\documentclass[english,11pt]{article}
\pdfoutput=1
\usepackage[sort,comma]{natbib}
\usepackage[pagebackref=false,colorlinks]{hyperref}
\hypersetup{pdffitwindow=true,linkcolor=blue,citecolor=blue,urlcolor=blue}
\usepackage{amsmath}
\usepackage{bbm}
\usepackage{multirow}

\usepackage{amsfonts}
\usepackage{mathdots}
\usepackage{amsthm}

\usepackage{multirow}
\usepackage{centernot}
\usepackage{todonotes}
\usepackage[ruled,linesnumbered]{algorithm2e}
\usepackage{float}
\usepackage{optidef}
\usepackage{dsfont}

\usepackage{lmodern}
\usepackage{amssymb,amsmath}
\usepackage{ifxetex,ifluatex}
\usepackage{fixltx2e} % provides \textsubscript
\ifnum 0\ifxetex 1\fi\ifluatex 1\fi=0 % if pdftex
  \usepackage[T1]{fontenc}
  \usepackage[utf8]{inputenc}
\else % if luatex or xelatex
  \ifxetex
    \usepackage{mathspec}
  \else
    \usepackage{fontspec}
  \fi
  \defaultfontfeatures{Ligatures=TeX,Scale=MatchLowercase}
\fi
\IfFileExists{upquote.sty}{\usepackage{upquote}}{}
\IfFileExists{microtype.sty}{%
	\usepackage{microtype}
	\UseMicrotypeSet[protrusion]{basicmath} % disable protrusion for tt fonts
}{}
\usepackage{hyperref}
\hypersetup{unicode=true,
            pdftitle={pcaNet example output},
            pdfborder={0 0 0},
            breaklinks=true}
\usepackage[margin=1in]{geometry}
\usepackage{color}
\usepackage{fancyvrb}

\DefineVerbatimEnvironment{Highlighting}{Verbatim}{commandchars=\\\{\}}
\usepackage{framed}

\SetKwInOut{Input}{Input}
\SetKwInOut{Output}{Output\,}
\SetKwInOut{Data}{Data}

\DeclareMathOperator*{\argmin}{arg\,min}

\definecolor{shadecolor}{RGB}{248,248,248}

\usepackage[font={footnotesize,it}, sc, bf]{caption}
\renewenvironment{abstract}{%
  \noindent\textbf\abstractname .\hspace{1pt}
}{
  \endlist \par\bigskip\bigskip
}

\usepackage{lastpage}
\RequirePackage[hyperpageref]{backref}
\renewcommand*{\backref}[1]{} 
\renewcommand*{\backrefalt}[4]{
    \ifcase #1
       No referred.
    \or
       \emph{Referred to on page #2.}
    \else
       \emph{Referred to on pages #2.}
    \fi}

\usepackage[shortlabels]{enumitem}

\usepackage[titletoc,title]{appendix}

\binoppenalty=\maxdimen
\relpenalty=\maxdimen

\usetikzlibrary{bayesnet}

\usepackage{xcolor}
\definecolor{cambridgebluecore}{RGB}{0, 176, 185}
\definecolor{cambridgebluedark}{RGB}{17, 94, 103}
\definecolor{cambridgebluelight}{RGB}{133, 176, 154}

\definecolor{cambridgeredcore}{RGB}{213, 0, 50}
\definecolor{cambridgereddark}{RGB}{138, 21, 56}
\definecolor{cambridgeredlight}{RGB}{232, 156, 174}

\definecolor{cambridgeblue2core}{RGB}{0, 114, 206}
\definecolor{cambridgeblue2dark}{RGB}{0, 60, 113}
\definecolor{cambridgeblue2light}{RGB}{108, 172, 228}

\definecolor{cambridgeorangecore}{RGB}{232, 119, 34}
\definecolor{cambridgeorangedark}{RGB}{190, 77, 0}
\definecolor{cambridgeorangelight}{RGB}{241, 190, 72}

\definecolor{cambridgegreencore}{RGB}{100, 167, 111}
\definecolor{cambridgegreendark}{RGB}{78, 91, 49}
\definecolor{cambridgegreenlight}{RGB}{183,	191, 16}

\usepackage{amsthm}
\usepackage{thmtools}
\newtheorem{mydef}{Definition}[section]

\newtheorem{mytheo}{Theorem}[section]

\newtheorem{myprop}{Proposition}[section]

\usepackage[]{graphicx}
\usepackage{caption}
\usepackage{subcaption}
\usepackage{epstopdf}
\usepackage{tikz}
\tikzset{font={\fontsize{9pt}{8}\selectfont}}
\usetikzlibrary{positioning}

\title{Kernel learning approaches for summarising and combining posterior similarity matrices}
\author{Alessandra Cabassi, Sylvia Richardson, Paul D. W. Kirk}

\begin{document}

\begin{center}
{\LARGE\bf Kernel learning approaches for summarising and combining posterior similarity matrices}
\end{center}
\medskip
\begin{center}
{\large Alessandra Cabassi$^{1}$, Sylvia Richardson$^{1}$, and Paul D. W. Kirk$^{1,2}$ \\[15pt]
\emph{$^{1}$MRC Biostatistics Unit}\\
\emph{$^{2}$Cambridge Institute of Therapeutic Immunology \& Infectious Disease\\
University of Cambridge, U.K.}\\
}
\end{center}

\bigskip

\begin{center}
Preprint, \today
\end{center}
\bigskip\bigskip

\begin{abstract}\newline
\textbf{Summary:} When using Markov chain Monte Carlo (MCMC) algorithms to perform inference for Bayesian clustering models, such as mixture models, the output is typically a sample of clusterings (partitions) drawn from the posterior distribution.  In practice, a key challenge is how to summarise this output.
Here we build upon the notion of the {\em posterior similarity matrix} (PSM) in order to suggest new approaches for summarising the output of MCMC algorithms for Bayesian clustering models.  A key contribution of our work is the observation that PSMs are positive semi-definite, and hence can be used to define probabilistically-motivated kernel matrices that capture the clustering structure present in the data.  This observation enables us to employ a range of kernel methods to obtain summary clusterings, and otherwise exploit the information summarised by PSMs. For example, if we have multiple PSMs, each corresponding to a different dataset on a common set of statistical units, we may use standard methods for combining kernels in order to perform {\em integrative clustering}.  We may moreover embed PSMs within predictive kernel models in order to perform {\em outcome-guided} data integration. We demonstrate the performances of the proposed methods through a range of simulation studies as well as two real data applications.\\
\textbf{Availability:} \textsf{R} code is available at \url{https://github.com/acabassi/combine-psms}.\\
\textbf{Contact:} \href{alessandra.cabassi@mrc-bsu.cam.ac.uk}{alessandra.cabassi@mrc-bsu.cam.ac.uk}, \href{paul.kirk@mrc-bsu.cam.ac.uk}{paul.kirk@mrc-bsu.cam.ac.uk}\\
\end{abstract}

\section{Introduction}
\label{sec:introduction}

Clustering techniques aim to partition a set of statistical units in such a way that items in the same group are similar and items in different groups are dissimilar.
The definition of similarity between observations varies according to the application at hand. In biomedical applications, for example, clustering can be used to define groups of patients who have similar genotypes or phenotypes and therefore are more likely to respond similarly to a certain treatment and/or have similar prognoses. Statistical methods for clustering can be divided into two main categories: heuristic approaches as, for instance, $k$-means \citep{hartigan1979algorithm} and hierarchical clustering \citep{kaufman2009finding}, and model-based techniques such as mixture models \citep{mclachlan2004finite}. Here we are interested in particular in Bayesian mixture models. Inference on the parameters of these models can be performed either via deterministic approximate inference methods based on variational inference \citep{blei2006variational, bishop2007pattern} or using Markov chain Monte Carlo (MCMC) schemes to sample from the posterior distribution \citep{gelfand1990sampling}.
One of the advantages of Bayesian model-based clustering is that, if infinite mixtures are used, one does not need to specify the number of clusters a priori \citep{rasmussen2000infinite}.
On the other hand, summarising the output of MCMC algorithms can be challenging, as it includes a large number of partitions of the data sampled from the posterior distribution.
The labels of each mixture component can change at each iteration of the MCMC, because the model likelihood is invariant under permutations of the indices. This phenomenon is known as \emph{label switching} \citep{celeux2000computational}.
One way of summarising the cluster allocations that circumvents this problem is to compute a posterior similarity matrix (PSM) containing the probabilities of each pair of observations of belonging to the same cluster (more details about Bayesian mixture models and PSMs can be found in Section \ref{sec:background}). 
In addition to that, however, one is often interested in finding one clustering of the data that best matches the information contained in the PSM \citep{fritsch2009improved}.  

We propose a new method to summarise the PSMs derived from the MCMC output of Bayesian model-based clustering, showing that they are valid kernel matrices.  Consequently, we are able to use kernel methods such as kernel $k$-means to find a summary clustering, and can make use of machine learning algorithms developed for pattern analysis (see e.g. \citealp{shawe2004kernel}) to combine the PSMs obtained from different datasets. 
We assume that we have a similarity matrix for each dataset, from initial clustering analyses performed for each dataset independently.
We show how these multiple kernel learning (MKL) techniques allow us to find a global clustering structure and assign different weights to each PSM, depending on how much information each provides about the clustering structure.
We also show how we may include a response variable, if it is available, in order to perform outcome-guided integrative clustering.
In particular, we show that, in the unsupervised framework, we can use the localised multiple kernel $k$-means approach of \cite{gonen2014localized}.
We further demonstrate that, if a response variable is available for each data point,  it is possible to incorporate this information using the \emph{simpleMKL} algorithm of \cite{rakotomamonjy2008SimpleMKL} in order to perform outcome-guided integrative clustering.
Both algorithms assign a weight to each PSM, that is output together with the global cluster assignment.  

This work therefore contributes to the many ways of summarising the clusterings sampled from the posterior distribution that have already been proposed \citep{fritsch2009improved, wade2018bayesian}. Our approach performs equally well in the case of one dataset and has the advantage of being easily extended to the case of multiple data sources.

From a different perspective, this work also suggests a new rational way to define kernels that are appropriate when the data are believed to possess a clustering structure.
\citet{lanckriet2004statistical} applied MKL methods to the problem of genomic data fusion, trying different kernels for each data source. 
However, how to define a kernel in general remained an open problem. \citet{cabassi2020multiple} suggested the use of methods based on multiple kernel learning to summarise the output of multiple similarity matrices, each resulting from applying consensus clustering \citep{monti2003consensus} to a different dataset.
Similarly, using PSMs as kernels ensures that the similarities between data points that we consider correctly reflect the clustering structure present in the data.

The manuscript is organised as follows. In Section \ref{sec:background} we introduce the problem of summarising PSMs. In Section \ref{sec:methods} we recall the theory of kernel methods, prove that PSMs are valid kernel matrices and explain how this allows us to apply kernel methods to the problems mentioned above. 
We also present the method that we use to choose the best number of clusters in the final summary.
In Section \ref{sec:simulation-examples} we present some simulation studies demonstrating the performances of kernel methods to summarise PSMs and to integrate multiple datasets in an unsupervised and outcome-guided fashion. 
Finally, in Section \ref{sec:results-psms} we introduce our motivating examples and illustrate how our methodology can be applied to it.

\section{Background}
\label{sec:background}

%\commentPDWK{More to be added.}
In this section we briefly recall the concept of Bayesian mixture models (Section \ref{sec:bayesian-mixture-models}) and we expose the problem of summarising the posterior distribution on the cluster allocations (Section \ref{sec:posterior-similarity-matrices}).

\subsection{Bayesian mixture models}
\label{sec:bayesian-mixture-models}
Mixture models assume that the data are drawn from a mixture of distributions:
\begin{equation}
p(\boldsymbol{x}) = \sum_{k=1}^K \pi_k f_{\boldsymbol{x}}(\boldsymbol{x} | \phi_k).
\end{equation}
where $f_{\boldsymbol{x}}$ is a parametric density that depends on the parameter(s) $\phi_k$, and $\pi_k$ are the mixture weights, which must satisfy $0 \leq \pi_k \leq 1$ and $\sum_{k=1}^K \pi_k = 1$.
In the Bayesian framework, we assign a prior distribution to the set of all parameters, $\Pi = [\pi_1, \dots, \pi_k]$ and $\Phi  = [\phi_1, \dots, \phi_k]$. %This allows us to take into account the uncertainty around...
In the finite mixtures case, methods to estimate the posterior distributions when the true number of mixture components is unknown include the MCMC-based algorithms proposed by \citet{ferguson1973bayesian} and \citet{richardson1997bayesian}.  MCMC approaches also exist for so-called infinite mixture models \citep{rasmussen2000infinite}, such as Dirichlet process mixtures and their generalisations  \citep{antoniak1974mixtures}.
%In this work, we make use of the \verb|MDI-GPU| package in \textsf{C} \citep{
%mason2016mdi}, the \textsf{R} package \verb|PreMiuM| of \citet{liverani2015premium}, and the \verb|DPMSysBio| Matlab toolbox \citep{zurauskiene2016graph} in order to fit Bayesian mixture models.
%
\subsection{Posterior similarity matrices}
\label{sec:posterior-similarity-matrices}
When using MCMC methods in order to perform Bayesian clustering on a dataset $X = [\boldsymbol{x}_1, \dots, \boldsymbol{x}_N] $, one obtains a vector of cluster assignments $\boldsymbol{c}^{(b)} = [c_1^{(b)}, \dots, c_N^{(b)}]$ from the posterior distribution for each iteration  $b = 1, \dots, B$ of the algorithm (see, for example, \citealp{neal2000markov}). From this, it is possible to obtain a Monte Carlo estimate of the probability that observations $i$ and $j$ belong to the same cluster as follows:
\begin{equation}\label{eq:psm}
P(c_i = c_j | X)  \approx \frac{1}{B} \sum_{b=1}^{B} \mathbb{I}_{\{c_i^{(b)} = c_j^{(b)} \}} = \vcentcolon \Delta_{ij} .
\end{equation}
We denote by $\Delta$  the posterior similarity matrix that is the matrix that has $ij$th entry $\Delta_{ij}$ equal to the right hand side of Equation \eqref{eq:psm}. 

Many ways to find a final clustering using the PSM $\Delta$ have been proposed \citep{binder1978bayesian, wade2018bayesian, dahl2006modelbased, fritsch2009improved, medvedovic2002bayesian}. 
A simple solution is to choose, among the $\boldsymbol{c}^{(b)}$, the one that maximises the posterior density. The problem with this approach is that many clusterings are associated with very similar posterior densities \citep{fritsch2009improved}. A more principled approach is to define a loss function $L(\boldsymbol{c}, \hat{\boldsymbol{c}})$ measuring the loss of information that occurs when estimating the true clustering $\boldsymbol{c}$ with $\hat{\boldsymbol{c}}$ \citep{binder1978bayesian}. The optimal clustering $\boldsymbol{c}^*$ is then defined as the one minimising the posterior expected loss:
\begin{equation}
\boldsymbol{c}^* =  \argmin_{\hat{\boldsymbol{c}}} \mathbb{E}  \left[ L(\boldsymbol{c}, \hat{\boldsymbol{c}}) | X \right]  =  \argmin_{\hat{\boldsymbol{c}}} \sum_{\boldsymbol{c}} L(\boldsymbol{c}, \hat{\boldsymbol{c}}) p\left( \boldsymbol{c} | X \right).
\end{equation}
\citet{binder1978bayesian}, for instance, suggested choosing the clustering $\hat{\boldsymbol{c}}$ that minimises the loss function
\begin{equation}
 L_{\text{Binder}} (\boldsymbol{c}, \hat{\boldsymbol{c}}) = \sum_{i < j} \left[ l_1 \mathbb{I}(c_i = c_j)\mathbb{I}(\hat{c}_i \not= \hat{c}_j) + l_2 \mathbb{I}(c_i \not= c_j) \mathbb{I}(\hat{c}_i = \hat{c}_j) \right],
\end{equation}
where $l_1$ and $l_2$ are positive constants determining whether assigning observations that belong to the same clusters to different clusters is penalised more highly than assigning observations that belong to different clusters to the same cluster ($l_1/l_2>1$) or vice versa ($l_1/l2<1$) and $\mathbb{I}$ is the indicator function. If $l_1 = l_2$, then
\begin{equation}\label{eq:bindersloss}
\boldsymbol{c}^*_{\text{Binder}} = \argmin_{\hat{\boldsymbol{c}}}\sum_{i < j} \lvert \mathbb{I}_{\{\hat{c}_i = \hat{c}_j\}} - \Delta_{ij} \rvert.
\end{equation}
More recently, \citet{wade2018bayesian} proposed an alternative to Binder's loss function based on the {\em variation of information} of \citet{meilua2007comparing}:
\begin{equation}
L_{\text{VI}} (\boldsymbol{c}, \hat{\boldsymbol{c}}) = H(\boldsymbol{c}) + H(\hat{\boldsymbol{c}}) - 2I(\boldsymbol{c}, \hat{\boldsymbol{c}}),
\label{eq:loss-vi}
\end{equation}
where $H(\boldsymbol{c})$ and $H(\hat{\boldsymbol{c}})$ represent the entropy of clusterings $\boldsymbol{c}$ and $\hat{\boldsymbol{c}}$ respectively, $I(\boldsymbol{c}, \hat{\boldsymbol{c}})$ is the mutual information between clusterings $\boldsymbol{c}$ and $\hat{\boldsymbol{c}}$. Defining by $C_i$ the set of elements that belong to cluster $i$ in clustering $\boldsymbol{c}$, $\hat{C}_j$ the set of elements that belong to cluster $j$ in clustering $\hat{\boldsymbol{c}}$, $n_{ij} = | C_i \cap \hat{C}_j |$, $n_{i+} = \sum_j n_{ij}$, $n_{+j} = \sum_i n_{ij}$, this is
\begin{equation}
L_{\text{VI}} = - \sum_{i = 1}^{K_N} \frac{n_{i+}}{N} \log_2 \left( \frac{n_{i+}}{N} \right) - \sum_{j = 1}^{\hat{K}_N} \frac{n_{+j}}{N} \log_2 \left( \frac{n_{+j}}{N} \right) - 2 \sum_{i = 1}^{K_N} \sum_{j = 1}^{\hat{K}_N} \frac{n_{ij}}{N} \log_2 \left(\frac{n_{ij} N}{n_{i+}n_{+j}}\right),
\end{equation}
where $K_N$ and $\hat{K}_N$ represent the number of clusters in $\boldsymbol{c}$ and $\hat{\boldsymbol{c}}$ respectively.

\citet{dahl2006modelbased} advanced the idea to choose, among all the clustering vectors $\boldsymbol{c}^{(b)}$, the one that minimises the least-squared distance to the PSM:
\begin{equation}
\boldsymbol{c}^*_{\text{Dahl}} = \argmin_{\hat{c}} \sum_{i < j} \left(\mathbb{I}_{\{\hat{c}_i = \hat{c}_j\}} - \Delta_{ij} \right)^2.
\end{equation}
This turned out to be equivalent to minimising Binder's loss function (Equation \ref{eq:bindersloss}).
\cite{fritsch2009improved} improved on the methods of Binder and Dahl by maximising the posterior expected adjusted Rand index of \citet{hubert1985comparing}.
%TODO cite also: \cite{Hurn2003}

Moreover, \citet{medvedovic2002bayesian} applied the complete linkage approach of \citet{everitt1993cluster} to the matrix of pseudo-distances $1- \Delta$, while \citet{molitor2010bayesian} used the partitioning around medoids algorithm of \citet{kaufman2009finding}. %\textcolor{red}{[\citet{wang2018optimal}?]}

All of these methods are only applicable with one PSM.
In what follows we describe a new way to find a clustering using posterior similarity matrices that also allows us to summarise multiple similarity matrices $\Delta_m$ and find a global clustering. 

\section{Methods}
\label{sec:methods}

In this section we explain how the output of the MCMC algorithms for Bayesian mixture models can be modified in order to obtain valid kernel matrices (Section \ref{sec:kernels}). After that, in Section \ref{sec:summarising-psms}, we show that this allows us to (i) use the kernel $k$-means algorithm to summarise a PSM and find a summary clustering of the data; (ii) combine multiple PSMs to perform integrative clustering of multiple datasets using an extension of kernel $k$-means; (iii) use an external response variable to determine the weight of each dataset in the integrative setting, by using predictive kernel methods such as Support Vector Machines (SVMs). In the last part of the section we explain how we choose the number of clusters in the final clustering.

\subsection{Kernels}
\label{sec:kernels}
Kernel methods allow non-linear relationships between data points with low computational complexity to be modelled, thanks to the so-called \emph{kernel trick} \citep{shawe2004kernel}.
For this reason, kernel methods have been widely used to extend many traditional algorithms to the non-linear framework, such as the PCA \citep{scholkopf1998nonlinear}, linear discriminant analysis \citep{mika1999fisher, roth2000nonlinear, baudat2000generalized} and ridge regression \citep{friedman2001elements, shawe2004kernel}.
%Moreover, mapping the data into some high dimensional space makes it easier to separate clusters of data \citep{Bishop2007}.

%We introduce now the notion of a kernel function.
Kernel methods proceed by embedding the observations into a higher-dimensional feature space $\mathcal{X}$ endowed with an inner product $\langle \cdot, \cdot \rangle_{\mathcal{X}}$ and induced norm $\lVert \cdot\rVert_{\mathcal{X}}$, making use of a map $\phi: \mathbb{R}^P \to \mathcal{X}$.
\begin{mydef}
	A \emph{positive definite kernel} or, more simply, a \emph{kernel} $\delta$ is a symmetric map $\delta : \mathbb{R}^P \times \mathbb{R}^P \to \mathbb{R}$ for which for all $\mathbf{x}_1, \mathbf{x}_2, \dots, \mathbf{x}_N \in \mathbb{R}^P$, the matrix $\Delta$ with entries $\Delta_{ij} = \delta(\mathbf{x}_i, \mathbf{x}_j)$ is positive semi-definite. The matrix $\Delta$ is called the \emph{kernel matrix} or \emph{Gram matrix}.
\end{mydef}
(The domain of the map $\phi$ is set to be $\mathbb{R}^P$ for simplicity, but the definition also holds for more general sets of departure for $\phi$.)
Using this definition, it is possible to prove that any inner product of feature maps gives rise to a kernel:
\begin{myprop}
	The map $\delta: \mathbb{R}^P \times \mathbb{R}^P \to \mathbb{R}$ defined by $\delta(\mathbf{x}, \mathbf{x}') = \langle \phi(\mathbf{x}), \phi(\mathbf{x}') \rangle_{\mathcal{X}}$ is a kernel. 
\end{myprop}

Moreover, using Mercer's theorem, it can be shown that for any positive semi-definite kernel function, $\delta$, there exists a corresponding feature map, $\phi$ (see e.g. \citealp{vapnik1998statistical}). That is,

\begin{mytheo}
	For each kernel $\delta$, there exists a feature map $\phi$ taking value in some inner product space $\mathcal{X}$ such that $\delta(\mathbf{x},\mathbf{x}') = \langle \phi(\mathbf{x}), \phi(\mathbf{x}') \rangle_{\mathcal{X}}$.
\end{mytheo}

In practice, it is therefore often sufficient to specify a positive semi-definite kernel matrix, $\Delta$, in order to allow us to apply kernel methods such as those presented in the following sections.
%perform kernel $k$-means clustering (or to apply other kernel methods).  
For a more detailed discussion of kernel methods, see e.g. \cite{shawe2004kernel}.
\subsection{Identifying PSMs as kernel matrices}
\label{sec:summarising-psms}

It has been shown elsewhere that co-clustering matrices are valid kernel matrices \citep{cabassi2020multiple}. We show here that this result also holds for PSMs, and hence they can be used as input for any kernel-based model.
PSMs are convex combinations of co-clustering matrices $C^{(b)}$, where each matrix $C^{(b)}$ is defined as follows:
\begin{equation*}
C_{ij}^{(b)} = \left\lbrace
\begin{array}{ll}
1 & \text{if } c^{(b)}_i = c^{(b)}_j ,\\
0 & \textit{otherwise},
\end{array}
\right.
\end{equation*}
where $C_{ij}^{(b)}$ indicates whether the statistical units $i$ and $j$ are assigned to the same cluster at iteration $b$ of the MCMC chain. 
Indicating by $K$ the total number of clusters and reordering the rows and column, each $C_{ij}^{(b)}$ can be written as a block-diagonal matrix where every block is a matrix of ones:
\begin{equation}\label{eq:block-diagonal-matrix}
C^{(b)} = \left[ 
\begin{array}{c c c c c}
J_1 & 0 & 0 & \dots & 0 \\
0 & J_2 & 0 & \dots & 0 \\
\vdots   & &  & \ddots & \vdots \\
0 & 0 & 0 & \dots &  J_K 
\end{array}
\right]
\end{equation}
where $J_k$ is an $n_k \times n_k$ matrix of ones, with $n_k$ being the number of items in cluster $k$.
The eigenvalues of a block diagonal matrix are simply the eigenvalues of its blocks, which, in this case, are nonnegative. Therefore all $C^{(b)}$, with $b = 1, \dots, B$, are positive semidefinite.
Now, if $\lambda$ is a nonnegative scalar, and $C$ is positive semidefinite, then $\lambda C$ is also positive semidefinite. Moreover, the sum of positive semidefinite matrices is a positive semidefinite matrix. 
Therefore, given any set of nonnegative $\lambda_b$, $b= 1, \dots, B$, $\sum_{b=1}^{B} \lambda_b C^{(b)}$ is positive semidefinite. We can conclude that any PSM is positive definite.  

\subsection{Summarising and combining PSMs using kernel methods}

Here we show that the fact that all posterior similarity matrices are valid kernels allows us to use kernel methods to find a clustering of the data that summarises a sample of clusterings $\boldsymbol{c}^{(1)}, \dots, \boldsymbol{c}^{(B)}$ from the posterior distribution of an MCMC algorithm for Bayesian clustering. To do this we suggest to use an extension of the well-known $k$-means algorithm that only needs as input a kernel matrix. 

Moreover, this method can be easily extended to allow us to combine multiple PSMs. This can be a useful feature under many circumstances. For instance, as it is the case in our motivating examples, one could have different types of information relative to the same statistical observations. In this situation, it may be appropriate to define and fit different mixture models on each data type, and then summarise the two posterior samples of clusterings at a later stage. This can be achieved by using multiple kernel $k$-means algorithms, that allow us to combine multiple kernels to find a global clustering. On top of that, these techniques also assign different weights to each kernel. These can be used to assess how much each dataset contributed to the final clustering and therefore to get an idea of how much information is present in each data type about the clustering structure.

The problem with combining multiple kernels is, however, that it is not always clear whether they all have the same clustering structure. To overcome this issue, we also propose an \textit{outcome-guided} algorithm to summarise multiple PSMs. The idea is that, instead of choosing the weight of each kernel in an unsupervised way, if we have a variable available which is closely related to the outcome of interest, we should weight more highly the kernels in which statistical units that have similar outcomes are closer to each other. In mathematical terms, this corresponds to using support vector machines to find the kernel weights, where the response variable is our proxy for the outcome.

\subsubsection{Summarising PSMs}

In order to illustrate our method for summarising PSMs, first we recall the main ideas behind $k$-means clustering and then we present its extension to the kernel framework.

\paragraph{$k$-means clustering}
$k$-means clustering is a widely used clustering algorithm, first introduced by \citet{steinhaus1956sur}. Let $\boldsymbol{x}_1, \dots, \boldsymbol{x}_N$ indicate the observed dataset, with $\boldsymbol{x}_n \in \mathbb{R}^p$ and $z_{nk}$ be the corresponding cluster labels, where $\sum_k z_{nk} = 1$ and 
\begin{equation}
z_{nk} = \left\{
\begin{array}{l l}
1, & \text{if } \boldsymbol{x}_n \text{ belongs to cluster }k, \\
0, & \text{otherwise}.\\
\end{array}
\right.
\end{equation}
We denote by Z the $N \times K$ matrix with $ij$th element equal to $z_{ij}$.
The goal of the $k$-means algorithm is to minimise the sum of all squared distances between the data points $\boldsymbol{x}_n$ and the corresponding cluster centroid $\boldsymbol{m}_k$. The optimisation problem is 
\begin{mini!}[2]
	{Z}{\sum_{n} \sum_{k} z_{nk} \lVert \boldsymbol{x}_n - \boldsymbol{m}_k \rVert^2_2 \label{eq:of_kmeans}}
	{\label{opt:kmeans}}
	{}
	\addConstraint{\sum_k z_{nk}}{=1, \; \forall n}
	\addConstraint{ N_k = }{\sum_n z_{nk}, \; \forall k}
	\addConstraint{\boldsymbol{m}_k }{= \frac{1}{N_k} \sum_n z_{nk} \boldsymbol{x}_n, \; \forall k.}
\end{mini!}
where $\lVert \cdot \rVert_2$ indicates the Euclidean norm.
\paragraph{Kernel $k$-means clustering}
\label{sec:kernel-k-means}

Now we can show how the kernel trick works in the case of the $k$-means clustering algorithm \citep{girolami2002mercer}.
Redefining the objective function of Equation \eqref{eq:of_kmeans} based on the distances between observations and cluster centres in the feature space $\mathcal{X}$, the optimisation problem becomes:
\begin{mini!}[2]
	{Z}{\sum_n \sum_k z_{nk} \lVert \phi(\boldsymbol{x}_n) - \tilde{\boldsymbol{m}}_k \rVert^2_{\mathcal{X}} \label{eq:of_kernelkmeans}}
	{\label{opt:kernelkmeans}}
	{}
	\addConstraint{\sum_k z_{nk}}{=1, \; \forall n}
	\addConstraint{ N_k = }{\sum_n z_{nk}, \; \forall k}
	\addConstraint{\tilde{\boldsymbol{m}}_k }{= \frac{1}{N_k} \sum_n z_{nk} \phi(\boldsymbol{x}_n), \; \forall k.}
\end{mini!}
where we indicated by $\tilde{\boldsymbol{m}}_k$ the cluster centroids in the feature space $\mathcal{X}$.  
Using this kernel, each term of the sum in Equation \eqref{eq:of_kernelkmeans} can be written as
\begin{align}
\lVert \phi(\boldsymbol{x}_n) - \tilde{\boldsymbol{m}}_k \rVert^2_{\mathcal{X}} & = \langle \phi(\boldsymbol{x}_n) - \tilde{\boldsymbol{m}}_k, \phi(\boldsymbol{x}_n) - \tilde{\boldsymbol{m}}_k \rangle_{\mathcal{X}}\\
& = \langle \phi(\boldsymbol{x}_n), \phi(\boldsymbol{x}_n) \rangle_{\mathcal{X}} - \frac{2}{N_k} \sum\limits_{i=1}^N z_{ik} \langle \phi(\boldsymbol{x}_n),\phi(\boldsymbol{x}_i) \rangle_{\mathcal{X}} \\
& \quad + \frac{1}{N_k^2} \sum\limits_{i=1}^N \sum\limits_{j=1}^N z_{ik} z_{jk} \langle \phi(\boldsymbol{x}_i), \phi(\boldsymbol{x}_j) \rangle_{\mathcal{X}} \\
& = \delta(\boldsymbol{x}_n, \boldsymbol{x}_n) - \frac{2}{N_k} \sum\limits_{i=1}^N z_{ik} \delta(\boldsymbol{x}_n, \boldsymbol{x}_i)  + \frac{1}{N_k^2} \sum\limits_{i=1}^N \sum\limits_{j=1}^N z_{ik} z_{jk} \delta(\boldsymbol{x}_i, \boldsymbol{x}_j).
\end{align}
Therefore, we do not need to evaluate the map $\phi$ at every point $\boldsymbol{x}_i$ to compute the objective function of Equation \eqref{eq:of_kernelkmeans}. Instead, we just need to know the values of the kernel evaluated at each pair of data points $\delta(\boldsymbol{x}_i,\boldsymbol{x}_j)$, $i,j = 1, \dots, N$. This is what is commonly referred to as the kernel trick.

We have seen how kernels can be used to perform $k$-means clustering. Now, if we have a sample of clusterings from the posterior, we can easily exploit this technique to find a summary clustering. Once we compute our PSM $\Delta$, this will be our kernel matrix. The clustering of interest will then be the one given by kernel $k$-means in the form of $z_{nk}$, $n=1, \dots, N$, $k=1, \dots, K$. The number of clusters $K$ is chosen as explained in Section \ref{sec:choose-k}.

\subsubsection{Combining PSMs to perform integrative clustering}
\label{sec:combining-psms}

To combine multiple PSMs relative to the same statistical units, all we need to do is to use the extension of kernel $k$-means to the case of multiple kernels.

\paragraph{Multiple kernel $k$-means clustering}
\label{sec:multiple-kernel-kmeans}
\cite{gonen2014localized} extended the kernel $k$-means approach to the case of multiple kernels. We consider multiple datasets $X_1, \dots, X_M$ each with a different mapping function $\phi_m: \mathbb{R}^P \to \mathcal{X}_m$ and corresponding kernel $\delta_m(\boldsymbol{x}_i, \boldsymbol{x}_j) = \langle \phi_m(\boldsymbol{x}_i), \phi_m(\boldsymbol{x}_j) \rangle_{\mathcal{X}_m} $ and kernel matrix $\Delta_m$. Then, if we define 
\begin{equation}
\phi_{\Theta} (\boldsymbol{x}_i) = [\theta_{i1}\phi_1(\boldsymbol{x}_i)', \theta_{i2}\phi_2(\boldsymbol{x}_i)', \dots, \theta_{iM}\phi_M(\boldsymbol{x}_i)']',
\end{equation}
where $\Theta \in \mathbb{R}^{N \times M}_{+}$ is a vector of kernel weights such that  $\theta_{im}$ is the weight of observation $\boldsymbol{x}_i$ in dataset $m$ and $\sum_m \theta_{im} = 1$ for all $i = 1, \dots, N$ and $\theta_{im} \geq 0$, the kernel function of this multiple feature problem is a convex sum of the single kernels:
\begin{align}
\delta_{\Theta}(\boldsymbol{x}_i, \boldsymbol{x}_j) & = \langle \phi_{\Theta}(\boldsymbol{x}_i), \phi_{\Theta}(\boldsymbol{x}_j) \rangle_{\mathcal{X}_m} \\ 
& = \sum_{m=1}^{M} \theta_{im} \theta_{jm} \langle \phi_m(\boldsymbol{x}_i), \phi_m(\boldsymbol{x}_j) \rangle_{\mathcal{X}_m}\\
& = \sum_{m=1}^{M}  \theta_{im} \theta_{jm} \delta_m(\boldsymbol{x}_i, \boldsymbol{x}_j).
\end{align}
We denote the corresponding kernel matrix by $\Delta_{\Theta}$. 
The optimisation strategy proposed by \cite{gonen2014localized} is based on the idea that, for some fixed vector of weights $\boldsymbol{\theta}$, the problem is equivalent to the one of Equation \eqref{eq:of_kernelkmeans}, where we had only one kernel. Therefore, they develop a two-step optimisation strategy: (1) given a fixed matrix of weights $\Theta$, solve the optimisation problem as in the case of one kernel, with kernel matrix $\delta_{\Theta}$ and then (2) minimise the objective function with respect to the kernel weights, keeping the assignment variables fixed. This is a convex quadratic programming (QP) problem that can be solved with any standard QP solver up to a moderate number of kernels $M$.

Similarly to before, once we have defined the kernels $\Delta_{m}$ to be equal to each of our PSMs,  the labels found through multiple kernel $k$-means constitute the clustering that we are looking for. Moreover, the kernel weights give us an indication of how each PSM contributed to the final clustering.

%This kernel integration strategy was first used for multi-omic studies by \citet{cabassi2020multiple}. In that case, however, kernels were obtained via consensus clustering.

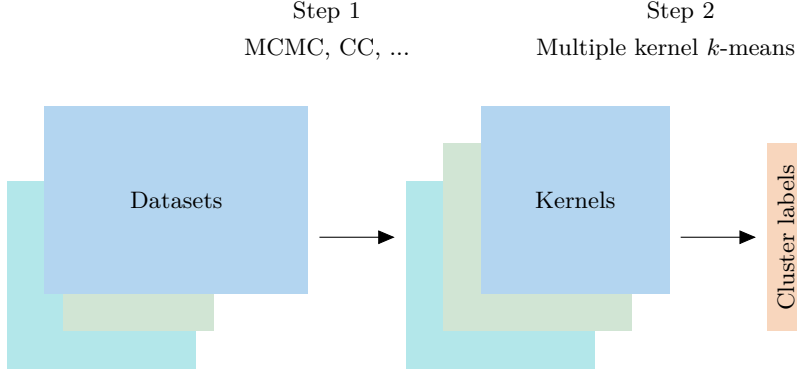
\begin{figure}[h!]
	\centering
	\begin{tikzpicture}
	\node (rect1) [minimum width=2.5cm,minimum height=2.5cm, fill=cambridgebluecore!30, cambridgebluecore!30] at (0,0) {};
	\node (rect2) [minimum width=2cm,minimum height=2.5cm, fill=cambridgegreencore!30, cambridgegreencore!30] at (0.5,0.5) {};
	\node (rect3) [minimum width=3.5cm,minimum height=2.5cm, fill=cambridgeblue2core!30, cambridgeblue2core!30, text = black] at (1,1) {Datasets};
	
	\node at (3,3.5) {Step 1};
	%	\draw[thick] (2.8,2.5)--(3.3,2.5)--(3.3,-1.5)--(2.8,-1.5);
	\draw [->] (2.9,0.5)--(3.9,0.5);
	\node at (3,3) {MCMC, CC, ...};
	
	\node (psm1) [minimum width=2.5cm,minimum height=2.5cm, fill=cambridgebluecore!30, cambridgebluecore!30] at (5.3,0) {};
	\node (psm2) [minimum width=2.5cm,minimum height=2.5cm, fill=cambridgegreencore!30, cambridgegreencore!30] at (5.8,0.5) {};
	\node (psm3) [minimum width=2.5cm,minimum height=2.5cm, fill=cambridgeblue2core!30, cambridgeblue2core!30, text = black] at (6.3,1) {Kernels};
	
	\node at (7.7,3.5) {Step 2};
	%	\draw[thick] (7.5,2.5)--(8,2.5)--(8,-1.5)--(7.5,-1.5);
	\draw [->] (7.7,0.5)--(8.7,0.5);
	\node at (7.5,3) {Multiple kernel $k$-means};
	
	\node (cluster) [minimum width=2.5cm,minimum height=.5cm, fill=cambridgeorangecore!30, cambridgeorangecore!30, text = black, rotate = 90] at (9.1,.5) {Cluster labels};
	\end{tikzpicture}
	\caption{Schematic representation of the MKL-based integrative clustering approach. Each colour indicates a different dataset/kernel. First, a mixture model is fit using MCMC on each dataset separately. The resulting PSMs are valid kernels that can be used as input to kernel $k$-means to find a global clustering of the data. Note that, as shown elsewhere \citep{cabassi2020multiple}, other similarity matrices, such as the similarity matrices given by consensus clustering (CC), define valid kernels. For this reason, they can be combined with PSMs via MKL.}
	\label{fig:theory-unsupervised}
\end{figure}

\subsubsection{Embedding PSMs within predictive kernel machines to perform outcome-guided integration}
\label{sec:outcome-guided-clustering}

\begin{figure}[h!]
	\centering
	\begin{tikzpicture}[x=0.75cm,y=0.75cm]
	\node (rect1) [minimum width=2.5cm,minimum height=2cm, fill=cambridgebluecore!30, cambridgebluecore!30] at (0,0) {};
	\node (rect2) [minimum width=2cm,minimum height=2cm, fill=cambridgegreencore!30, cambridgegreencore!30] at (0.5,0.5) {};
	\node (rect3) [minimum width=3cm,minimum height=2cm, fill=cambridgeblue2core!30, cambridgeblue2core!30, text = black] at (1,1) {Datasets};
	
	\node at (3.5,4) {Step 1};
%	\draw[thick] (3.3,3)--(3.8,3)--(3.8,-2)--(3.3,-2);
	\node at (3.5,3.5) {MCMC, CC, ...};
	\draw [->] (3.4,0.5)--(4.4,0.5);
	
	\node (psm1) [minimum width=2.5cm,minimum height=2.5cm, fill=cambridgebluecore!30, cambridgebluecore!30] at (6.2,0) {};
	\node (psm2) [minimum width=2.5cm,minimum height=2.5cm, fill=cambridgegreencore!30, cambridgegreencore!30] at (6.7,0.5) {};
	\node (psm3) [minimum width=2.5cm,minimum height=2.5cm, fill=cambridgeblue2core!30, cambridgeblue2core!30, text = black] at (7.2,1) {Kernels};
	\node (response) [minimum width=2.5cm,minimum height=.5cm, fill=cambridgeredcore!30, cambridgeredcore!30, text = black, rotate = 90] at (9.3,0.5) {Response};
	
	\node at (10.3,4)[black] {Step 2};
	\node at (10.3,3.5) {SimpleMKL};
%	\draw[thick] (9.9,3)--(10.4,3)--(10.4,-2)--(9.9,-2);
	\draw [->] (10,0.5)--(11,0.5);
	
	\node (weightedpsm) [minimum width=2.5cm,minimum height=2.5cm, fill=cambridgeredlight!30, cambridgeblue2light!30, text = black, text width = 2.5cm, align = center] at (13,0.5) {Weighted kernel};
	
	\node at (15.2,4)[rectangle, black] {Step 3};
%	\draw[thick] (15.1,3)--(15.6,3)--(15.6,-2)--(15.1,-2);
	\node at (15.2,3.5) {Kernel $k$-means};
	\draw [->] (15.2,0.5)--(16.2,0.5);
	
	\node (cluster) [minimum width=2.5cm,minimum height=.5cm, fill=cambridgeorangecore!30, cambridgeorangecore!30, text=black, rotate = 90] at (16.7,.5) {Cluster labels};
	\end{tikzpicture}
	\caption{Schematic representation of the MKL-based outcome-guided integrative clustering approach. Each colour indicates a different dataset/kernel. First, a mixture model is fit on each dataset separately. The resulting PSMs are valid kernels that can be used as input to simpleMKL, if a response variable is available, to find a global clustering of the data.}
	\label{fig:theory-outcome-guided}
\end{figure}
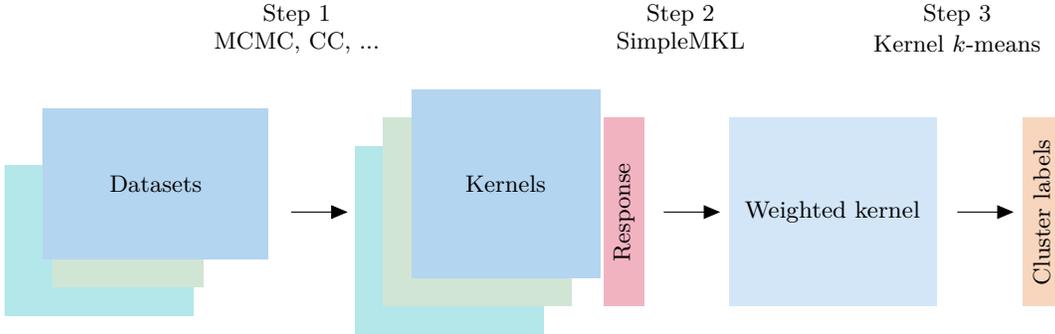

Suppose now that, in addition to the posterior similarity matrices $\Delta_1, \dots, \Delta_M$, we also have a categorical response variable $y_n$ associated with each observation $\boldsymbol{x}_n$. As we explained above, we would like to use this information to guide our clustering algorithm. We can use the \emph{simpleMKL} algorithm described in the remainder of this section to find the kernel weights $\theta_1, \dots, \theta_M$ and then use kernel $k$-means on the weighted kernel $\Delta = \sum_m \theta_m \Delta_m$ to find the final clustering (Figure \ref{fig:theory-outcome-guided}).

\paragraph{Support vector machines}
We briefly recall here the concept of the support vector machine \citep{boser1992training} that is widely used for solving problems in classification and regression \citep{scholkopf2001learning, bishop2007pattern}. 

In its simplest form, this method is applied to a binary classification problem, in which the data points $\boldsymbol{x}_1, \dots, \boldsymbol{x}_N \in \mathbb{R}^P$ in the training set are assigned to two classes indicated by the target values $y_n \in \{-1,1\}, n = 1, \dots, N$.
We consider a feature map $\phi: \mathbb{R}^P \to \mathcal{X}$  and the associated kernel $\delta(\cdot, \cdot): \mathcal{X} \times \mathcal{X} \to \mathbb{R}$ such that $\delta(\boldsymbol{x}_i, \boldsymbol{x}_j) = \langle \phi(\boldsymbol{x}_i), \phi(\boldsymbol{x}_j) \rangle_{\mathcal{X}}$. Suppose that there exist some values of $\alpha_n$ and $b$ such that 
\begin{equation}
\label{eq:decision-boundary-svm}
f(\boldsymbol{x} ) = \sum_{n=1}^{N} \alpha_n \delta(\boldsymbol{x}, \boldsymbol{x}_n)
\end{equation}
satisfies $f(\boldsymbol{x}_n) +b  > 0$ if $y_n = 1$ and $f(\boldsymbol{x}_n) +b< 0$ otherwise. $f$ is a function that lives in a function space $\mathcal{H}$ endowed with the norm $\lVert \cdot \rVert_{\mathcal{H}}$.
Then, this function can be used to classify new data points $\boldsymbol{x}$ according to the sign of $f(\boldsymbol{x}) + b$. For support vector machines, the parameters $\alpha_n$ and $b$ are chosen so as to maximise the \textit{margin}, i.e. the distance between the decision boundary given by Equation \eqref{eq:decision-boundary-svm} and the point $\boldsymbol{x}_n$ that is closest to the boundary. It can be shown that this can be achieved by solving the quadratic programming problem (see e.g. \citealp{bishop2007pattern,rakotomamonjy2008SimpleMKL})
\begin{mini!}[2]
	{f,b}{ \frac{1}{2} \lVert f \rVert^2_{\mathcal{H}}  \label{eq:of_SVM}}
	{\label{op:SVM}}{}
	\addConstraint{ y_n \big[f(\boldsymbol{x}_n) + b\big] }{\geq 1, \quad \forall n.}
\end{mini!}

However, in real applications, it is usually not possible to separate the two classes perfectly. Hence, in order to take into account misclassifications, it is necessary to introduce a penalty term that is linear with respect to the distance of the misclassified points to the classification boundary \citep{bennett1992robust}.
To this end, we define a variable $\xi_n$ (known as a {\em slack variable}) for each data point such that
\begin{equation}
 \xi_n =  \left\{
 \begin{array}{l l}
 0, & \text{if } \boldsymbol{x}_n \text{ is correctly classified,} \\
 \lvert y_n - f(\boldsymbol{x}_n) \rvert, & \text{otherwise}.
 \end{array}
 \right.
\end{equation} 
The optimisation problem of Equation \eqref{op:SVM} then becomes
\begin{mini!}[2]
	{f,b,\{\xi_n\}}{ \frac{1}{2} \lVert f \rVert^2_{\mathcal{H}} + \lambda \sum_n \xi_n \label{eq:of_SVMpenalised}}
	{\label{op:SVMpenalised}}{}
	\addConstraint{ y_n \big[f(\boldsymbol{x}_n) + b\big] }{\geq 1 - \xi_n, \quad \forall n}
	\addConstraint{\xi_n }{\geq 0, \quad \forall n,}
\end{mini!}
where $\lambda>0$ is a parameter that controls the penalisation of misclassifications.
The objective functions \eqref{eq:of_SVM} and \eqref{eq:of_SVMpenalised} are quadratic, so any local optimum is also a global optimum. 
%However, this optimisation problem is very computationally demanding with a complexity of $O(N^3)$ and therefore hard to solve directly. 
One of the most popular approaches to solve this type of problems is \emph{sequential minimal optimisation} \citep{platt1999fast}. 
%This method has a complexity between $O(N)$ and $O(N^2)$ depending on the specific application.}
For more details about SVMs see e.g. \cite{bishop2007pattern}.
\paragraph{Multiple kernel learning for SVMs}
\label{sec:MKL}
In the multiple kernel learning framework for SVMs, we consider $M$ different feature representations, with mapping functions $\phi_m$ and corresponding kernel functions $\delta_m$ and feature spaces $\mathcal{X}_m$. We substitute the kernel $\delta$  of Equation \eqref{eq:decision-boundary-svm} with a convex combination of kernels $\delta_m$ \citep{lanckriet2004statistical}:
\begin{equation}
	f(\boldsymbol{x}) + b = \sum_{m=1}^{M} \theta_m f_m(\boldsymbol{x}) + b
\end{equation}
where $\theta_m \geq 0$, $\sum_m \theta_m = 1$ and $f_m = \sum_n \delta_m (\boldsymbol{x}, \boldsymbol{x}_n)$.
\cite{rakotomamonjy2007more} proposed then to solve the optimisation problem 
\begin{mini!}[2]
	{\{f_m\}, b, \{\xi_n\}, \{\theta_m\}}{J(\boldsymbol{\theta}) \vcentcolon = \frac{1}{2} \sum_m \frac{1}{\theta_m} \lVert f_m \rVert^2_{\mathcal{H}_m} + \lambda \sum_n \xi_n \label{eq:of_multipleSVM}}
   {\label{op:multipleSVM}}{}
   \addConstraint{y_n \bigg[ \sum_m f_m(\boldsymbol{x}_n) +  b \bigg] }{\geq 1 - \xi_n, \quad \forall n}
   \addConstraint{\xi_n  }{\geq 0, \quad \forall n}
   \addConstraint{\sum_m \theta_m }{= 1}
   \addConstraint{\theta_m }{\geq 0, \quad \forall m}
\end{mini!}
using the convention that $x/0 = 0$ if $x = 0$ and $\infty$ otherwise. The algorithm of Rakotomamonjy and Bach takes the name of \emph{simpleMKL} and is based on the idea that one can iteratively solve a standard SVM problem \eqref{eq:of_SVMpenalised} for a fixed value of $\boldsymbol{\theta}$ and then update the vector of weights $\boldsymbol{\theta}$ using the gradient descent method on the objective function $J(\boldsymbol{\theta})$. Since the objective function is smooth and differentiable with Lipschitz gradient, it can be easily optimised with the reduced gradient algorithm (\citealp{luenberger1984linear}, Chapter 11). If the standard SVM problem is solved exactly at each iteration, then convergence to the global optimum is guaranteed \citep{luenberger1984linear}. 
% However, if the SVM algorithm is stopped when a criterion on the duality gap is satisfied, convergence is not guaranteed by standard arguments.
%{\color{blue}[I think you need to say more about the optimisation here and elsewhere -- e.g. are you guaranteed to find a global optimum?  What kind of stopping criterion is used?]}. 
%
\paragraph{Multiclass multiple kernel learning} 
%{\color{blue}[References needed in this subsection]}
SVMs can be used also when the target value $y_n$ takes more than two different values. The  most commonly used approaches are called \textit{one-versus-one} \citep{knerr1990singlelayer} and \textit{one-versus-the-rest} \citep{vapnik1998statistical}. In the first one, we consider in turn each class as the ``positive'' case, and all the others as the ``negative'' cases. This way, we construct $K$ different classifiers and then assign a new observation $\boldsymbol{x}$ using 
\begin{equation}
y(\boldsymbol{x}) = \max\limits_{k \in \{1, \dots, K\}} y_k(\boldsymbol{x})
\end{equation}
%to 
%\begin{equation}
%k = \argmax\limits_{k \in \{1, \dots, K\}} y_k({\bf x}).
%\end{equation}
The second approach is to train one SVM for each pair of classes and then assign a point $\boldsymbol{x}$ to the class to which it is assigned more often. 

\cite{rakotomamonjy2008SimpleMKL} extended the \emph{simpleMKL} algorithm to the case of a response with $K > 2$ classes.
Both these approaches can be used with the simpleMKL algorithm, defining a new cost function $J(\boldsymbol{\theta})$ as the sum of all the cost functions of the partial SVMs $J_s(\boldsymbol{\theta})$:
\begin{equation}
	J(\boldsymbol{\theta}) = \sum_{s \in \mathcal{S}} J_s({\boldsymbol{\theta}})
\end{equation}
where $\mathcal{S}$ indicates the set of all partial SVMs and each $J_s$ is defined as in Equation \eqref{eq:of_multipleSVM}.

\vspace{1em}
It is important to note that none of these approaches explicitly rely on the fact that the $\Delta_m$ are posterior similarity matrices. Hence, any other type of matrix $\Delta_m$ can be used, as long as it is symmetric, positive semi-definite and the entries $\Delta^m_{ij}$ can be interpreted as some measure of the similarity between $\boldsymbol{x}_i$ and $\boldsymbol{x}_j$. 

\subsection{Choosing the number of clusters in the final summary}
\label{sec:choose-k}
Many possible approaches have been proposed to choose the number of clusters (see e.g. \citealp{milligan1985examination, tibshirani2001estimating, yeung2001validating, dudoit2002predictionbased}).
We focus on the so-called \textit{silhouette}, a measure of compactness of the clustering structure, proposed by \citet{rousseeuw1987silhouettes}. There are two ways of defining the silhouette of a cluster, based respectively on the similarities and dissimilarities between the data. Here we briefly explain the former.

Given some cluster assignment labels $\boldsymbol{c} = [c_1, \dots, c_N]$ and some measure of the dissimilarity between the data points $\Delta_{ij}$ for all $i, j = 1, \dots, N$, we can define the following quantities: $a_n$ is the average similarity of $\boldsymbol{x}_n$ to all objects of cluster $c_n$ and, for each $c_i \not = c_n$, $\Delta_{n, c_i}$ is the average similarity of $n$ to all objects belonging to cluster $c_i$.  Moreover, let us indicate by $b_n$ the maximum $\Delta_{n, c_i}$ over all $i$ such that $c_i \not = c_n$. Then, for each observation $n = 1, \dots, N$, we can calculate
\begin{equation}
s_n = \left\{
\begin{array}{ll}
1 - a_n/b_n, & \text{if } a_n < b_n,\\
0, & \text{if } a_n = b_n,\\
a_n/b_n - 1, & \text{if } a_n > b_n.
\end{array}
\right.
\end{equation}
This takes values between $-1$ and $1$, with higher values indicating that $\boldsymbol{x}_i$ is {\em well-clustered} and negative values suggesting that $\boldsymbol{x}_i$ has been misclassified. 

Thus, we run our algorithms for combining the posterior similarity matrices with different number of clusters from $K_{\text{min}}$ to $K_{\text{max}}$. We consider the overall average silhouette width $\bar{s} = \frac{1}{N}\sum_{n=1}^N s_n$ as a measure of the compactness of clusters and we choose the value of $K$ that gives the highest value of $\bar{s}$. 

%Each time, we compute the average silhouette based on the similarities $\Delta_{ij}$, indicated with $\bar{s}_s$, the average silhouette based on the dissimilarities $1-\Delta_{ij}$, indicated with $\bar{s}_d$, and the following quantity
%%we also look at the silhouette calculated with the dissimilarities between the data $\bar{d}$ and the so-called \emph{double silhouette}, calculated as follows:
%\begin{equation}
%\min \{\text{sign}(s_s), \text{sign}(s_d)\} |s_s s_d|,
%\end{equation}
%that we refer to as the \emph{double silhouette}.
%%%%

\section{Simulation examples}
\label{sec:simulation-examples}
Here we show how the methods presented above perform in practice.
In Section \ref{sec:synthetic-datasets} we explain how we generate the synthetic datasets for the simulation studies. 
In Section \ref{sec:single-datasets} we show that the kernel $k$-means approach applied to a posterior similarity matrix derived from a single dataset performs similarly to standard clustering methods.
Additionally, to assess the MKL-based integrative clustering approaches described in Sections \ref{sec:combining-psms} and \ref{sec:outcome-guided-clustering}, we perform a range of simulation studies; the results are presented in Section \ref{sec:simulations-integrative-clustering}. 
\subsection{Synthetic datasets}
\label{sec:synthetic-datasets}
We generate four synthetic datasets, each composed of data belonging to six different clusters of equal size. 
Each observation $\boldsymbol{x}_n^{(k)} \in \{0,1,2\}^{10}$ belonging to cluster $k$ is drawn from a multivariate categorical distribution such that, for each covariate $ j = 1, \dots, 10$,
\begin{equation}
x_{nj}^{(k)} \sim \text{Categorical} (\pi_{1k}, \pi_{2k}, \pi_{3k}),
\end{equation}
where $\pi_{ik}$, $i = 1,2,3$ are such that 
$\pi_{ik} = w \rho_{ik} + (1-w)/3$, with $[\rho_{1k}, \rho_{2k}, \rho_{3k}] \sim \text{Dirichlet}(0.01)$.
Each dataset has a different value of $w \in [0,1]$. Higher values of $w$ give clearer clustering structures.
The response variable is binary, with $P(y_n = 1 | z_{n} = k) = \theta_k$, where $\theta_k \in \left\lbrace 0.01, 0.1, 0.15, 0.85, 0.9, 0.99 \right\rbrace$.
We repeat each experiment 100 times.
For each synthetic dataset, we use the MCMC algorithm for Dirichlet process mixture models implemented in the \textsf{R} package \verb|PreMiuM| of \citet{liverani2015premium} to obtain the PSMs. We use discrete mixtures \citep[Section 3.2]{liverani2015premium} except in one setting (detailed below) where the profile regression model of \citet{molitor2010bayesian} is used, using a discrete mixture with categorical response. The idea of profile regression is that, if a response $y_n$ is available for each $n=1, \dots, N$, the observations $d_n = (x_n, y_n)$ are jointly modelled as the product of the response model and a covariate model. In both cases we use the default hyperparameters, which we found to work well in practice.

\begin{figure}[h]
	\centering
	\includegraphics[width=.96\textwidth]{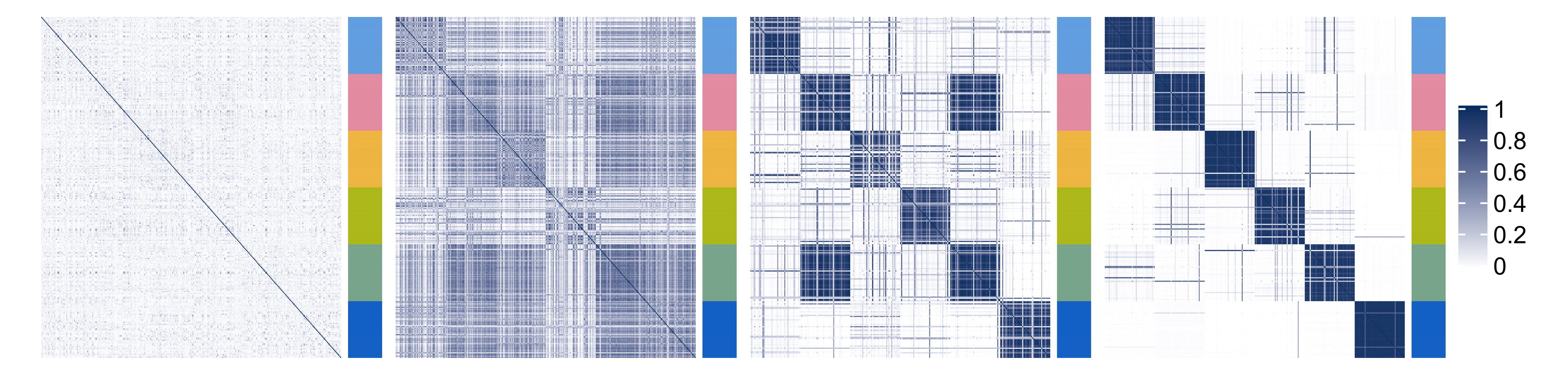}
	\caption{PSMs of the datasets used for setting A. The rows and columns correspond to the statistical units. The coloured bar on the right of each PSM represents the true clusters. Higher probabilities of belonging to the same cluster are indicated in blue. The values of $w$ used to generate these matrices are, from left to right, 0.2, 0.4, 0.6, and 0.8.}
	\label{fig:psms}
\end{figure}

We consider four different simulation settings:
\begin{enumerate}[(A)]
	\item The clustering structure in every dataset is the same and is related to the outcome of interest. %(Figures \ref{fig:psms-simulations}a-d);
	\item As in setting A, the clustering structure in each dataset is the same. In this case, however, each dataset contains some additional covariates that have no clustering structure.
		\item The dataset with highest cluster separability has a clustering structure that is unrelated to the response variable, all the other datasets are the same as in setting A. % (Figures \ref{fig:psms-simulations}a-c and 4e).
	\item This is the same as setting C, but profile regression is used to derive the PSMs.
\end{enumerate}
One set of PSMs used for setting A is shown in Figure \ref{fig:psms}.

In this case we know that higher values of $w_d$ are associated with higher levels of cluster separability. However, in general, we do not know how dispersed the elements in each matrix are. So, we define another way to score how strong the signal is in each dataset. We use the \emph{cophenetic correlation coefficient}, a measure of how faithfully hierarchical clustering would preserve the pairwise distances between the original data points. Given a dataset $X= [\boldsymbol{x}_1, \boldsymbol{x}_2, \dots, \boldsymbol{x}_N]$ and a similarity matrix $\Delta \in \mathbb{R}^{N\times N}$, we define the \emph{dendrogrammatic distance} between $\boldsymbol{x}_i$ and $\boldsymbol{x}_j$ as the height of dendrogram at which these two points are first joined together by hierarchical clustering and we denote it by  $\eta_{ij}$. The cophenetic correlation coefficient $\rho$ is calculated as 
\begin{equation}
\rho = \frac{\sum_{i<j} (\Delta_{ij} - \bar{\Delta})(\eta_{ij} - \bar{\eta})}{\sqrt{\sum_{i < j} (\Delta_{ij} - \bar{\Delta}) \sum_{i<j} (\eta_{ij} - \bar{\eta})}},
\end{equation}
where $\bar{\Delta}$ and $\bar{\eta}$ are the average values of $\Delta_{ij}$ and $\eta_{ij}$ respectively.

\subsection{Summarising PSMs}
\label{sec:single-datasets}
%{\color{blue} [I think this section needs some tidying up, to make things more precise.]}

%{\color{blue}[I think you need to say somewhere - possibly here, possibly in Section \ref{sec:single-datasets}  - that the point of considering the single dataset case is to show that kernel k means does as good a job of summarising the MCMC output as other methods, but has the additional advantage that you can then combine the resulting kernel matrix with others, using MKL.  Moreover, this simulation study allows you to check that the transformation you are doing to the PSM in order to obtain a kernel matrix is fine, and does not have a negative impact on the quality of the clustering (compared to other methods that just use the PSM without transformation)].  } 

Before using the PSMs with the MKL-based integrative clustering methods, we carry out some simulation studies to ensure that the kernel $k$-means algorithm performs equally well at summarising the MCMC output as other methods from the literature. 
The advantage is that the kernel $k$-means can be extended to combine multiple datasets.
%Moreover, this simulation study allows us to check that the transformation we apply to the PSMs in order to obtain valid kernel matrices does not have a negative impact on the quality of clustering, since we compare our results to the ones of methods who use the PSMs without transformation.

We use the adjusted Rand index \citep[ARI;][]{hubert1985comparing} as a measure of the similarity between the output of each clustering method and the true partition of the data. %{\color{blue} [I think that perhaps a lot of the text about the ARI could be relegated to an appendix -- this can be sorted out at the end.]}  

To compute the ARI given two partitions $U = \{U_1, \dots, U_K\}$ and $V = \{V_1, \dots, V_L\}$, we summarise the overlapping between each pair of subsets $U_i$ and $V_j$ by the values $\nu_{ij} = |U_i \cap V_j|$, $i = 1, \ldots, K$, $j = 1, \dots, L$.
The ARI is then calculated as 
\begin{equation}
ARI = \frac{
	\sum_{k,l} \binom{\nu_{kl}}{2} - \sum_k \binom{\nu_{k \cdot}}{2} \sum_l \binom{\nu_{\cdot l}}{2} \big/ \binom{\nu}{2}
}{
	\frac{1}{2} \big[ \sum_k \binom{\nu_{k \cdot}}{2} + \sum_l \binom{\nu_{\cdot l}}{2} \big] - \sum_k \binom{\nu_{k\cdot}}{2} \sum_l \binom{\nu_{\cdot l}}{2} \big/ \frac{\nu}{2}
}.
\end{equation}
This is the corrected-for-chance version of the Rand index \citep{rand1971objective} that is simply the number $n_s$ of pairs of elements that are in the same subsets in both partitions $U$ and $V$,  plus the number $n_d$ of pairs of elements that are in different subsets in both $U$ and $V$, divided by the total number of pairs $N^2$
\begin{equation}
RI = \frac{n_s+n_d}{N^2}.
\end{equation}

We use the synthetic datasets described in setting A of Section \ref{sec:synthetic-datasets} and compare the kernel $k$-means algorithm to the methods implemented in the \textsf{R} package \verb|mcclust| \citep{fritsch2009improved}.
All these methods take a PSM $\Delta$ as input and find the clustering $\boldsymbol{c}^*$ that maximises the posterior expected adjusted Rand index (PEAR).
This is achieved by choosing the clustering $\boldsymbol{c}^*$ that maximises the following quantity:
\begin{equation}\label{eq:PEAR}
\frac{
	\sum_{i<j} \mathbb{I}_{\{c^*_i = c^*_j\} } \Delta_{ij} - \sum_{i<j} \mathbb{I}_{\{c_i^*=c_j^*\}}\sum_{i<j}\Delta_{ij} \big/ \binom{n}{2}
}{
	\frac{1}{2} \big[ \sum_{i<j} \mathbb{I}_{\{c_i^* = c_j^*\}} + \sum_{i<j} \Delta_{ij} \big] - \sum_{i<j} \mathbb{I}_{\{c_i^* = c_j^*\}} \sum_{i<j} \Delta_{ij}  \big/ \frac{n}{2}
}.
\end{equation}
The clusterings $\boldsymbol{c}^*$ taken into consideration by these methods can be chosen in different ways.
We use hierarchical clustering with $1 - \Delta$ as distance matrix with average and complete linkage.
The {\tt maxpear} function tries all the possible number of clusters between one and a maximum number of cluster $K_{\max}$ specified by the user. We consider both 6 and 20 as values for $K_{\max}$. 
We also use the {\tt maxpear} function to try with all the clusterings in the MCMC output and take the one that maximises the quantity of Equation \eqref{eq:PEAR}. We repeat this procedure using the {\tt minVI} function of the \textsf{R} package {\tt mcclust.ext} of \citet{wade2018bayesian}, where the selected $\boldsymbol{c^*}$ is the one that minimises the lower bound to the posterior expected variation of information (Equation \ref{eq:loss-vi}) from Jensen's inequality.

\begin{figure}[h!]
	\centering
	\includegraphics[width=.8\linewidth]{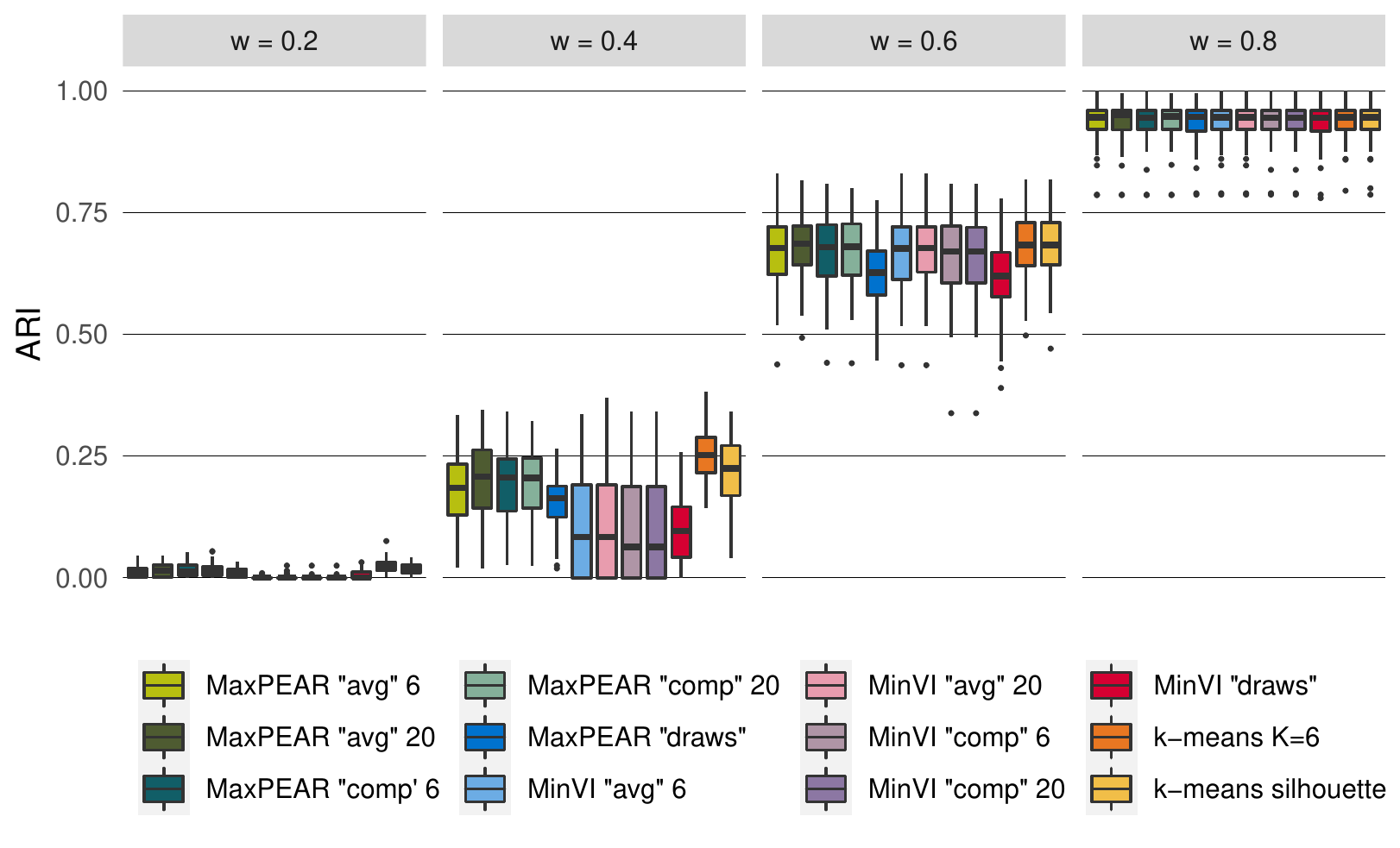}
	\caption{ARI  for the kernel $k$-means applied to one dataset at a time, for different values of $\rho$ compared to maximising the PEAR as suggested by \citet{fritsch2009improved} and to minimising the VI as suggested by \citet{wade2018bayesian}. For both methods, we try different settings, namely: performing hierarchical clustering on the matrix $1-\Delta$  with average (avg) and complete (comp) linkage, with maximum number of clusters equal to either 6 or 20, as well as considering all the clusterings samples that are appear in the MCMC output (draws). For kernel $k$-means, the results obtained fixing the number of clusters to six and choosing it via the silhouette are presented.}
	\label{fig:singleDatasets_densityPlot_ARI}
\end{figure}

Figure \ref{fig:singleDatasets_densityPlot_ARI} shows the boxplots of the adjusted Rand index obtained repeating the experiment with 100 different sets of synthetic data calculated for each of the considered clustering algorithms and for all values of $w$. 
We can see that the kernel $k$-means on average performs better than the other methods when the number of clusters is known and that it performs similarly to the others when the number of clusters is chosen to maximise the average silhouette. %This demonstrates that we can use kernel $k$-means to summarise the output of MCMC algorithms and that the transformation applied to the PSMs in order to create valid kernel matrices does not affect the quality of the clustering.

\subsection{Integrative clustering}
\label{sec:simulations-integrative-clustering}

We assess the MKL-based integrative clustering approaches described in Sections \ref{sec:combining-psms} and \ref{sec:outcome-guided-clustering} in the four settings presented in Section \ref{sec:synthetic-datasets}.
For each setting we consider four different subsets of data, each combining three out of our four synthetic datasets generated with $w = 0.2, 0.4, 0.6, 0.8$. In what follows, we refer to the dataset generated with $w = 0.2$ as ``dataset 0'', the one generated with $w = 0.4$ as ``dataset 1'', and so on, for ease of reference. Moreover, we indicate by ``0+1+2'' the integration of the datasets generated with values of $w$ equal to 0.2, 0.4, and 0.6 respectively, and similarly for the other combinations of datasets. Here we show the ARI between the clusterings found via MKL integration and the true cluster labels, the weights assigned to each dataset in each setting are instead reported in the Supplementary Material.

\begin{figure}
	\centering 
	\includegraphics[width=.85\textwidth]{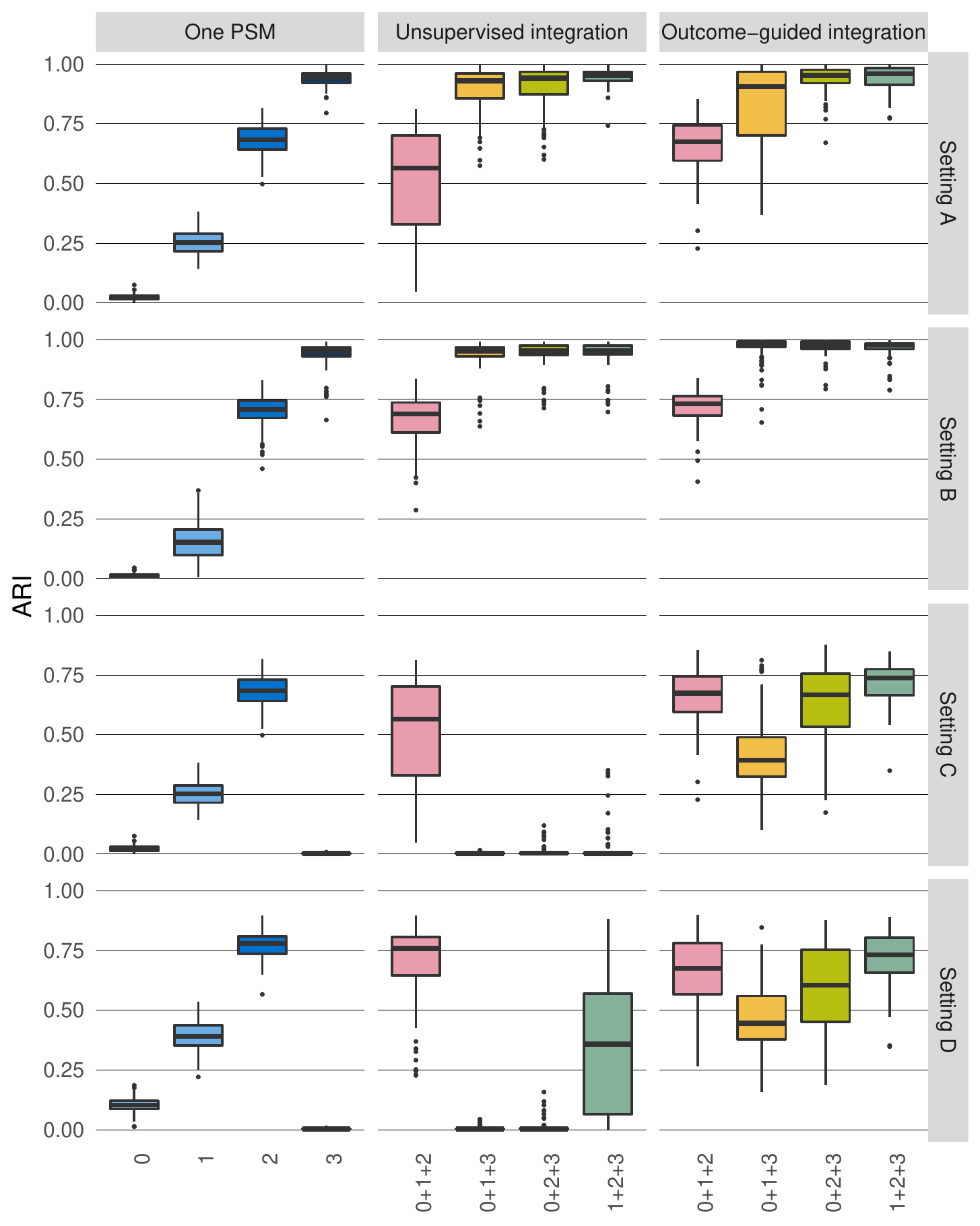}
	\caption{Simulation study, adjusted Rand index obtained by summarising the PSMs one at a time using kernel $k$-means (left), combining different subsets of three PSMs in an unsupervised fashion using localised multiple kernel $k$-means (centre), and combining the same subsets making use of a response variable and multi-class SVMs to determine each PSM's weight and using kernel $k$-means for the final clustering (right).}
	\label{fig:simulation-ari}
\end{figure}

\paragraph*{Setting A}
The ARI obtained by combining the datasets in the unsupervised and outcome-guided frameworks is shown in the first row of Figure \ref{fig:simulation-ari}. The values of the ARI obtained in the previous section on each dataset separately are also reported. In all settings we set the number of clusters to the true value, six. The unsupervised integration performed using localised multiple kernel $k$-means allows us to reach values of the ARI that are close to those of the ``best'' dataset (i.e. the dataset that has the highest value of cluster separability) among the three datasets in each subset. This is because the unsupervised MKL approach considered here assign higher weights to the datasets that give rise to kernels with higher values of $\rho$, which in this case correspond to higher values of $w$ \citep{cabassi2020multiple}. Moreover, even higher values of the ARI are achieved via outcome-guided integration thanks to the smarter weighting of the datasets. In this case, the kernels that help separate the classes in the response have higher weights than the others.

\paragraph{Setting B}
In the second row of Figure \ref{fig:simulation-ari} are shown the results obtained for setting B, where the PSMs are obtained exploiting (an adaptation of) the variable selection strategy of \citet{chung2009nonparametric} implemented in the \textsf{R} package {\tt PReMiuM}. Despite the fact that the ARI of dataset 2 is lower than in the previous case, the integration results are better than in Setting A. Again, this is due to the fact that most informative kernels are weighted more highly than the other ones.

\paragraph*{Setting C}
This simulation study helps us to show that the outcome-guided approach favours the clustering structures that agree with the structure in the response. For this reason we use a dataset with high cophenetic correlation coefficient whose clustering structure is not related to the response.
The results are presented in the third row of Figure \ref{fig:simulation-ari}.
Again, localised multiple kernel $k$-means assigns higher weights to the datasets that are more easily separable, i.e. datasets that give rise to kernels having higher cophenetic correlation coefficients. Note that here higher values of $w$ correspond to higher cophenetic correlation. In this situation, this causes the ARI of the subsets of kernels that include dataset 4 to drop to zero.
In the outcome-guided case, instead, the dataset that has the highest level of cluster separability but is not related to the outcome of interest has (almost) always weight equal to zero. 

\paragraph*{Setting D}
Lastly, we consider the case where the model used to generate the PSMs is profile regression (fourth row of Figure \ref{fig:simulation-ari}). We see that, as expected, the ARI is higher than in the previous cases for the clustering obtained with each dataset taken separately, except of course for dataset 4, that has a different clustering structure. This is reflected in an improvement of the ARI of the unsupervised and outcome-guided integration, for all considered subsets of data. In particular, the latter almost always allows us to retrieve the true clustering.

\section{Integrative clustering: biological applications}
\label{sec:results-psms}
We present results from applying our method to two exemplar integrative clustering problems from the literature. 
In Section \ref{sec:pancan-10} we perform integrative clustering on the dataset used for the analysis of 3,527 tumour samples from 12 different tumour types of \citet{hoadley2014multiplatform}.
In Section \ref{sec:yeast-dataset} we consider an example of {\em transcriptional module discovery}, to which a number of existing integrative clustering approaches have previously been applied \citep{savage2010discovering,kirk2012bayesian, cabassi2020multiple}.

\subsection{Multiplatform analysis of ten cancer types}
\label{sec:pancan-10}

We analyse the dataset of \citet{hoadley2014multiplatform}, which contains data regarding 3,527 tumour samples spanning 12 different tumour types. This dataset is particularly suitable for two purposes: (i) determining whether 'omic signatures are shared across different tumour types and (ii) discovering tumour subtypes. The 'omic layers available are: DNA copy number, DNA methylation, mRNA expression, microRNA expression and protein expression. \citet{hoadley2014multiplatform} used \emph{Cluster-Of-Clusters Analysis} (COCA; \citealp{cabassi2020multiple}) to cluster the data and found that the clusters where highly correlated with the tissue of origin of each tumour sample and were shown to be of clinical interest.

Here, we combine the data layers both in the unsupervised and outcome-guided frameworks. We make use of the \textsf{C} implementation \citep{mason2016mdi} of the \emph{multiple dataset integration} (MDI) method of \citet{kirk2012bayesian} to produce PSMs for each data layer separately. In order to be able to do so, we only include in our analysis the tumour samples that have no missing values; this reduces the sample size to 2,421 and the number of tumour types available for the analysis to ten. A mixture of Gaussians is used for the continuous layers (DNA copy number, microRNA, and protein expression), while the multinomial model is used for the methylation data, which are categorical. Due to the high number of features, it is not possible to produce a PSM for the full mRNA dataset, so we exclude it from the analysis presented here. In the Supplementary Material, however, we show how the variable selection method developed specifically for multi-omic data by \citet{cabassi2020penalised} can be employed in this case to reduce the size of each data layer and integrate all five data types.

\subsubsection{Unsupervised integration}

\begin{figure}
	\centering
	\begin{subfigure}[t]{\textwidth}
		\includegraphics[width=\textwidth]{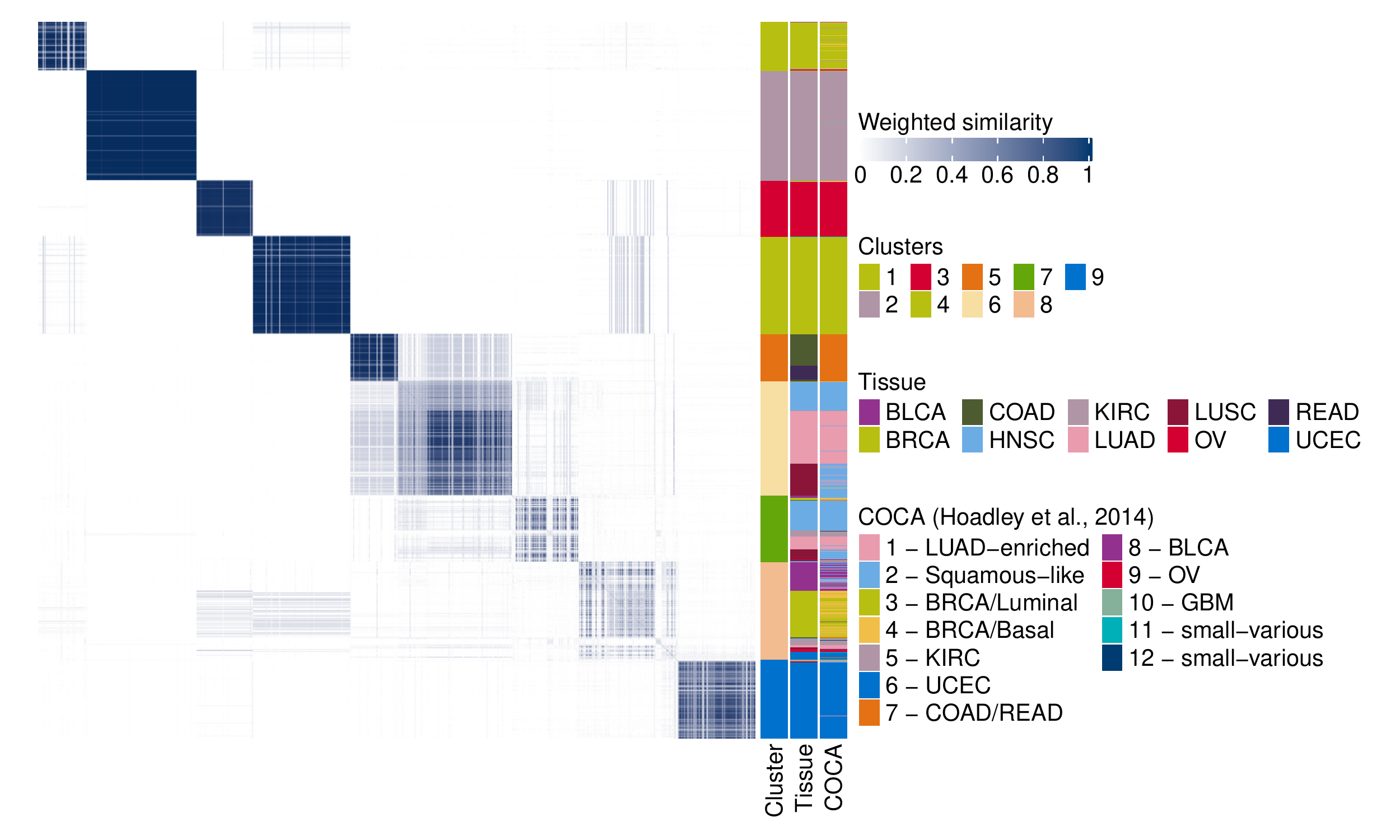}
		\caption{Clusters and weighted kernel.}
	\end{subfigure}
	\begin{subfigure}[t]{.48\textwidth}
		\includegraphics[width=\textwidth]{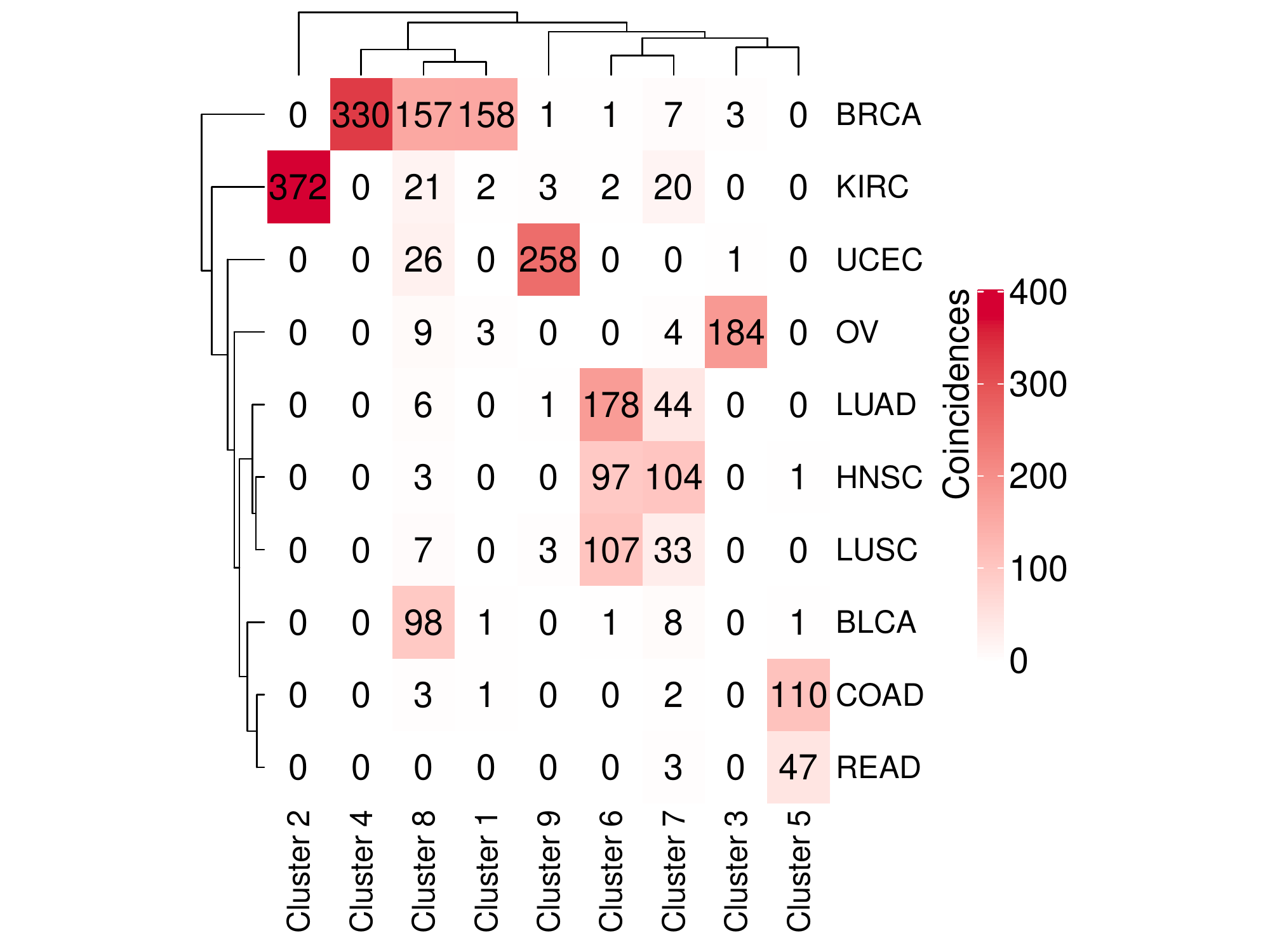}
		\caption{Coincidence matrix.}
	\end{subfigure}
	\caption{Unsupervised multiplatform analysis of ten cancer types. \textbf{(a)} Left: weighted kernel. The rows and columns correspond to cancer samples. Higher values of similarity between samples are indicated in blue. Right: final clusters, tissues of origin, and COCA clusters. \textbf{(b)} Coincidence matrix comparing the tissue of origin of the tumour samples (rows) with the clusters (columns).}
	\label{fig:unsupervised-pancan10}
\end{figure}

We combine the PSMs of the four data layers via multiple kernel $k$-means with number of clusters going from 2 to 50. We choose the number of clusters that maximises the silhouette, which is 9 (Supplementary Material). The resulting clusters are shown in Figure \ref{fig:unsupervised-pancan10}. Six out of the nine clusters contain almost exclusively samples from one tissue: most samples of renal cell carcinoma (KIRC) are in cluster 2, almost all statistical units in clusters 1 and 4 are breast cancer samples, most serous ovarian carcinoma (OV) samples are in cluster 3, bladder urothelial adenocarcinoma (BLCA) samples in cluster 8, and endometrial cancer (UCEC) samples in cluster 9. 
Cluster 5, instead, is formed by the colon and rectal adenocarcinoma samples together, and corresponds exactly to cluster 7 of Hoadley \emph{et al.} (COAD/READ). Moreover, lung squamous cell carcinoma (LUSC), lung adenocarcinoma (LUAD), and head and neck squamous cell carcinoma (HNSC) are divided into two clusters (6 and 7). Cluster 8 contains the remaining samples. The average weights assigned to each data layer are: 6.9\% to the copy number data, 7.5\% to the methylation data, 7\% to the microRNA expression data, and 78.7\% to the protein expression data.

\subsubsection{Outcome-guided integration}

\begin{figure}
	\centering
	\begin{subfigure}[t]{\textwidth}
		\includegraphics[width=\textwidth]{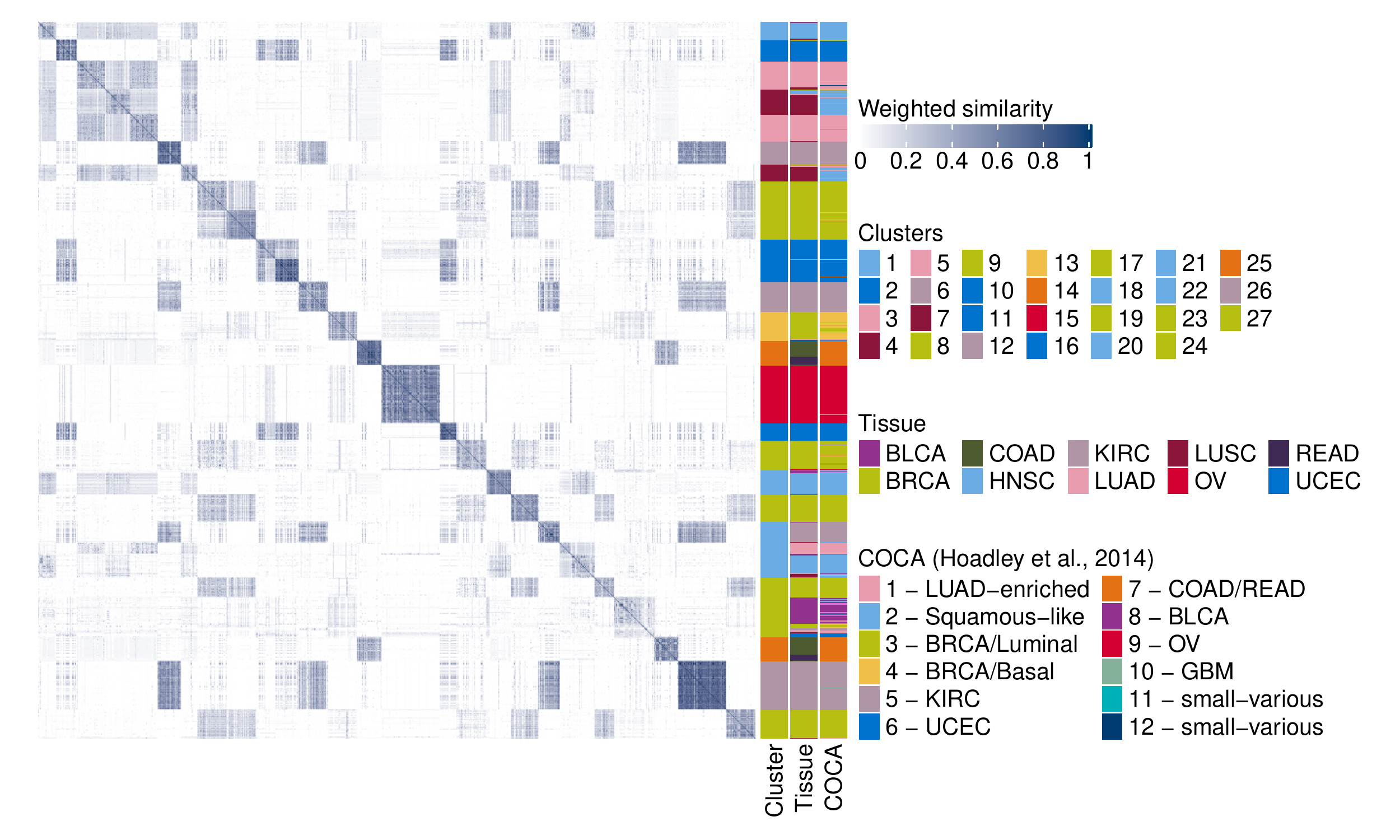}
		\caption{Clusters and weighted kernel.}
	\end{subfigure}
	\begin{subfigure}[t]{\textwidth}
		\includegraphics[width=\textwidth]{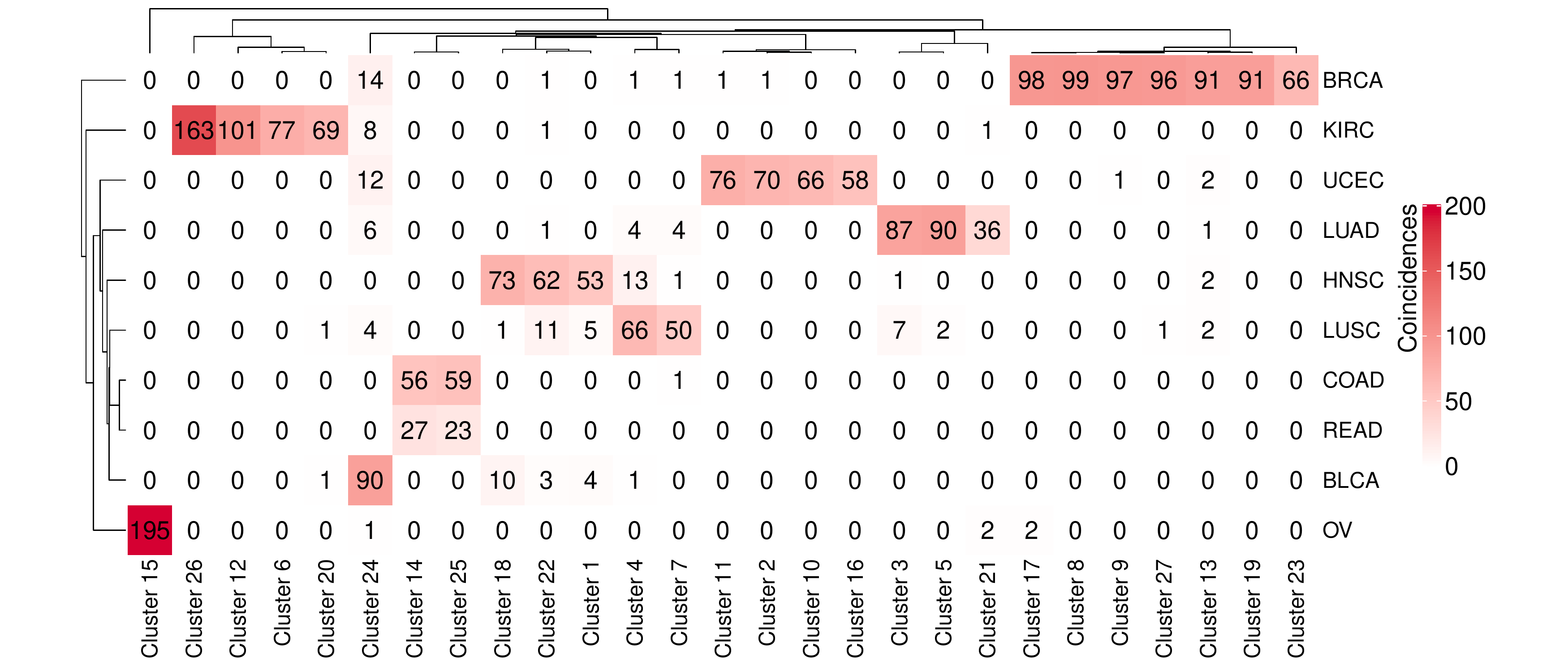}
		\caption{Coincidence matrix.}
	\end{subfigure}
	\caption{Outcome-guided multiplatform analysis of ten cancer types. \textbf{(a)} Left: weighted kernel. The rows and columns correspond to cancer samples. Higher values of similarity between samples are indicated in blue. Right: final clusters, tissues of origin, and COCA clusters. \textbf{(b)} Coincidence matrix comparing the tissue of origin of the tumour samples (rows) with the clusters (columns).}
	\label{fig:outcome-guided-pancan10}
\end{figure}

We obtain the weights for the outcome-guided integration via the SimpleMKL algorithm, which are as follows: DNA copy number 35.9\%, methylation 13.5\%, microRNA expression 33.8\%, and protein expression 16.8\%. We then cluster the data using kernel $k$-means with number of clusters going from 2 to 50. The silhouette is maximised at $K=27$ (Supplementary Material). The clusters obtained in this way are shown in Figure \ref{fig:outcome-guided-pancan10}. It is interesting to note that, in this case, each cluster contains almost exclusively tumour samples from the same tissue. The only exceptions are clusters 4 and 22, which contain both lung and head/neck squamous cell carcinoma samples, and clusters 14 and 25 in which colon and rectal adenocarcinomas are clustered together, like in the unsupervised case. Each tumour type, except for ovarian and bladder cancers, is divided into multiple subclusters. Further analysis would be required to assess whether these clusters are clinically relevant. Interestingly, we observe a distinction between luminal (i.e. estrogen receptor-positive and HER2-positive) and basal breast cancer samples (the former are in clusters 8, 9, 17, 19, 23, 24, 27, while the latter are in cluster 13). This was also observed by \citet{hoadley2014multiplatform}.

\subsection{Transcriptional module discovery}
\label{sec:yeast-dataset}

For this example we consider transcriptional module discovery for yeast (\emph{Saccharomyces cerevisiae}). The goal is to find clusters of genes that share a common biological function and are co-regulated by a common set of transcription factors. Previous studies have demonstrated that combining gene expression datasets with information about transcription factor binding can improve detection of meaningful transcriptional modules \citep{ihmels2002revealing, savage2010discovering}. 

We combine the ChIP-chip dataset of \citet{harbison2004transcriptional}, which provides binding information for 117 transcriptional regulators, with the expression dataset of \citet{granovskaia2010high}. The ChIP-chip data are discretised as in \citet{savage2010discovering} and \citet{kirk2012bayesian}. The measurements in the dataset of \citet{granovskaia2010high} represent the expression profiles of 551 genes at 41 different time points of the cell cycle.

Since the goal is to find clusters of genes, here the statistical units correspond to the genes and the covariates to 41 experiments of gene expression measurement for the expression dataset and to the 117 considered transcriptional regulators in the ChIP-chip dataset.
To produce the posterior similarity matrices for the two datasets, we use the \verb|DPMSysBio| Matlab package of \citet{zurauskiene2016graph}. For each dataset, we run 10.000 iterations of the MCMC algorithm and summarise the output into a PSM. The PSMs obtained in this way are reported in the Supplementary Material.

\begin{figure}[h]
	\centering
	\begin{subfigure}[b]{.49\textwidth}
		\centering
		\includegraphics[width =\textwidth]{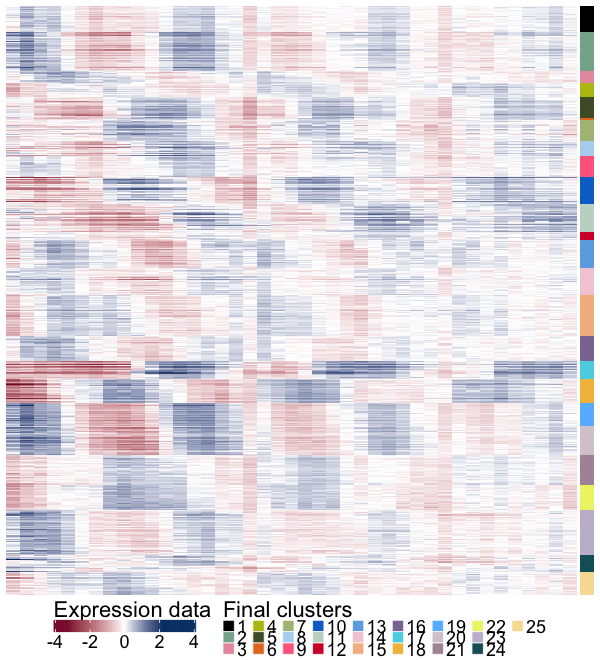}
		\caption{Expression data.}
	\end{subfigure}
	\begin{subfigure}[b]{.49\textwidth}
		\centering
		\includegraphics[width = \textwidth]{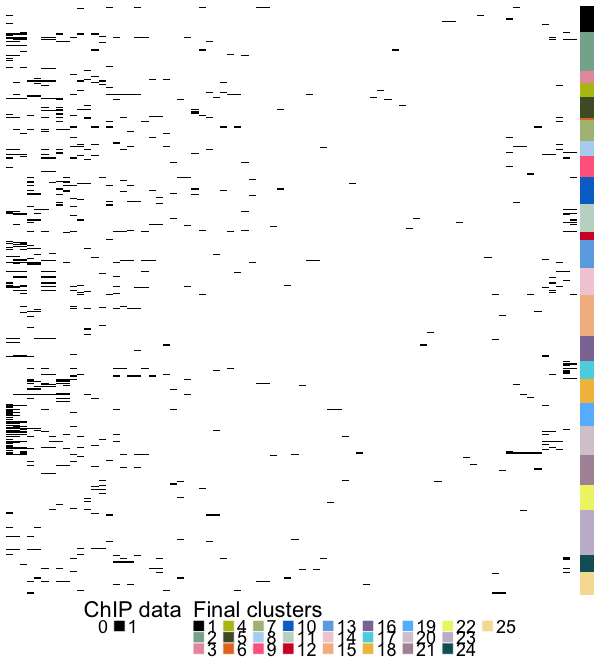}
		\caption{ChIP-chip data.}
	\end{subfigure}
	\begin{subfigure}[b]{.49\textwidth}
		\centering
		\includegraphics[width = \textwidth]{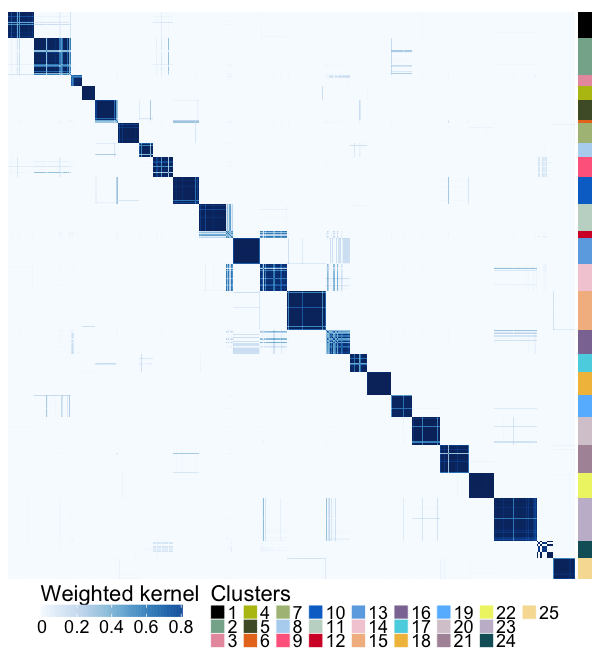}
		\caption{Weighted kernel.}
	\end{subfigure}
	\caption{Transcriptional module discovery, integration of the \citet{harbison2004transcriptional} and \citet{granovskaia2010high} datasets. \textbf{(a)} Expression data. Each row corresponds to a gene and each column to a different time point. \textbf{(b)} ChIP-chip data. Each row corresponds to a gene and each column to a transcriptional regulator. \textbf{(c)} Weighted kernel. The rows and columns correspond to the genes. Higher values of similarity between genes are indicated in blue. To the left of each plot is shown the final clustering, obtained by integrating the PSMs of the expression and ChIP-chip data via multiple kernel $k$-means.}
	\label{fig:transcriptional-psms-second-set-of-data}
\end{figure} 

We combine the PSMs using our unsupervised MKL approach. The average weights assigned by the localised multiple kernel $k$-means to each matrix and the values of the average silhouette for different numbers of clusters are reported in the Supplementary Material. %Interestingly, the PSM of the expression data has slightly higher cophenetic coefficient but lower average weight than the ChIP-chip data.
We set the number of clusters to 25, which is the value that maximises the silhouette.
The final clusters are shown in Figure \ref{fig:transcriptional-psms-second-set-of-data} next to the two datasets and the combined PSM.
In order to determine whether our clustering is biologically meaningful, we use the Gene Ontology Term Overlap (GOTO) scores defined by \citet{mistry2008gene}. Denoting by $\text{annot} g_i$ the set of all direct annotations for each gene and all of their associated parent terms, the GOTO similarity between two genes $g_i, g_j$ is the number of annotations that the two genes share:
\begin{equation}
\text{sim}_{\text{GOTO}} (g_i, g_j)= |\text{annot} g_i \cap \text{annot} g_j|.
\end{equation}
Details on how to compute the average overall GOTO score of a clustering are given in the Supplementary Material of \citet{kirk2012bayesian}.
The GOTO scores are reported in Table \ref{table:goto-scores-second-set-of-data}. The two datasets combined achieve higher GOTO scores than those of the clusters obtained using each dataset separately. We also compare these GOTO scores to those obtained with the two methods used in \citet{cabassi2020multiple}. The first one is COCA, a simple, unweighted algorithm for integrative clustering that is widely used in practice \citep{TCGA2012comprehensive, hoadley2014multiplatform}. The other integrative method considered here is KLIC (\emph{Kernel Learning Integrative Clustering}), by which we mean the integration of multiple kernels via multiple kernel $k$-means, where the kernels are generated via consensus clustering (Figure \ref{fig:theory-unsupervised}).
The clusters obtained with these two alternative methods have lower GOTO scores than the integration of PSMs. For COCA, this is result not unexpected, since the method is unweighted and has previously been shown to perform less well than KLIC. The difference between what is referred to as KLIC here and the unsupervised integration of PSMs, however, only lies in how the kernels are constructed. These scores therefore suggest that kernels generated from probabilistic models can lead to more accurate results than those built using consensus clustering.

\begin{table}[h]
\centering
\begin{tabular}{l c c c }
Dataset(s) &GOTO BP & GOTO MF & GOTO CC \\
\hline
ChIP data (Harbison {\em et al.}) & 6.18 & 0.97 & 8.54 \\
Expression data (Granovskaia {\em et al.}) & 7.07 & 1.04 & 8.90 \\
ChIP+Expression data: COCA & 5.74 & 0.90 & 8.19 \\
ChIP+Expression data: KLIC & 6.60 & 0.96 & 8.66 \\
ChIP+Expression data: integration of PSMs & \textbf{7.15} & \textbf{1.05} & \textbf{8.93} \\
\hline\\
\end{tabular}
\caption{Gene Ontology Term Overlap scores. ``BP'' stands for Biological Process ontology, ``MF'' for Molecular Function, and ``CC'' for Cellular Component.
The number of clusters used for every method is 25.
}
\label{table:goto-scores-second-set-of-data}
\end{table} 

\section{Conclusion}
We have presented a novel method for summarising a sample of clusterings from the posterior distribution of an MCMC algorithm for Bayesian clustering, based on kernel methods.
We have also extended this method to allow us to integrate multiple PSMs. This can be done either in an unsupervised or in an outcome-guided way. The former weights each PSM according to how well defined is the clustering structure that it shows, the latter gives more importance to the PSMs that better reflect the structure encountered in the response variable of choice.

We have used simulation examples to show that our method gives comparable performances in terms of proportion of correct co-clustering as the existing techniques. We have also demonstrated that  the integration of multiple datasets gives better results than using one dataset at a time. 
Additionally, we proved that, if a variable related to the output of interest is available, our method can assign higher weights to the PSMs that are more closely related to that. The simulation examples prove that this feature can be extremely useful when not all the PSMs have the same clustering structure.

Finally, we have applied the novel methods to two real data applications. The pancancer data analysis shows that the outcome-guided integration of multiple PSMs can potentially be used in the context of tumour subtype discovery. The yeast example demonstrates that the proposed method is able to identify groups of genes that are co-expressed and co-regulated that are more biologically meaningful than those determined via state-of-the-art integrative algorithms. 

\clearpage
\section*{Funding}
A. Cabassi and P.D.W. Kirk are supported by the MRC [MC\_UU\_00002/13]. This work was supported by the National Institute for Health Research [Cambridge Biomedical Research Centre at the Cambridge University Hospitals NHS Foundation Trust]. The views expressed are those of the authors and not necessarily those of the NHS, the NIHR or the Department of Health and Social Care. Partly funded by the RESCUER project. RESCUER has received funding from the European Union's Horizon 2020 research and innovation programme under grant agreement No. 847912.

\bibliographystyle{apalike}
\addcontentsline{toc}{section}{Bibliography}
\bibliography{main_arxiv}

\end{document}

\begin{center}
{\LARGE\bf Kernel learning approaches for summarising and combining posterior similarity matrices}
\end{center}
\medskip
\begin{center}
{\large Alessandra Cabassi$^{1}$, Sylvia Richardson$^{1}$, and Paul D. W. Kirk$^{1,2}$ \\[15pt]
\emph{$^{1}$MRC Biostatistics Unit}\\
\emph{$^{2}$Cambridge Institute of Therapeutic Immunology \& Infectious Disease\\
University of Cambridge, U.K.}\\
}
\end{center}

\bigskip

%\begin{center}
%Supplement, \today
%\end{center}
\bigskip\bigskip

\tableofcontents
\clearpage

\section{Simulation study}

\subsection{Data}

\begin{figure}[h]
	\centering
	\includegraphics[width=.8\textwidth]{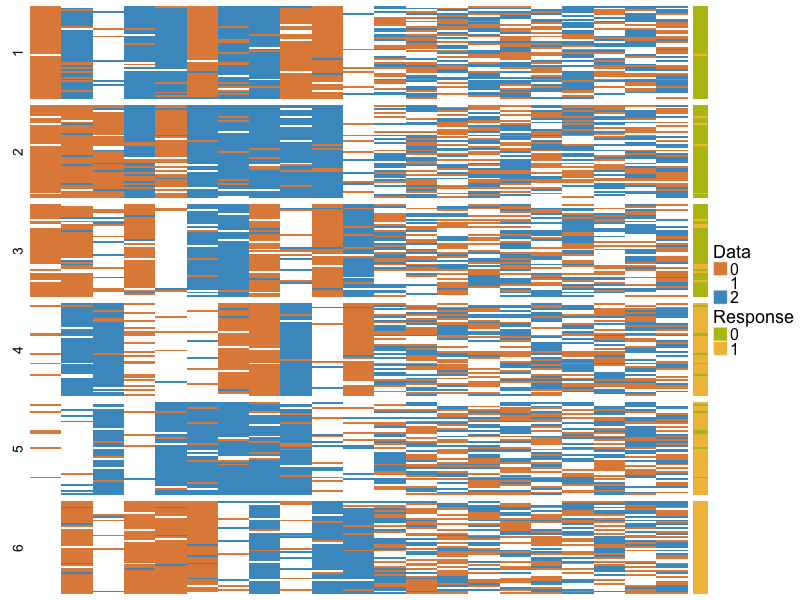}
	\caption{One of the datasets used for the simulation studies. The data are categorical, taking values 0, 1 or 2, the response is binary. The rows are separated by cluster, the cluster labels are indicated on the right of the data matrix.}
	\label{fig:data}
\end{figure}

In Figure \ref{fig:data} is shown one of the datasets used for the simulation studies, with value of $w$ set to 0.8 and number of covariates equal to 20. The first ten covariates determine six different clusters, the remaining covariates have no clustering structure. The response variable is generated as described in the main paper.

\clearpage
\subsection{Integrative clustering}

\begin{figure}[h]
	\centering 
	\includegraphics[width=.8\textwidth]{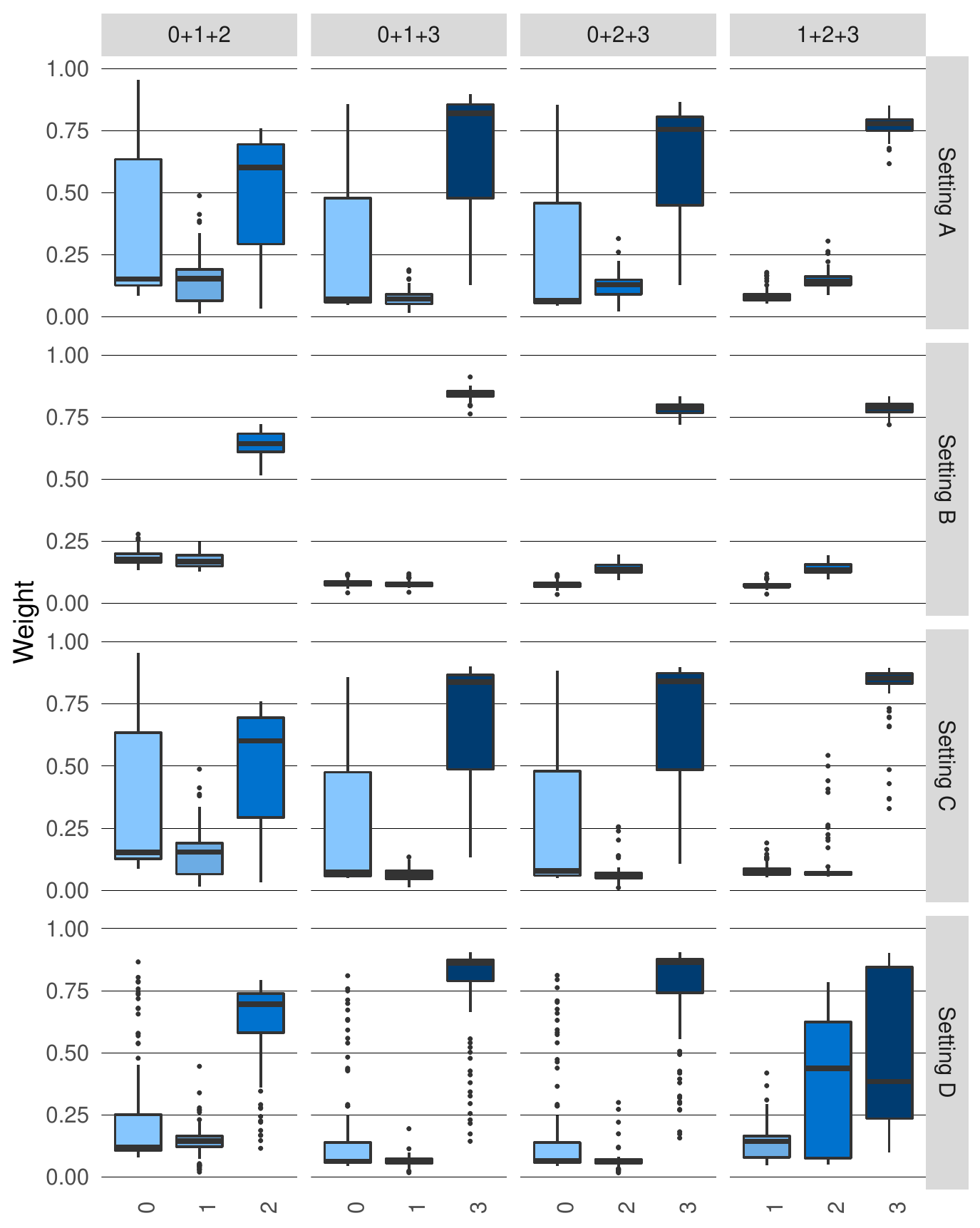}
		\caption{Weights assigned to each PSM for each subset of datasets by the unsupervised integration method.}
	\label{fig:simulation-weights-unsupervised}
\end{figure}

\begin{figure}[h]
	\centering 
	\includegraphics[width=.8\textwidth]{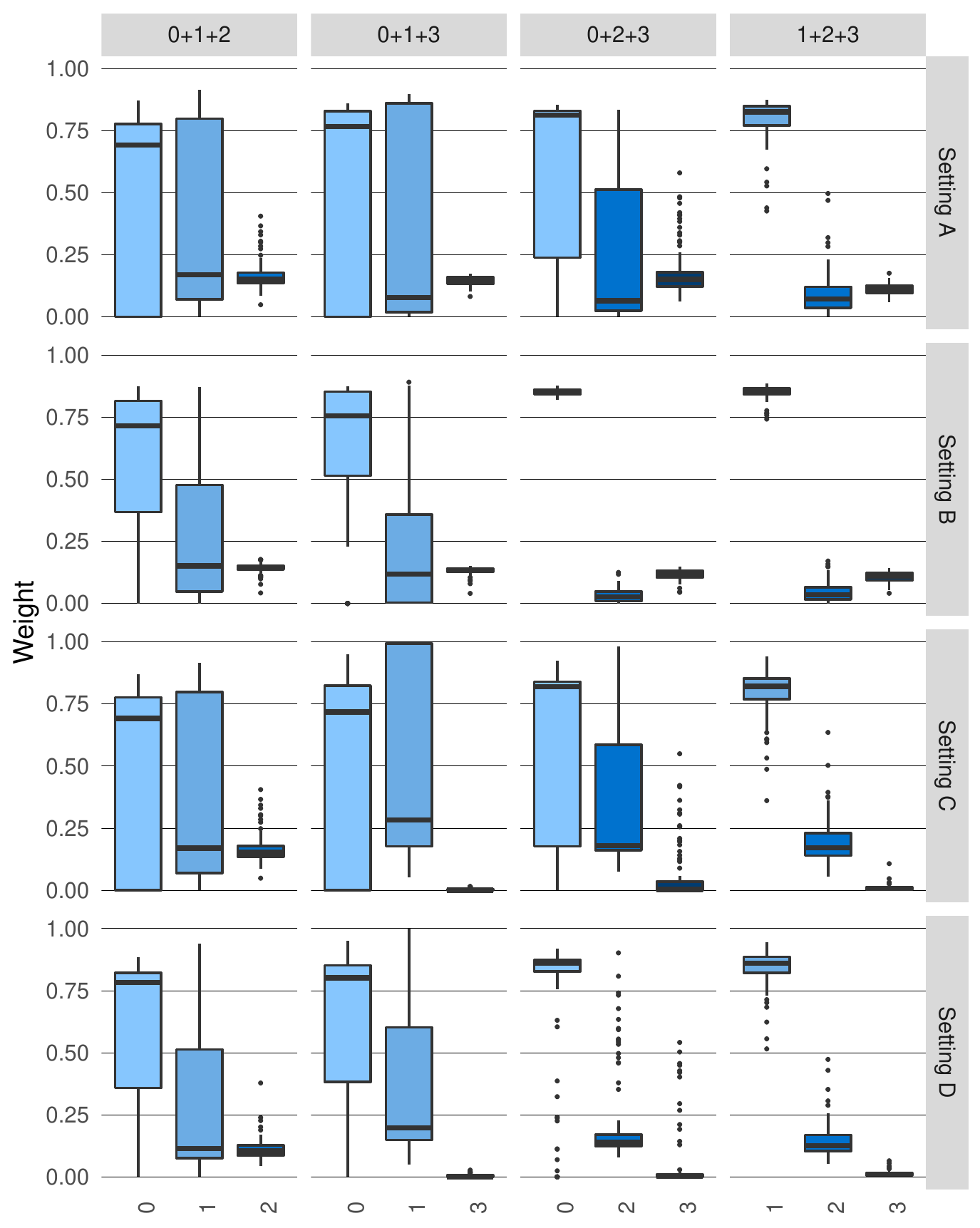}
		\caption{Weights assigned to each PSM for each subset of datasets by the outcome-guided integration methods.}
	\label{fig:simulation-weights-outcomeguided}
\end{figure}

Figures \ref{fig:simulation-weights-unsupervised} and \ref{fig:simulation-weights-outcomeguided} show the weights assigned to each dataset by the unsupervised and outcome-guided methods for the integration of multiple PSMs presented in the main paper.

\clearpage
\subsection{Additional simulation settings}
\label{sec:additional-simulation-settings}

\subsubsection{Different number of covariates}

\begin{figure}[h]
	\begin{subfigure}{\textwidth}
		\centering
		\includegraphics[width=.8\textwidth]{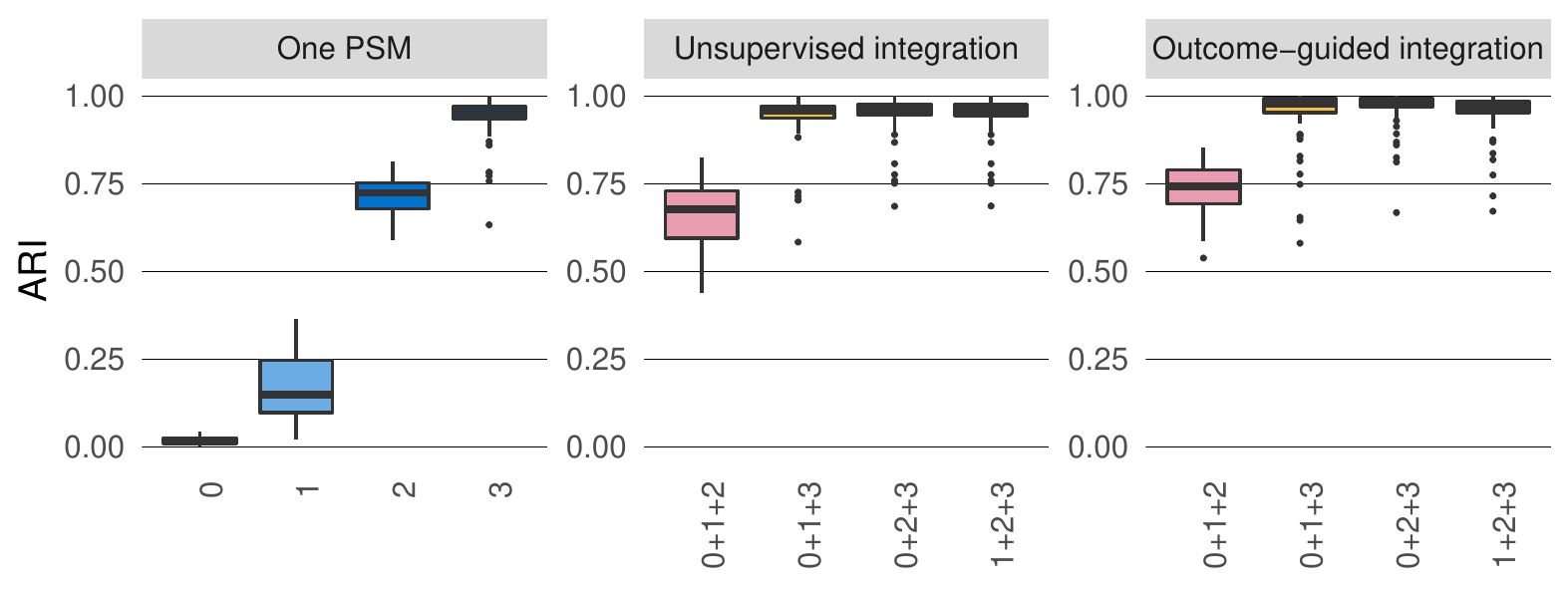}
		\caption{Setting B, 2 covariates without clustering structure.}
		\label{fig:simulation-b2-ari}
	\end{subfigure}
		\begin{subfigure}{\textwidth}
		\centering
		\includegraphics[width=.8\textwidth]{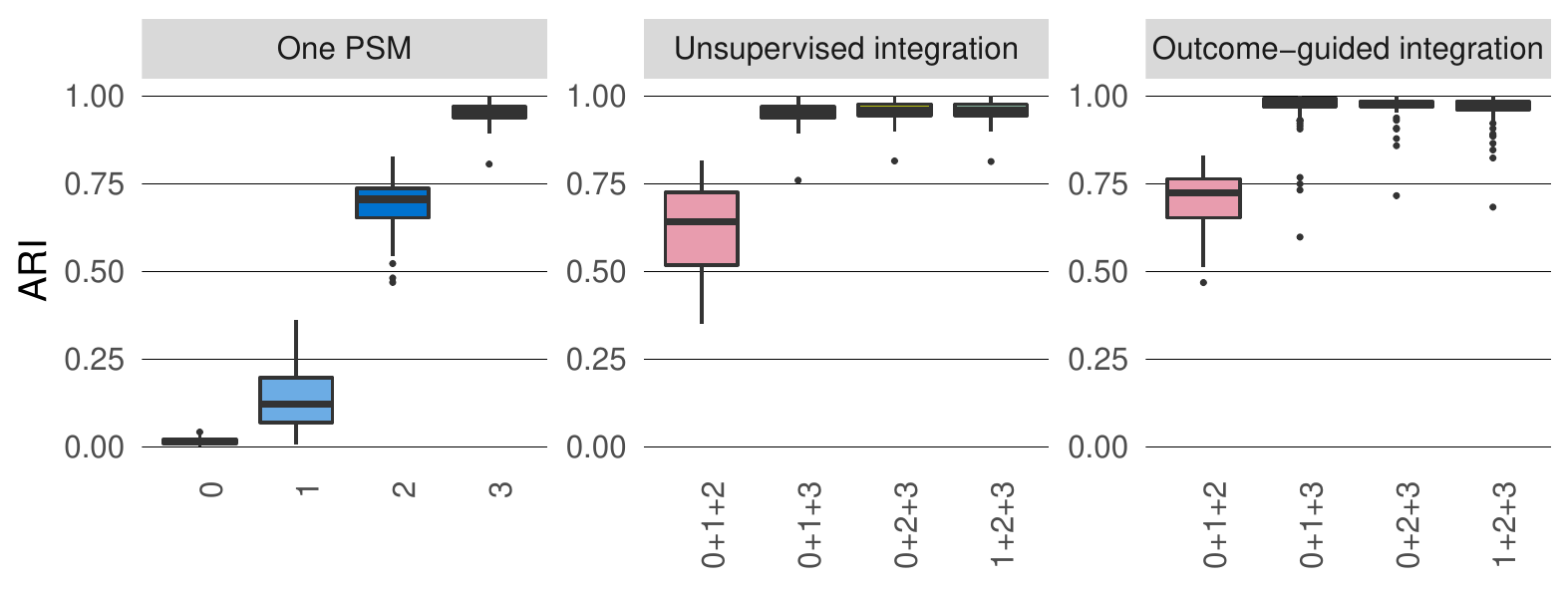}
		\caption{Setting B, 5 covariates without clustering structure.}
		\label{fig:simulation-b5-ari}
	\end{subfigure}
	\caption{Adjusted Rand index obtained by summarising the PSMs one at a time using kernel $k$-means (left), combining different subsets of three PSMs in an unsupervised fashion using localised multiple kernel $k$-means (centre), and combining the same subsets making use of a response variable and multi-class SVMs to determine each PSM's weight and using kernel $k$-means for the final clustering (right).}
	\label{fig:additional-simulations-b-ari}
\end{figure}

\begin{figure}
	\begin{subfigure}{\textwidth}
		\centering
		\includegraphics[width=.65\textwidth]{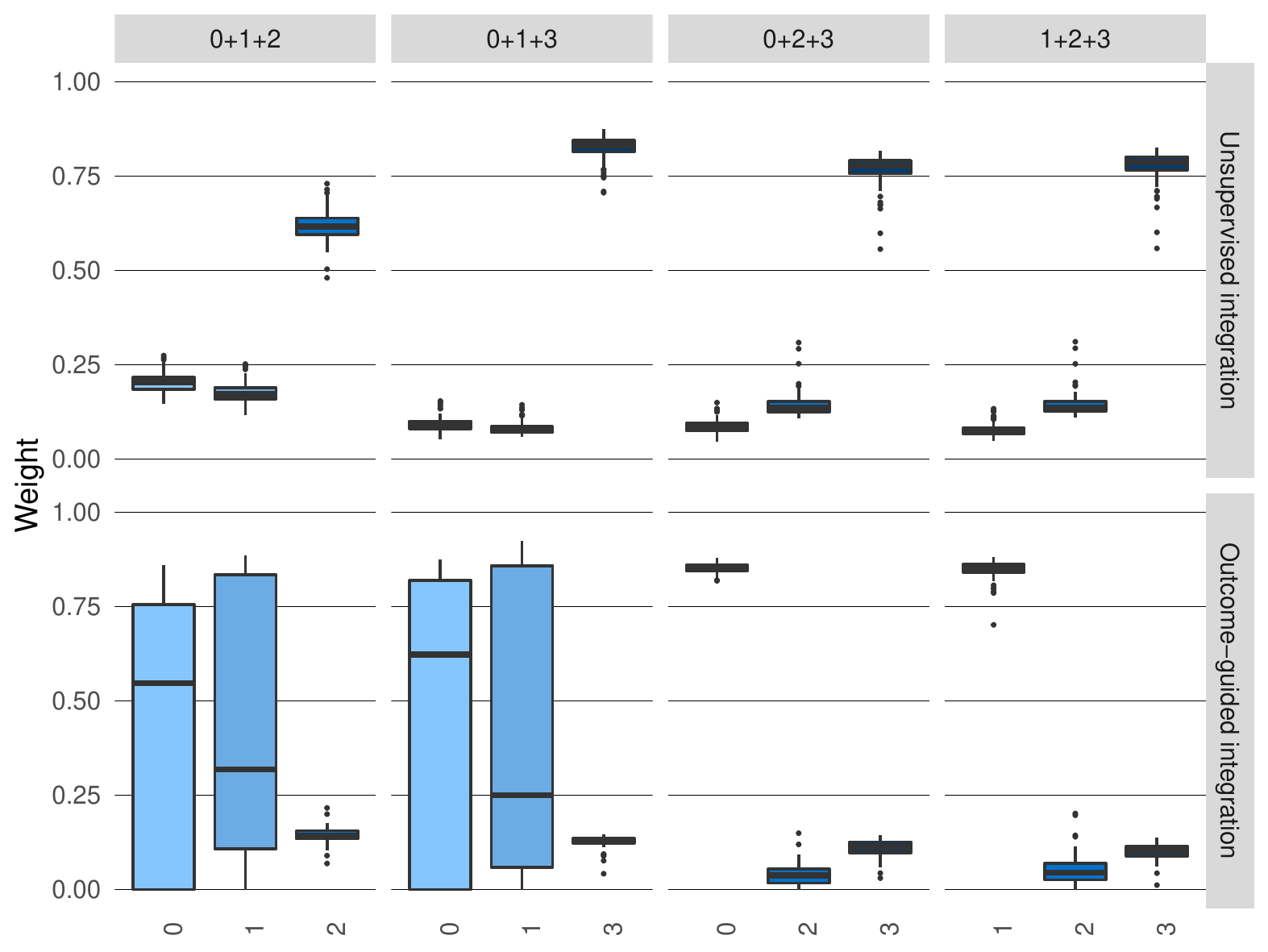}
		\caption{Setting B, 2 covariates without clustering structure.}
		\label{fig:simulation-b2-weights}
	\end{subfigure}
		\begin{subfigure}{\textwidth}
		\centering
		\includegraphics[width=.65\textwidth]{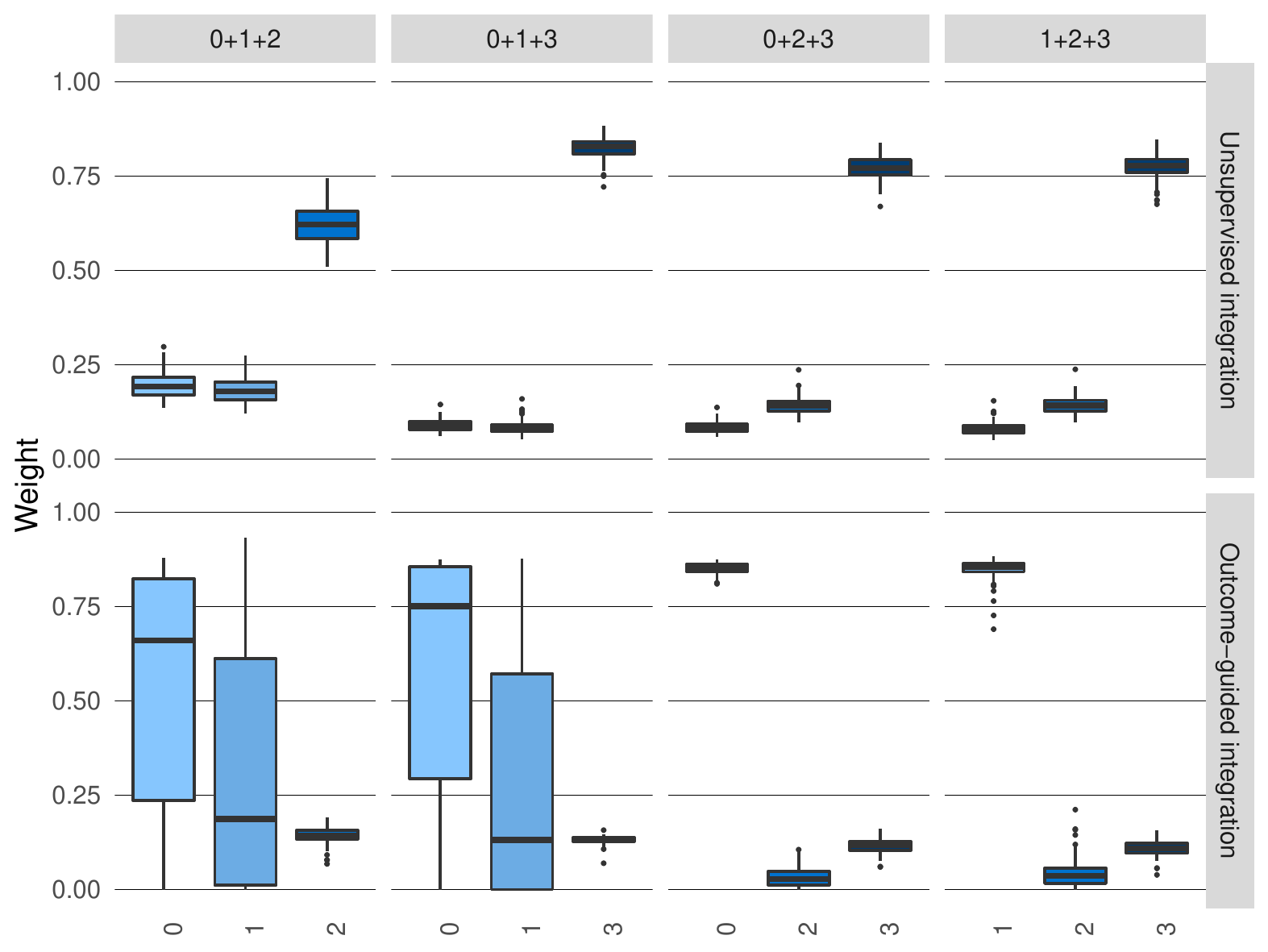}
		\caption{Setting B, 5 covariates without clustering structure.}
		\label{fig:simulation-b5-weights}
	\end{subfigure}
		\caption{Weights assigned to each PSM for each subset of datasets by the unsupervised (above) and outcome-guided (below) integration methods for PSMs.}
	\label{fig:additional-simulations-b-weights}
\end{figure}

We present here the results obtained for simulation setting B, with different numbers of irrelevant covariates. Figure \ref{fig:additional-simulations-b-ari} shows the adjusted Rand index and Figure \ref{fig:additional-simulations-b-weights} the weights assigned to each kernel.

\clearpage

\subsubsection{Using the true cluster labels as response variable}

We repeat the simulation study presented in the main paper, using the true cluster labels as the response variable, both for the outcome-guided integration (in all simulation settings) and to generate the PSMs with profile regression (in Setting D). Although the true cluster labels are not available in practice, this simulation study is used here to determine a putative upper bound on the performances of outcome-guided integration. 

\begin{figure}
	\centering 
	\includegraphics[width=.85\textwidth]{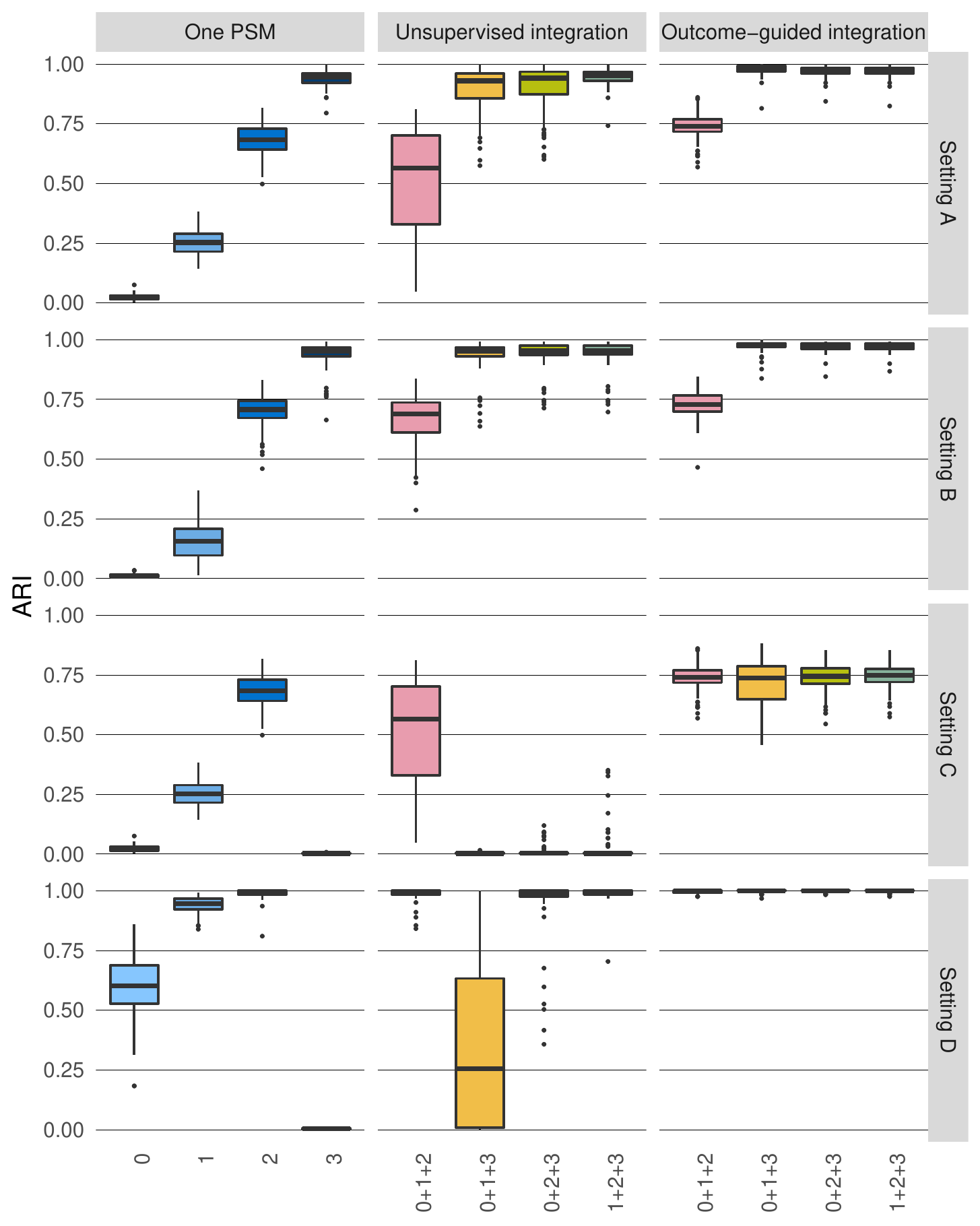}
	\caption{Simulation study where the response for each observation is given by its true cluster label. Adjusted Rand index obtained by summarising the PSMs one at a time using kernel $k$-means (left), combining different subsets of three PSMs in an unsupervised fashion using localised multiple kernel $k$-means (centre), and combining the same subsets making use of a response variable and multi-class SVMs to determine each PSM's weight and using kernel $k$-means for the final clustering (right).}
	\label{fig:simulation-ari-true-cl-labels}
\end{figure}

The ARI is reported in Figure \ref{fig:simulation-ari-true-cl-labels}. As expected, the outcome-guided integration has higher values of the ARI in all settings, compared to the case where the outcome is a binary variable. Moreover, in Setting D the ARI of each PSM taken individually is also higher here than in the other simulation study. 

\begin{figure}[h]
	\centering
	\includegraphics[width=.8\textwidth]{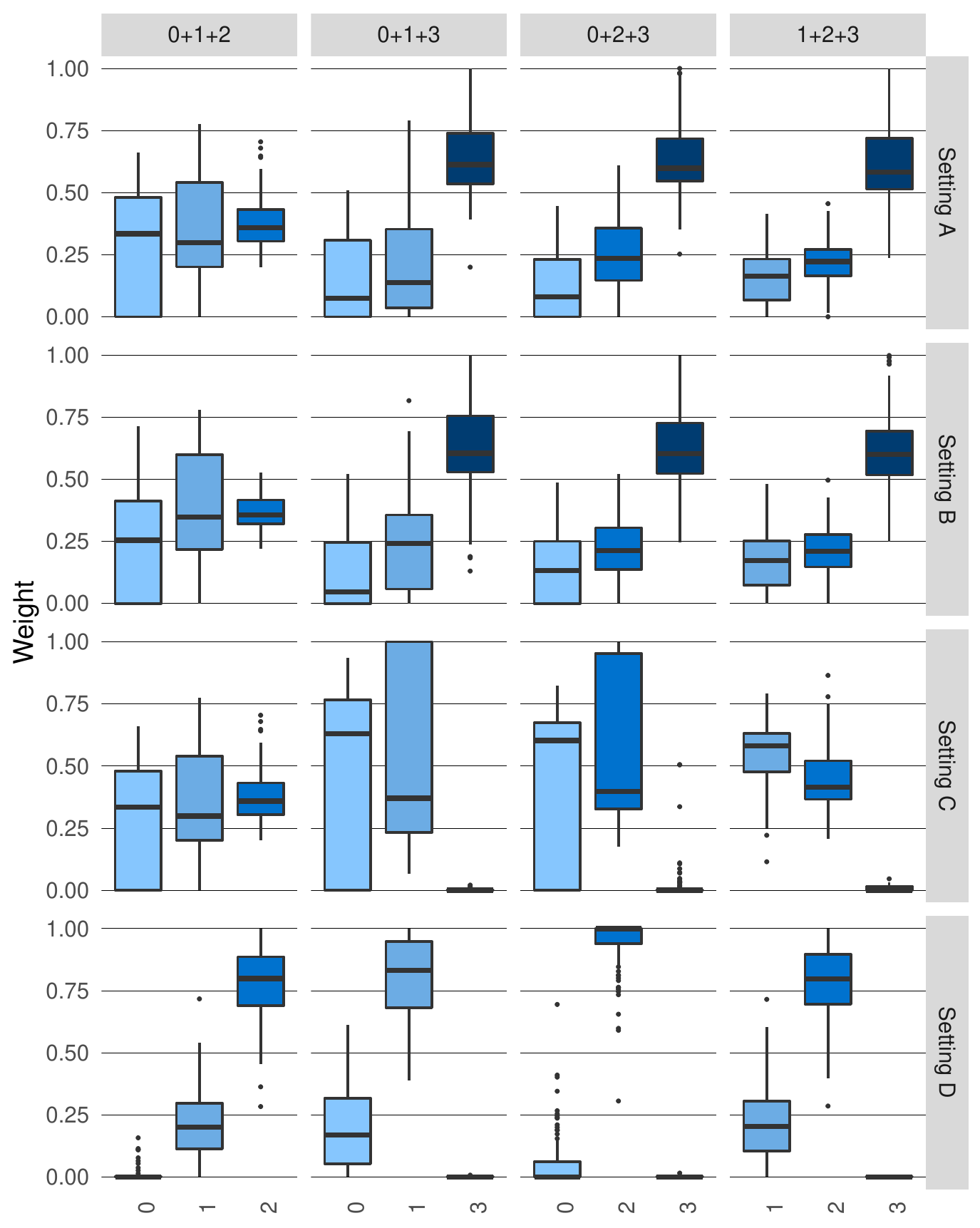}
		\caption{Weights assigned to each PSM for each subset of datasets by the unsupervised (above) and outcome-guided (below) integration methods for PSMs for the simulation study where the response for each observation is given by its true cluster label.}
	\label{fig:simulation-weights-outcomeguided-true-cl-labels}
\end{figure}

The weights assigned to each kernel matrix are shown in Figure \ref{fig:simulation-weights-outcomeguided-true-cl-labels}. These are of easier interpretation compared to those presented above (Figure \ref{fig:simulation-weights-outcomeguided}): on average, kernels originated from datasets with higher cluster separability have higher weights.

\clearpage
\section{Integrative clustering: biological applications}
\subsection{Multiplatform analysis of ten cancer types}

\subsubsection{Variable selection}

In Table \ref{table:selected-variables} are reported the number of variables measured in each layer and the number of selected variables using separate elastic-net (EN) on each layer as in \citet{cabassi2020penalised} and \citet{seyres2020transcriptional}.

\begin{table}[H]
\centering
\begin{tabular}{l c c c c}
Dataset & Full dataset & $\alpha=0.1$ & $\alpha=0.5$ & $\alpha=1$ \\
\hline
Protein expression & 131 & 131 & 131 & 124 \\
mRNA expression & 6000 & 1893 & 568 & 258 \\
Methylation & 2043 & 1439 &  623 & 322 \\
DNA copy number & 84 & 84 & 84 & 84 \\
microRNA expression & 51 & 51 & 51 & 50 \\
\hline\\
\end{tabular}
\caption{Number of selected variables in each dataset for different values of the EN parameter $\alpha$.}
\label{table:selected-variables}
\end{table} 

The full mRNA dataset is too large to be used as input to MDI and for this reason it is only used for the integration of the reduced datasets obtained via variable selection with values of $\alpha$ of 0.5 and 1.

\subsubsection{MCMC convergence assessment}

We run five MCMC chains for 50.000 iterations, with a burn-in period of 25.000 iterations and thinning of 5. For each set of five chains, we check the Vats-Knudson $\hat{R}$ \citep{vats2018revisiting} with parameters  $\epsilon = 0.1$ and $\alpha = 0.1$ to assess the convergence of the mass parameter. The PSMs obtained for the five chains are summarised into one by taking the average. 

\begin{figure}[H]
	\centering
	\includegraphics[width=.36\linewidth]{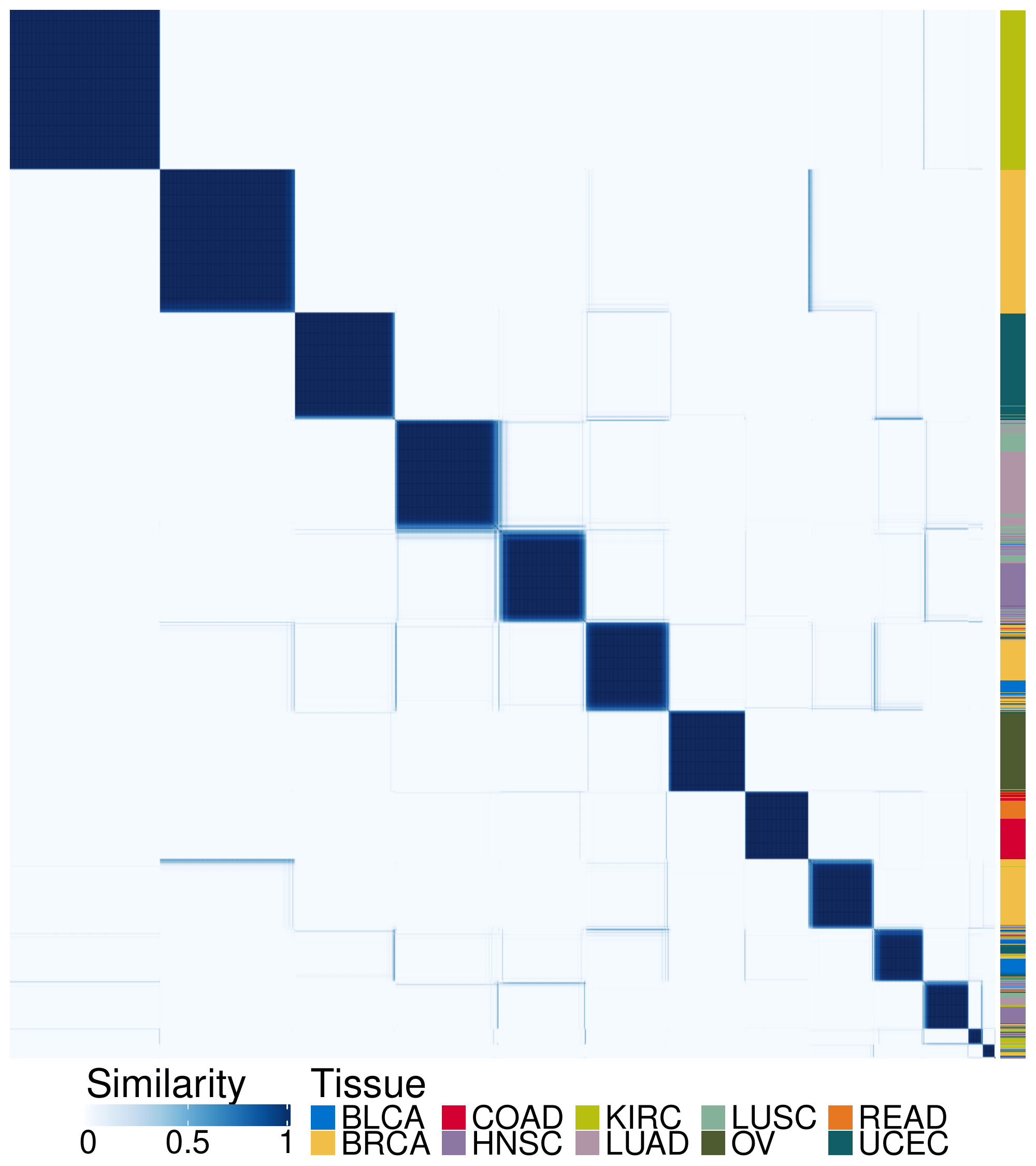}
	\includegraphics[width=.36\linewidth]{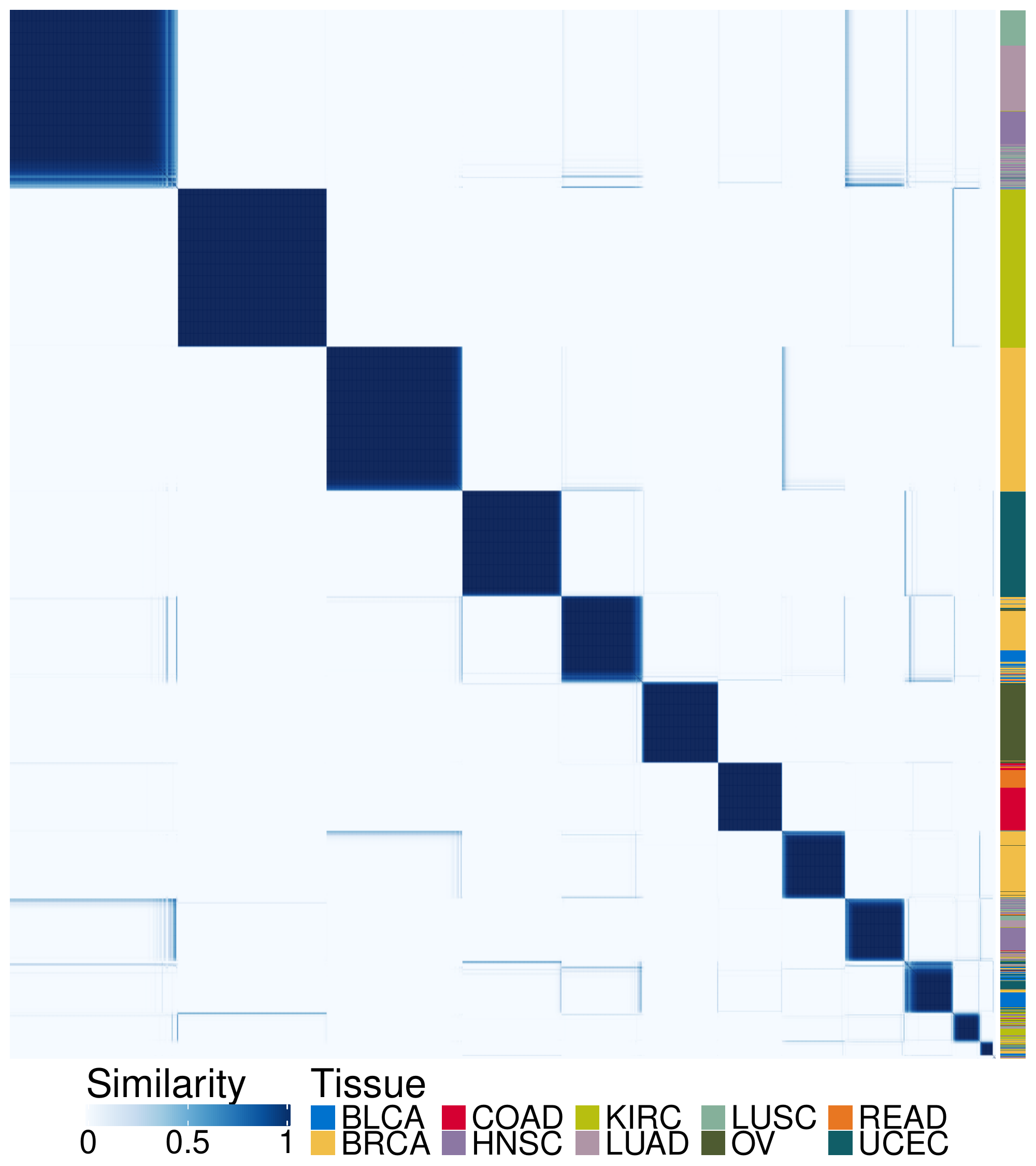}
	\includegraphics[width=.36\linewidth]{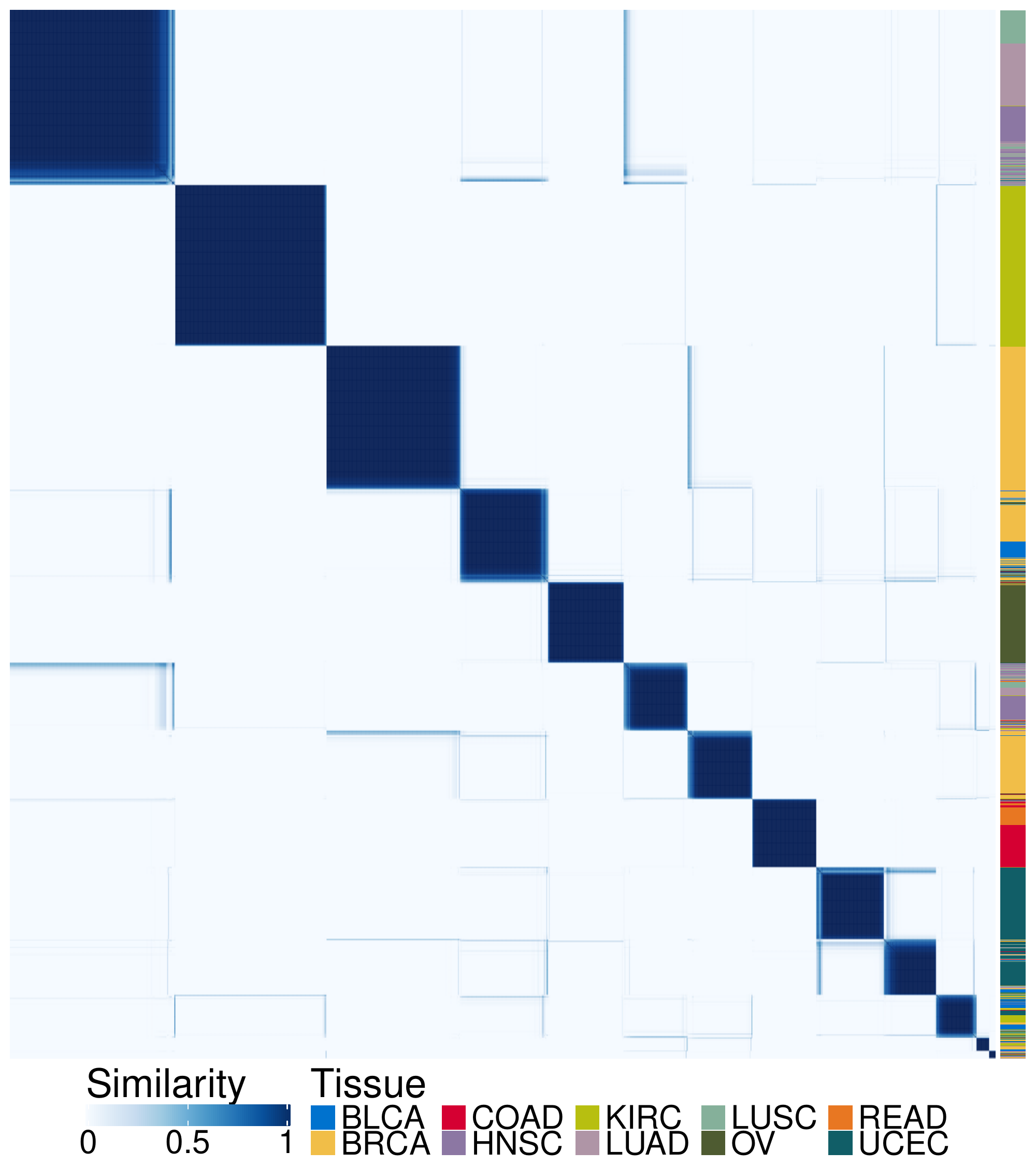}
	\includegraphics[width=.36\linewidth]{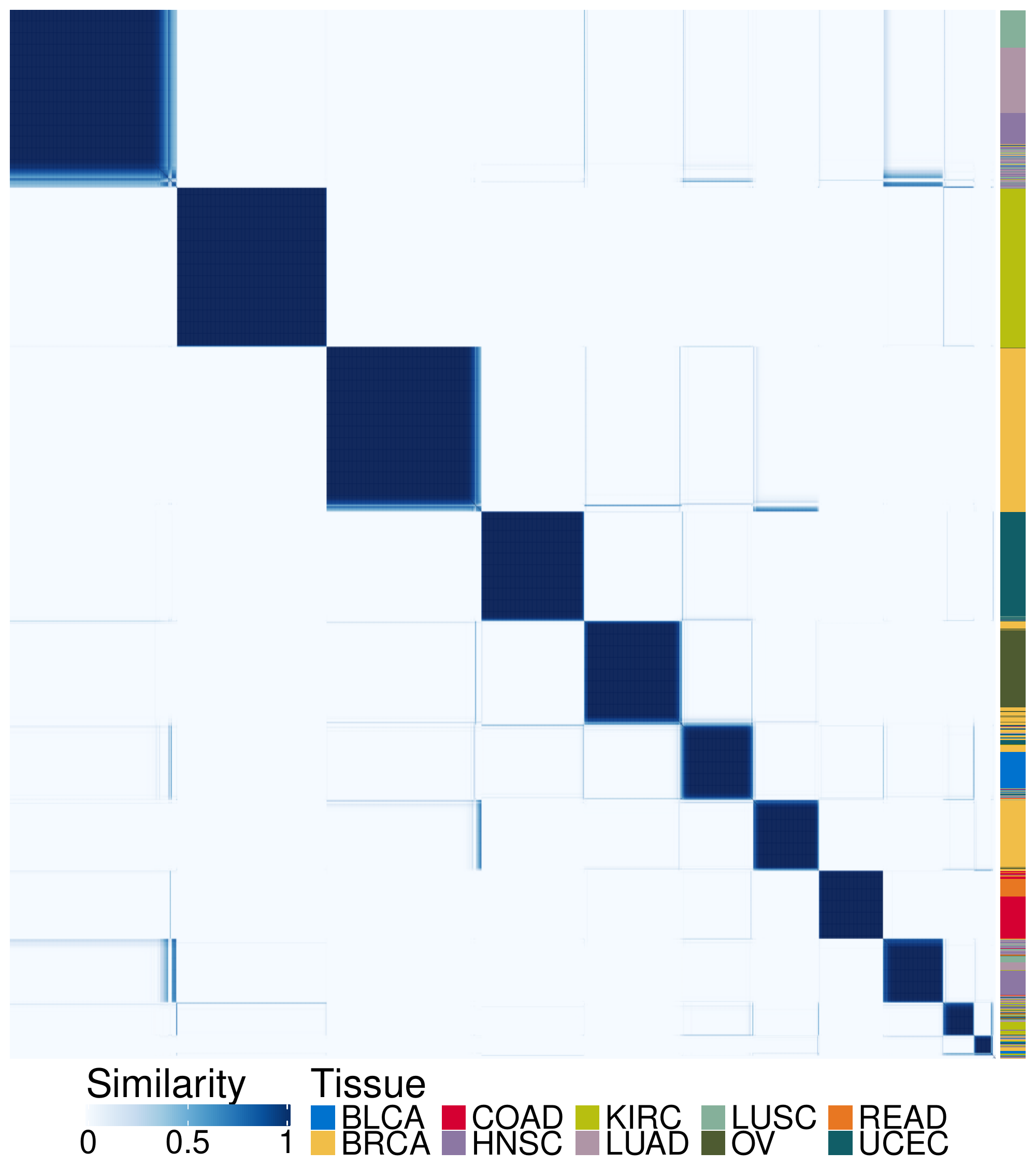}
	\includegraphics[width=.36\linewidth]{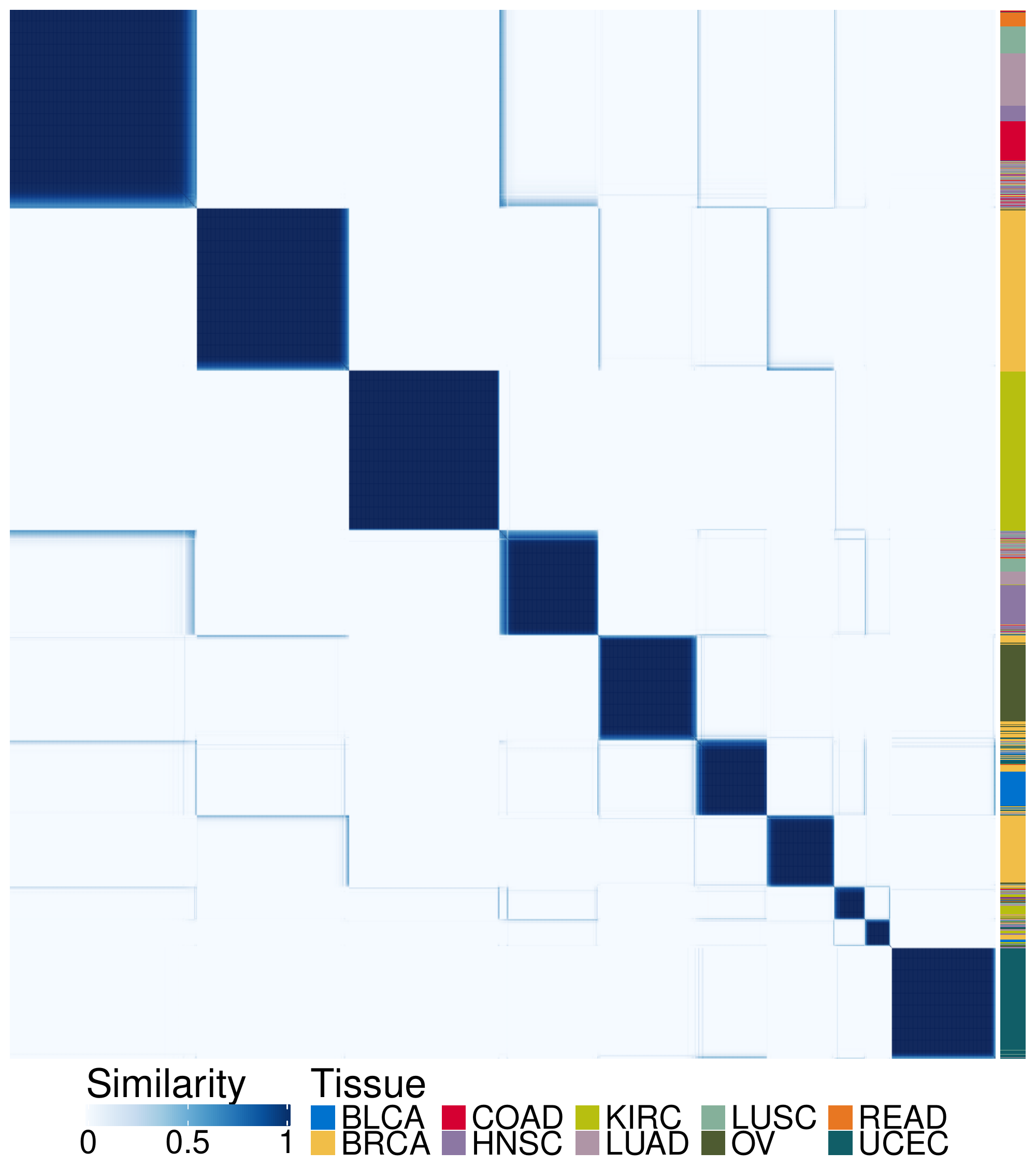}
	\includegraphics[width=.36\linewidth]{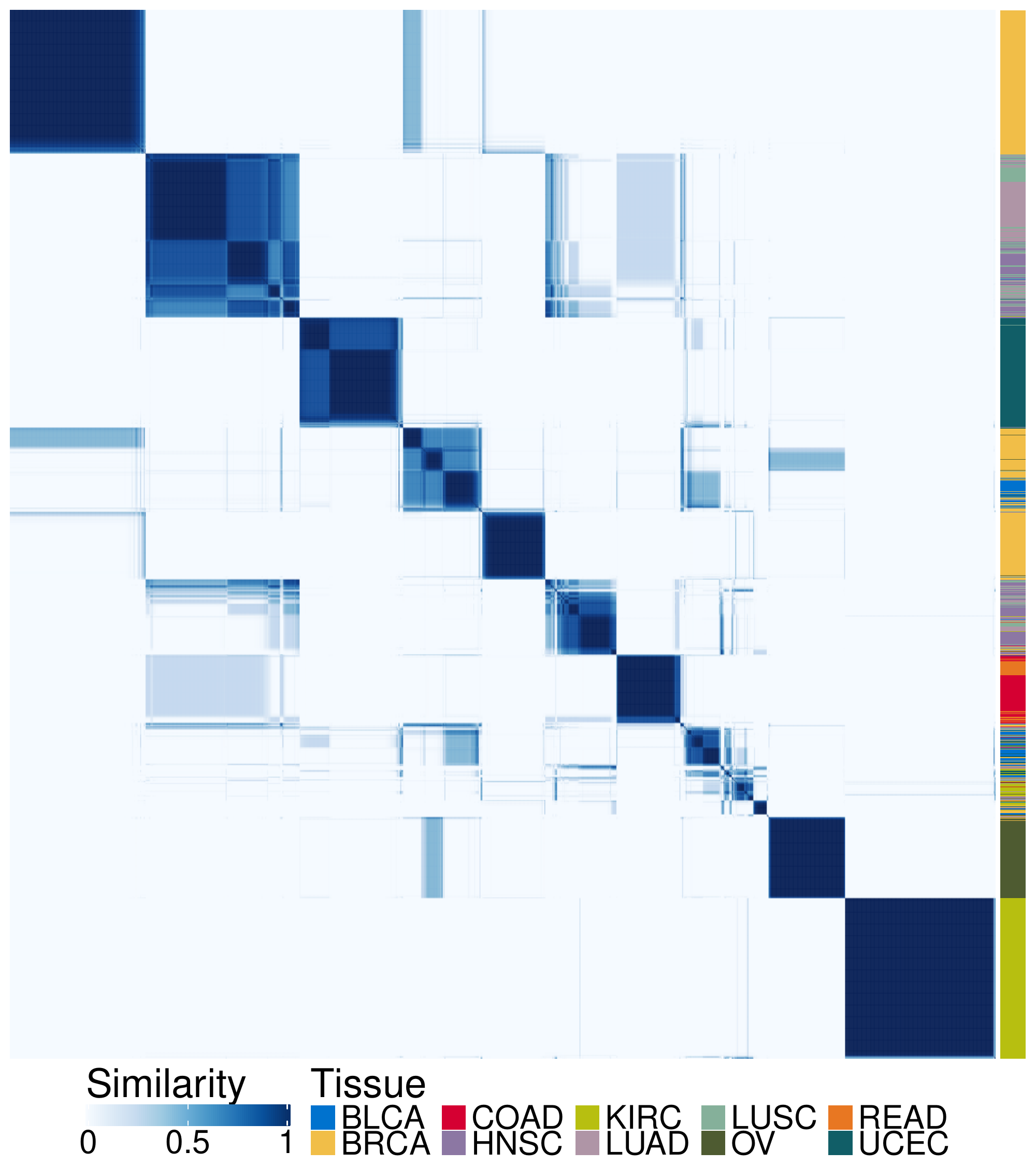}
	\caption{PSMs of the protein expression data. $\lambda=0$, $\alpha=0.1, 0.5$.}
\end{figure}

\begin{table}[H]
\centering
\begin{tabular}{l c c c c}
& \textbf{Chain 2} & \textbf{Chain 3} & \textbf{Chain 4} & \textbf{Chain 5} \\
\hline
\textbf{Chain 1} & 0.87 & 0.80 & 0.76 & 0.69 \\
\textbf{Chain 2} &1 & 0.91 & 0.87 & 0.73 \\
\textbf{Chain 3} && 1 & 0.82 & 0.68 \\
\textbf{Chain 4} && & 1 & 0.83 \\
\hline\\
\end{tabular}
\caption{ARI between the clusterings found on the PSMs of different chains with the number of clusters that maximises the silhouette.}
\end{table} 

\begin{figure}[H]
\centering
\includegraphics[width=.45\linewidth]{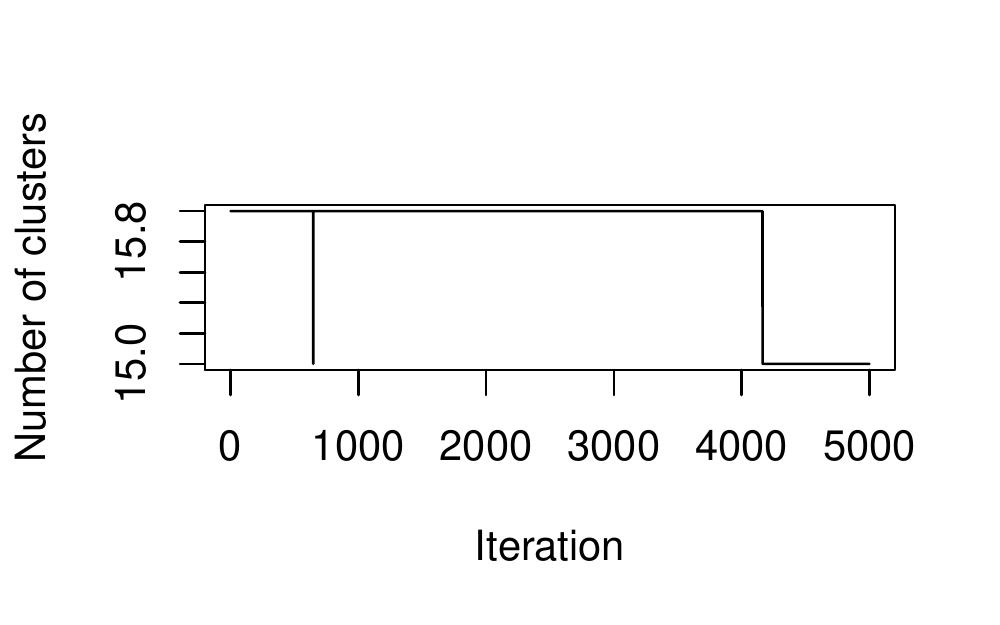}
\includegraphics[width=.45\linewidth]{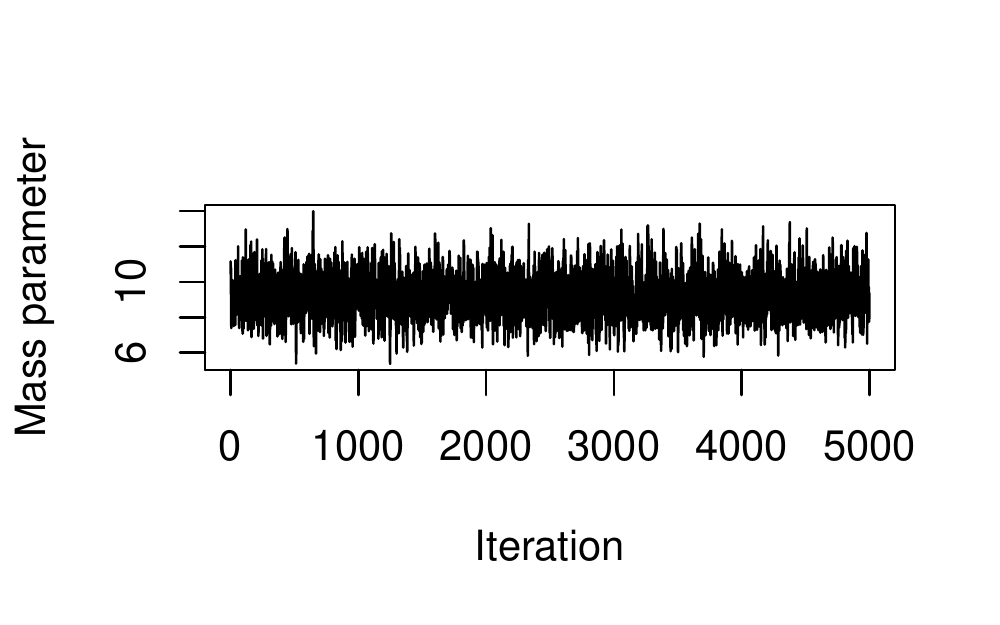}
\vspace{-1.6cm}

\includegraphics[width=.45\linewidth]{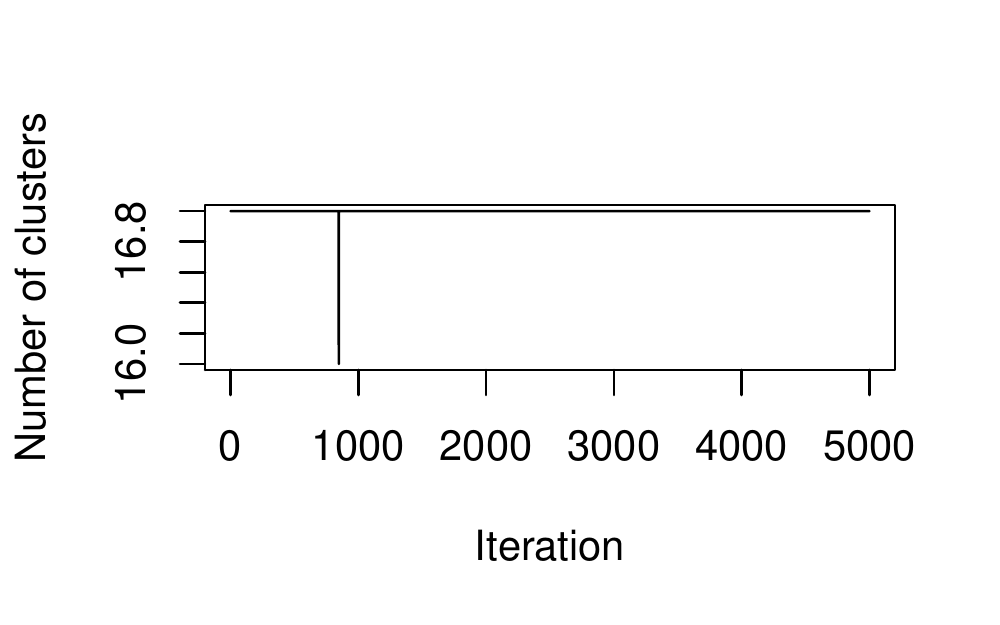}
\includegraphics[width=.45\linewidth]{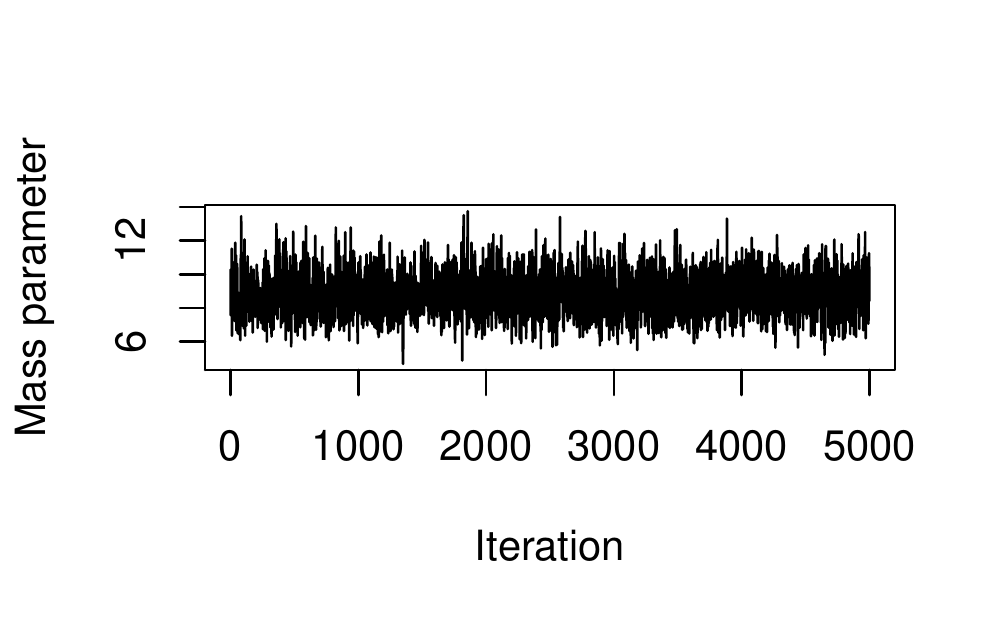}
\vspace{-1.6cm}

\includegraphics[width=.45\linewidth]{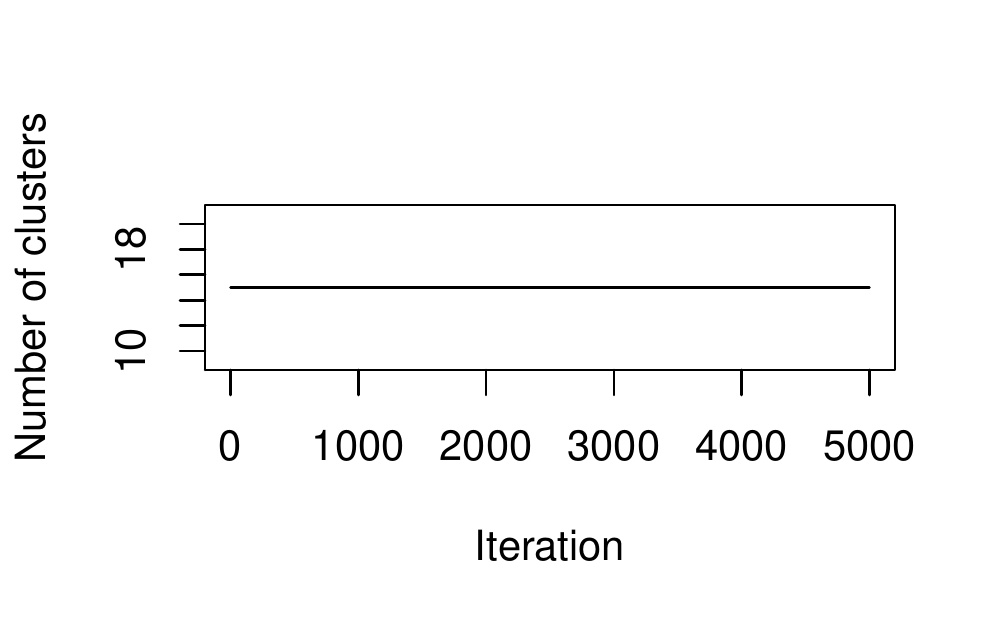}
\includegraphics[width=.45\linewidth]{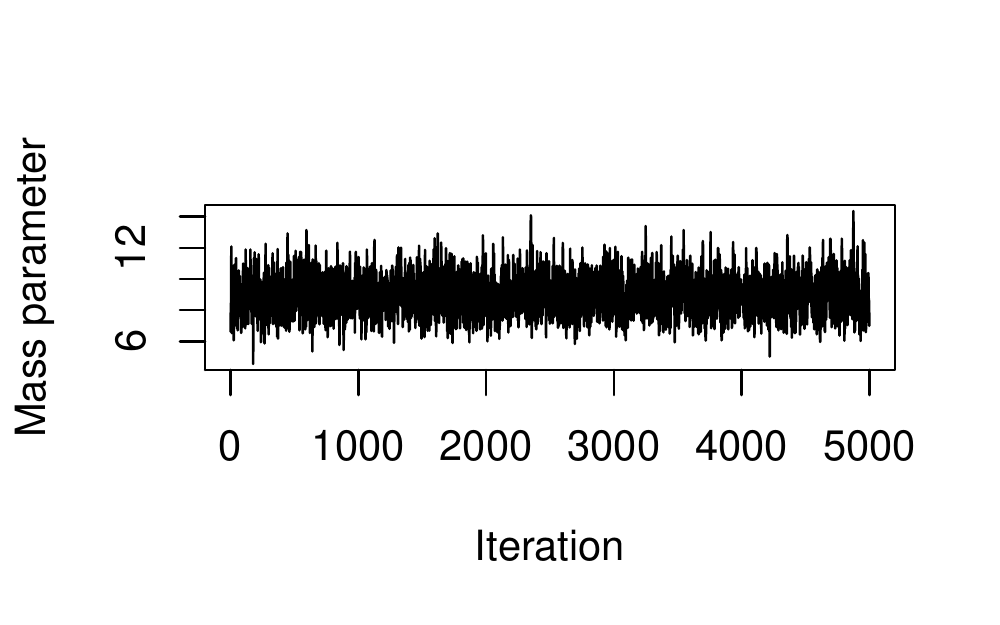}
\vspace{-1.6cm}

\includegraphics[width=.45\linewidth]{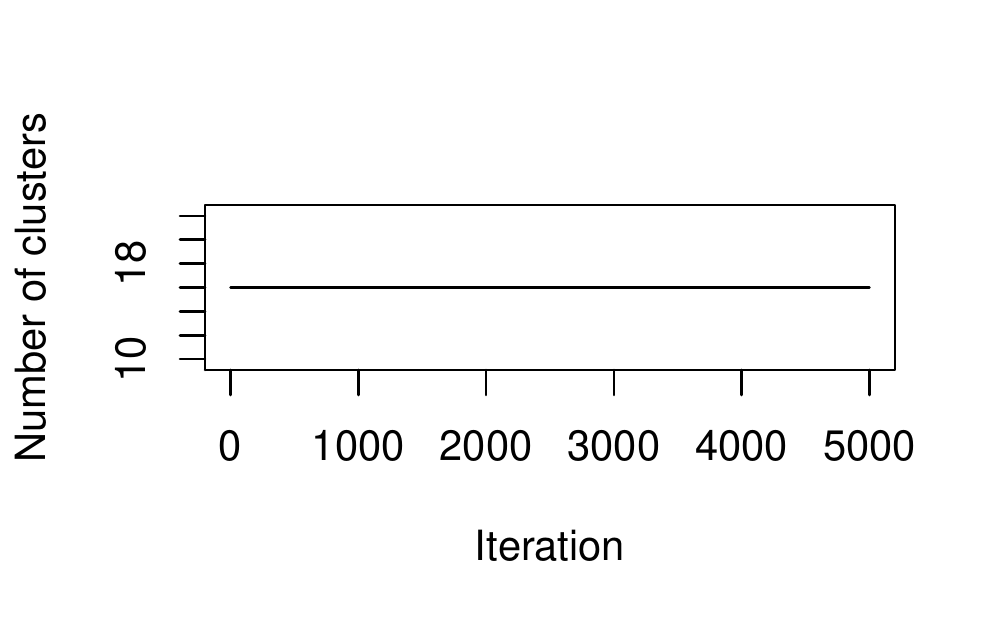}
\includegraphics[width=.45\linewidth]{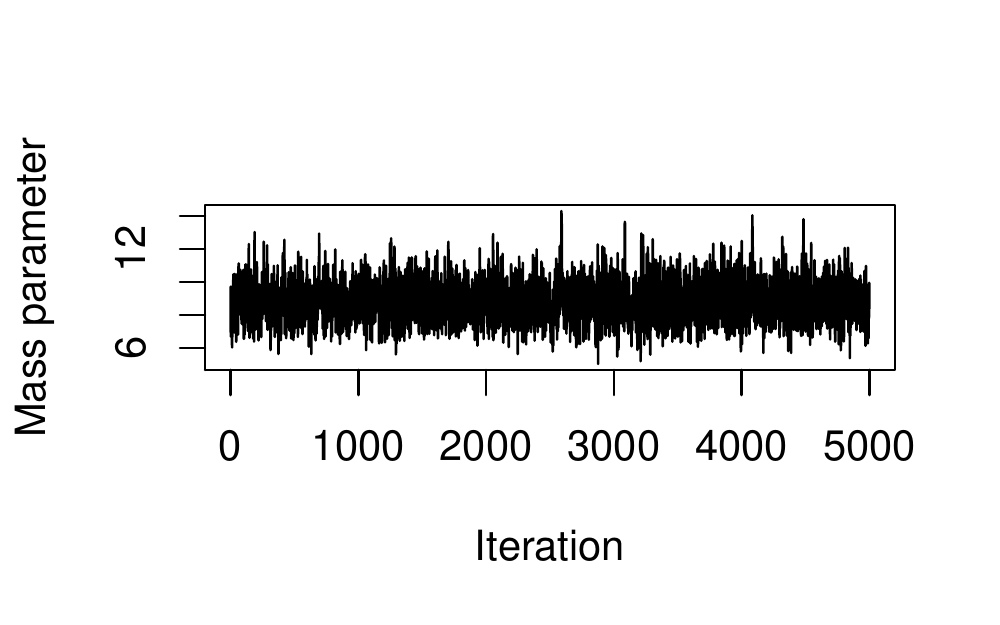}
\vspace{-1.6cm}

\includegraphics[width=.45\linewidth]{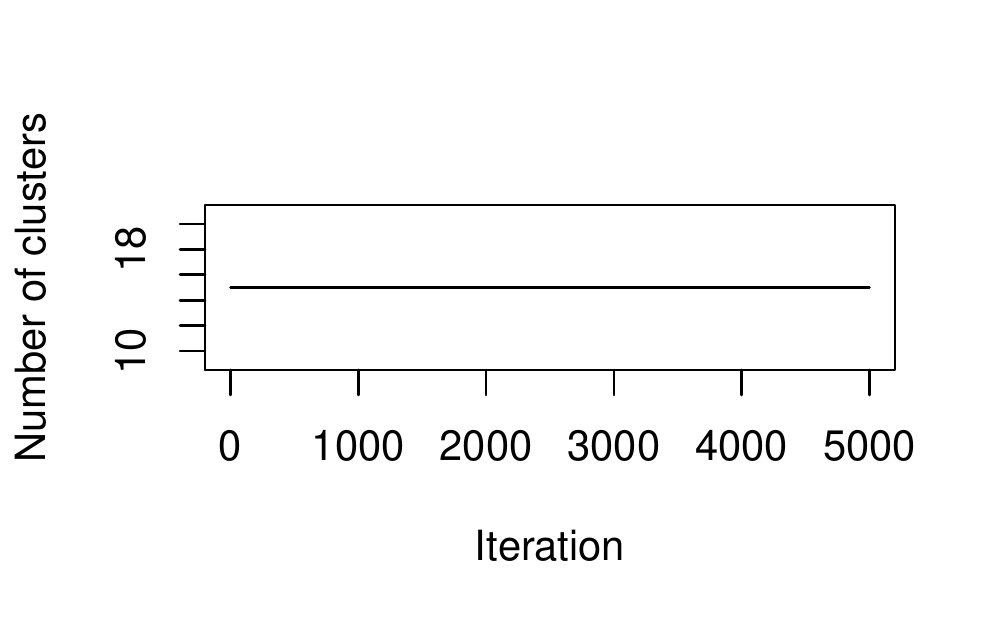}
\includegraphics[width=.45\linewidth]{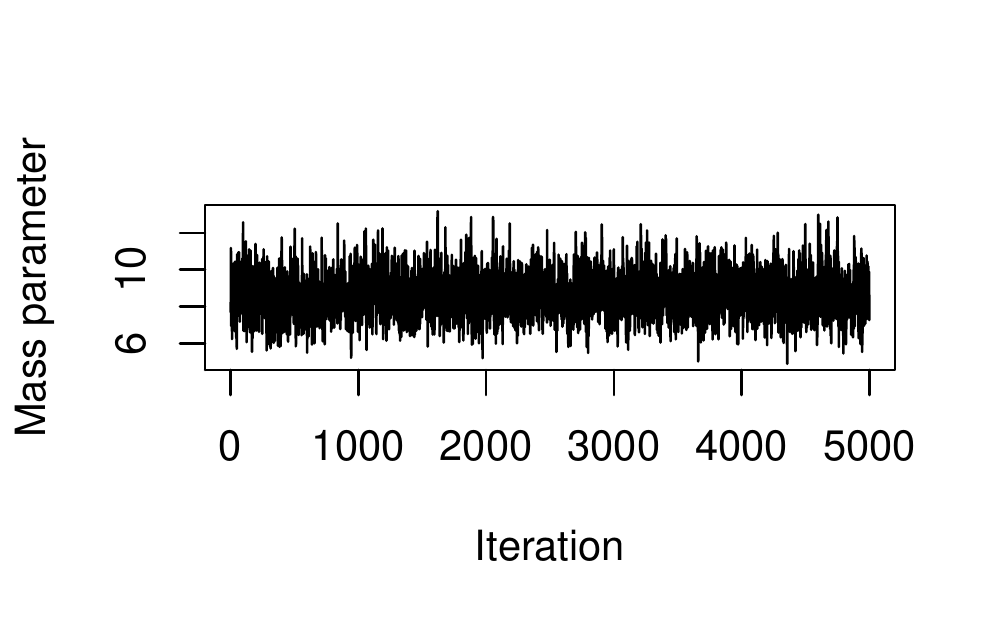}
\caption{MCMC convergence assessment, protein expression data. $\lambda=0$, $\alpha=0.1, 0.5$.}
\end{figure}

\clearpage

\begin{figure}[H]
	\centering
	\includegraphics[width=.36\linewidth]{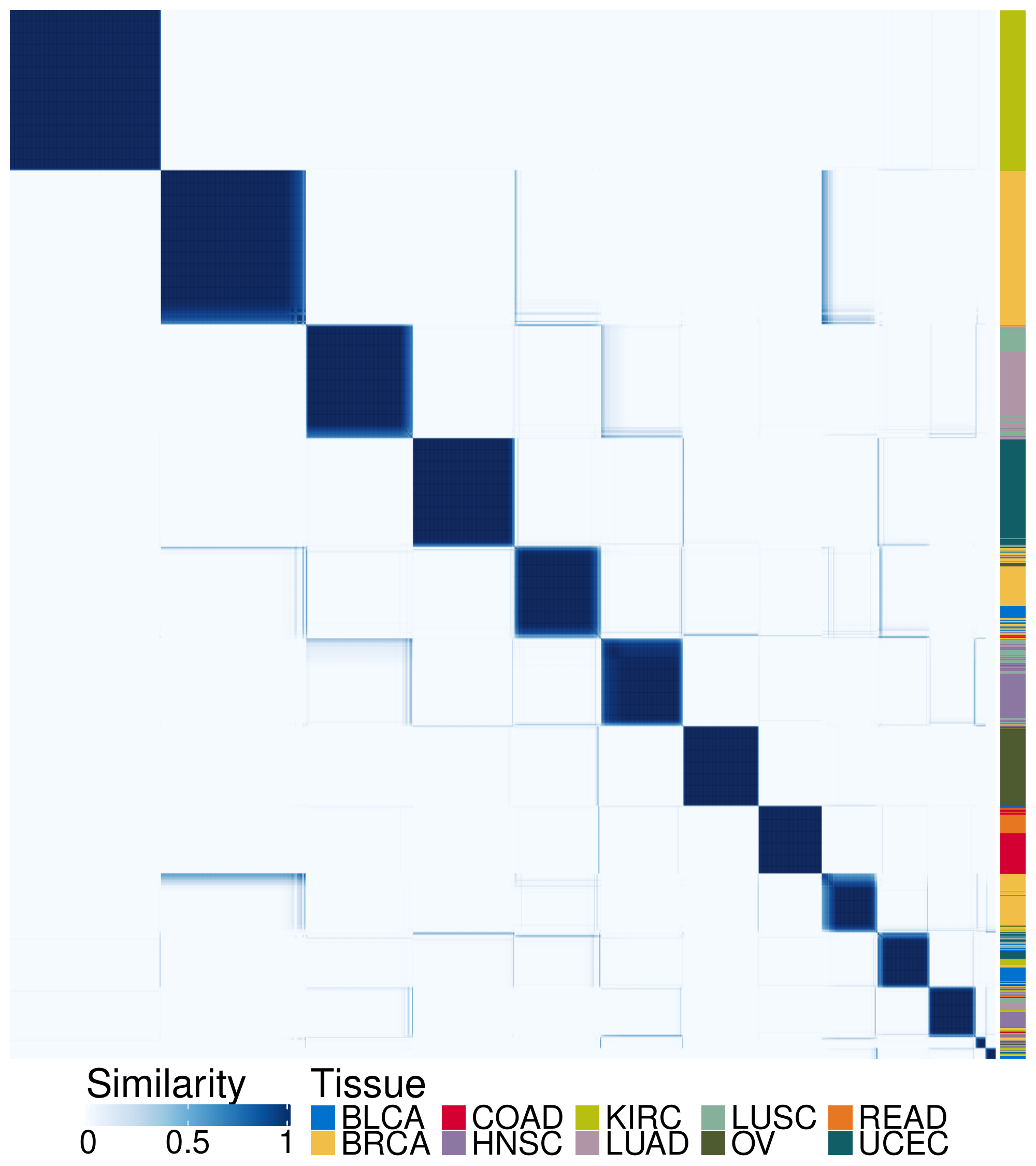}
	\includegraphics[width=.36\linewidth]{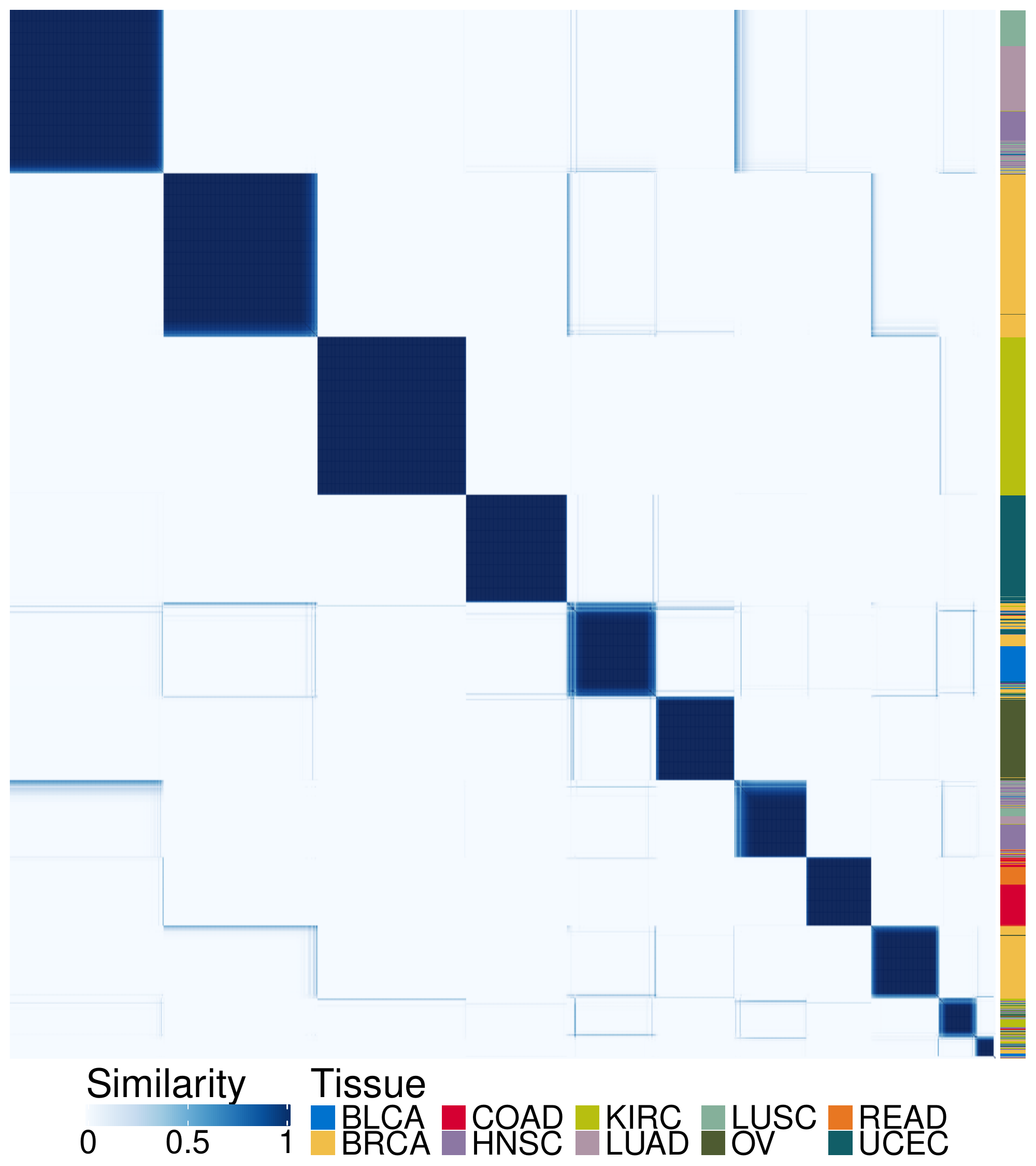}
	\includegraphics[width=.36\linewidth]{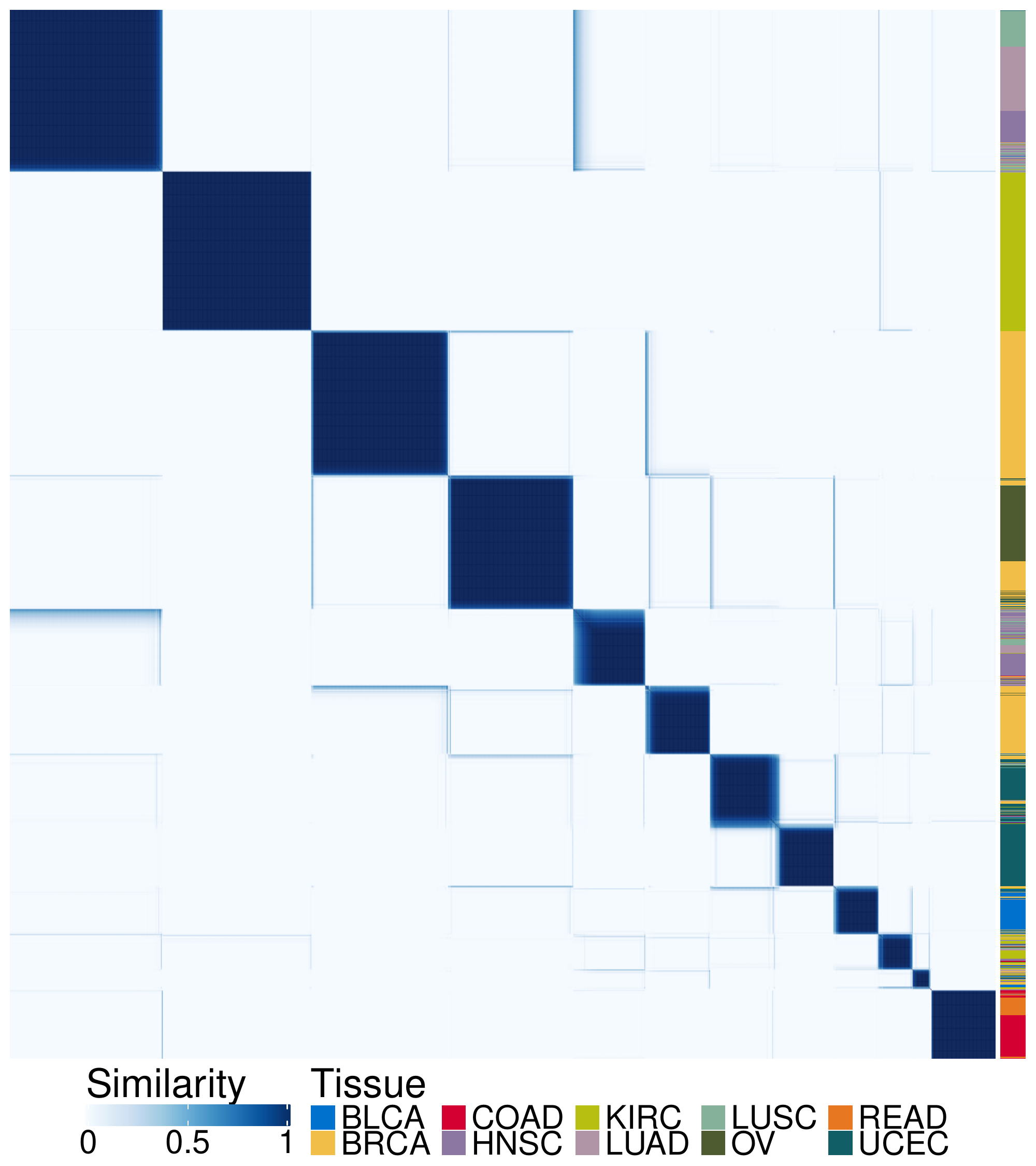}
	\includegraphics[width=.36\linewidth]{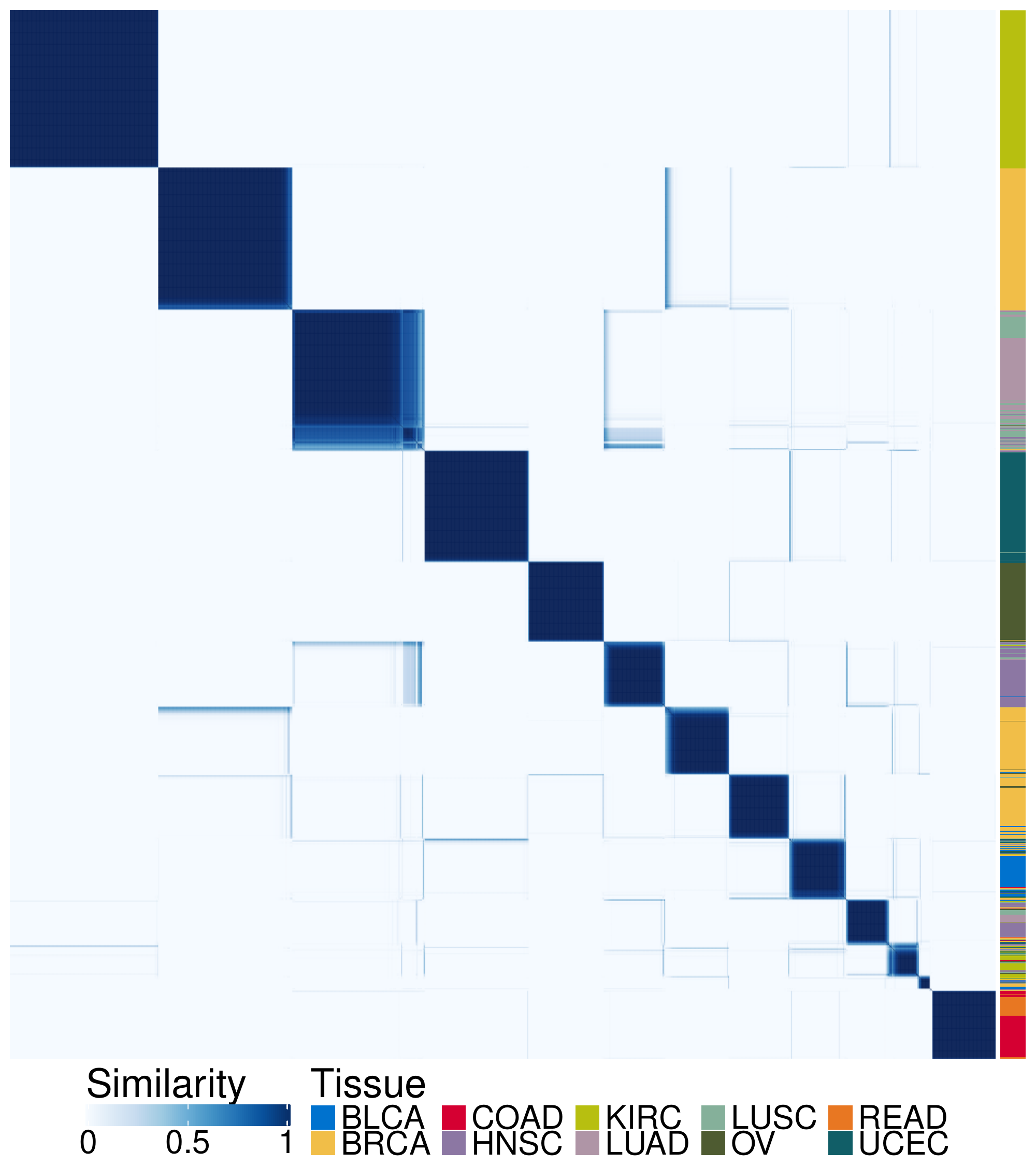}
	\includegraphics[width=.36\linewidth]{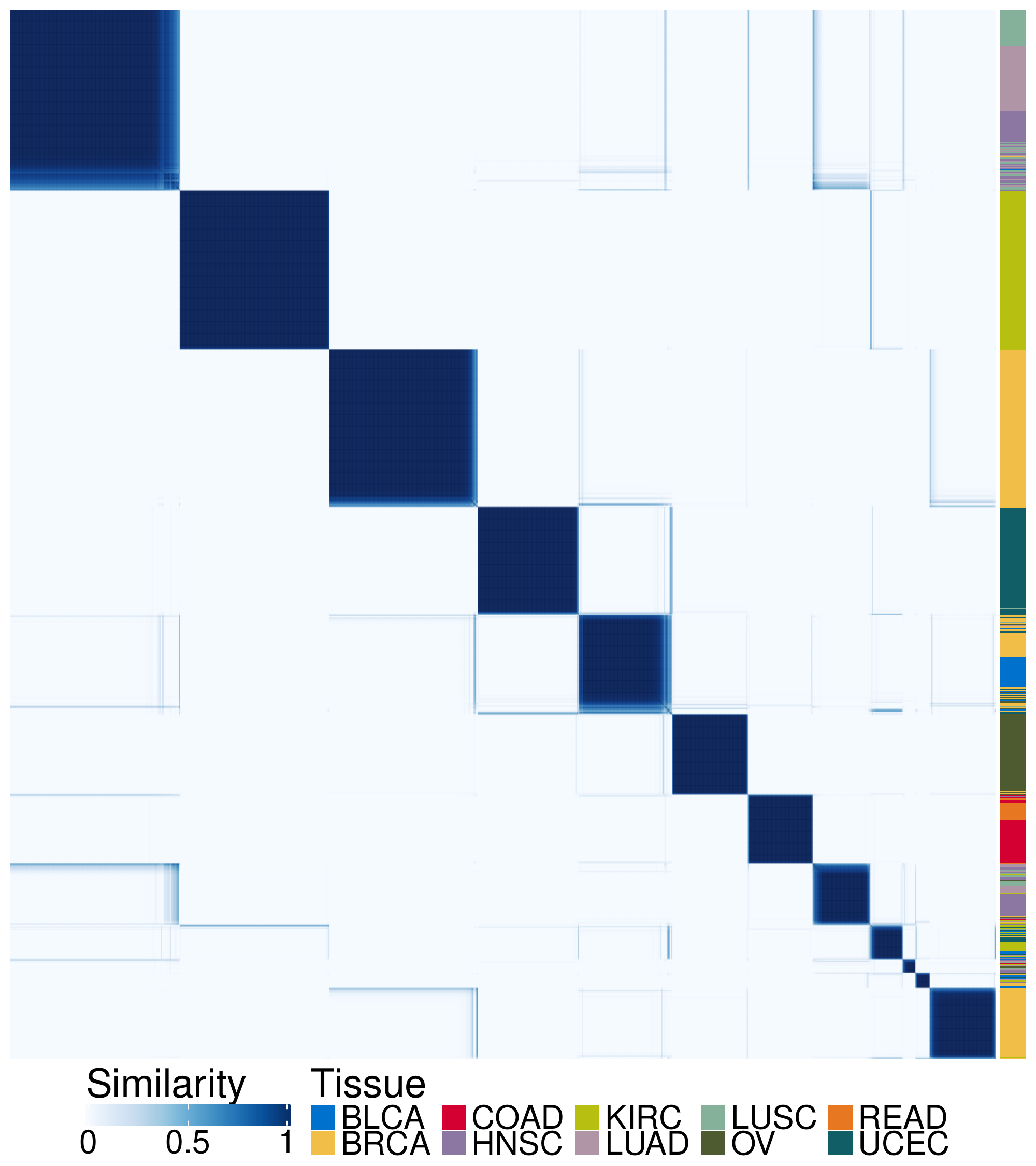}
	\includegraphics[width=.36\linewidth]{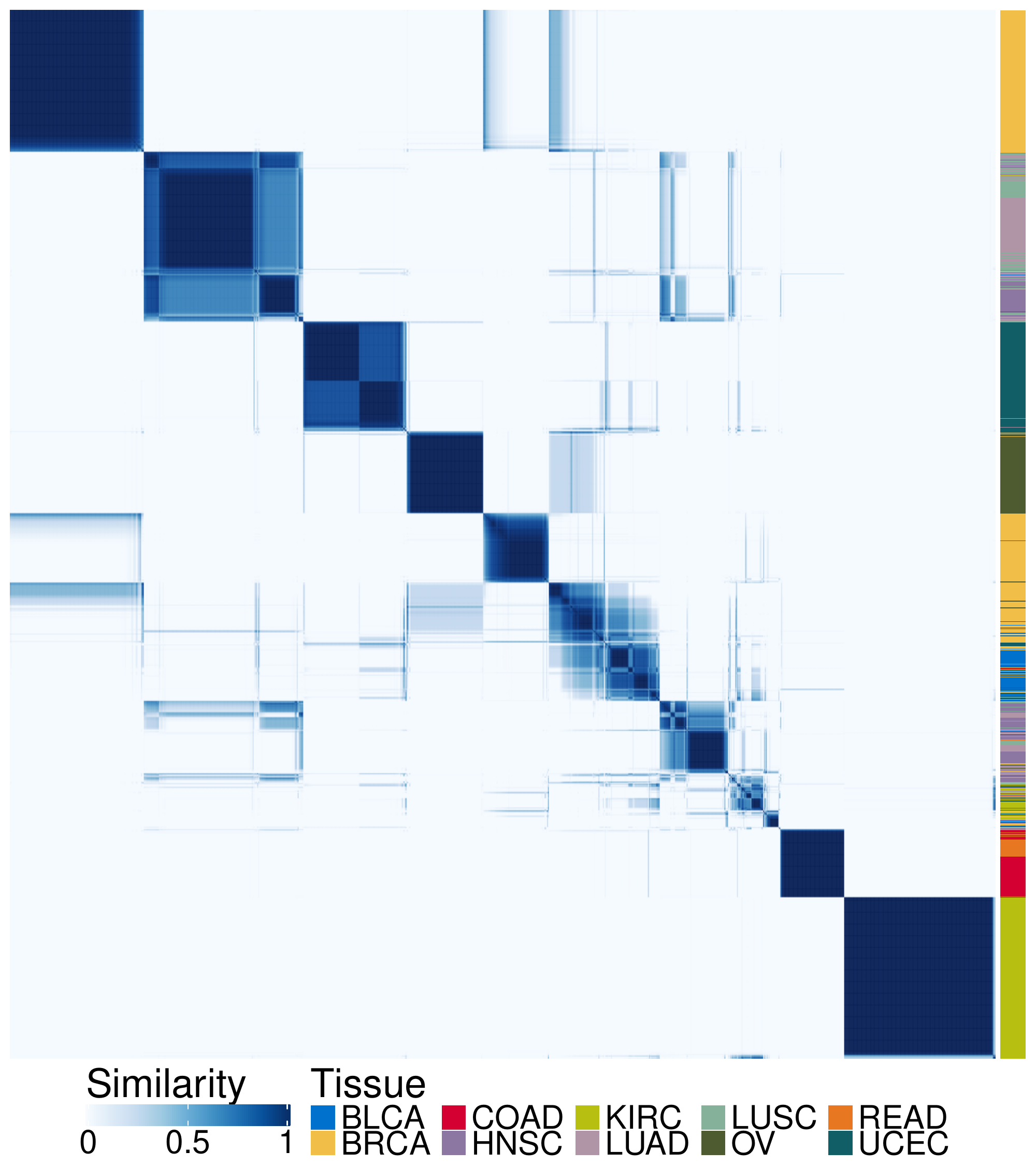}
	\caption{PSMs of the protein expression data. $\alpha=1$.}
\end{figure}

\begin{table}[H]
\centering
\begin{tabular}{l c c c c}
& \textbf{Chain 2} & \textbf{Chain 3} & \textbf{Chain 4} & \textbf{Chain 5} \\
\hline
\textbf{Chain 1} & 0.75 & 0.72 & 0.89 & 0.79 \\
\textbf{Chain 2} &1 & 0.82 & 0.80 & 0.92 \\
\textbf{Chain 3} && 1 &  0.76 & 0.81 \\
\textbf{Chain 4} && & 1 & 0.80 \\
\hline\\
\end{tabular}
\caption{ARI between the clusterings found on the PSMs of different chains with the number of clusters that maximises the silhouette.}
\end{table} 

\begin{figure}[H]
\centering
\includegraphics[width=.45\linewidth]{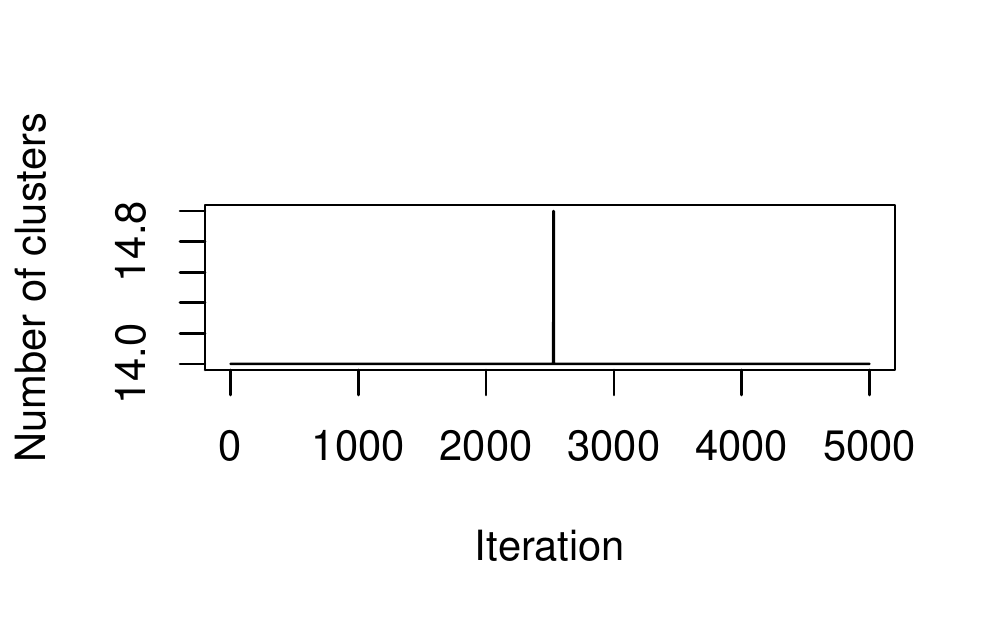}
\includegraphics[width=.45\linewidth]{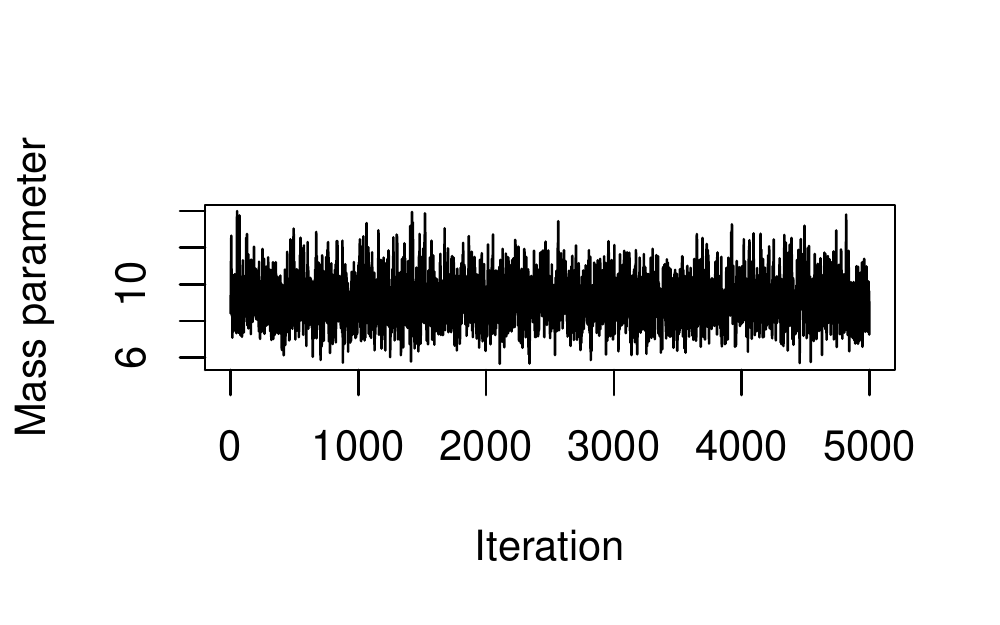}
\vspace{-1.6cm}

\includegraphics[width=.45\linewidth]{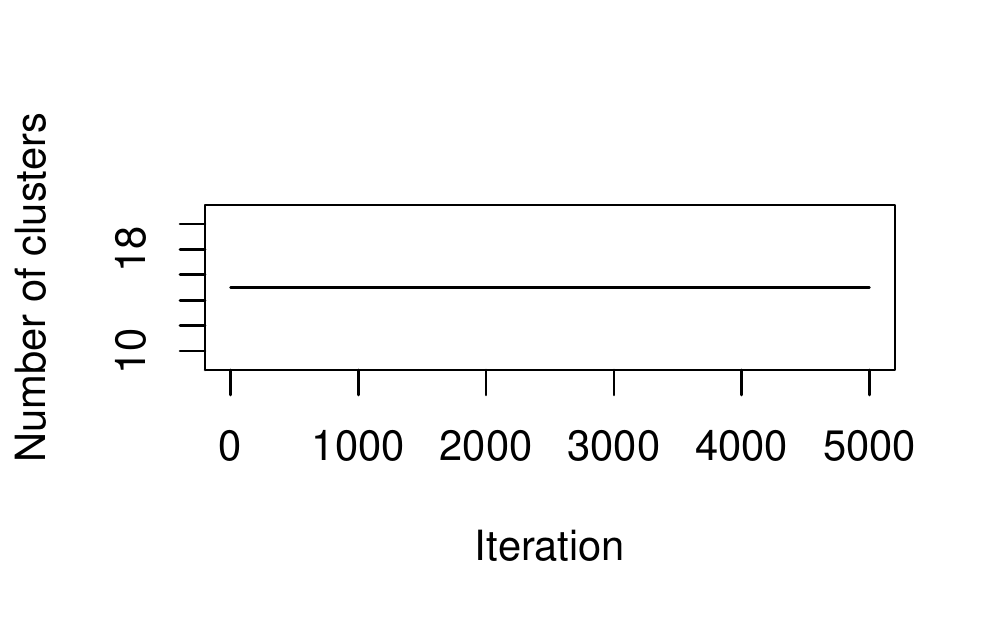}
\includegraphics[width=.45\linewidth]{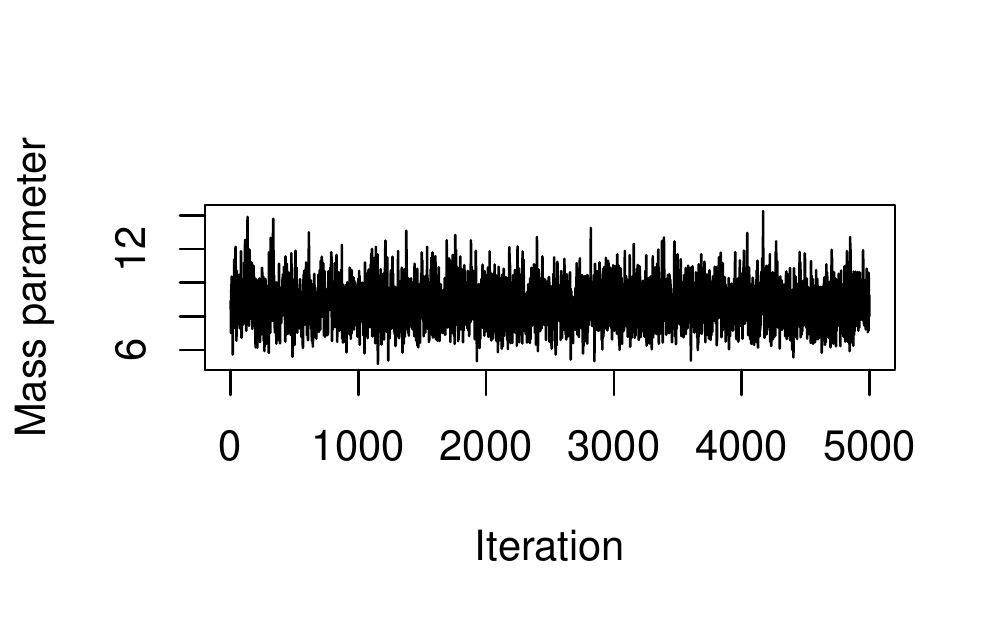}
\vspace{-1.6cm}

\includegraphics[width=.45\linewidth]{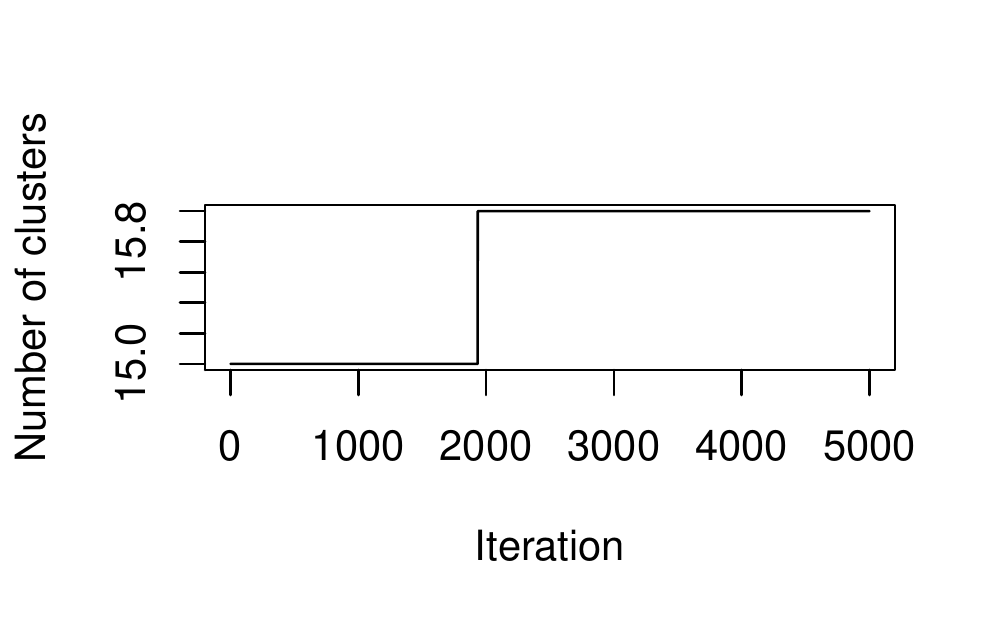}
\includegraphics[width=.45\linewidth]{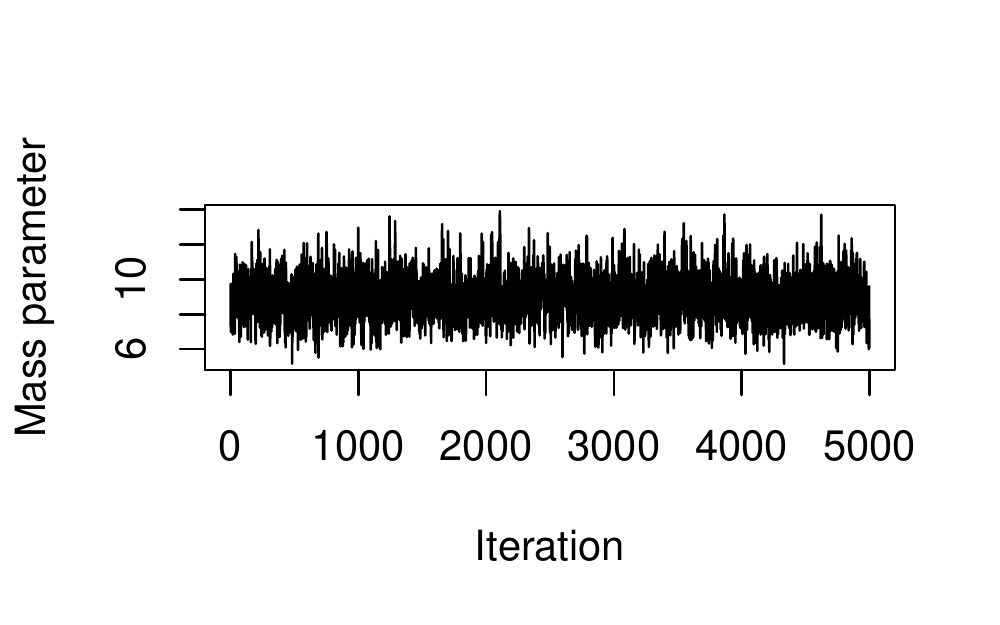}
\vspace{-1.6cm}

\includegraphics[width=.45\linewidth]{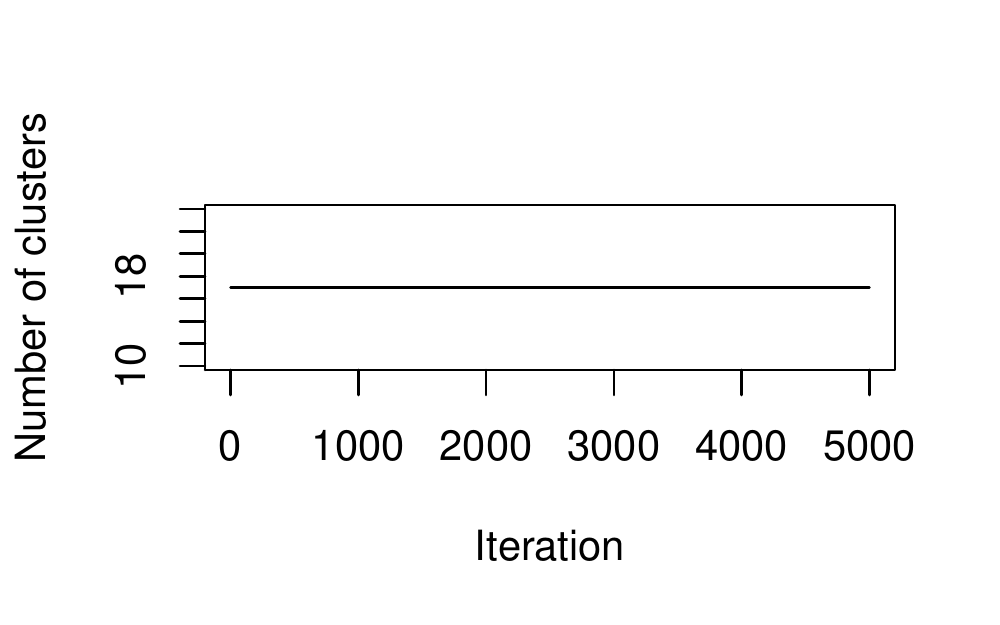}
\includegraphics[width=.45\linewidth]{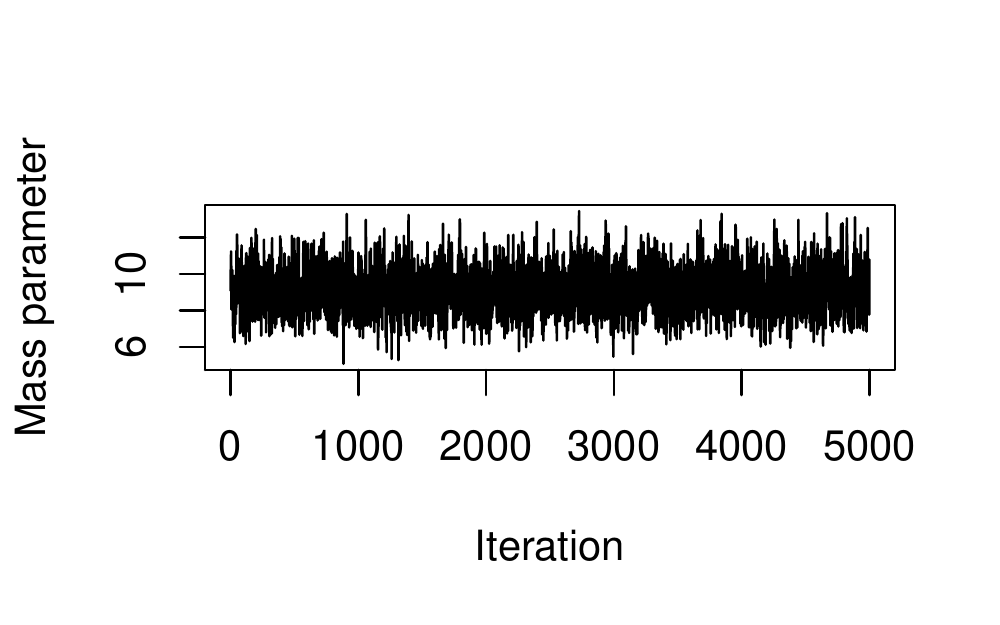}
\vspace{-1.6cm}

\includegraphics[width=.45\linewidth]{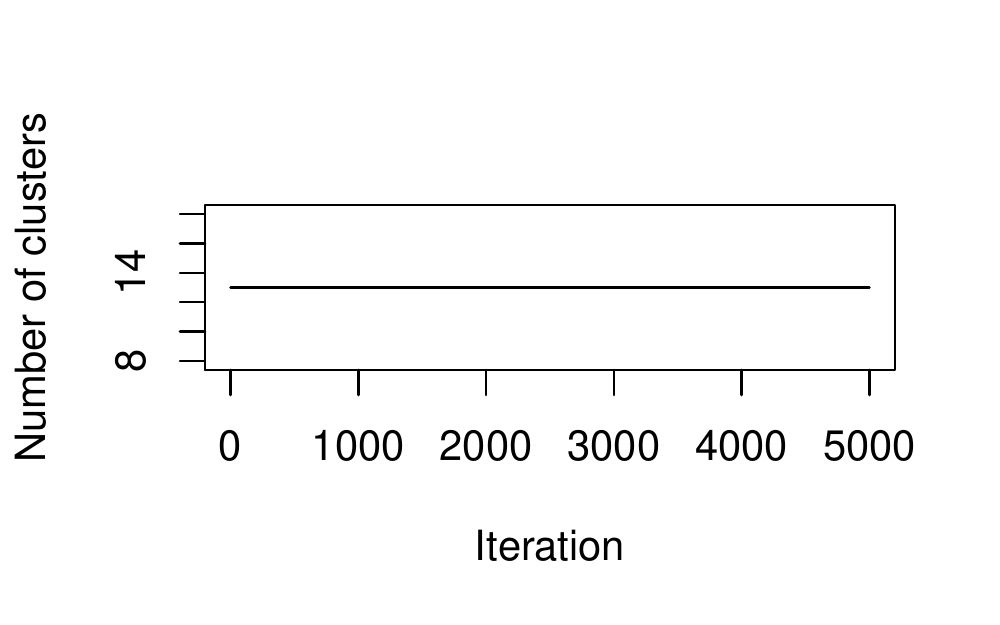}
\includegraphics[width=.45\linewidth]{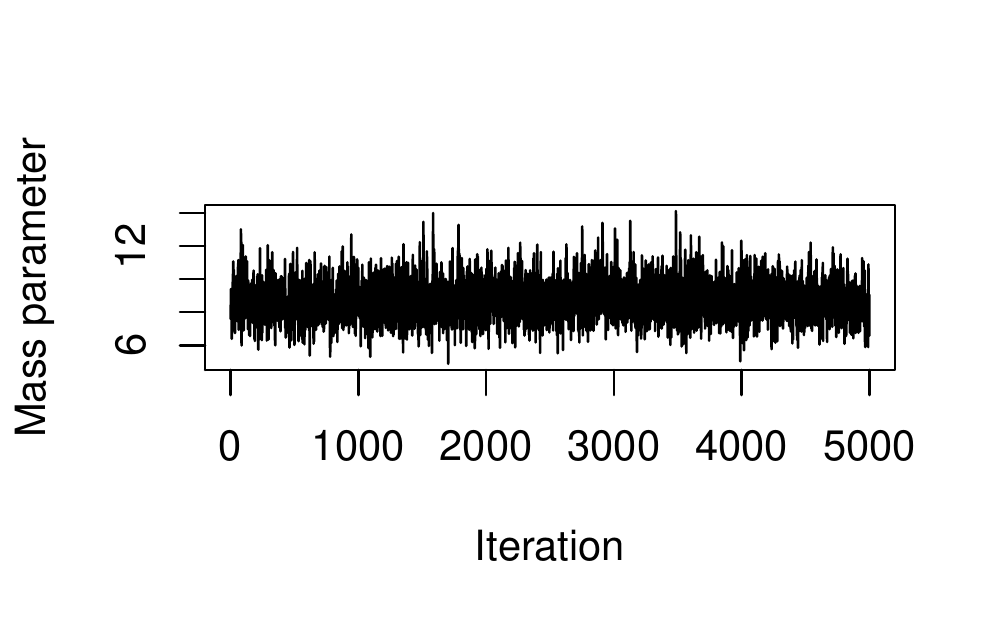}
\caption{MCMC convergence assessment, protein expression data. $\alpha=1$.}
\end{figure}

\clearpage

\begin{figure}[H]
	\centering
	\includegraphics[width=.36\linewidth]{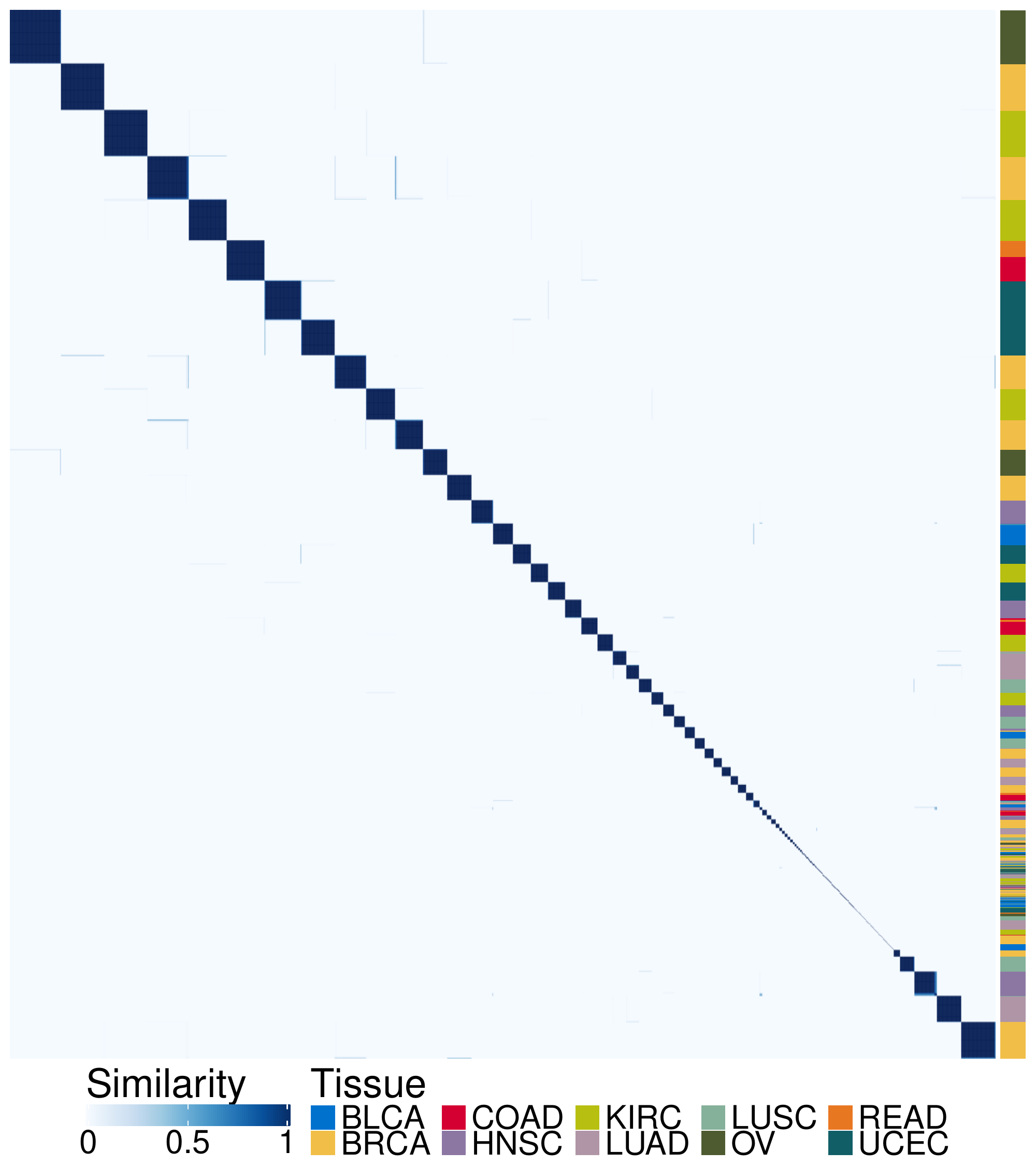}
	\includegraphics[width=.36\linewidth]{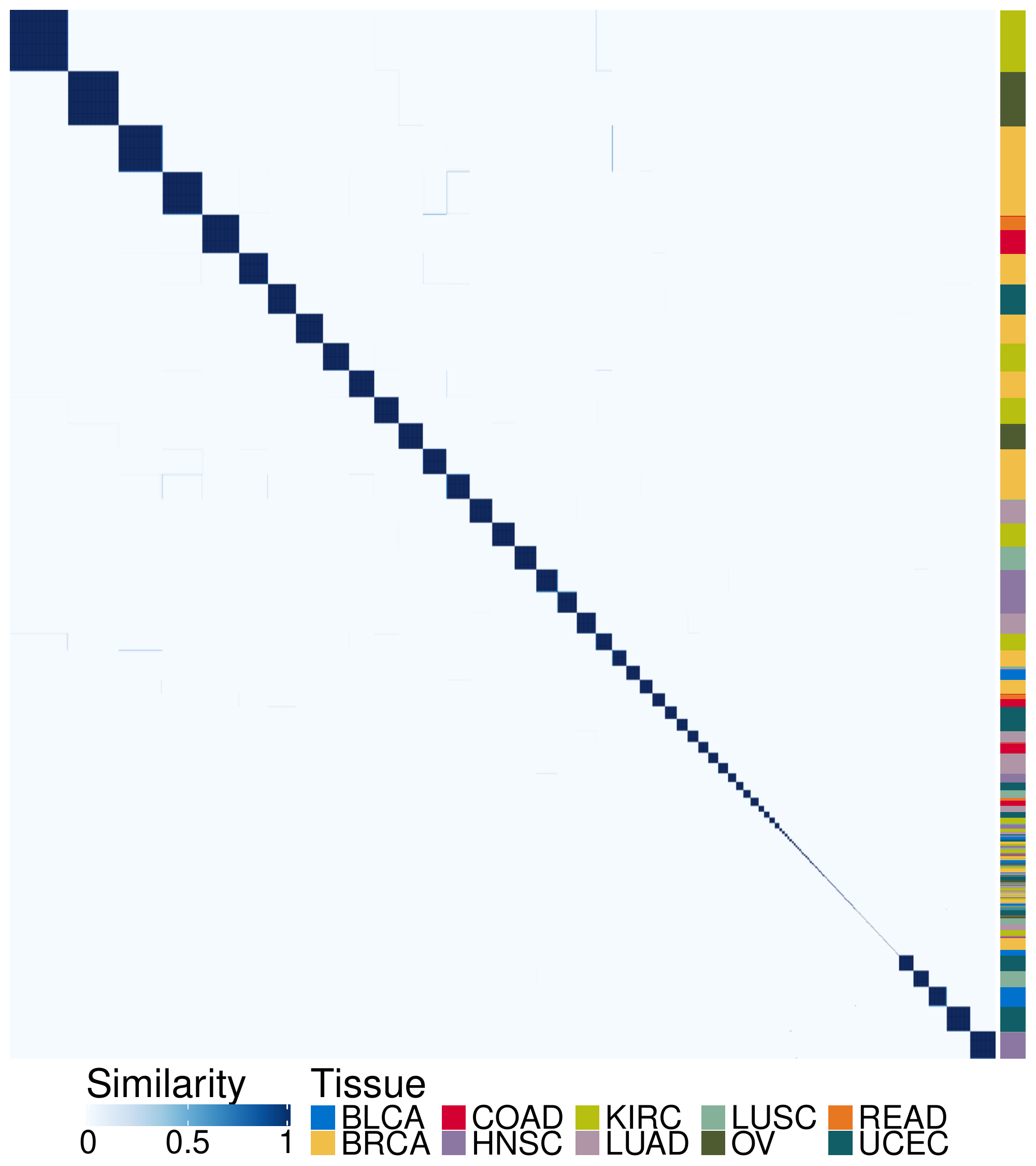}
	\includegraphics[width=.36\linewidth]{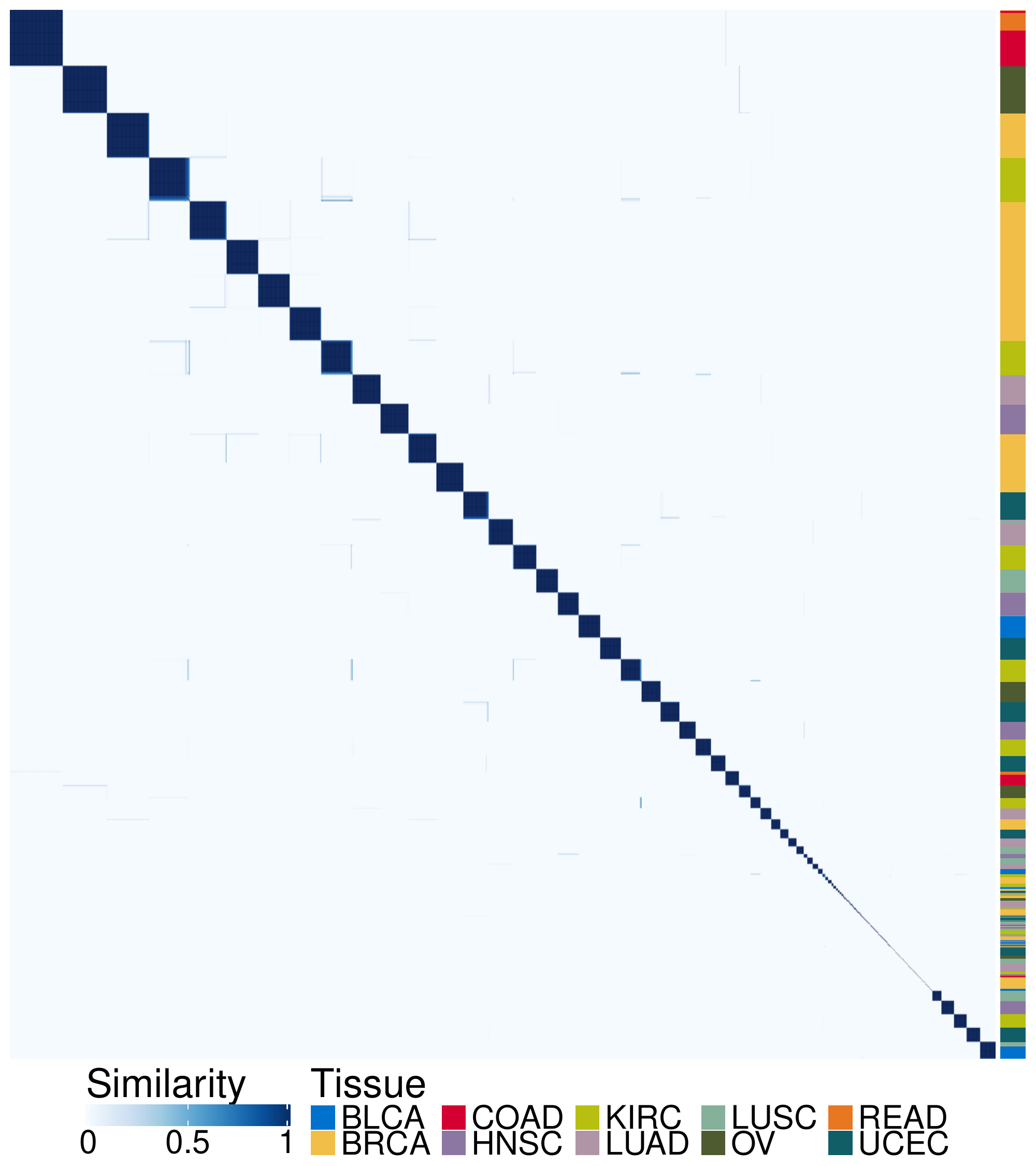}
	\includegraphics[width=.36\linewidth]{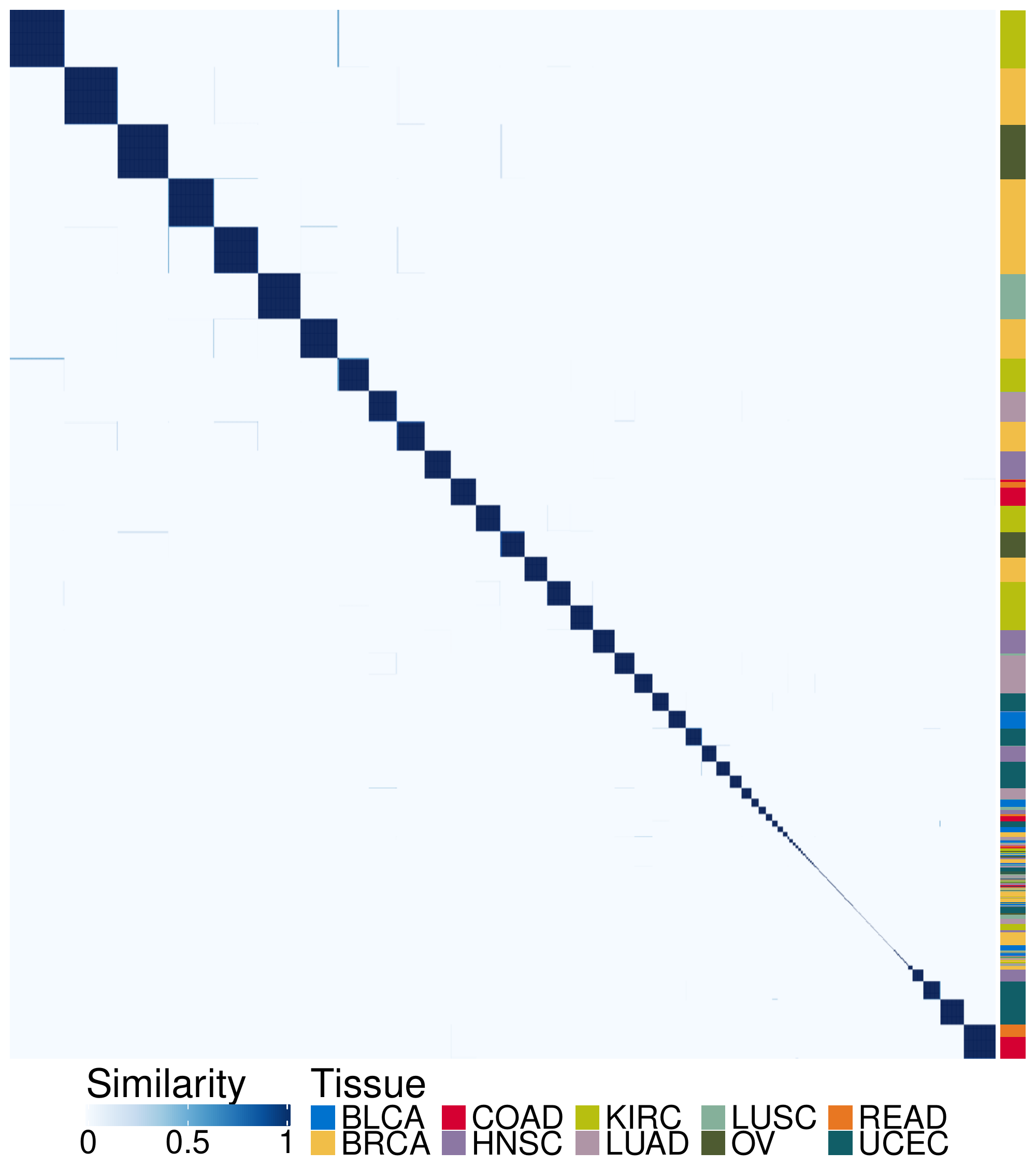}
	\includegraphics[width=.36\linewidth]{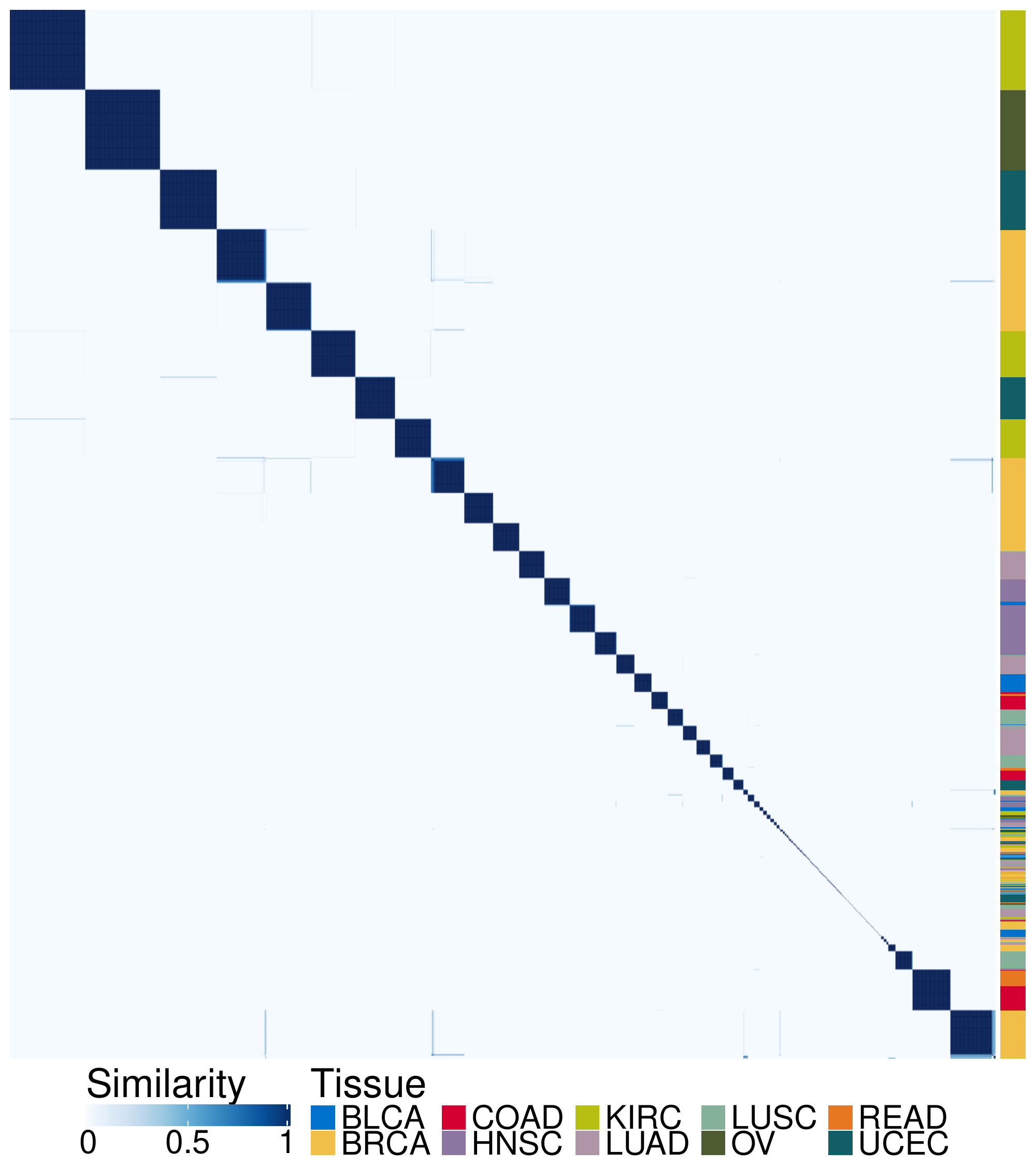}
	\includegraphics[width=.36\linewidth]{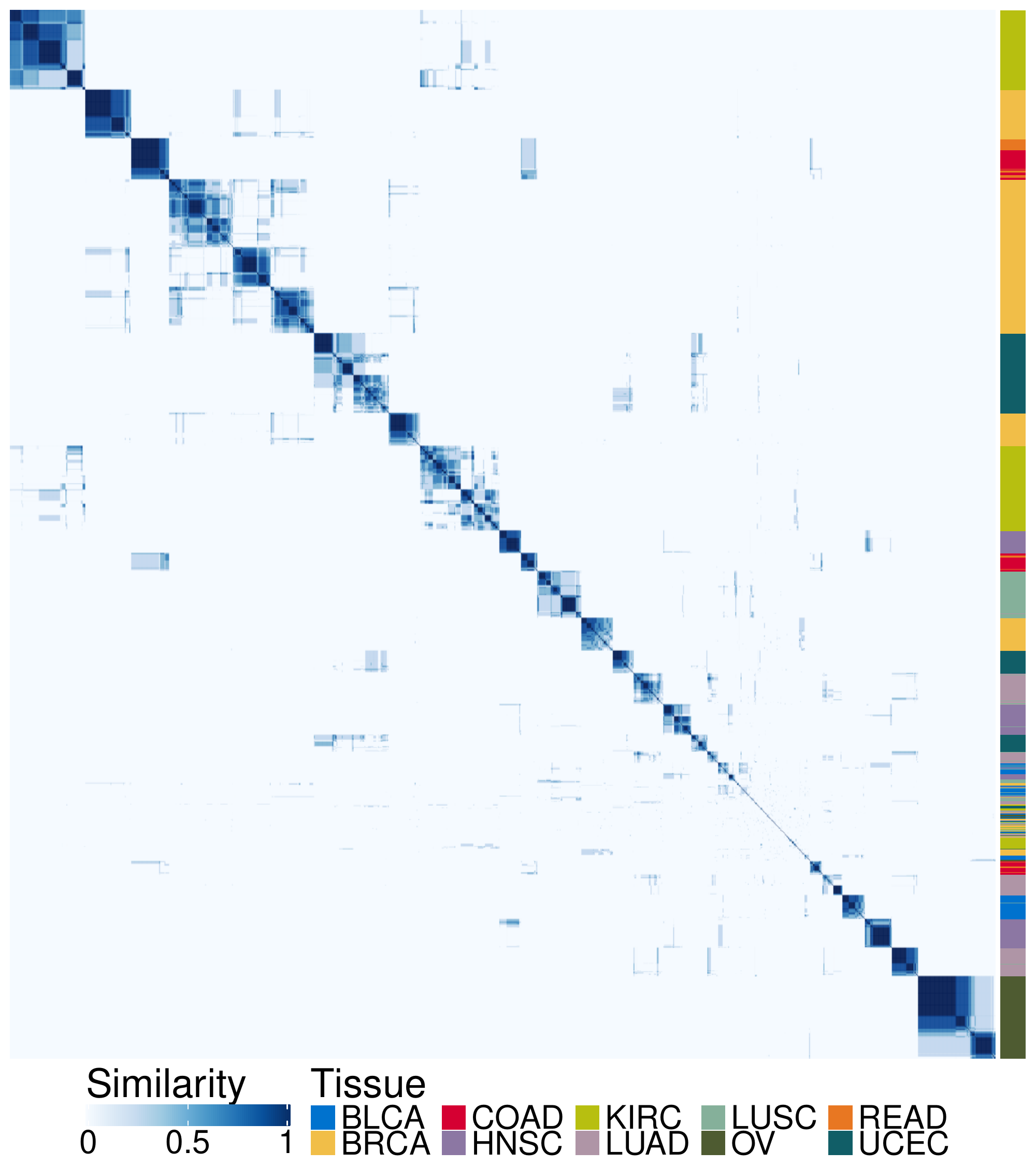}
	\caption{PSMs of the mRNA expression data. $\alpha=0.5$.}
\end{figure}

\begin{table}[H]
\centering
\begin{tabular}{l c c c c}
& \textbf{Chain 2} & \textbf{Chain 3} & \textbf{Chain 4} & \textbf{Chain 5} \\
\hline
\textbf{Chain 1} & 0.46 & 0.43 & 0.48 & 0.42 \\
\textbf{Chain 2} &1 & 0.42 & 0.52 & 0.42 \\
\textbf{Chain 3} && 1 &  0.43 & 0.41 \\
\textbf{Chain 4} && & 1 & 0.49 \\
\hline\\
\end{tabular}
\caption{ARI between the clusterings found on the PSMs of different chains with the number of clusters that maximises the silhouette.}
\end{table} 

\begin{figure}[H]
\centering
\includegraphics[width=.45\linewidth]{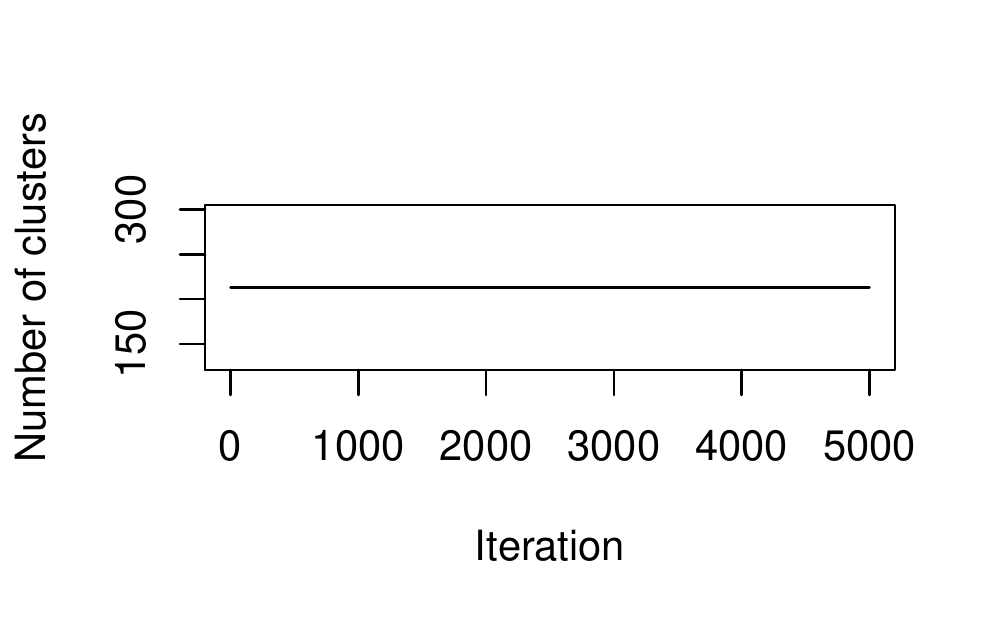}
\includegraphics[width=.45\linewidth]{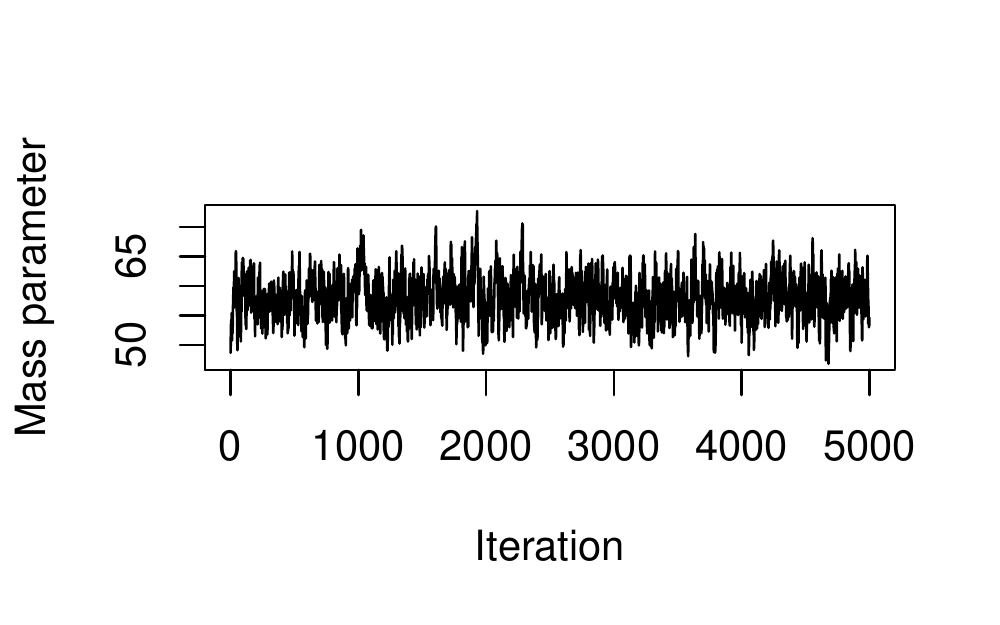}
\vspace{-1.6cm}

\includegraphics[width=.45\linewidth]{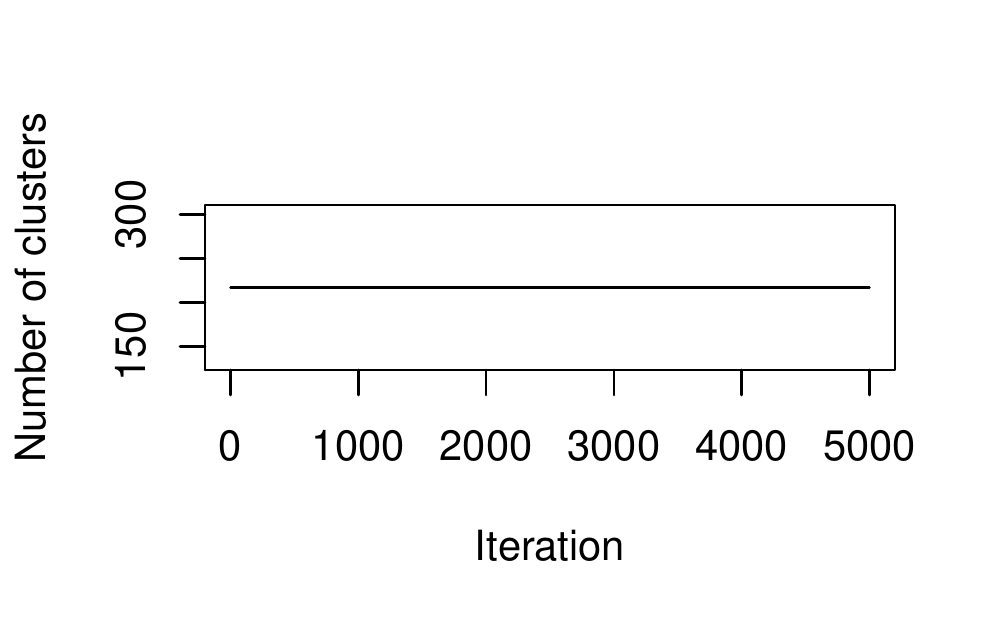}
\includegraphics[width=.45\linewidth]{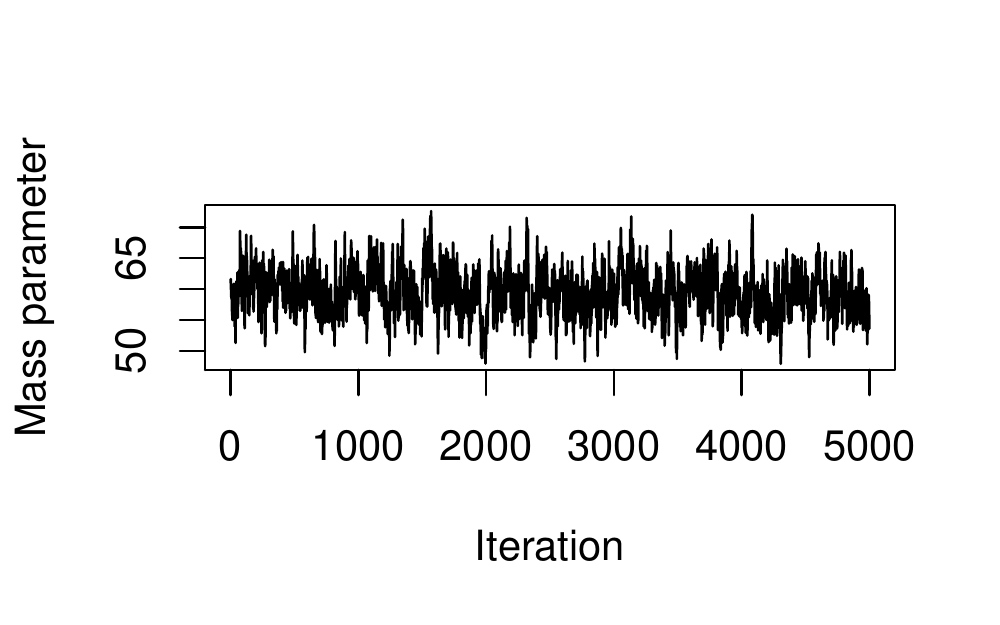}
\vspace{-1.6cm}

\includegraphics[width=.45\linewidth]{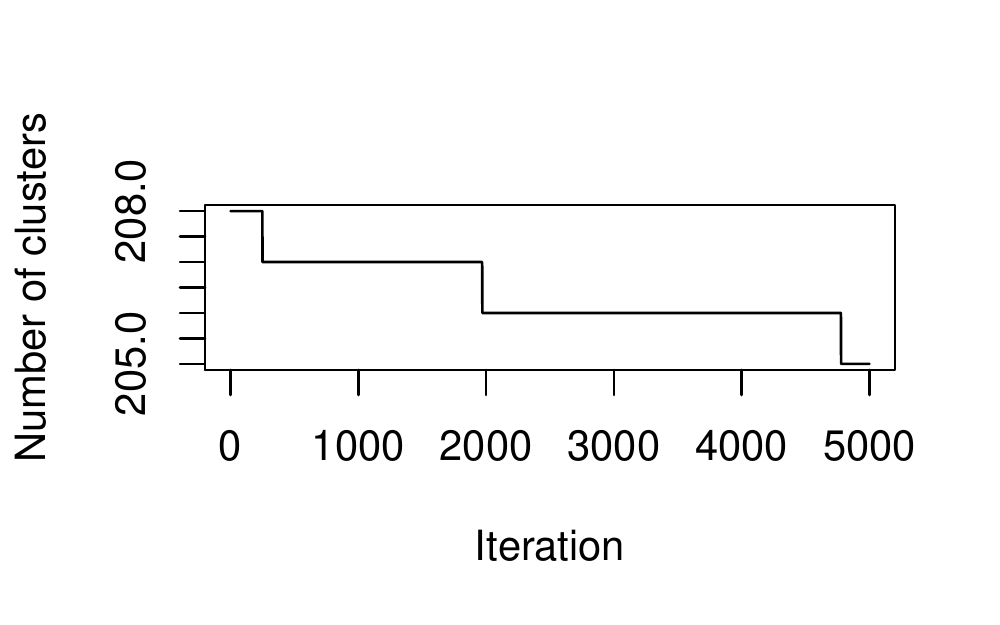}
\includegraphics[width=.45\linewidth]{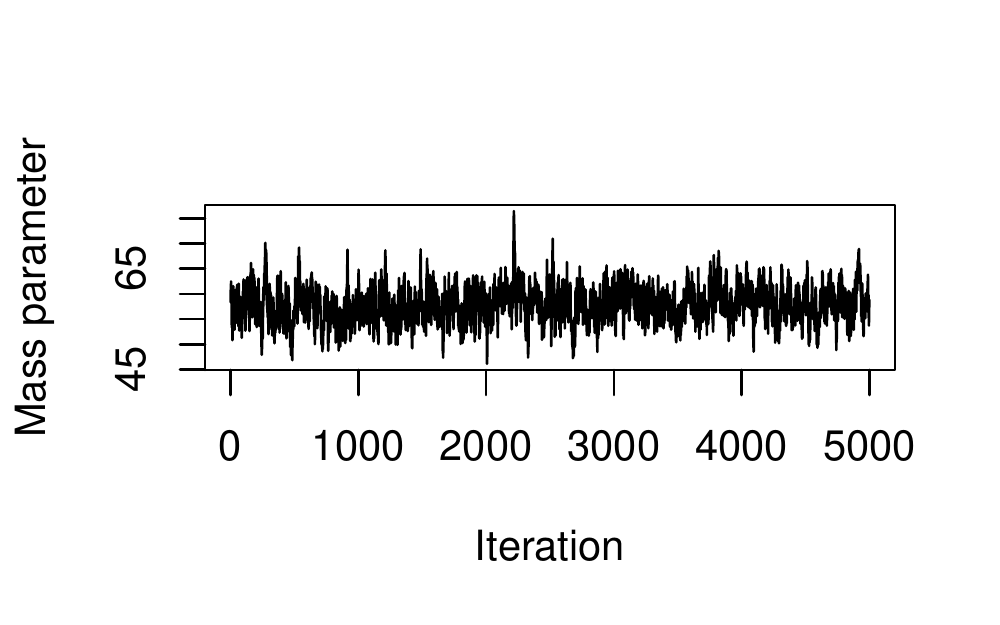}
\vspace{-1.6cm}

\includegraphics[width=.45\linewidth]{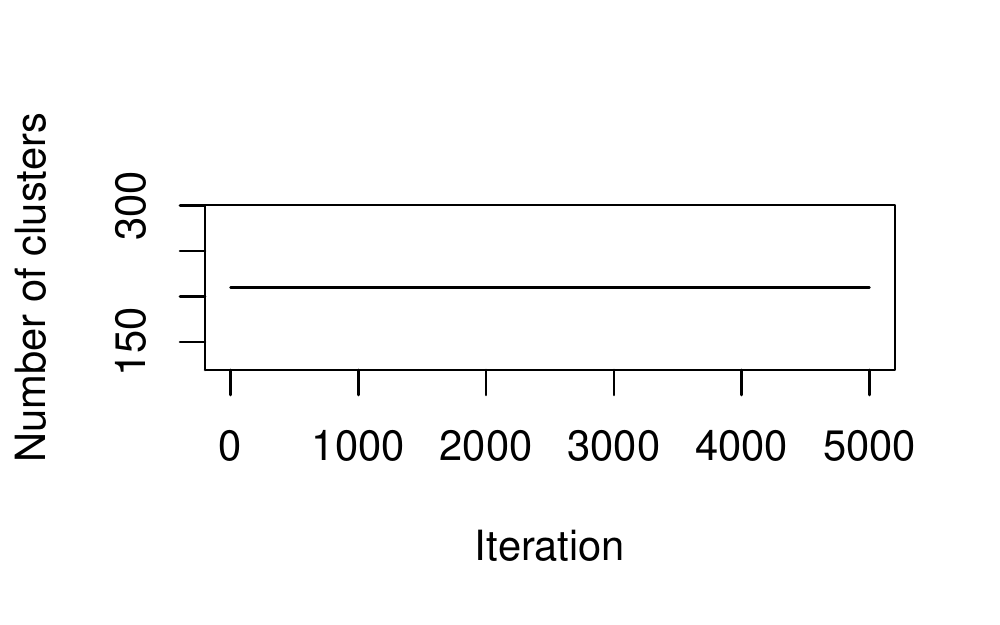}
\includegraphics[width=.45\linewidth]{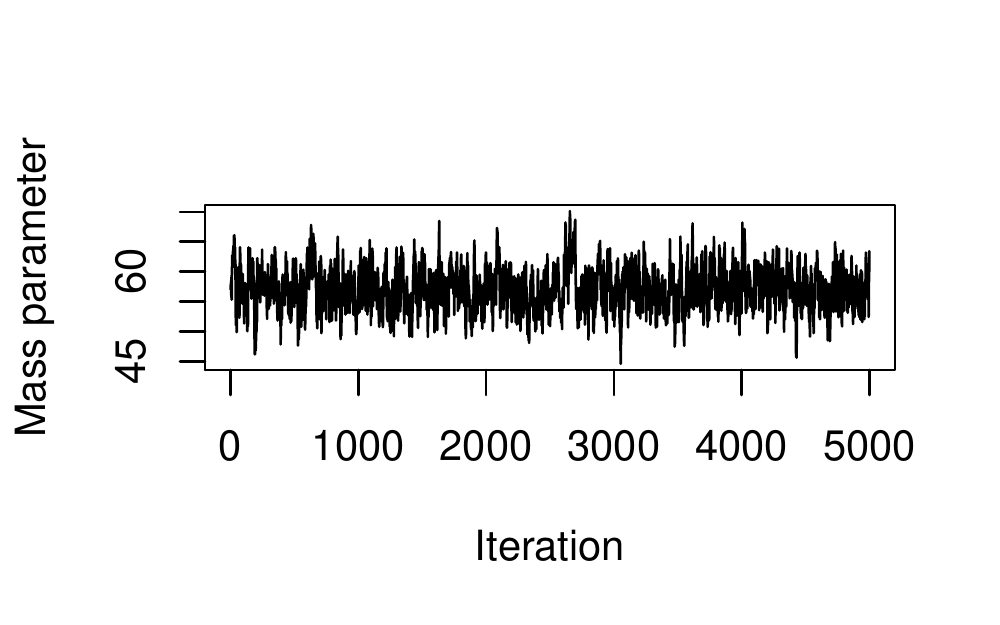}
\vspace{-1.6cm}

\includegraphics[width=.45\linewidth]{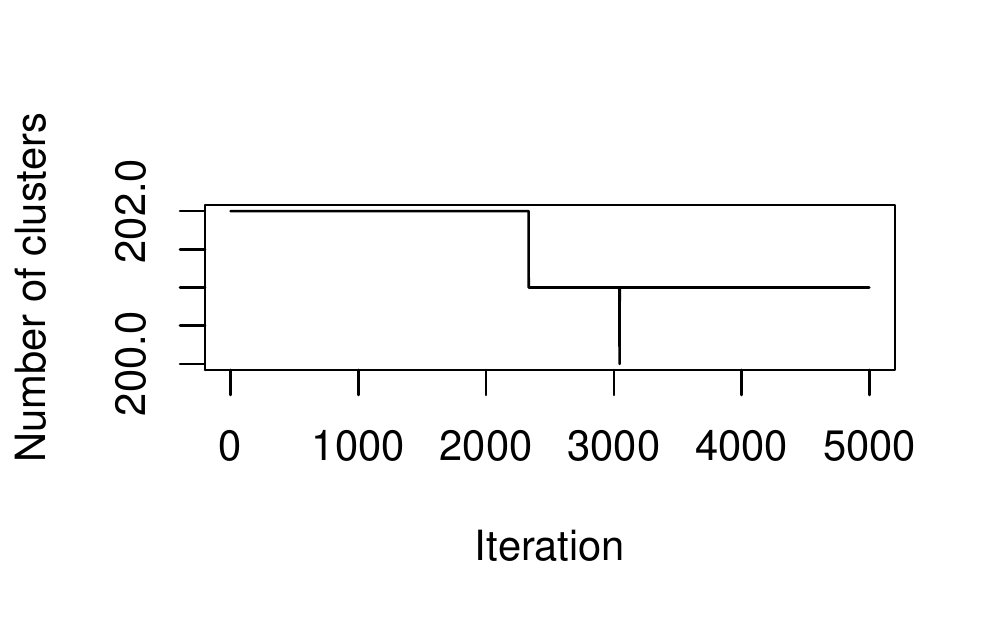}
\includegraphics[width=.45\linewidth]{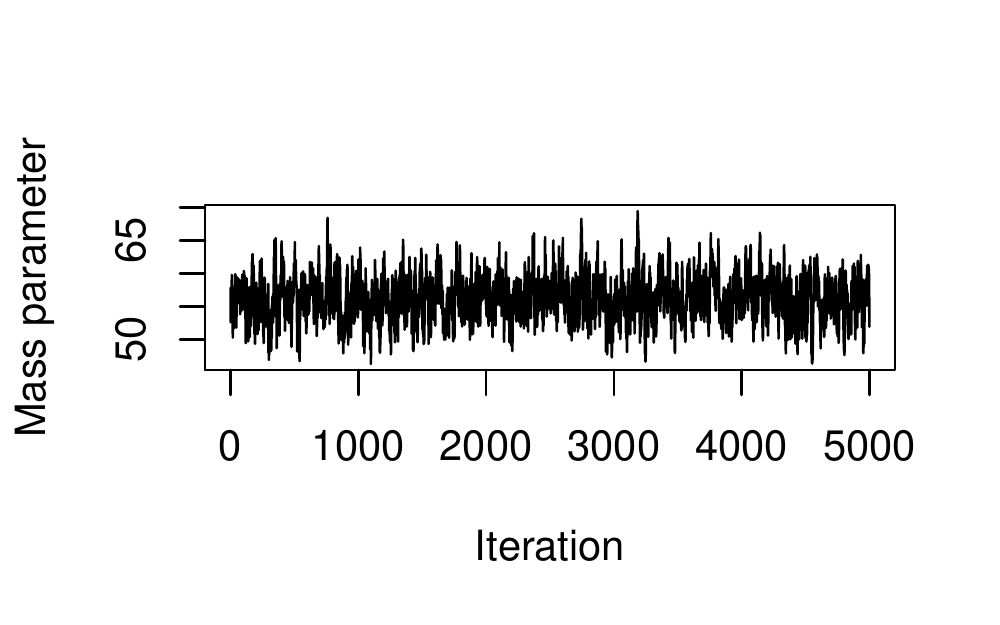}
\caption{MCMC convergence assessment, mRNA expression data. $\alpha=0.5$.}
\end{figure}

\clearpage

\begin{figure}[H]
	\centering
	\includegraphics[width=.36\linewidth]{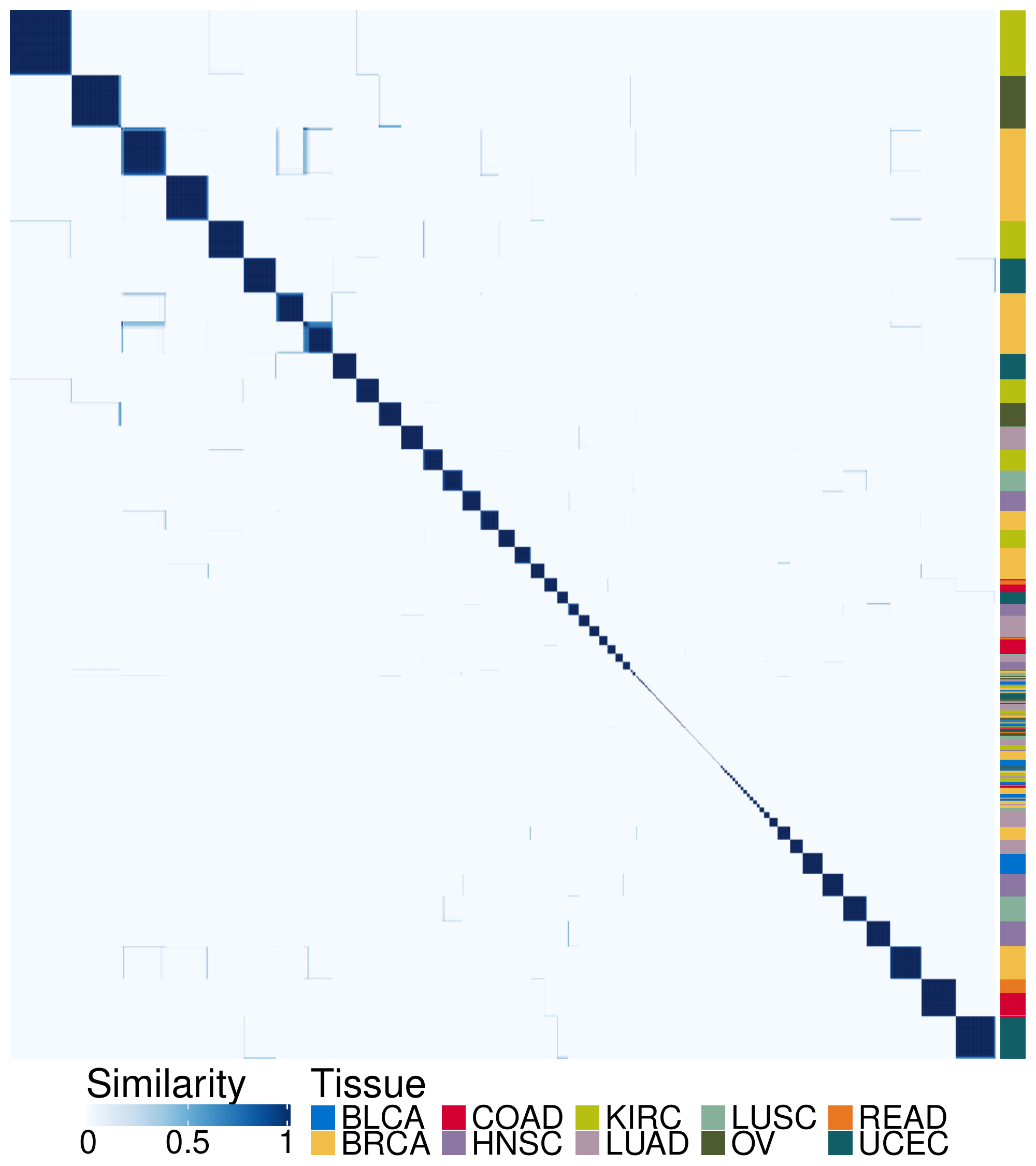}
	\includegraphics[width=.36\linewidth]{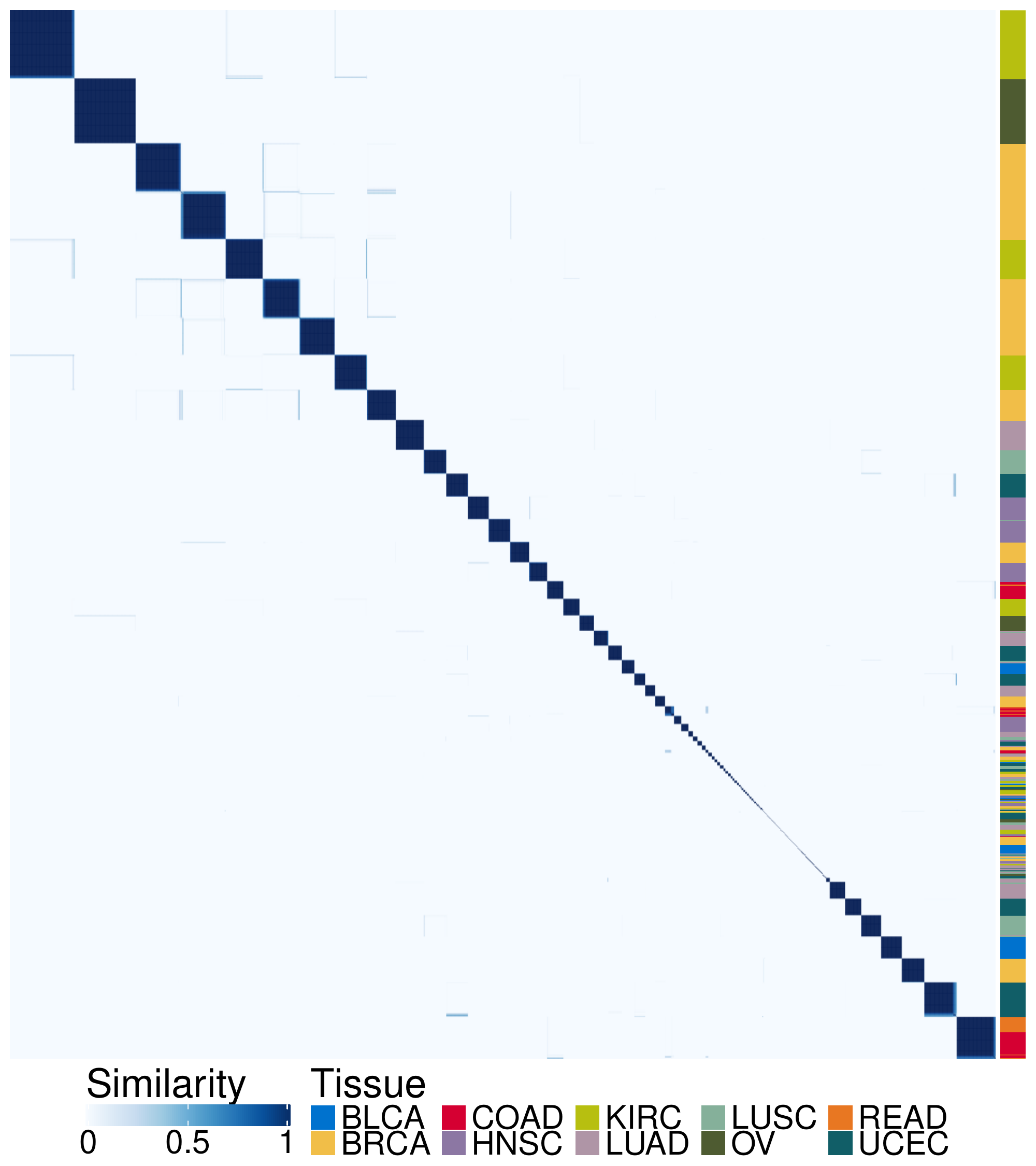}
	\includegraphics[width=.36\linewidth]{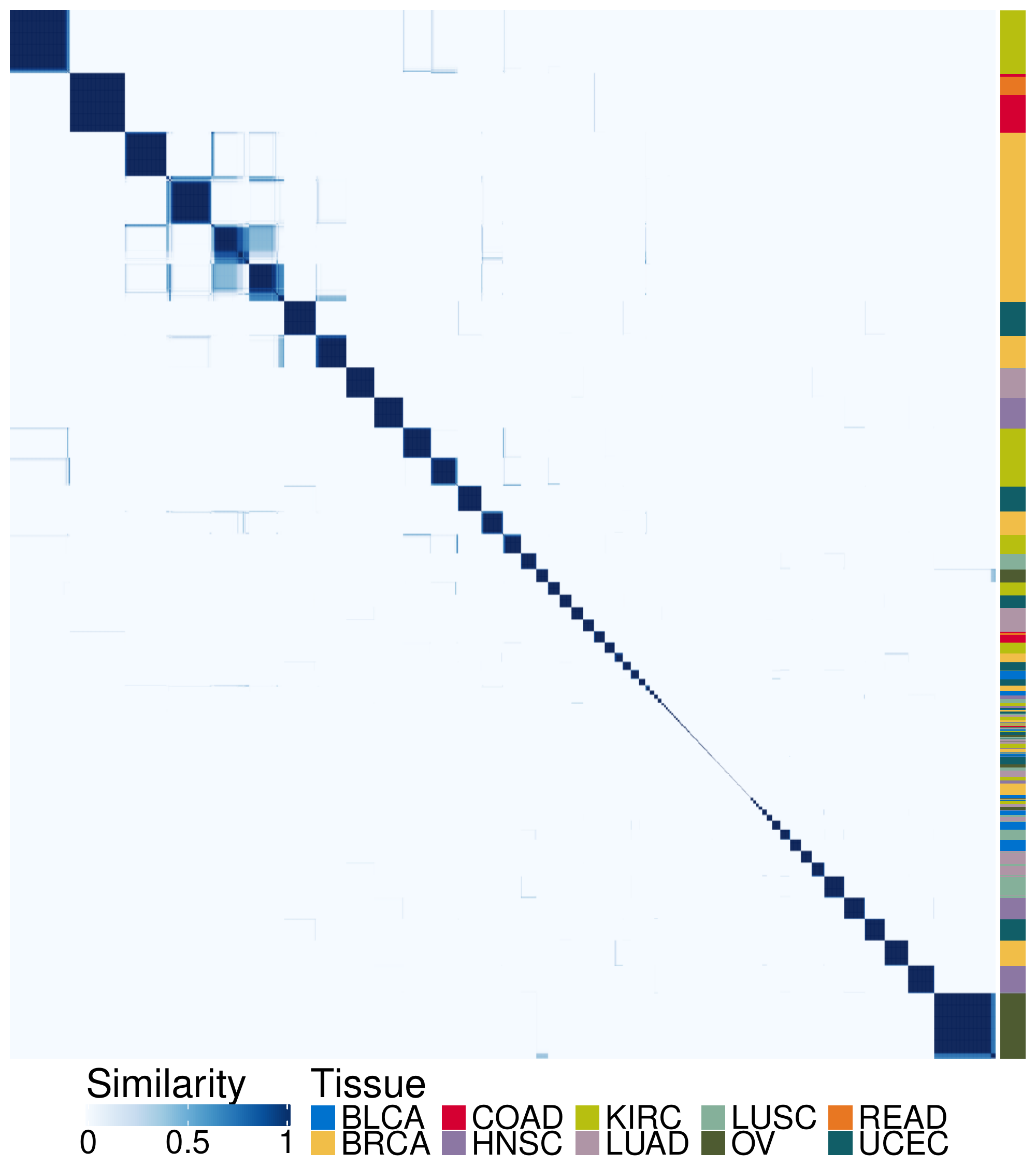}
	\includegraphics[width=.36\linewidth]{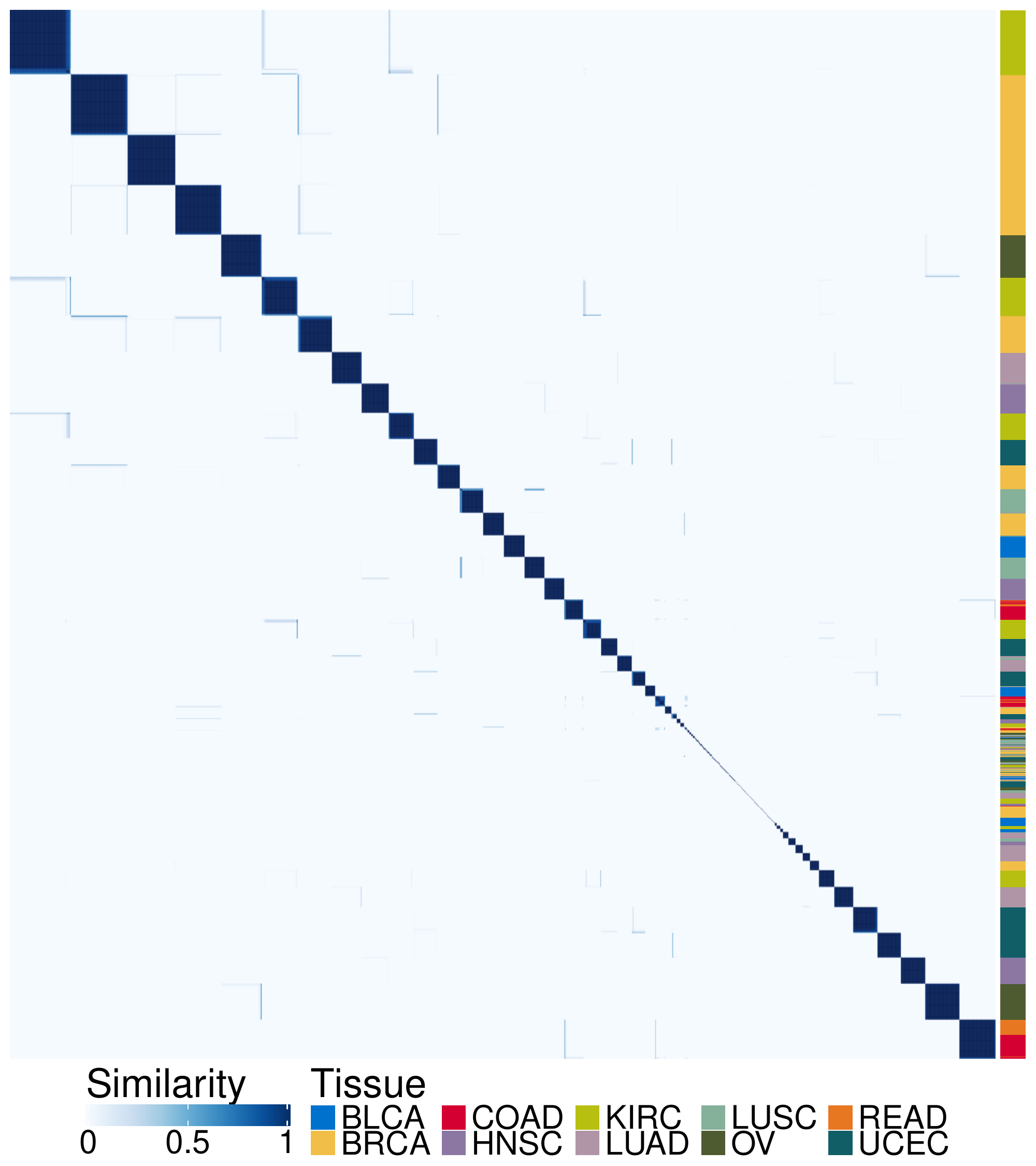}
	\includegraphics[width=.36\linewidth]{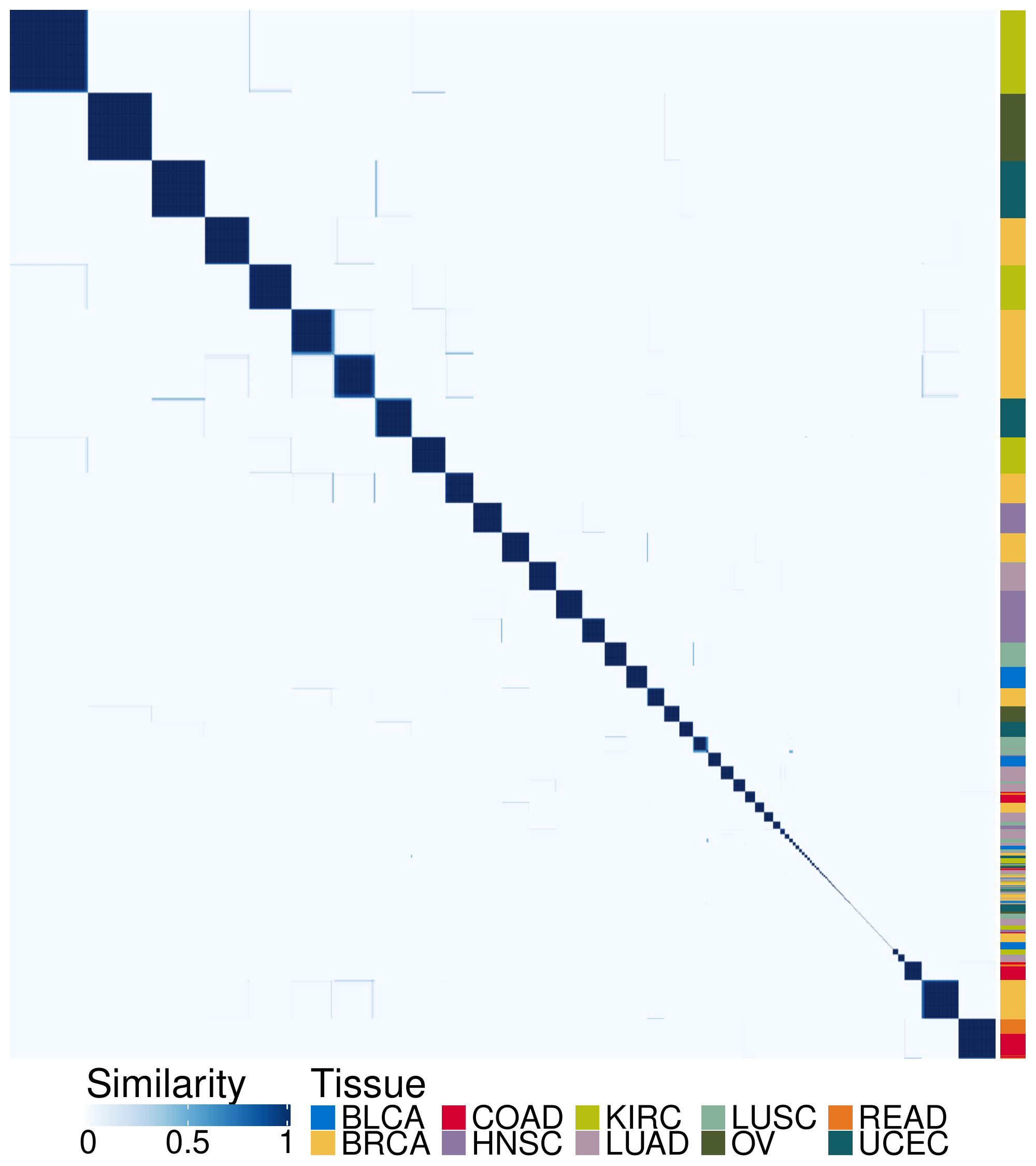}
	\includegraphics[width=.36\linewidth]{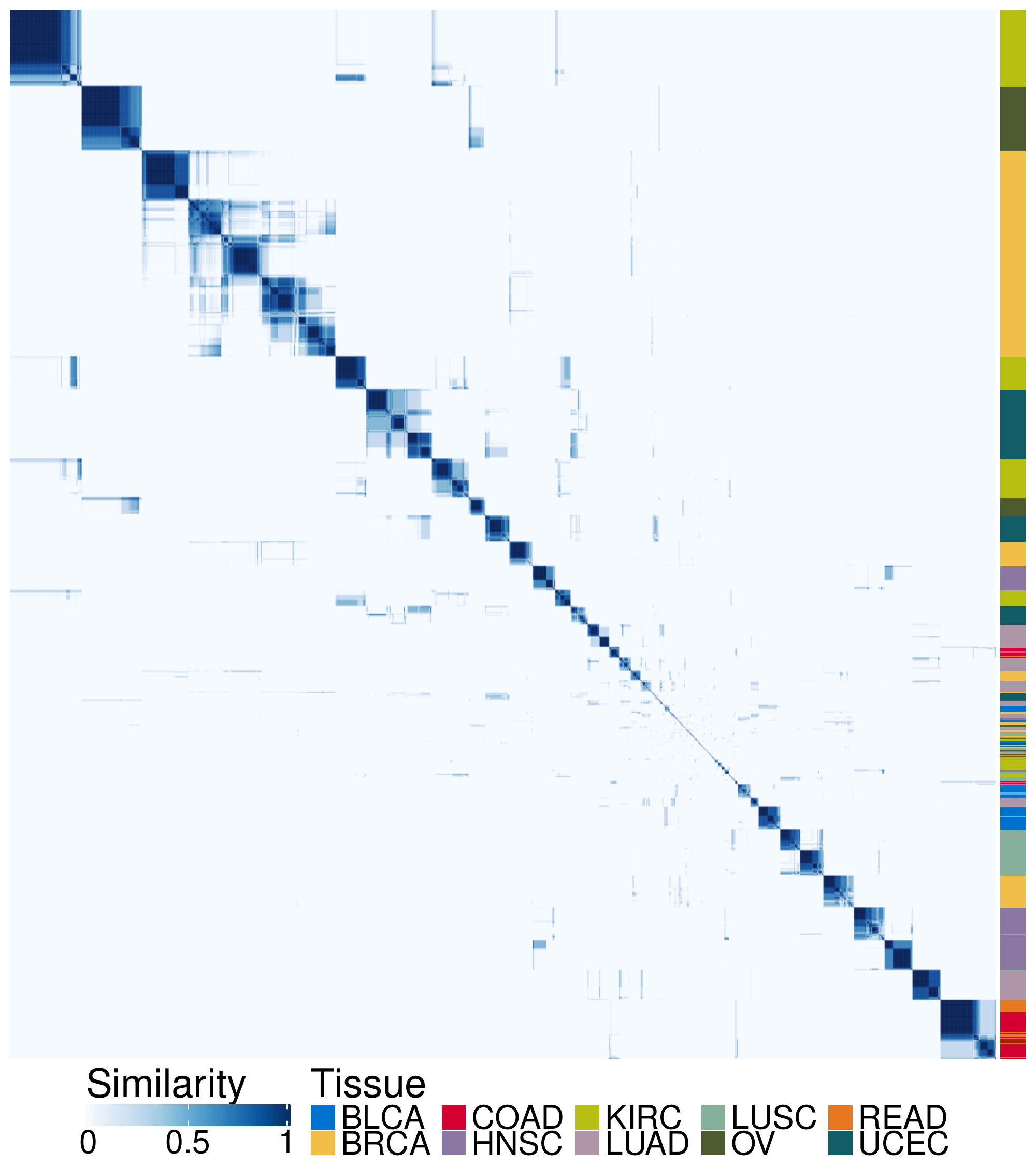}
	\caption{PSMs of the mRNA expression data. $\alpha=1$.}
\end{figure}

\begin{table}[H]
\centering
\begin{tabular}{l c c c c}
& \textbf{Chain 2} & \textbf{Chain 3} & \textbf{Chain 4} & \textbf{Chain 5} \\
\hline
\textbf{Chain 1} & 0.63 & 0.55 & 0.61 & 0.52 \\
\textbf{Chain 2} &1 & 0.58 & 0.64 & 0.63 \\
\textbf{Chain 3} && 1 &  0.59 & 0.58 \\
\textbf{Chain 4} && & 1 & 0.57 \\
\hline\\
\end{tabular}
\caption{ARI between the clusterings found on the PSMs of different chains with the number of clusters that maximises the silhouette.}
\end{table} 

\begin{figure}[H]
\centering
\includegraphics[width=.45\linewidth]{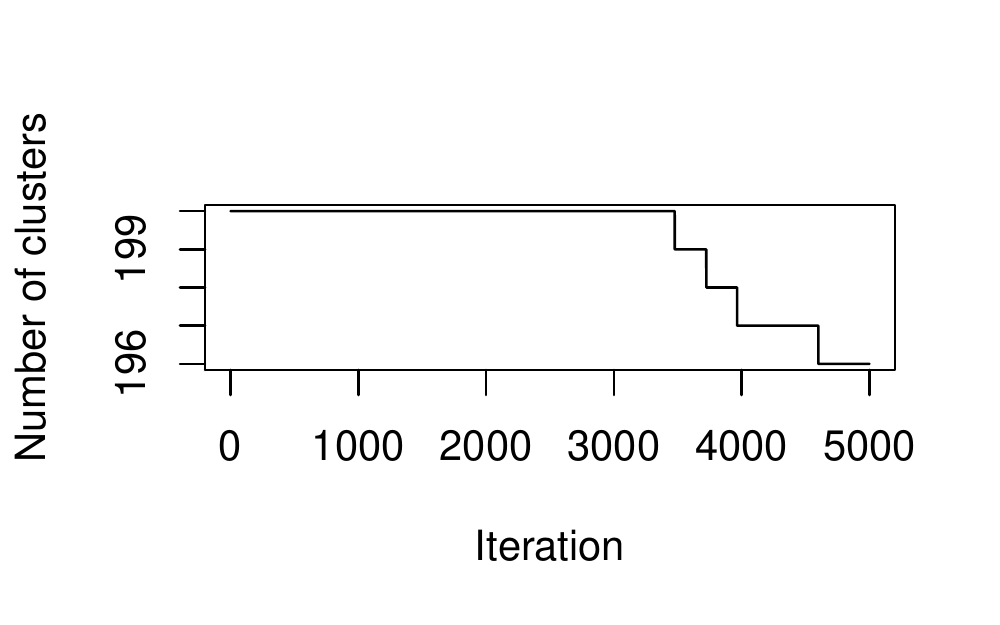}
\includegraphics[width=.45\linewidth]{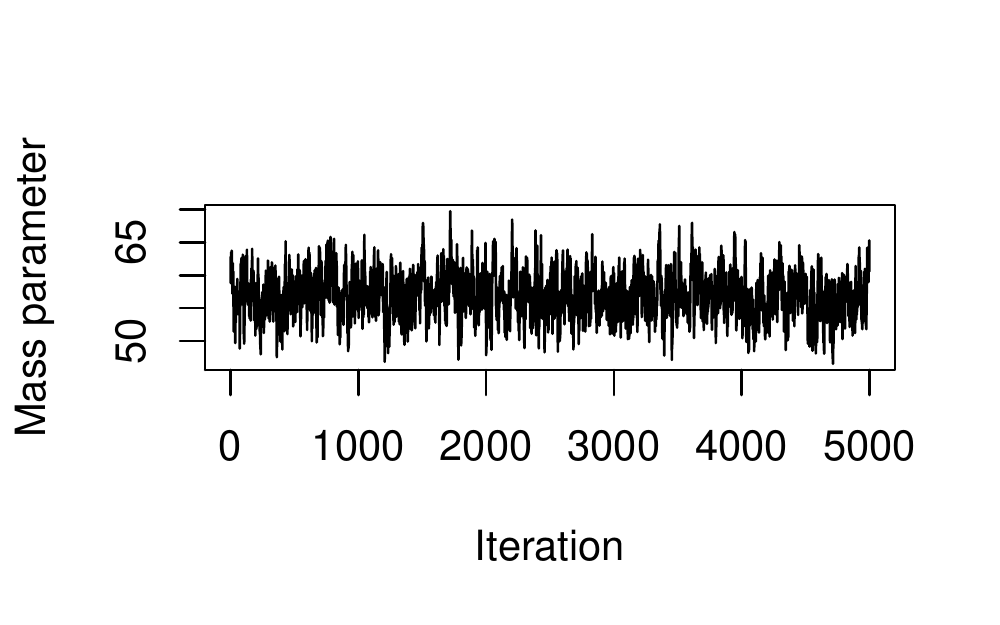}
\vspace{-1.6cm}

\includegraphics[width=.45\linewidth]{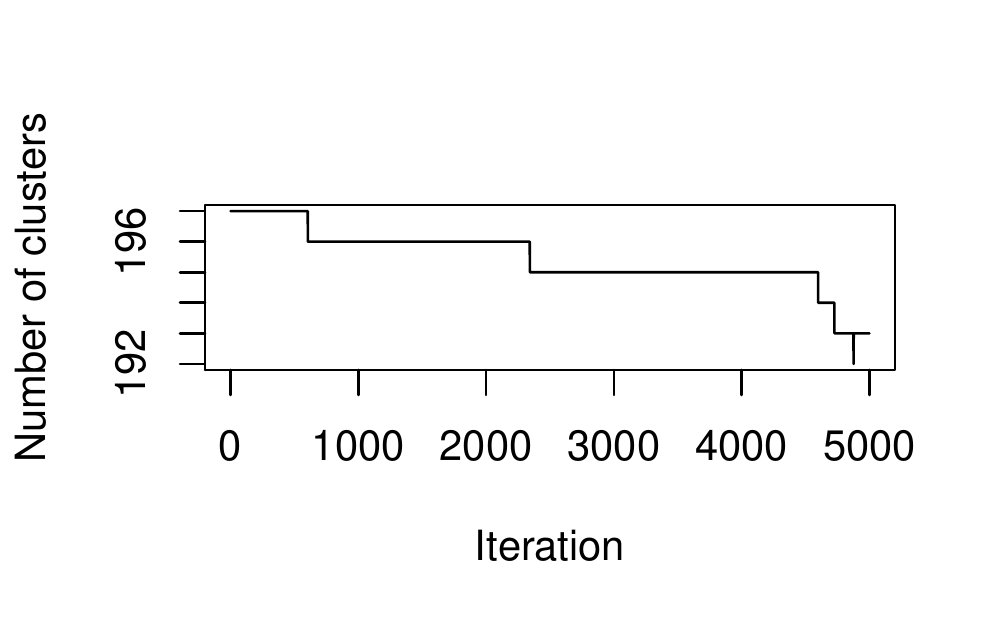}
\includegraphics[width=.45\linewidth]{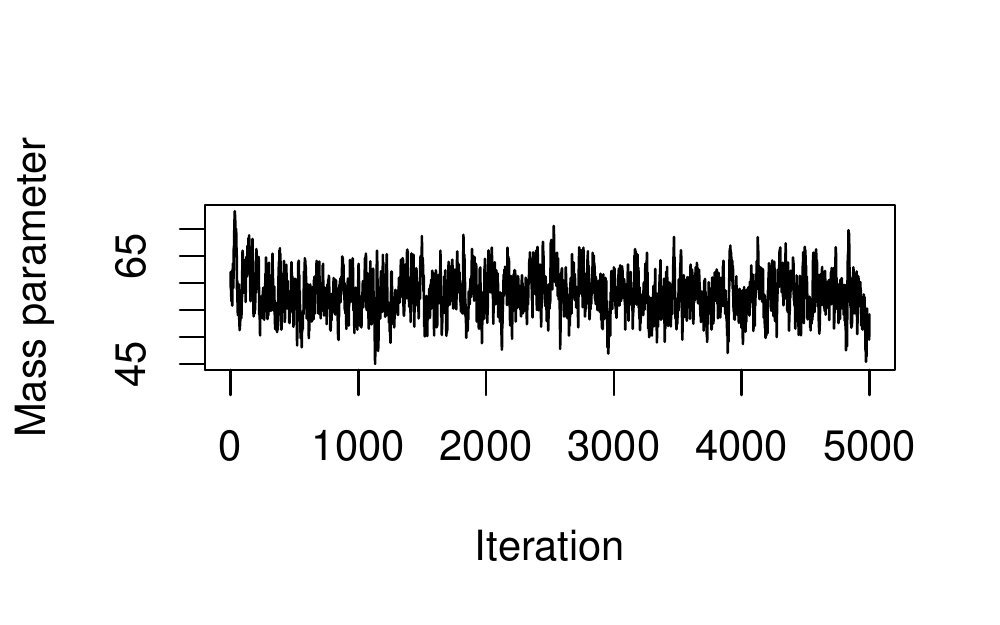}
\vspace{-1.6cm}

\includegraphics[width=.45\linewidth]{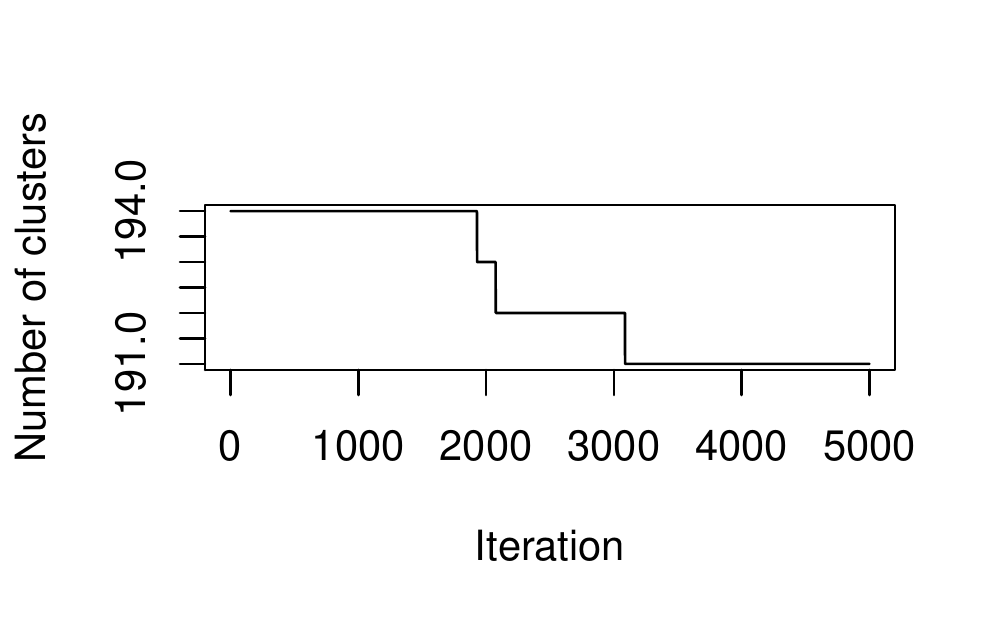}
\includegraphics[width=.45\linewidth]{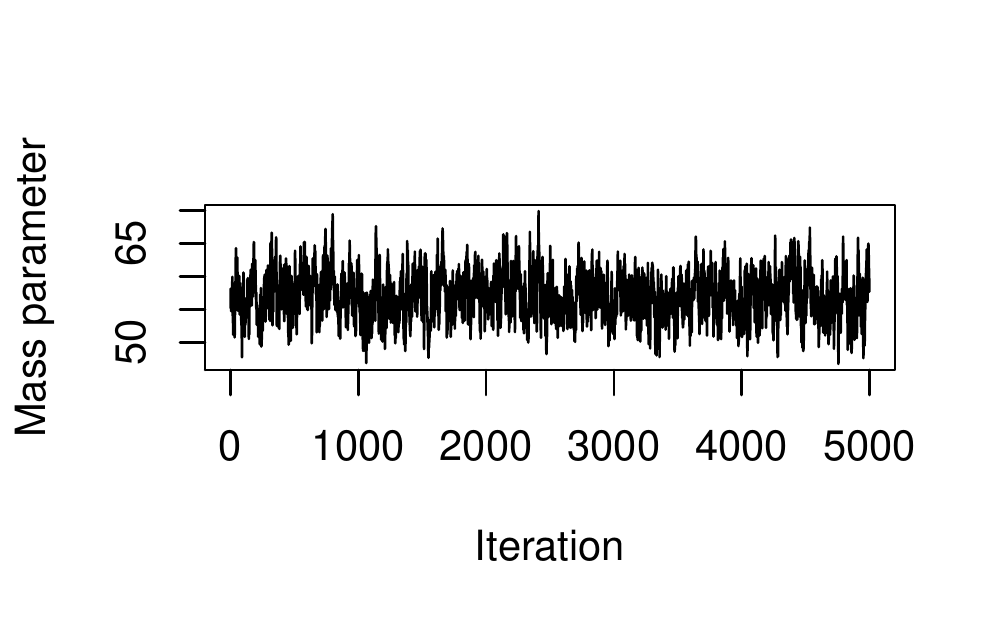}
\vspace{-1.6cm}

\includegraphics[width=.45\linewidth]{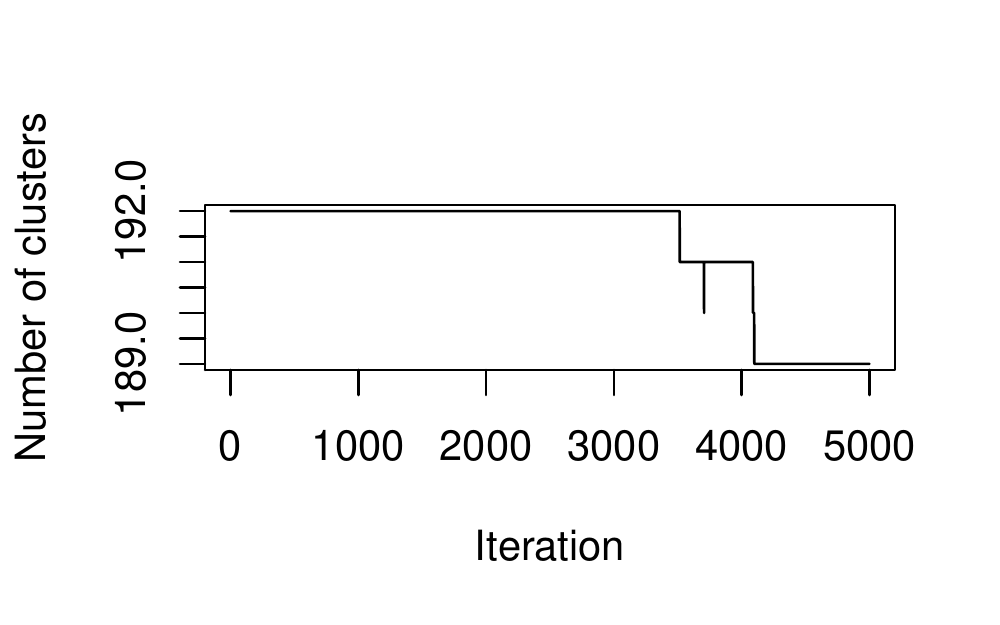}
\includegraphics[width=.45\linewidth]{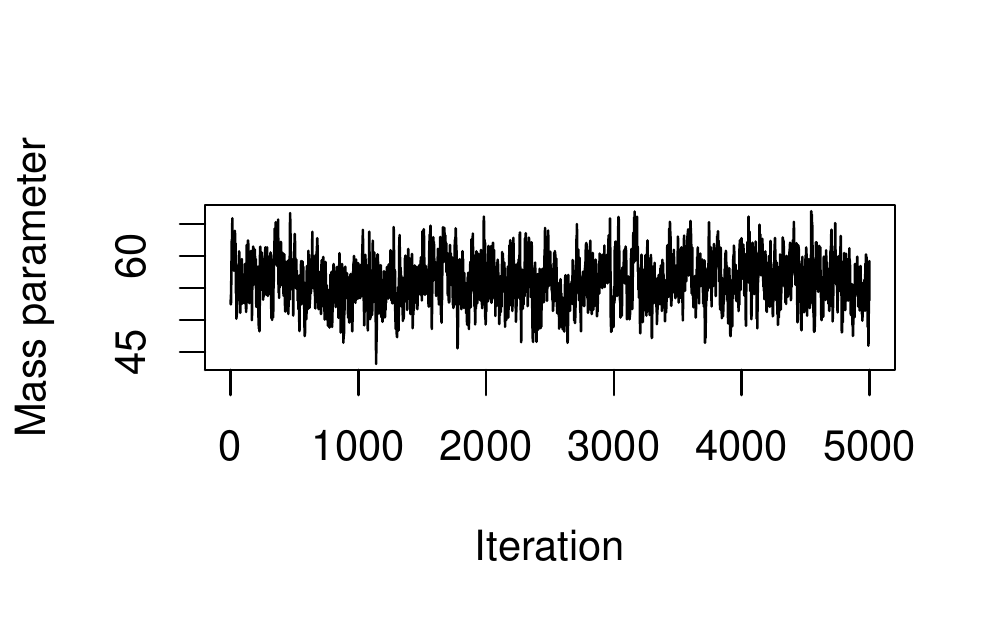}
\vspace{-1.6cm}

\includegraphics[width=.45\linewidth]{n_clusters_mRNA_chain5_alpha05_temp}
\includegraphics[width=.45\linewidth]{mass_parameter_mRNA_chain5_alpha05_temp}
\caption{MCMC convergence assessment, mRNA expression data. $\alpha=1$.}
\end{figure}

\clearpage

\begin{figure}[H]
	\centering
	\includegraphics[width=.36\linewidth]{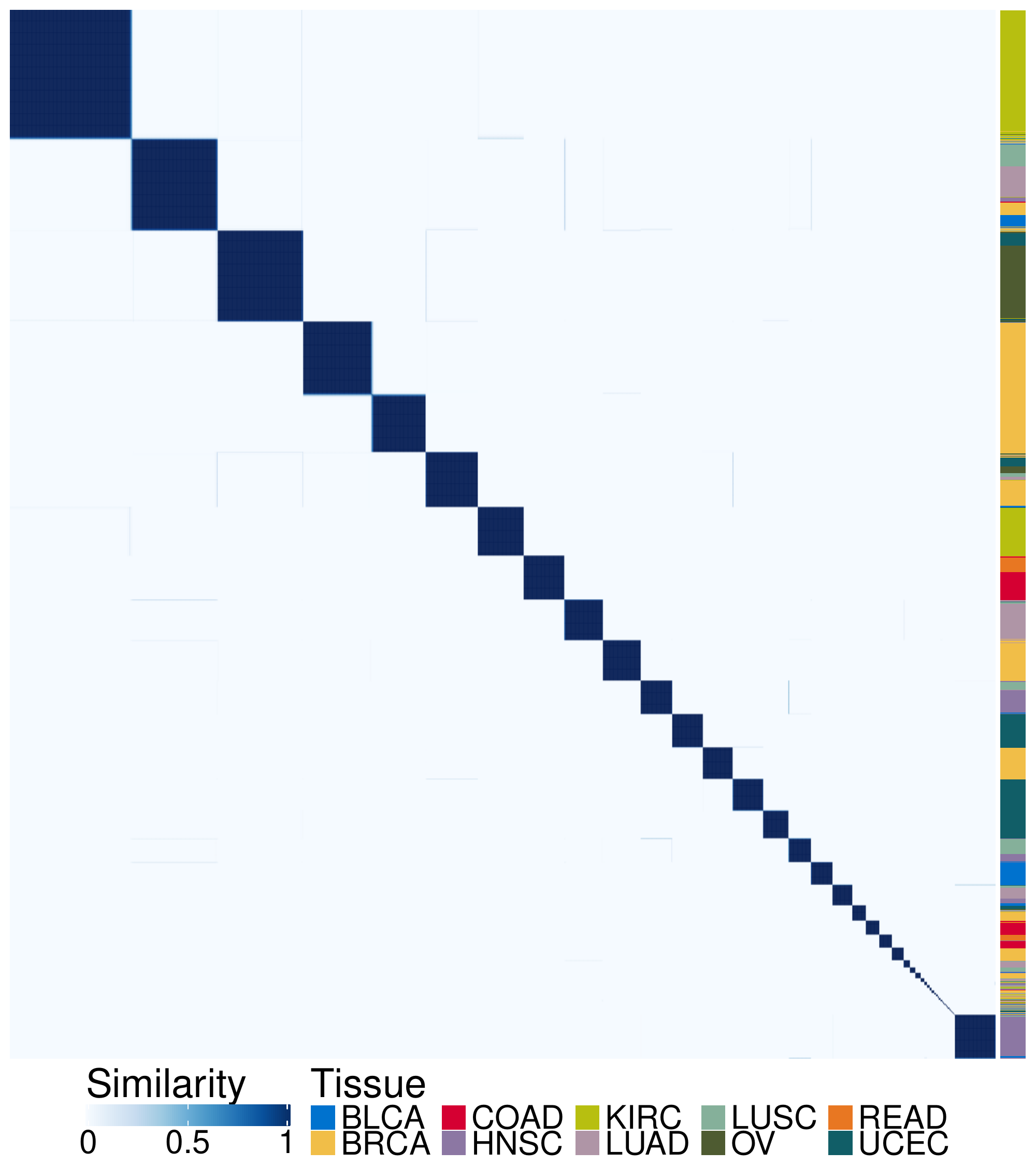}
	\includegraphics[width=.36\linewidth]{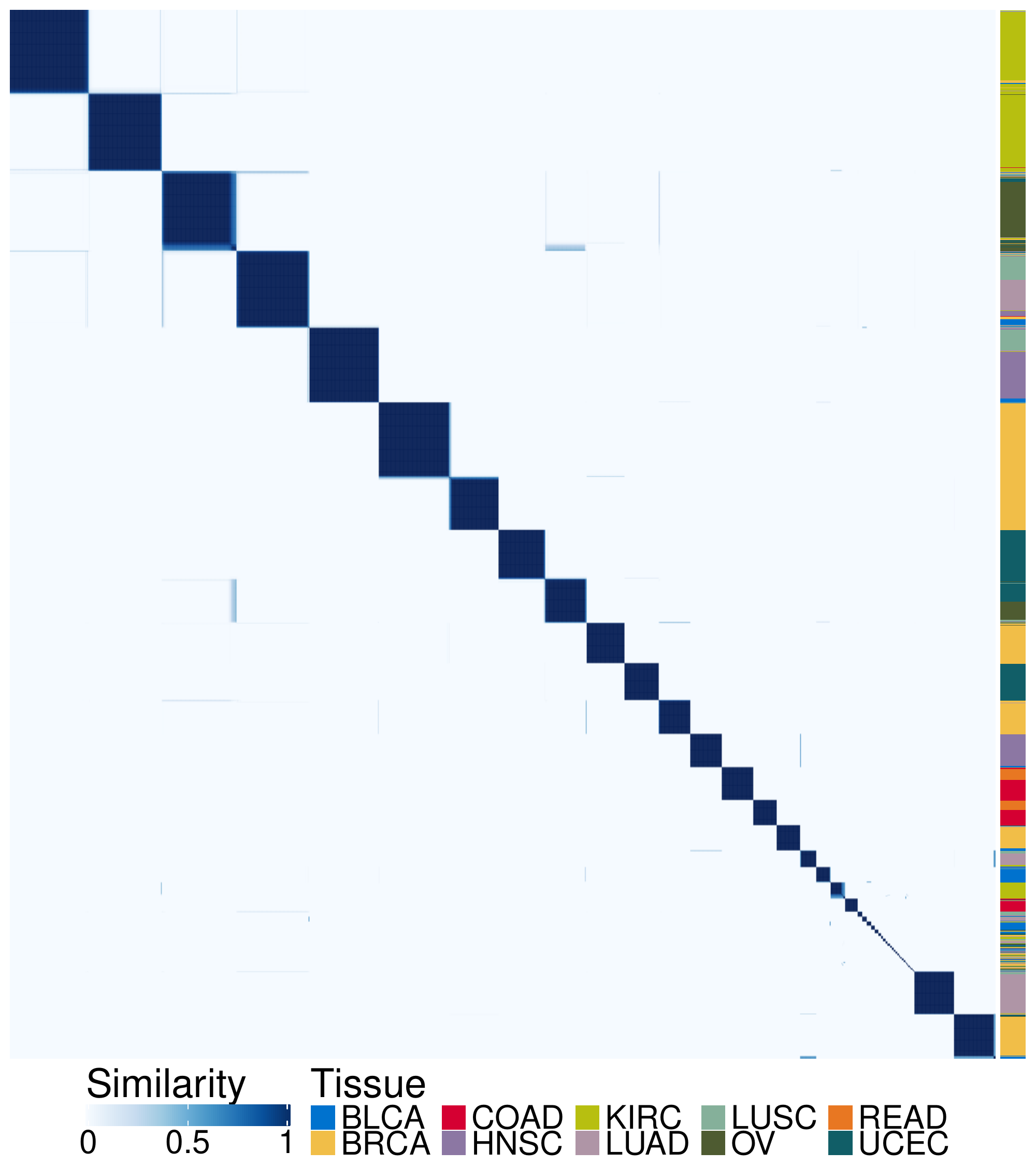}
	\includegraphics[width=.36\linewidth]{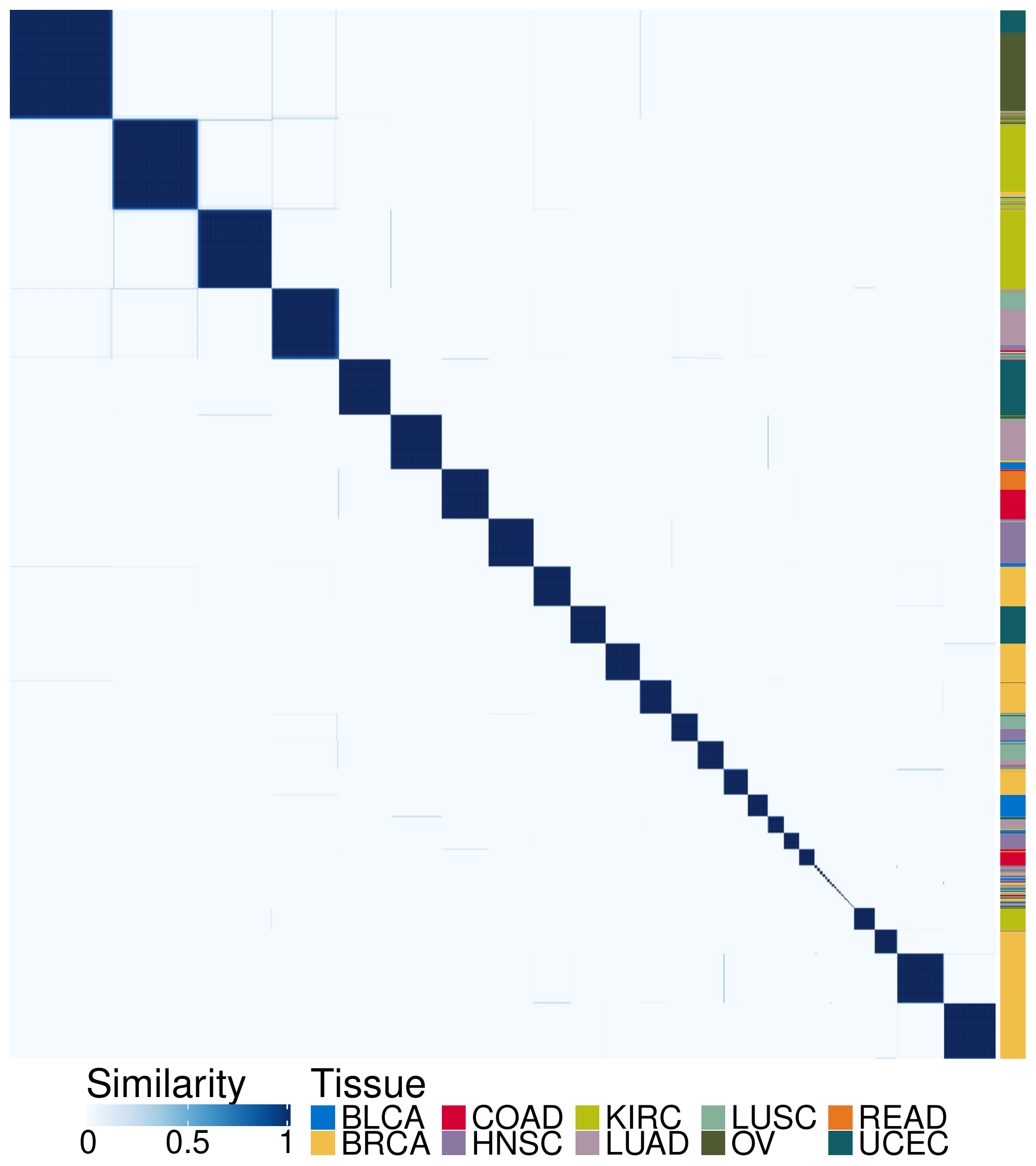}
	\includegraphics[width=.36\linewidth]{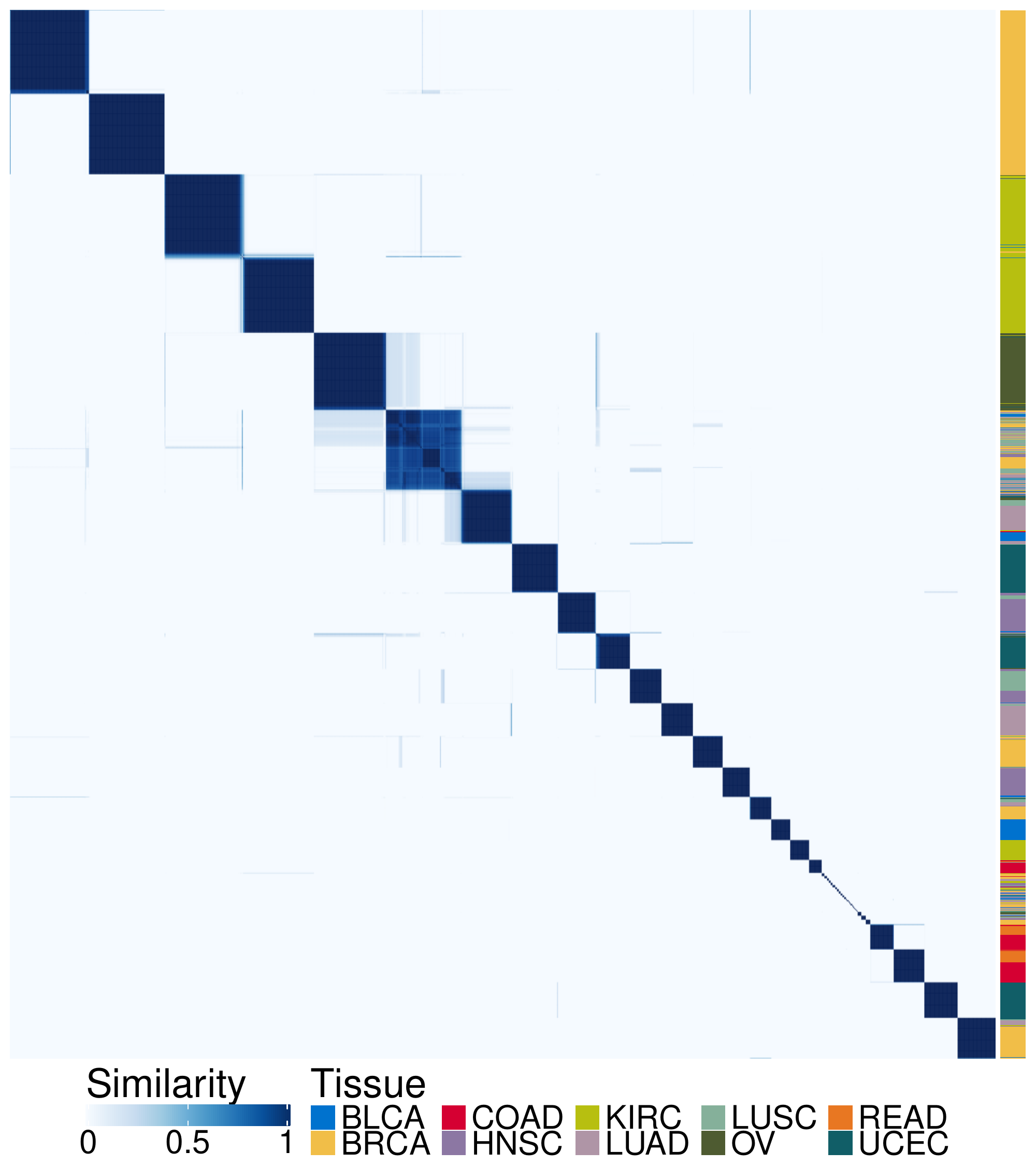}
	\includegraphics[width=.36\linewidth]{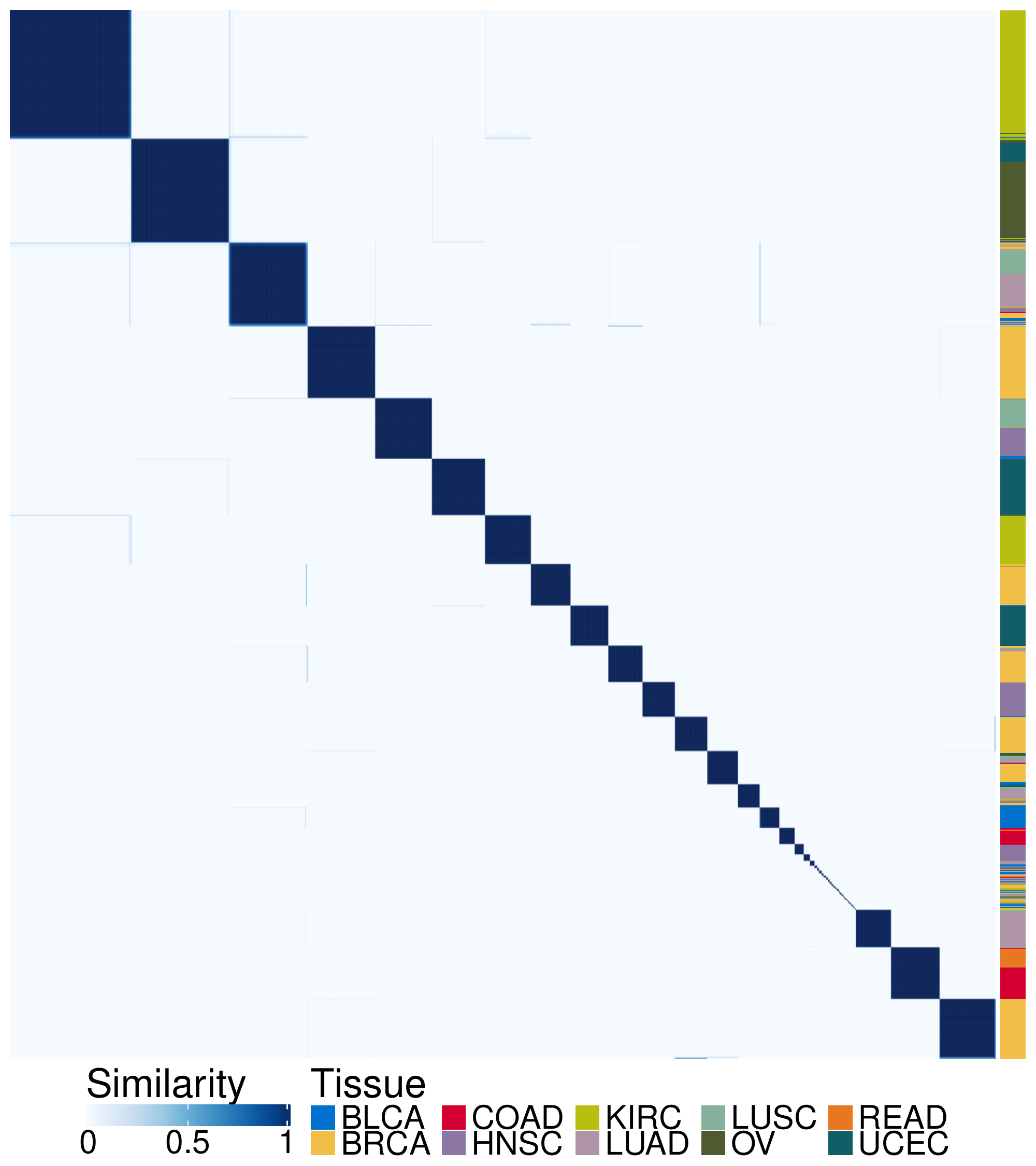}
	\includegraphics[width=.36\linewidth]{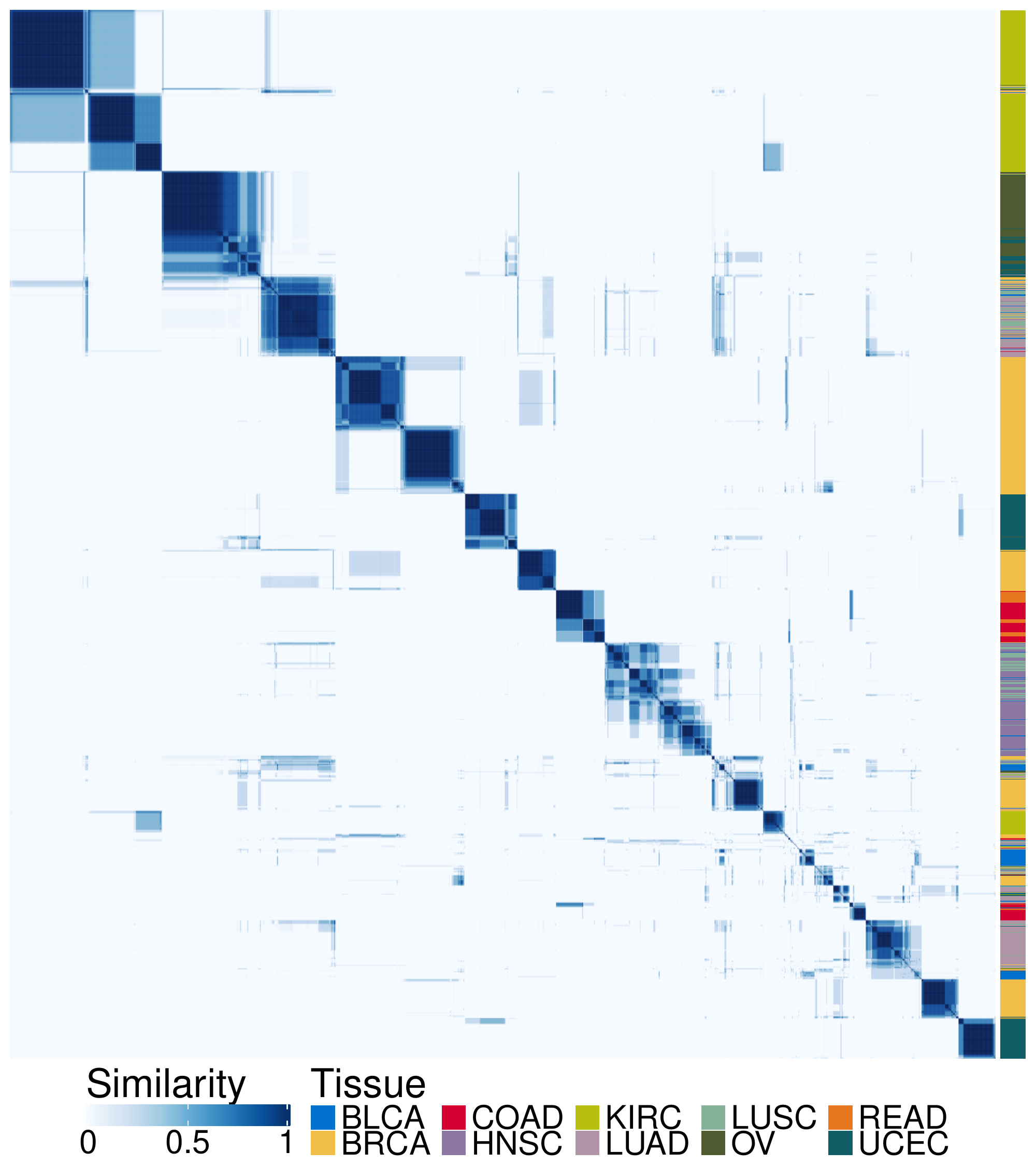}
	\caption{PSMs of the methylation data. $\lambda=0$.}
\end{figure}

\begin{table}[H]
\centering
\begin{tabular}{l c c c c}
& \textbf{Chain 2} & \textbf{Chain 3} & \textbf{Chain 4} & \textbf{Chain 5} \\
\hline
\textbf{Chain 1} & 0.60 & 0.57 & 0.53 & 0.79 \\
\textbf{Chain 2} &1 & 0.67 & 0.63 & 0.64 \\
\textbf{Chain 3} && 1 &  0.59 & 0.66 \\
\textbf{Chain 4} && & 1 & 0.54 \\
\hline\\
\end{tabular}
\caption{ARI between the clusterings found on the PSMs of different chains with the number of clusters that maximises the silhouette.}
\end{table} 

\begin{figure}[H]
	\centering
	\includegraphics[width=.45\linewidth]{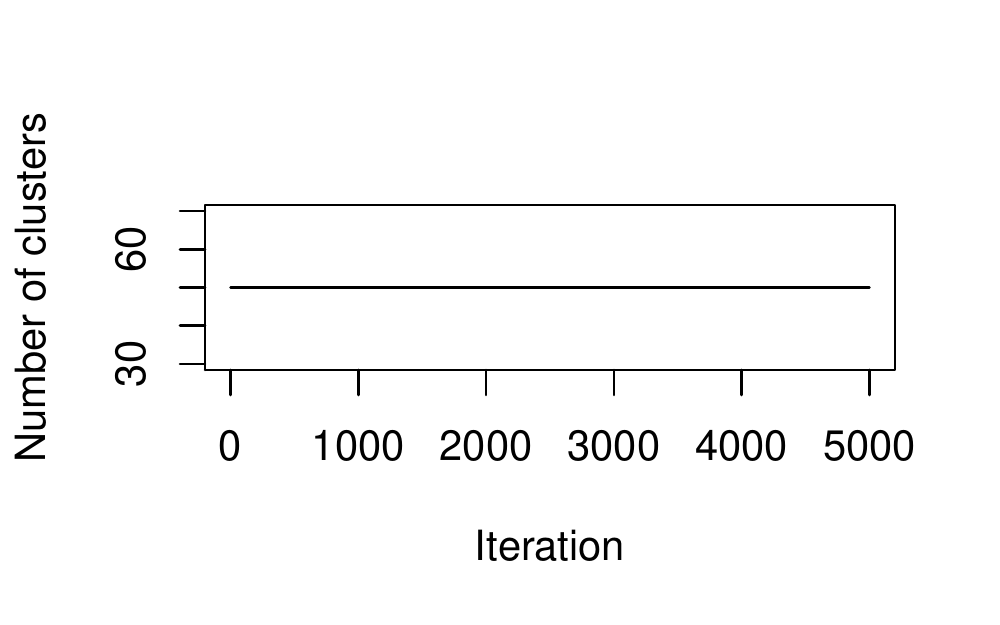}
	\includegraphics[width=.45\linewidth]{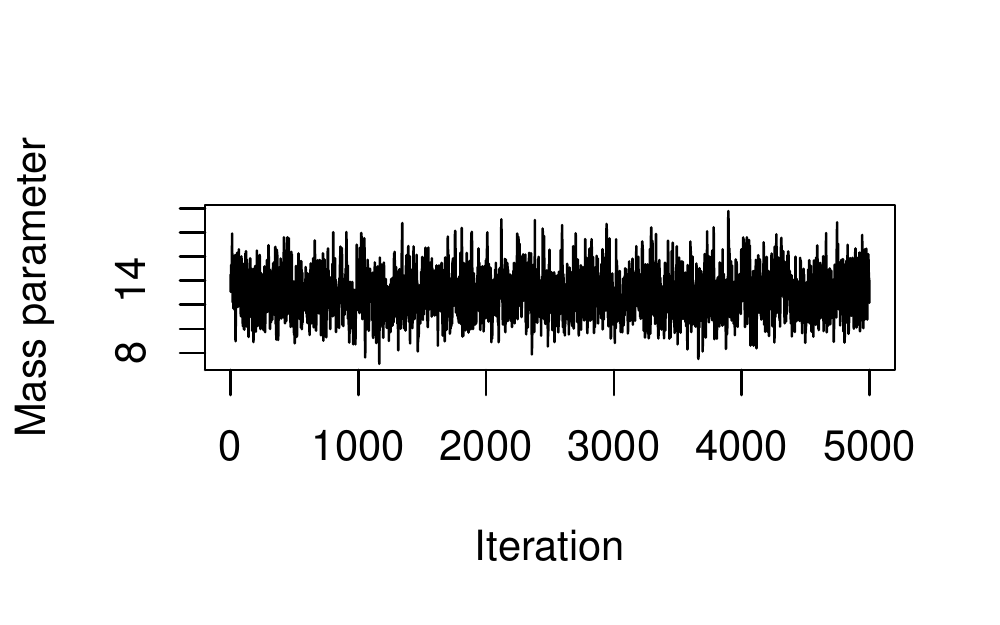}
	\vspace{-1.6cm}

	\includegraphics[width=.45\linewidth]{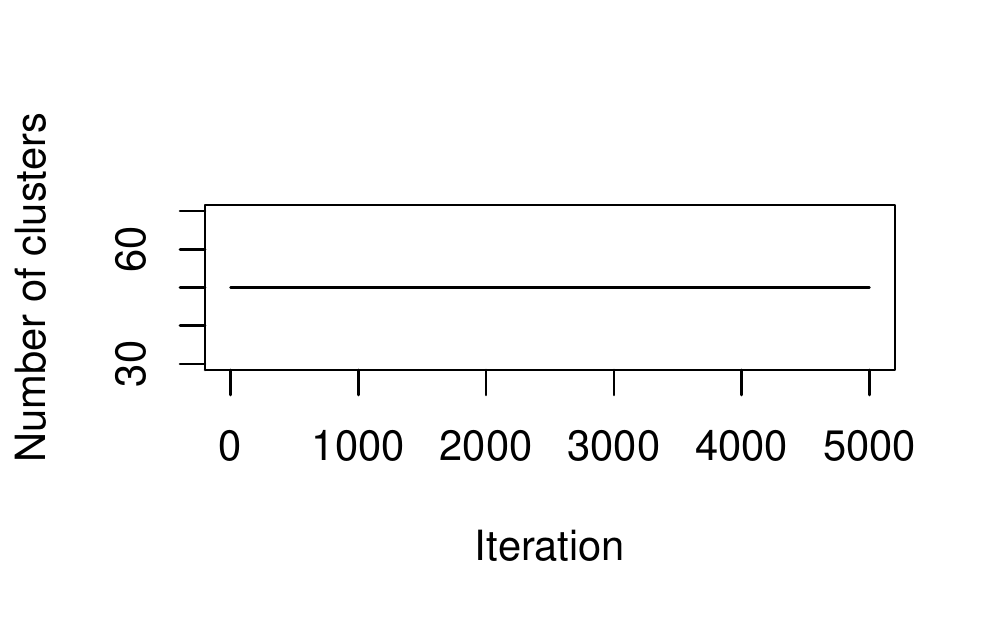}
	\includegraphics[width=.45\linewidth]{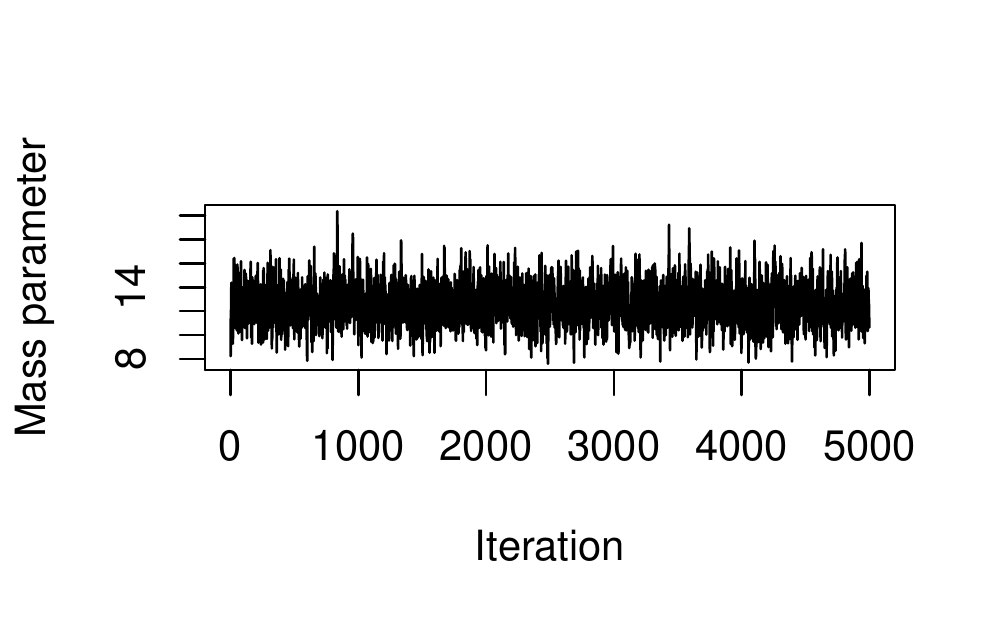}
	\vspace{-1.6cm}

	\includegraphics[width=.45\linewidth]{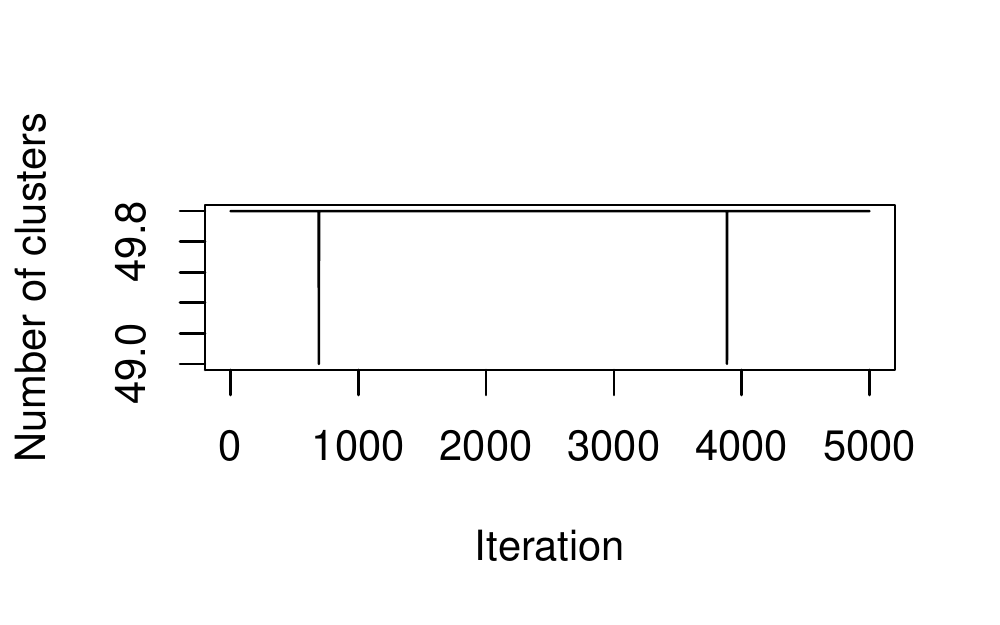}
	\includegraphics[width=.45\linewidth]{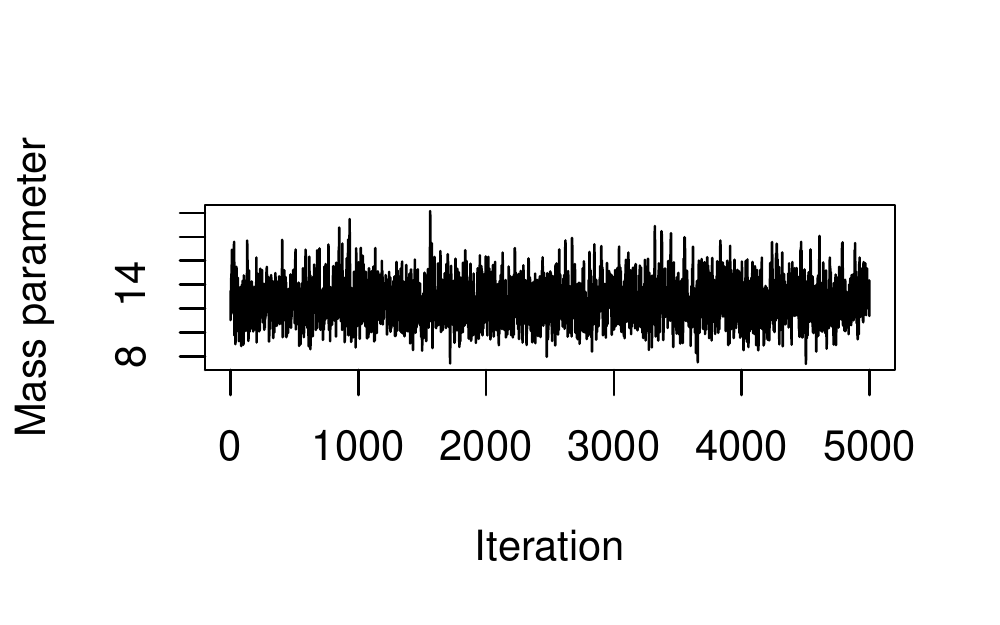}
	\vspace{-1.6cm}

	\includegraphics[width=.45\linewidth]{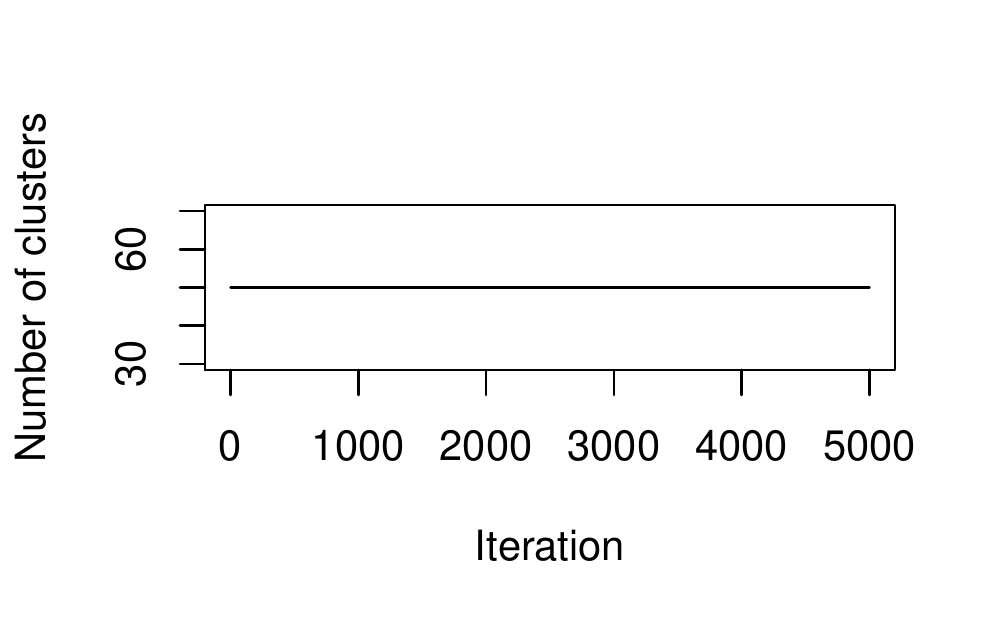}
	\includegraphics[width=.45\linewidth]{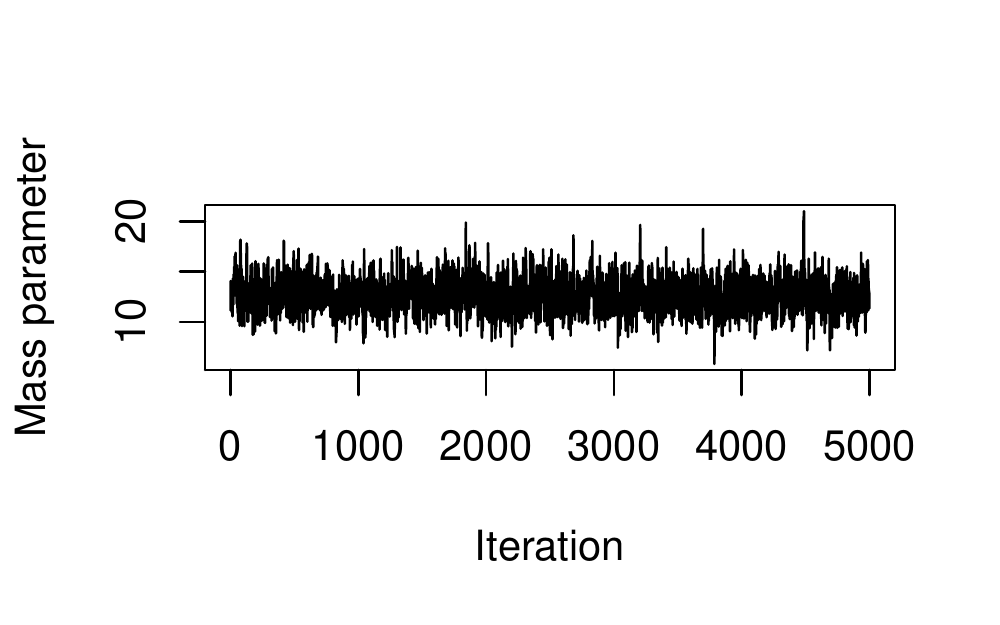}
	\vspace{-1.6cm}

	\includegraphics[width=.45\linewidth]{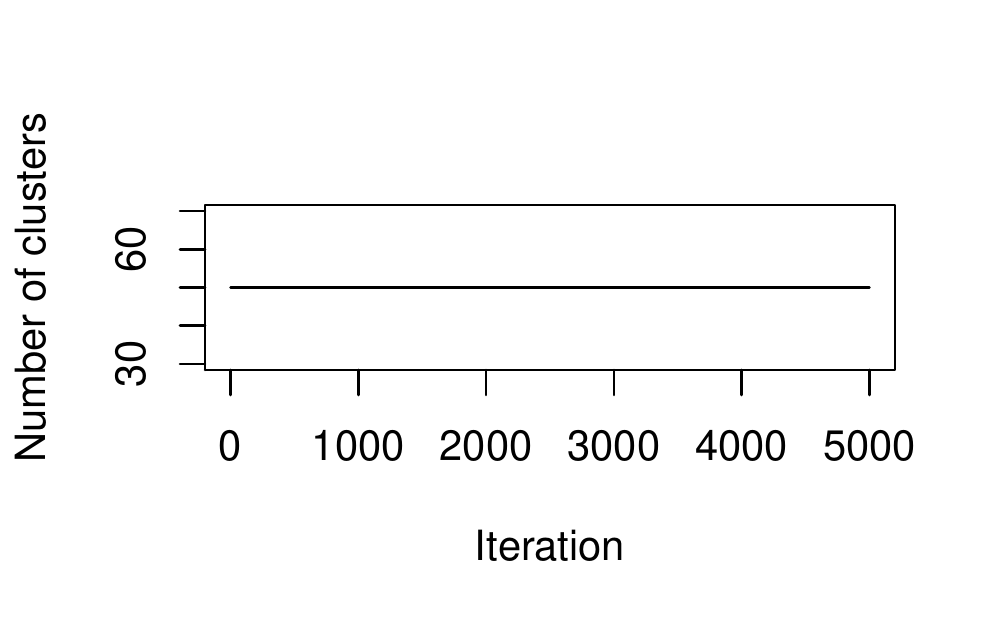}
	\includegraphics[width=.45\linewidth]{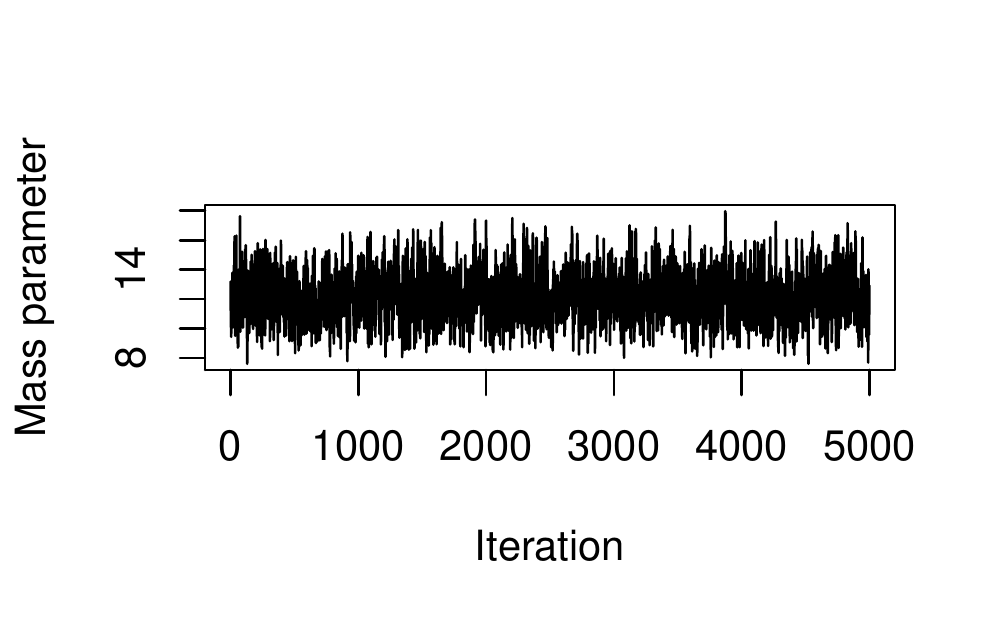}
	\caption{MCMC convergence assessment, methylation data. $\lambda=0$.}
\end{figure}

\clearpage

\begin{figure}[H]
	\centering
	\includegraphics[width=.36\linewidth]{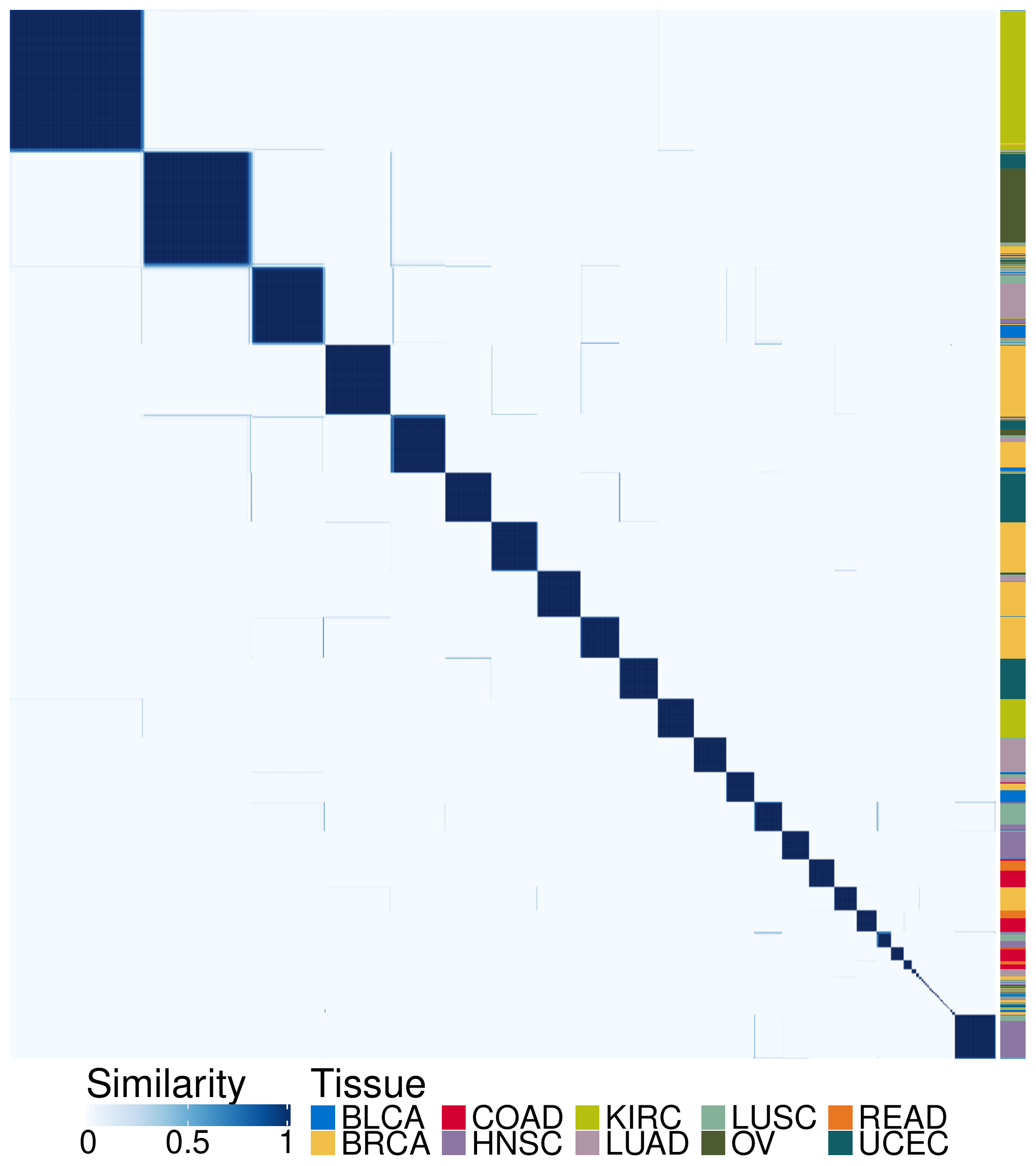}
	\includegraphics[width=.36\linewidth]{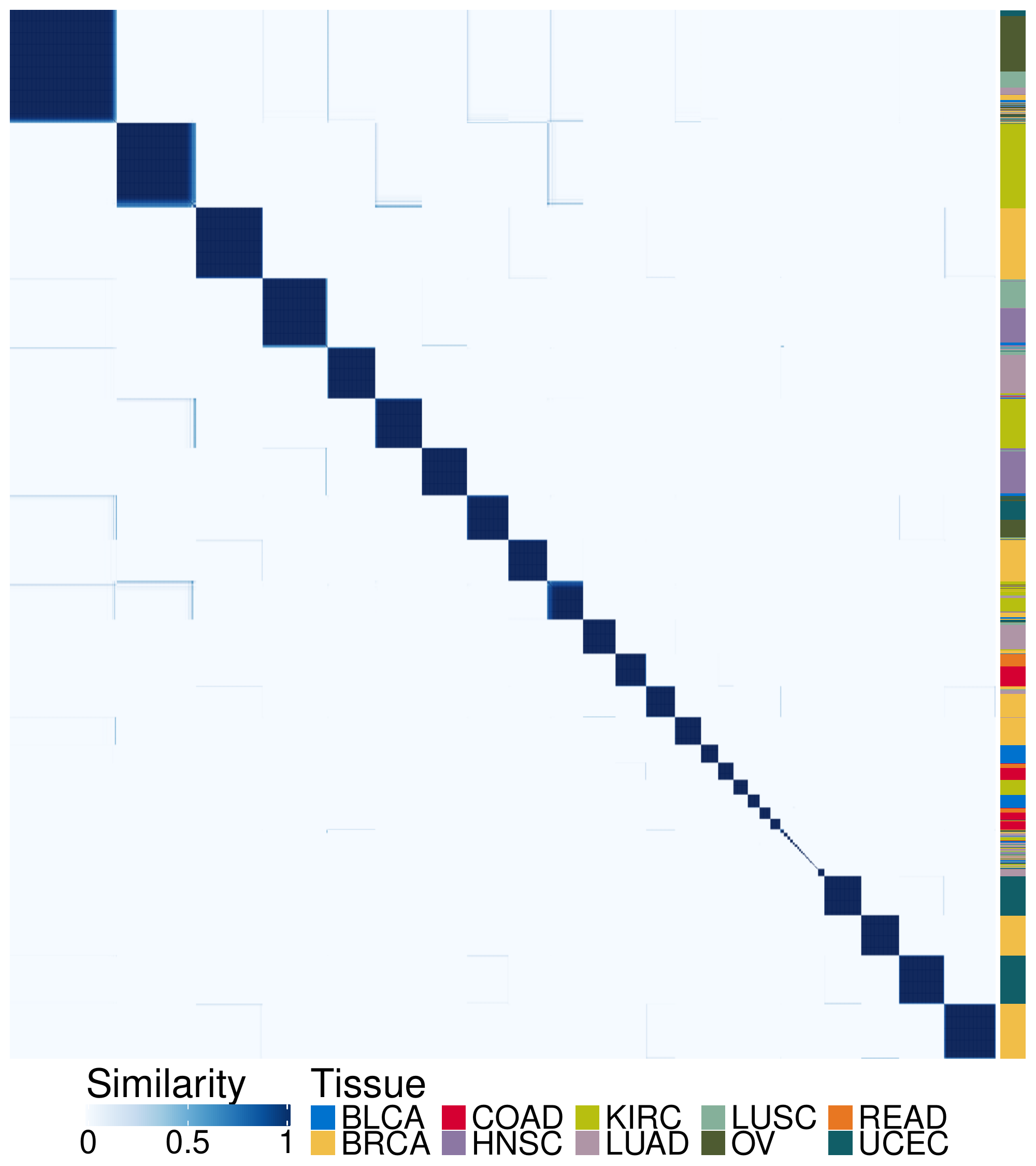}
	\includegraphics[width=.36\linewidth]{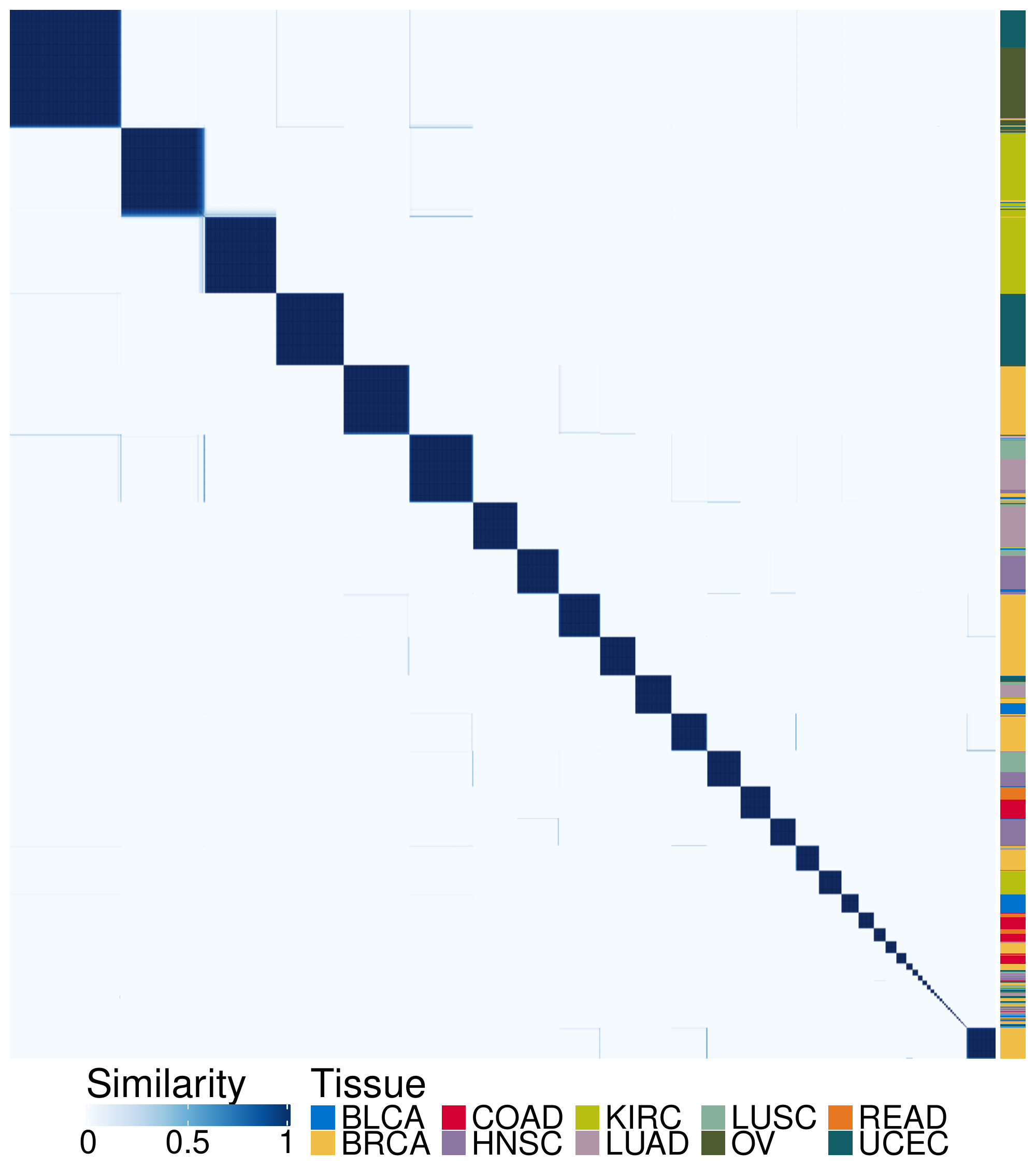}
	\includegraphics[width=.36\linewidth]{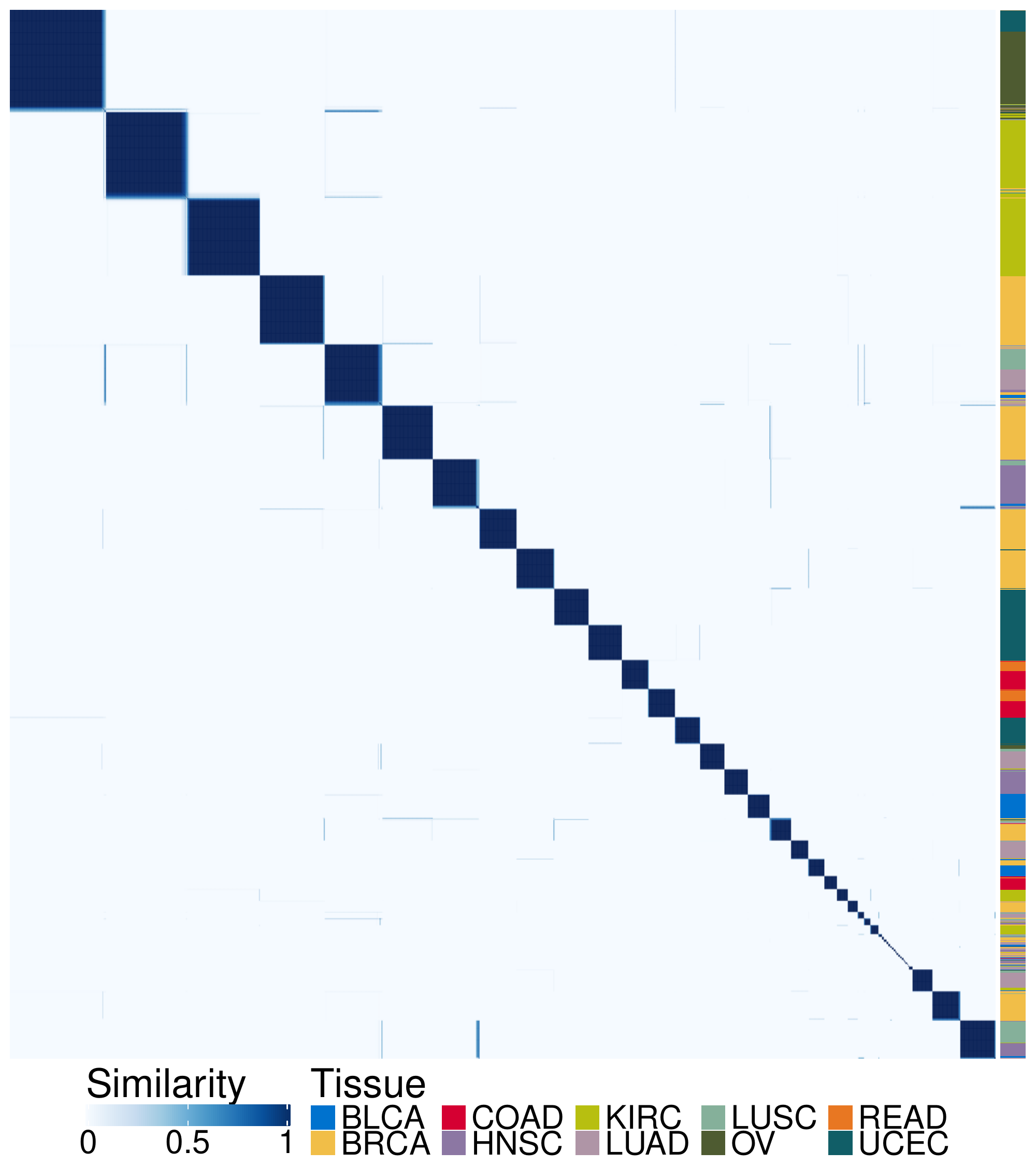}
	\includegraphics[width=.36\linewidth]{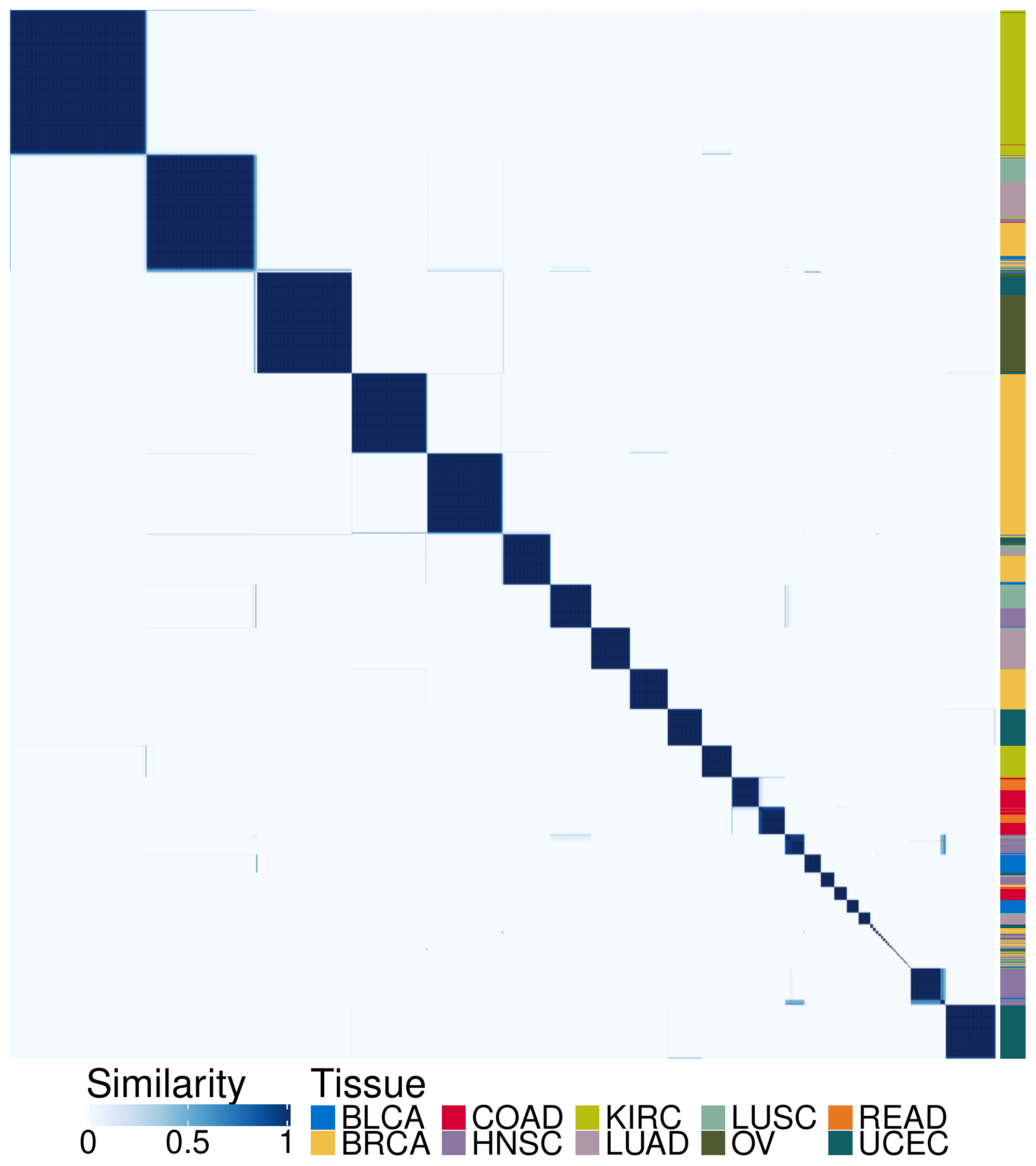}
	\includegraphics[width=.36\linewidth]{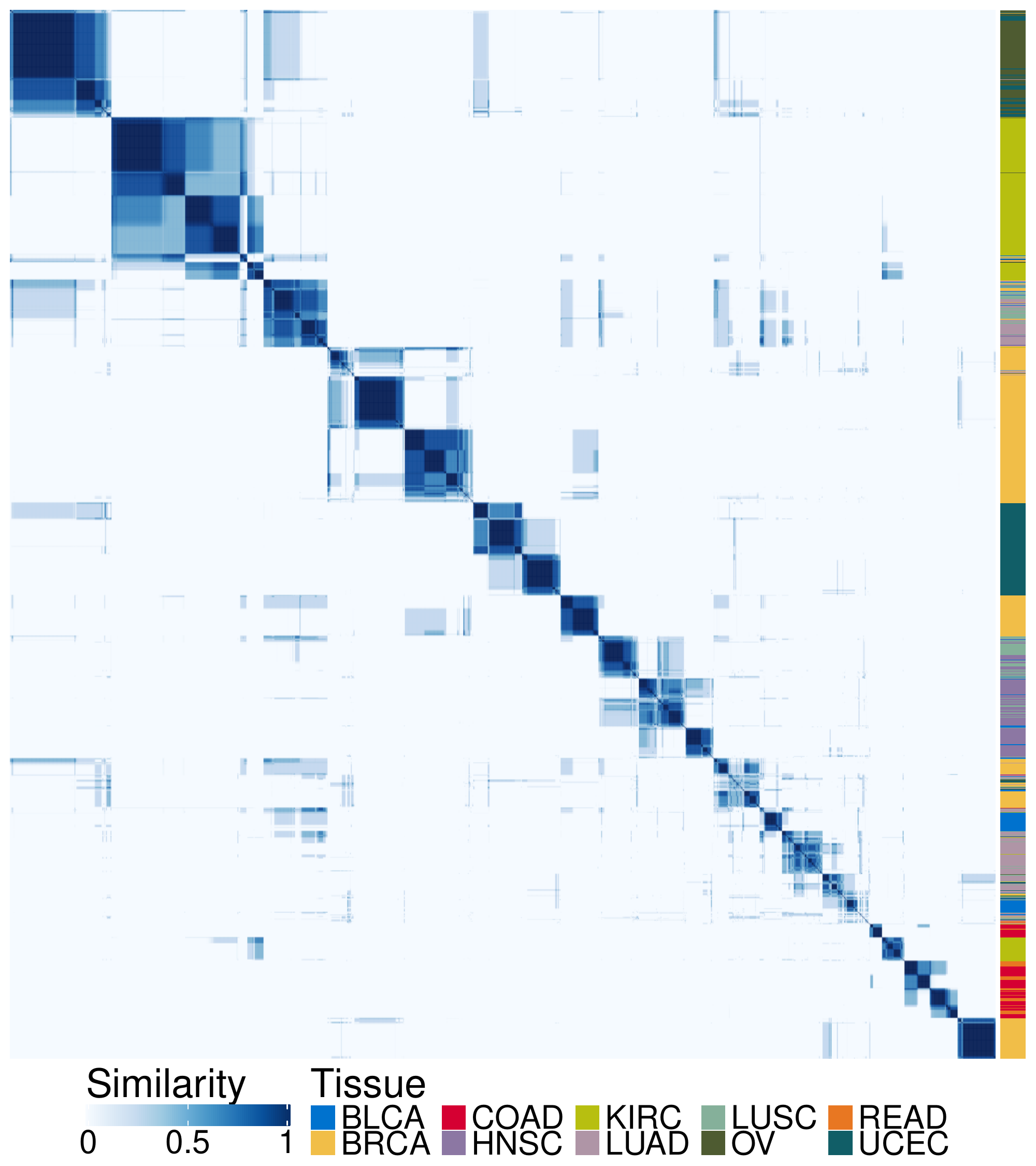}
	\caption{PSMs of the methylation data. $\alpha=0.1$.}
\end{figure}

\begin{table}[H]
\centering
\begin{tabular}{l c c c c}
& \textbf{Chain 2} & \textbf{Chain 3} & \textbf{Chain 4} & \textbf{Chain 5} \\
\hline
\textbf{Chain 1} & 0.55 & 0.53 & 0.61 & 0.64 \\
\textbf{Chain 2} &1 & 0.48 & 0.57 & 0.49 \\
\textbf{Chain 3} && 1 &  0.71 & 0.55 \\
\textbf{Chain 4} && & 1 & 0.57 \\
\hline\\
\end{tabular}
\caption{ARI between the clusterings found on the PSMs of different chains with the number of clusters that maximises the silhouette.}
\end{table} 

\begin{figure}[H]
	\centering
	\includegraphics[width=.45\linewidth]{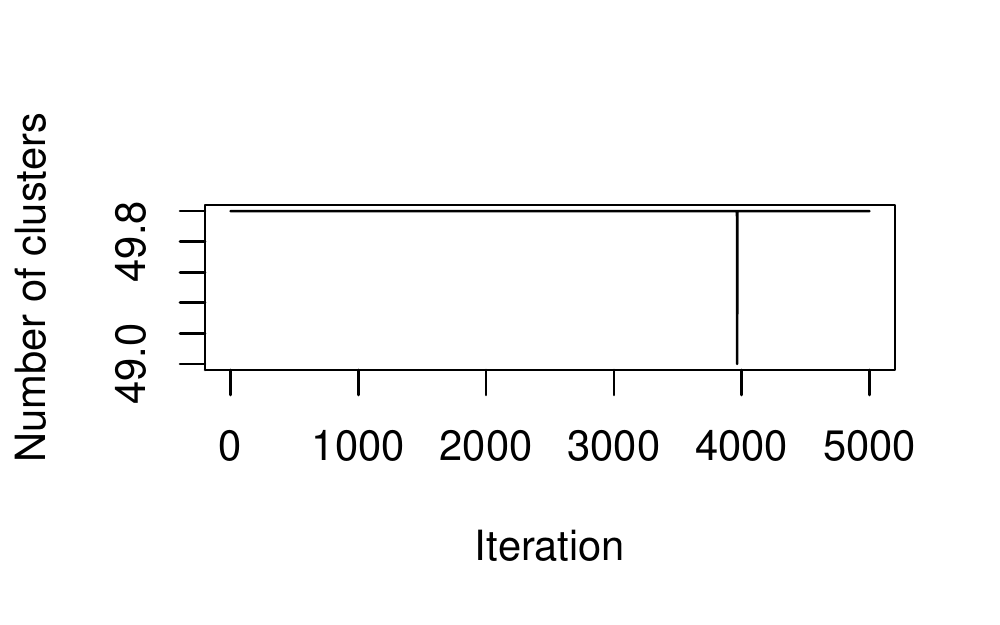}
	\includegraphics[width=.45\linewidth]{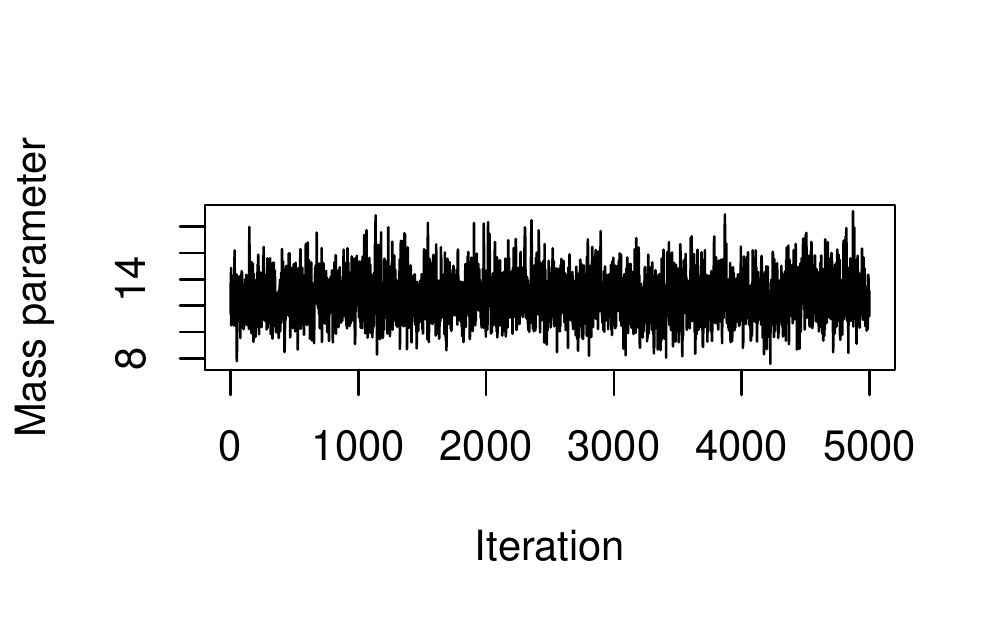}
	\vspace{-1.6cm}
		
	\includegraphics[width=.45\linewidth]{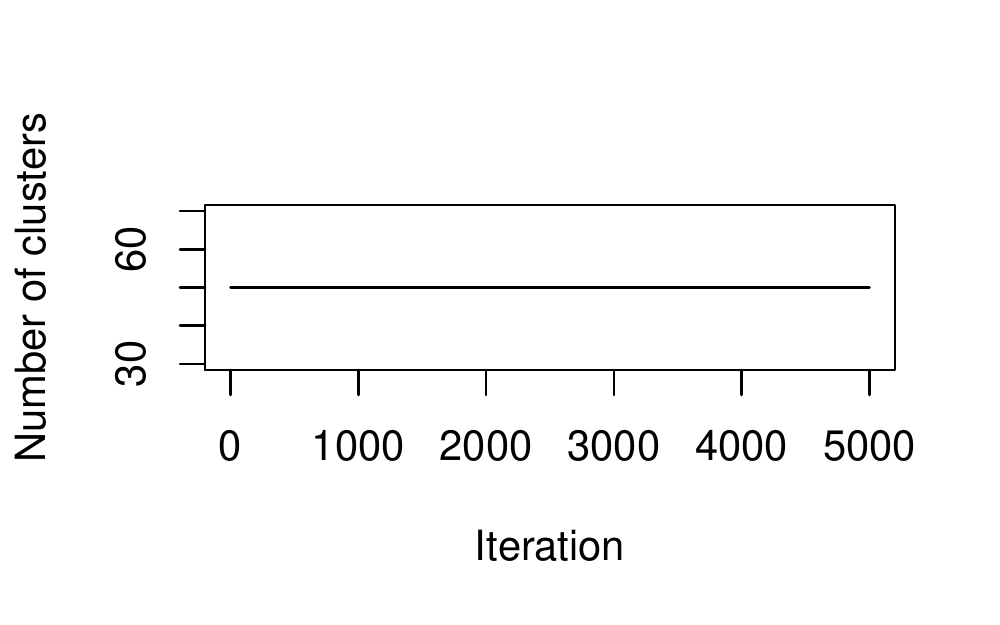}
	\includegraphics[width=.45\linewidth]{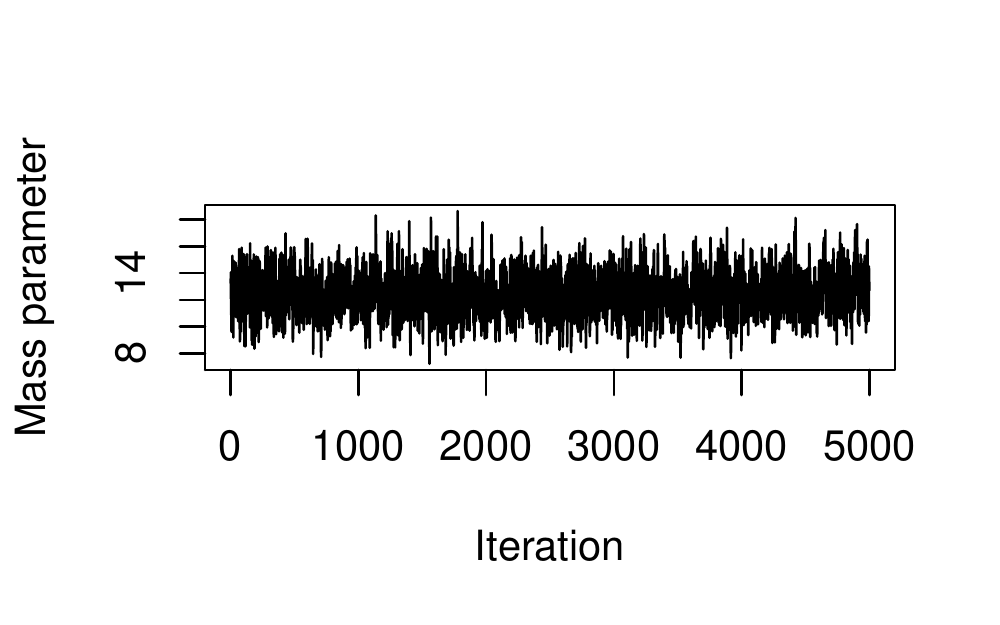}
	\vspace{-1.6cm}

	\includegraphics[width=.45\linewidth]{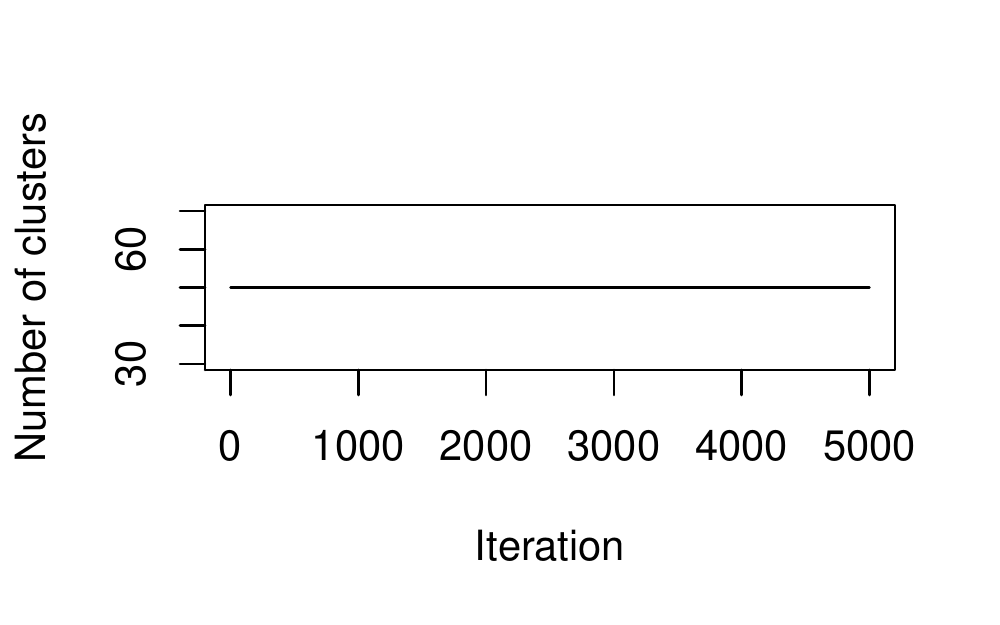}
	\includegraphics[width=.45\linewidth]{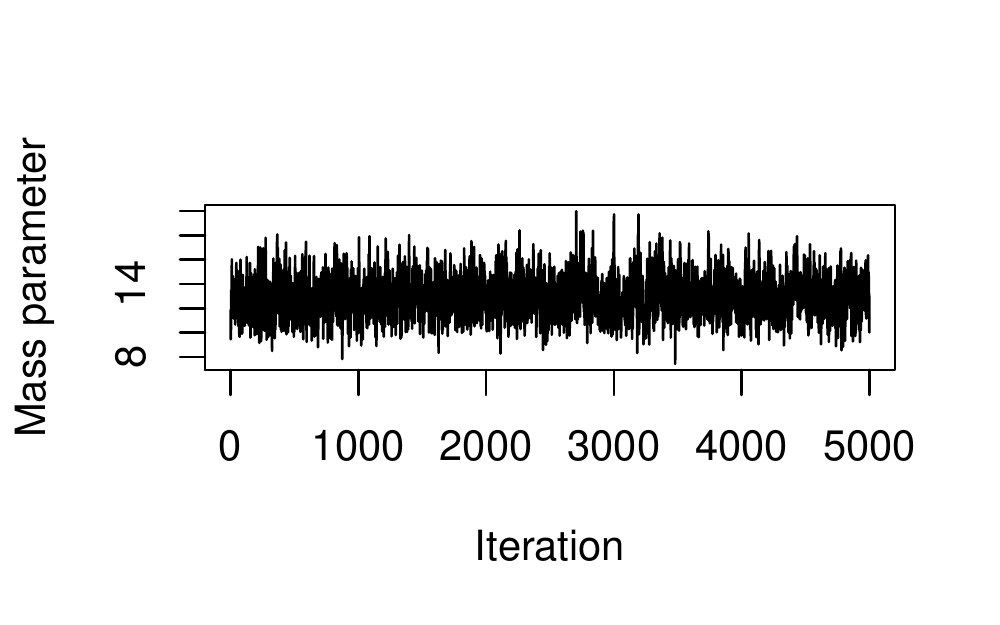}
	\vspace{-1.6cm}

	\includegraphics[width=.45\linewidth]{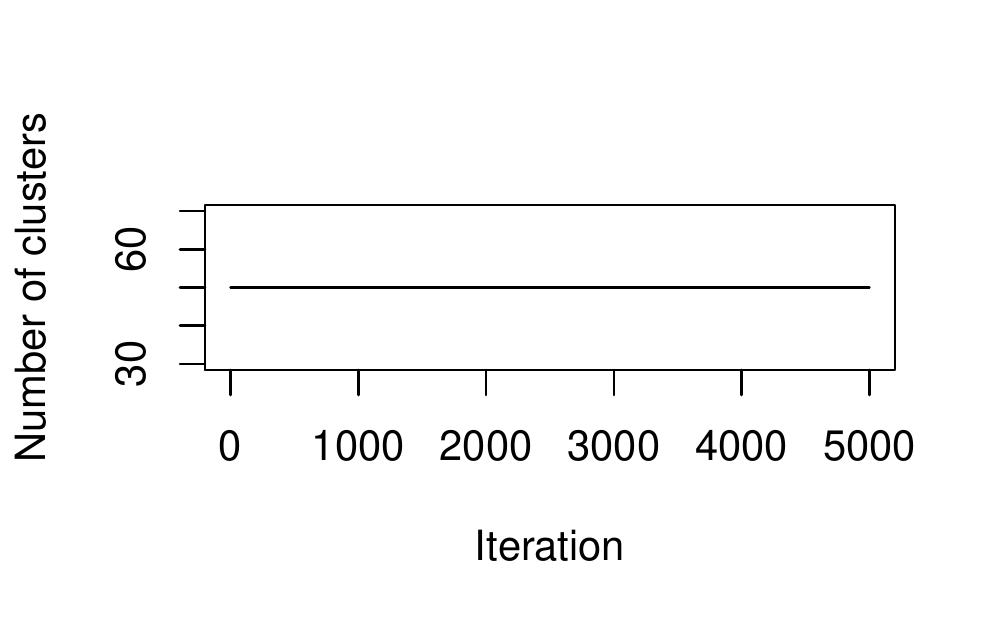}
	\includegraphics[width=.45\linewidth]{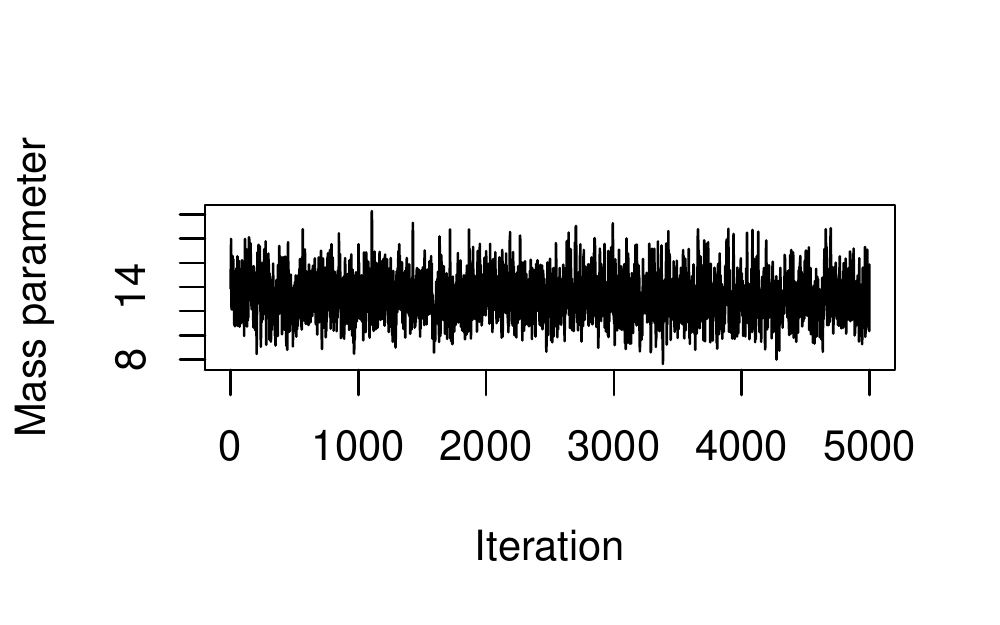}
	\vspace{-1.6cm}

	\includegraphics[width=.45\linewidth]{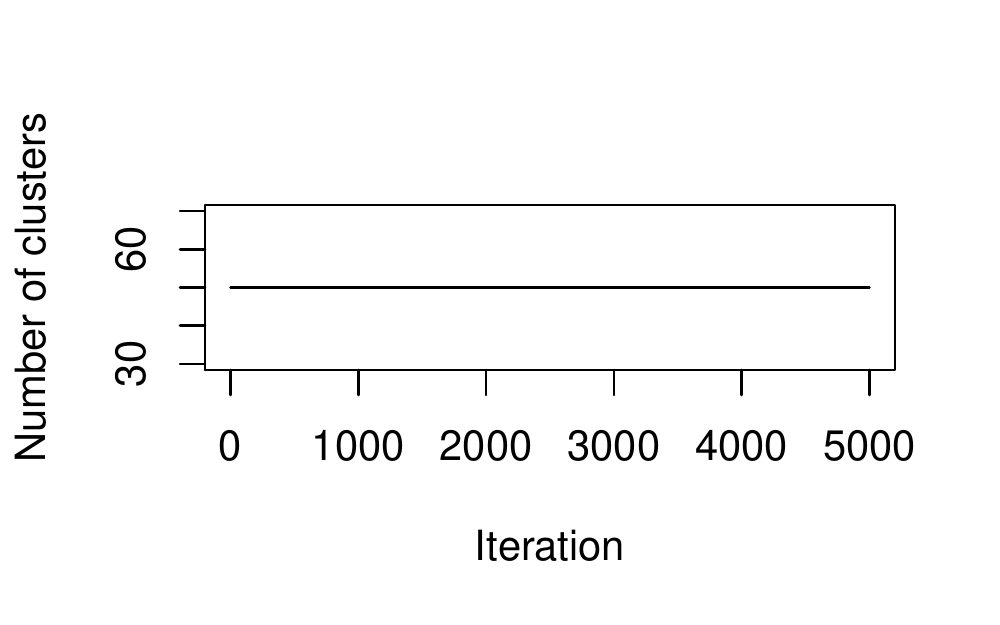}
	\includegraphics[width=.45\linewidth]{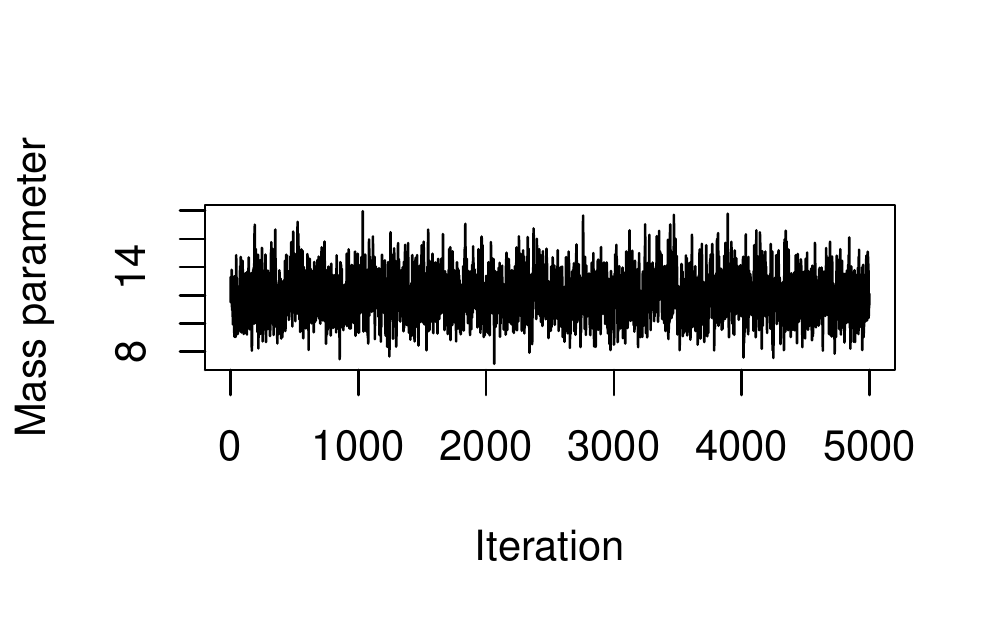}
	\caption{MCMC convergence assessment, methylation data. $\alpha=0.1$}
\end{figure}

\clearpage

\begin{figure}[H]
	\centering
	\includegraphics[width=.36\linewidth]{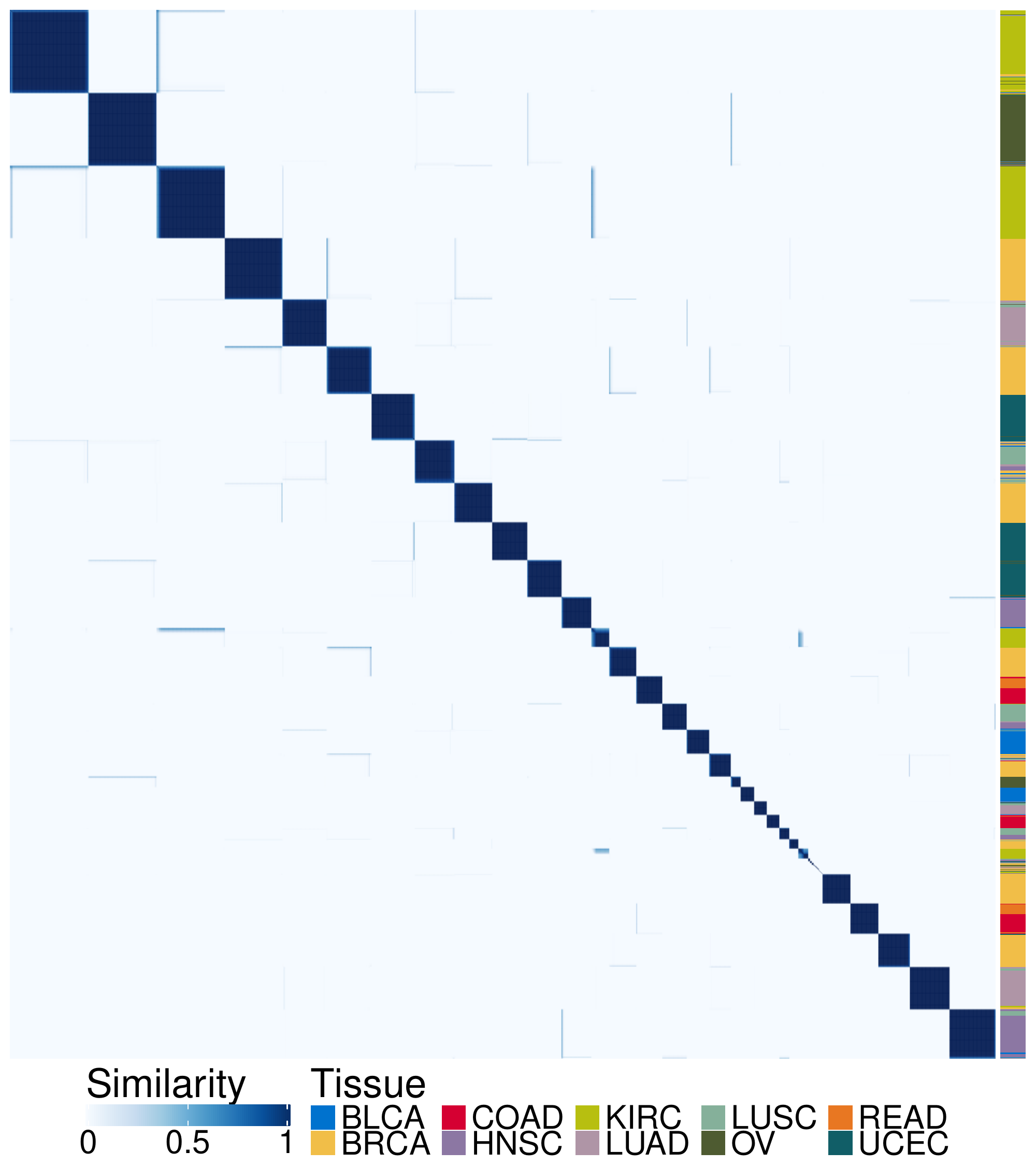}
	\includegraphics[width=.36\linewidth]{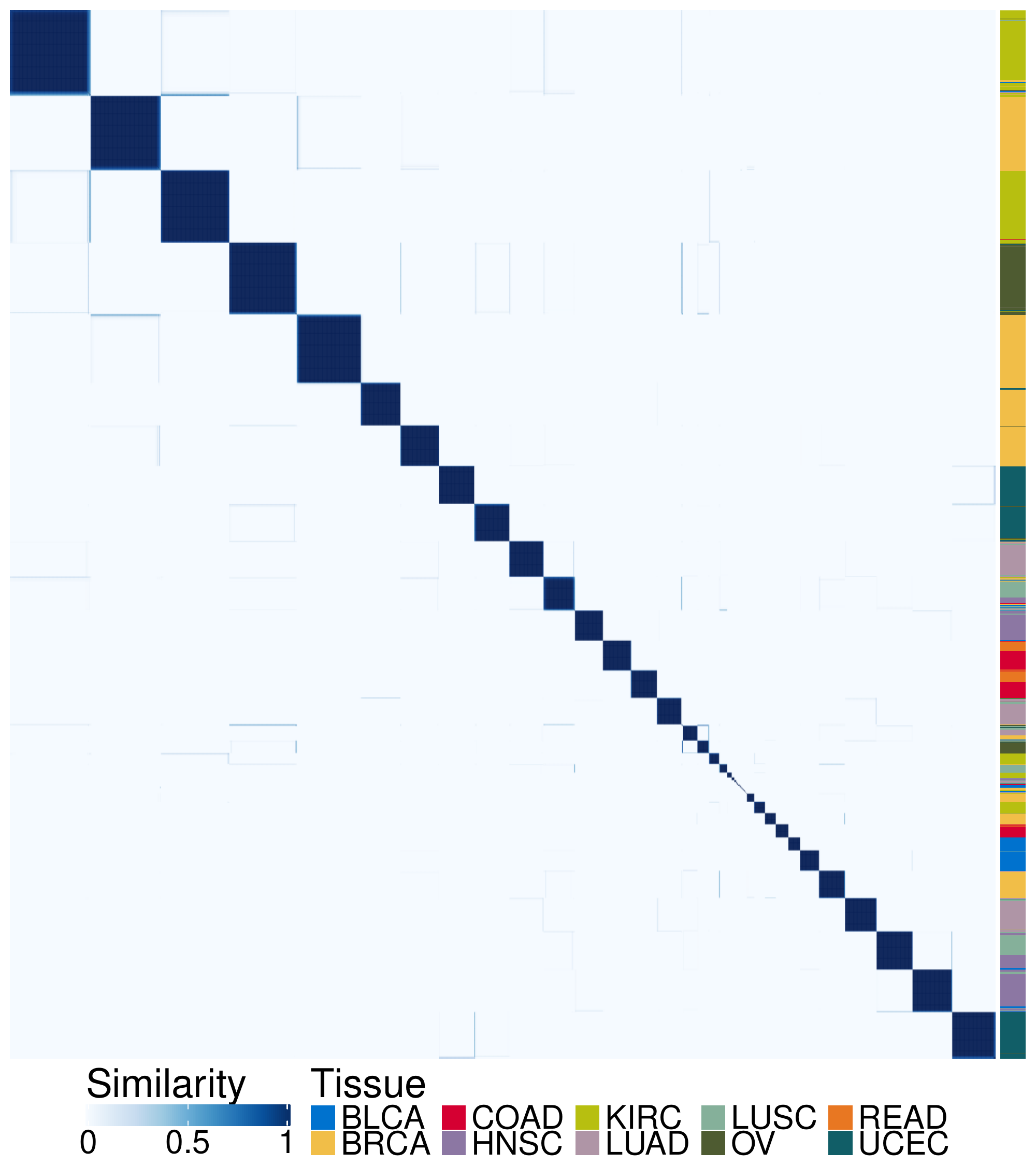}
	\includegraphics[width=.36\linewidth]{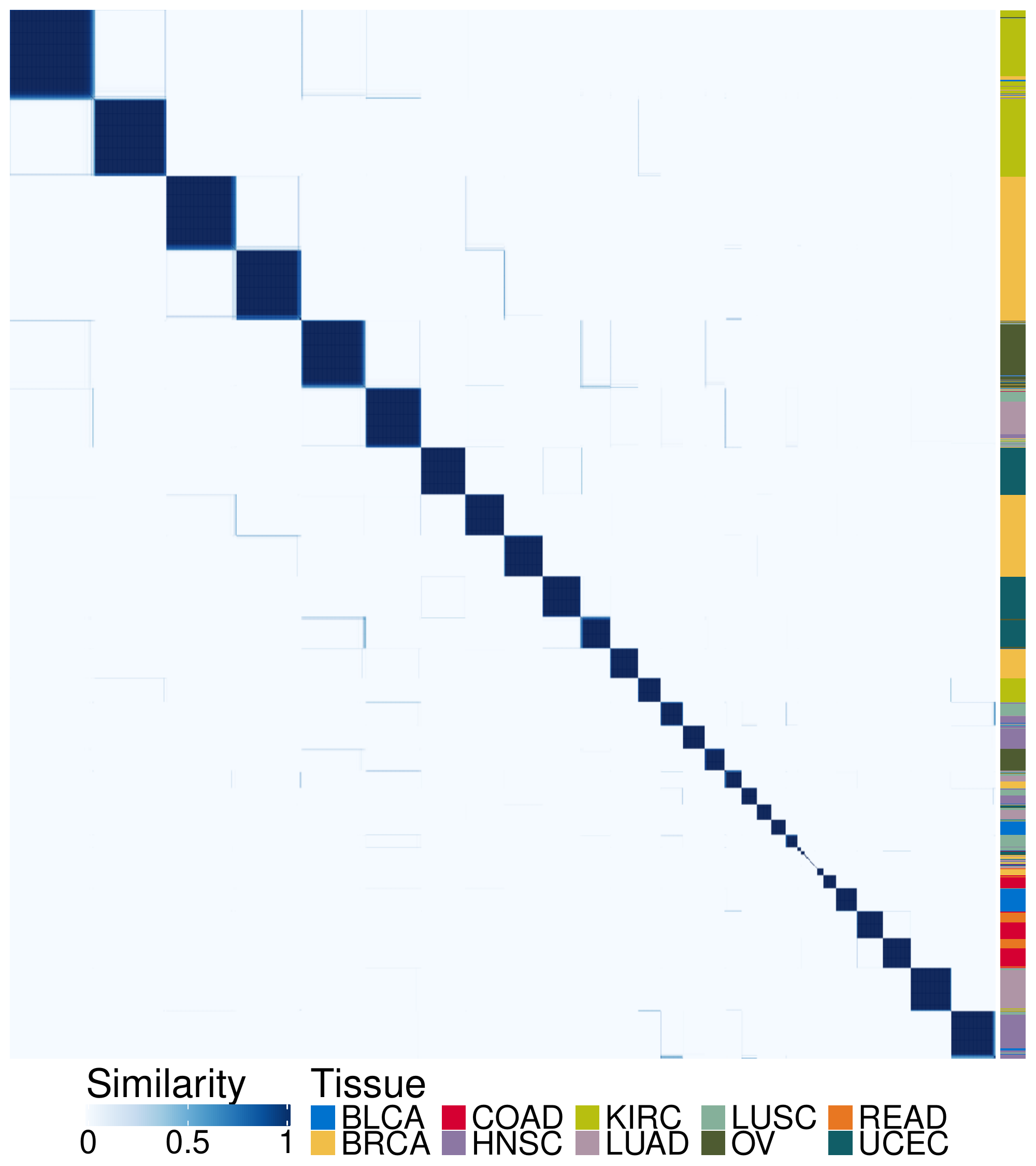}
	\includegraphics[width=.36\linewidth]{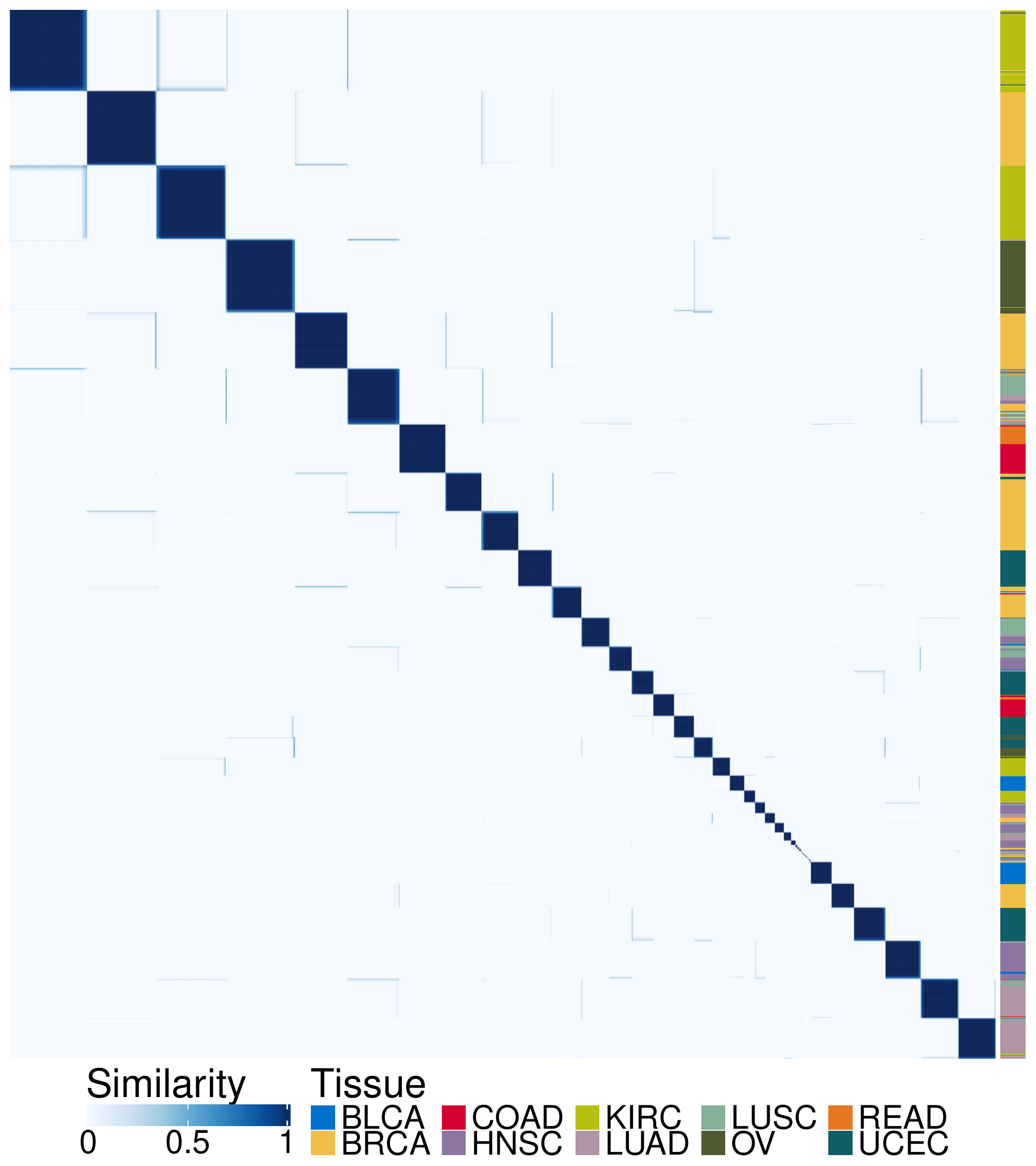}
	\includegraphics[width=.36\linewidth]{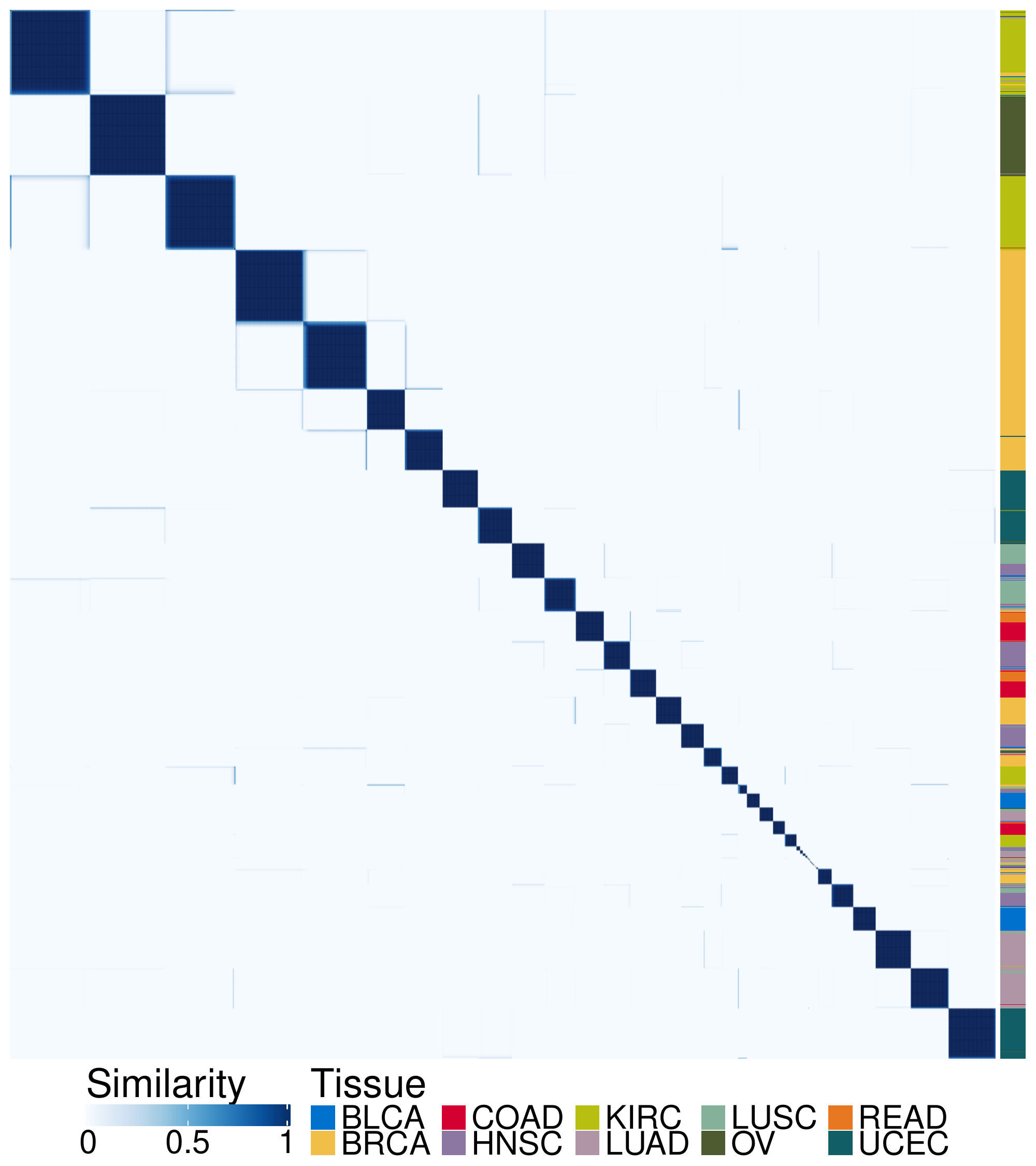}
	\includegraphics[width=.36\linewidth]{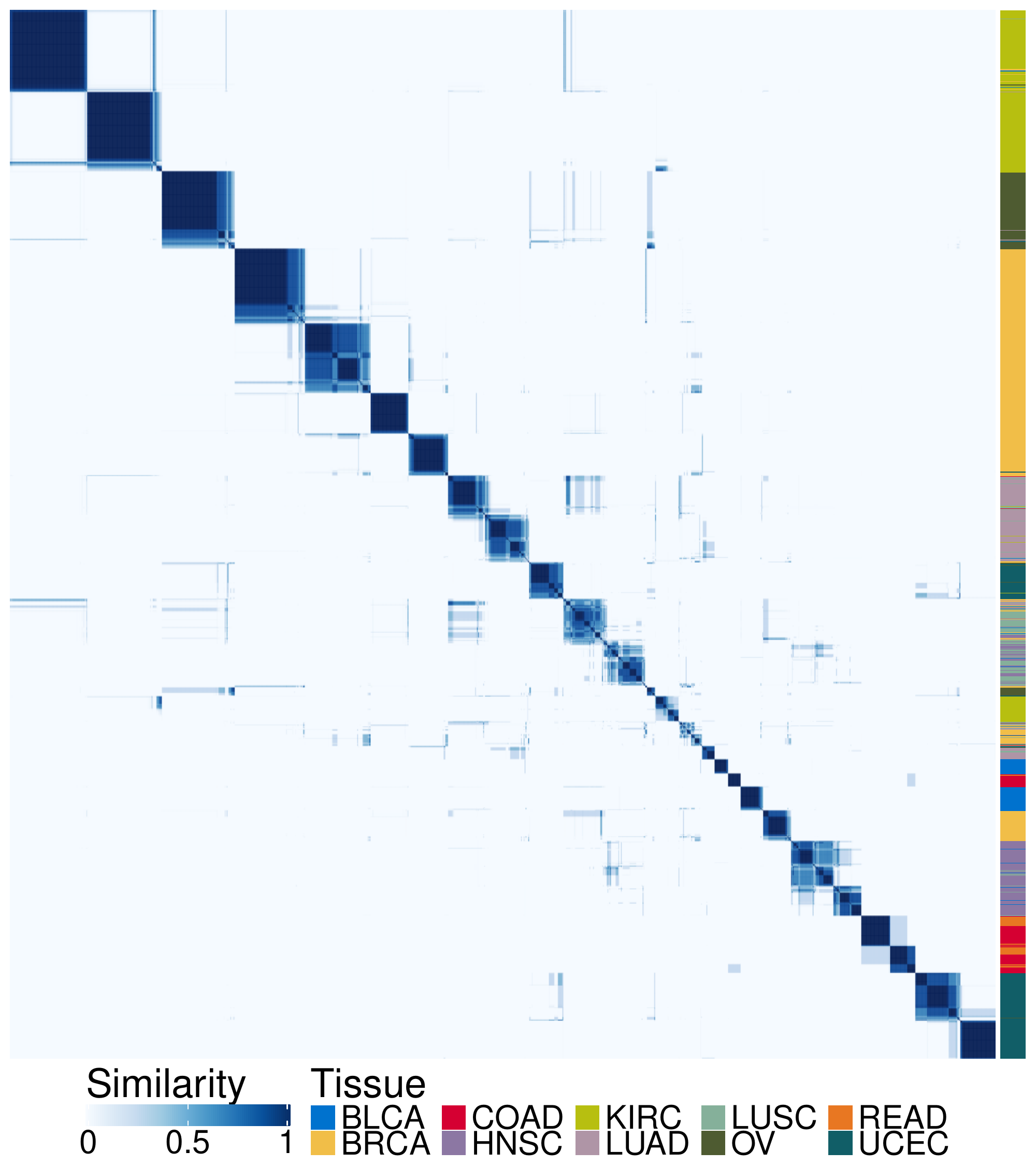}
	\caption{PSMs of the methylation data. $\alpha=0.5$.}
\end{figure}

\begin{table}[H]
\centering
\begin{tabular}{l c c c c}
& \textbf{Chain 2} & \textbf{Chain 3} & \textbf{Chain 4} & \textbf{Chain 5} \\
\hline
\textbf{Chain 1} & 0.79 & 0.74 & 0.74 & 0.78 \\
\textbf{Chain 2} &1 & 0.81 & 0.73 & 0.82 \\
\textbf{Chain 3} && 1 &  0.70 & 0.77 \\
\textbf{Chain 4} && & 1 & 0.75 \\
\hline\\
\end{tabular}
\caption{ARI between the clusterings found on the PSMs of different chains with the number of clusters that maximises the silhouette.}
\end{table} 

\begin{figure}[H]
	\centering
	\includegraphics[width=.45\linewidth]{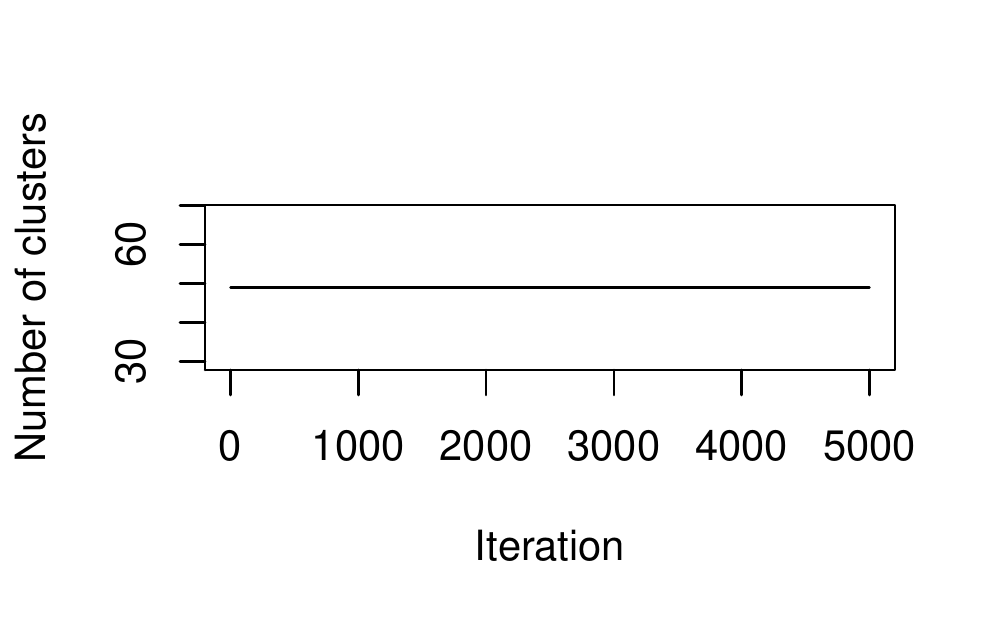}
	\includegraphics[width=.45\linewidth]{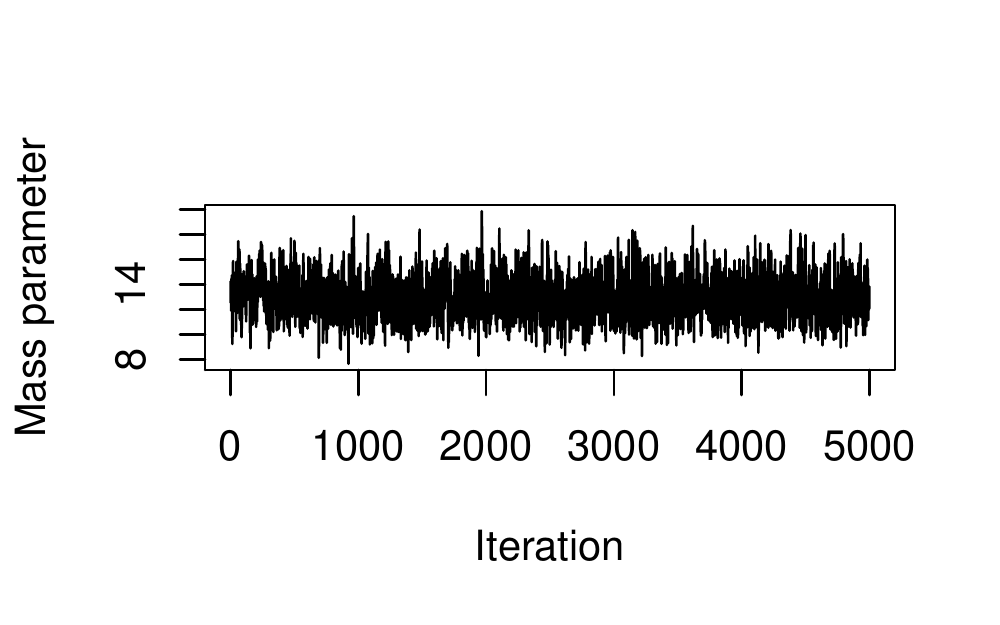}
	\vspace{-1.6cm}

	\includegraphics[width=.45\linewidth]{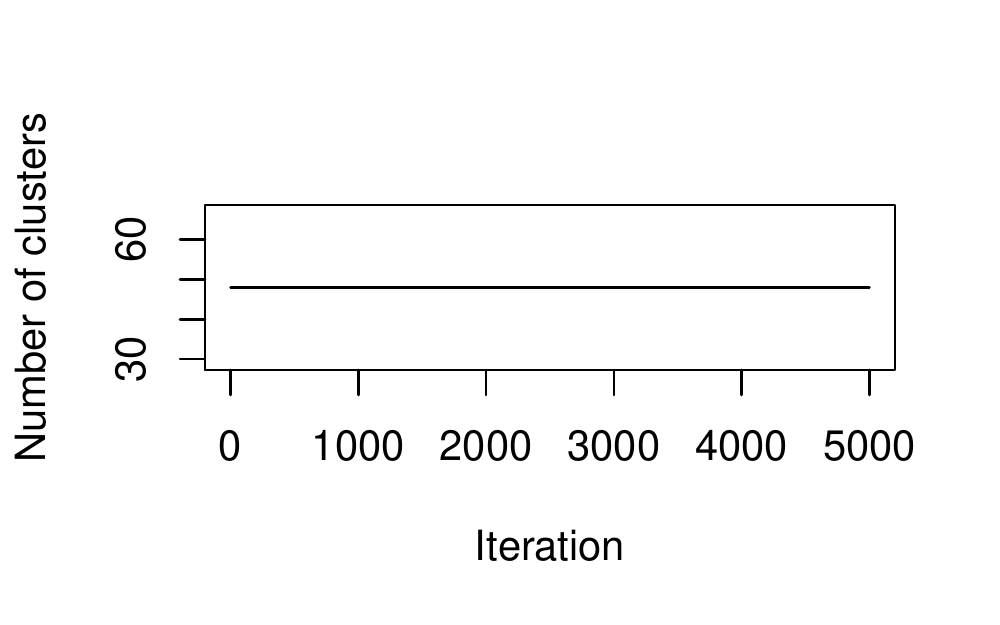}
	\includegraphics[width=.45\linewidth]{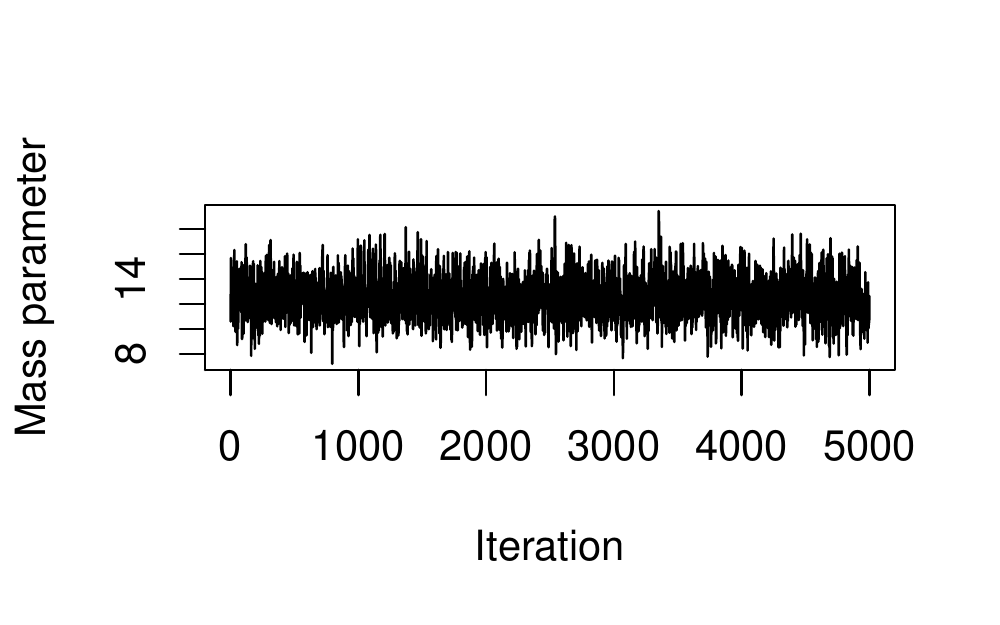}
	\vspace{-1.6cm}

	\includegraphics[width=.45\linewidth]{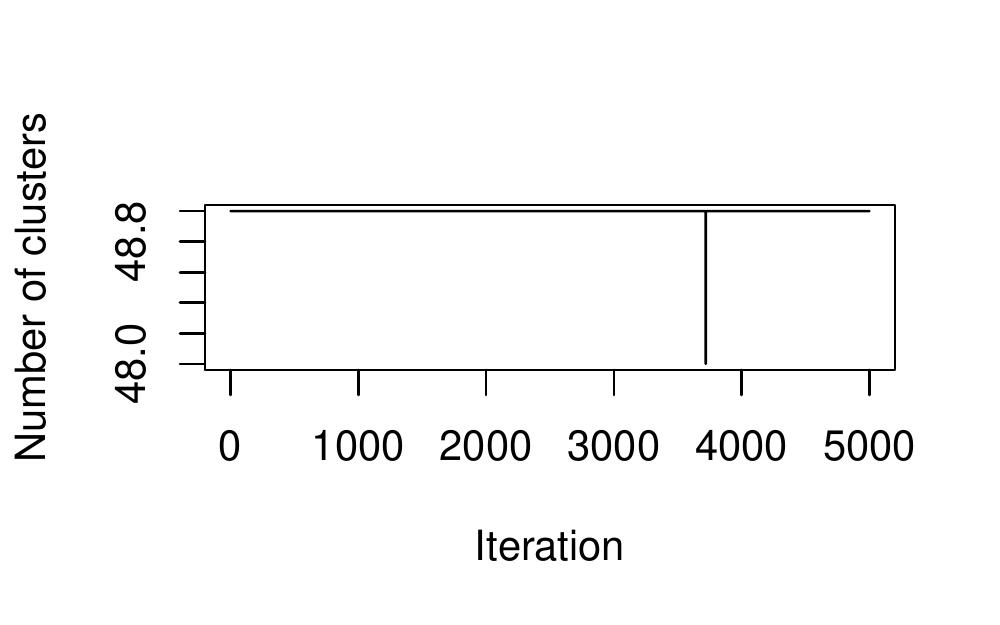}
	\includegraphics[width=.45\linewidth]{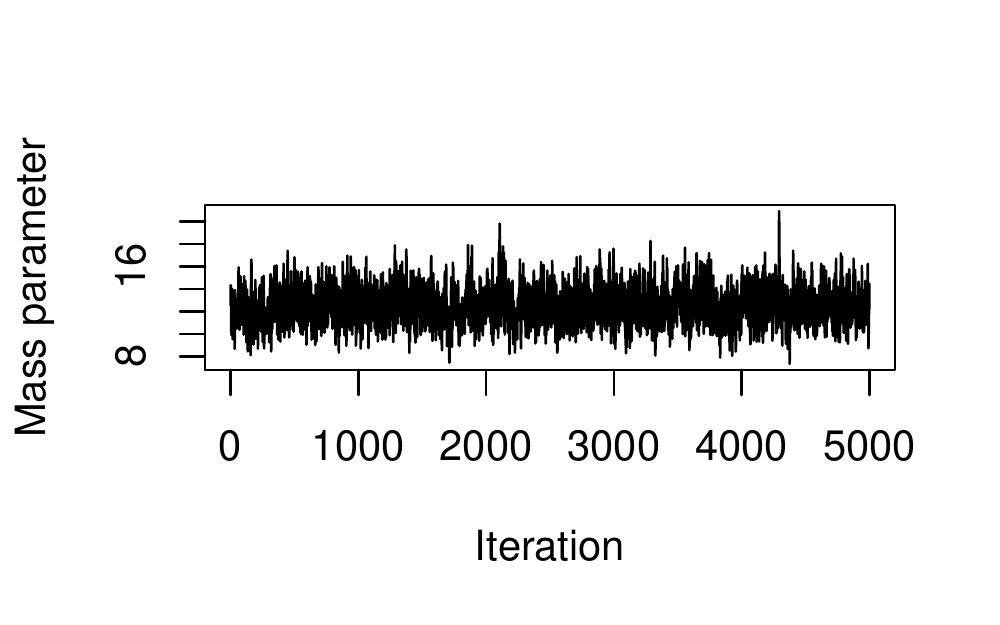}
	\vspace{-1.6cm}

	\includegraphics[width=.45\linewidth]{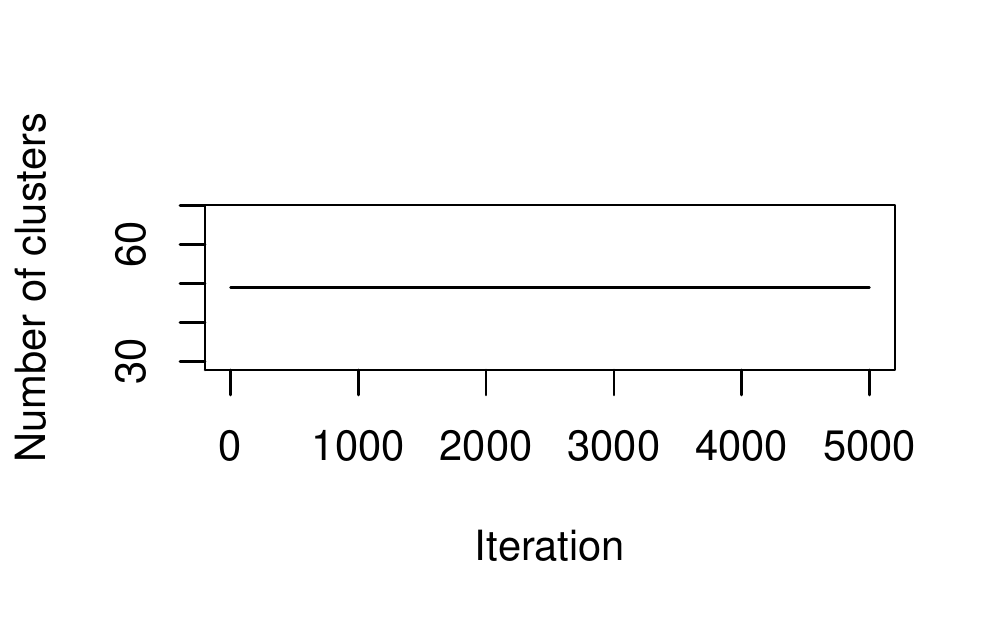}
	\includegraphics[width=.45\linewidth]{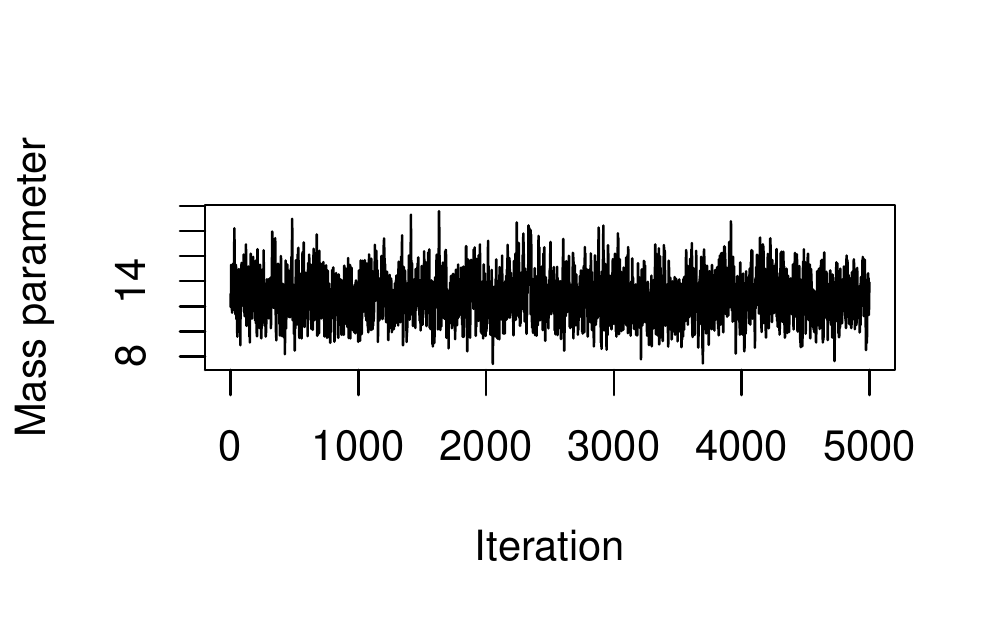}
	\vspace{-1.6cm}

	\includegraphics[width=.45\linewidth]{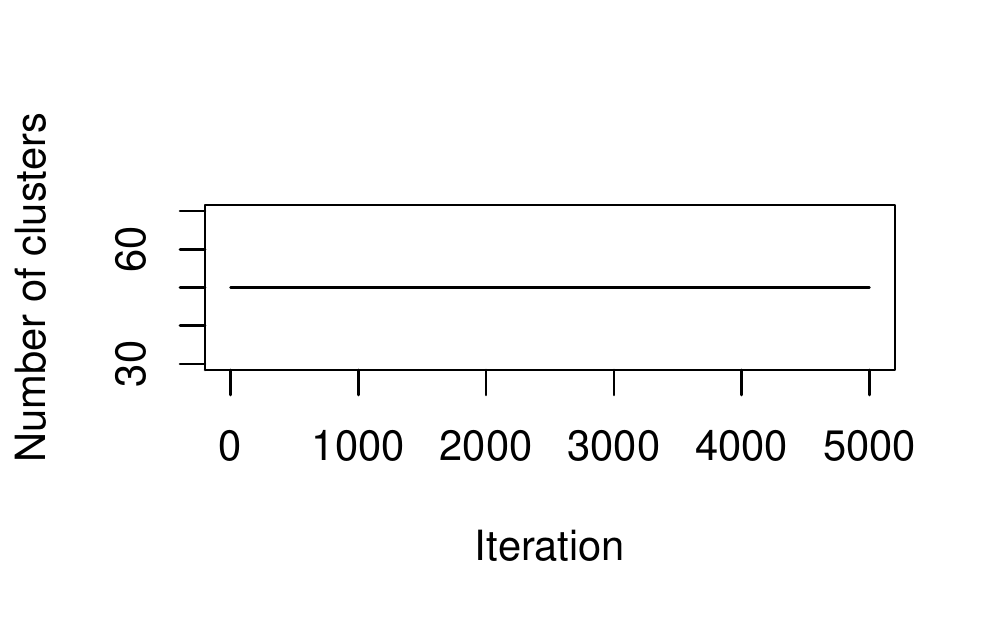}
	\includegraphics[width=.45\linewidth]{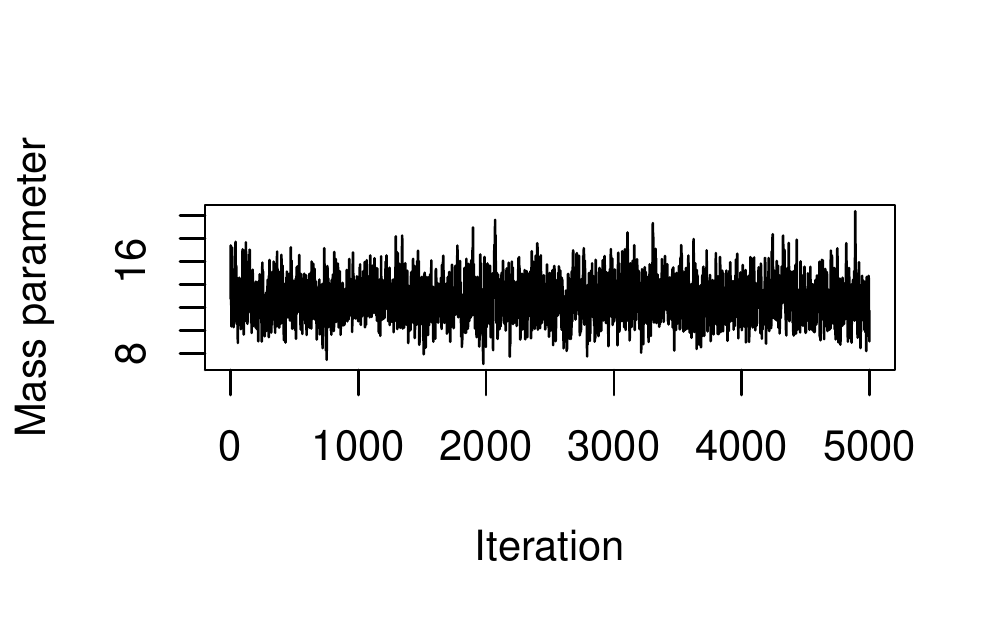}
	\caption{MCMC convergence assessment, methylation data.}
\end{figure}

\clearpage

\begin{figure}[H]
	\centering
	\includegraphics[width=.36\linewidth]{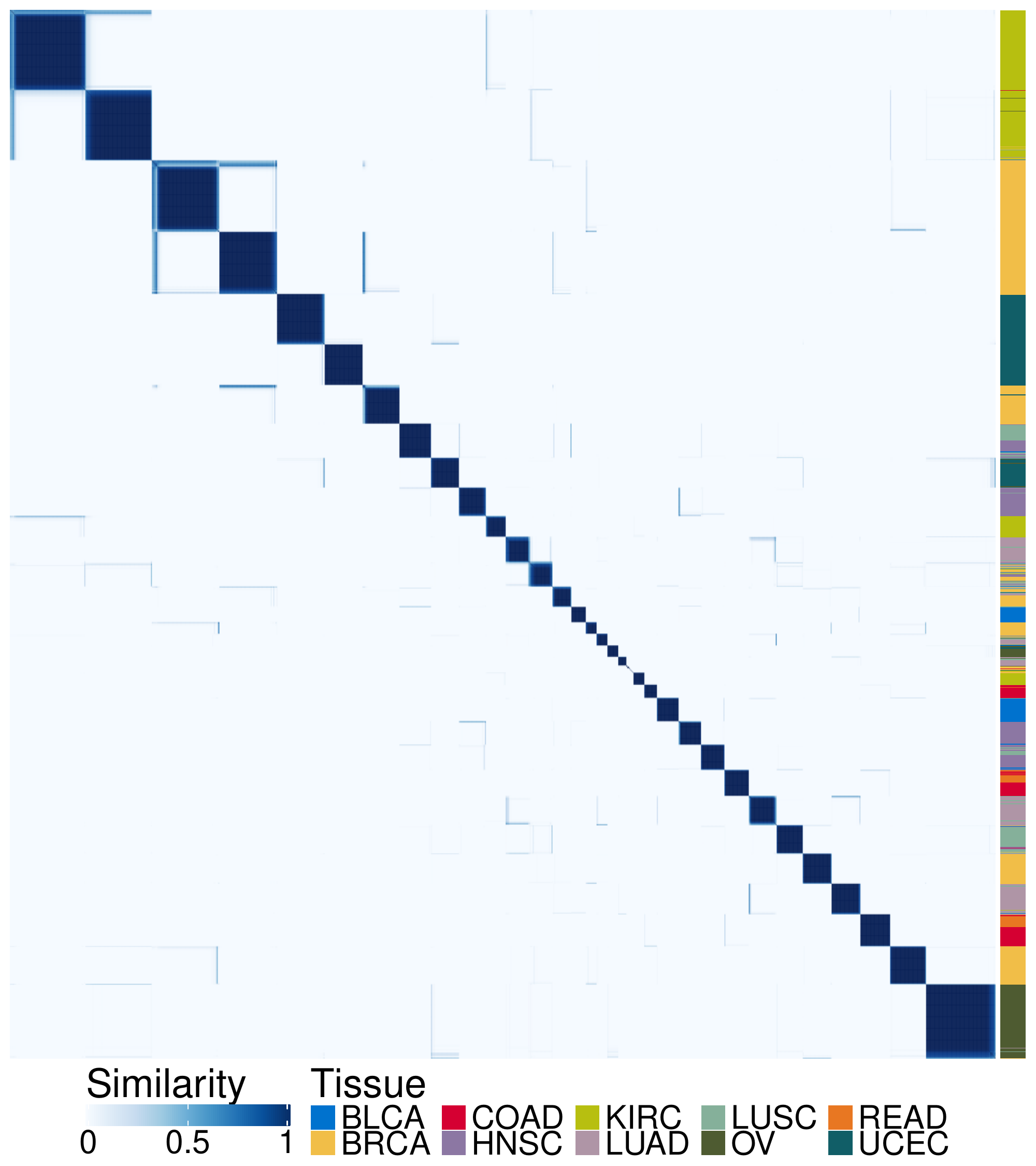}
	\includegraphics[width=.36\linewidth]{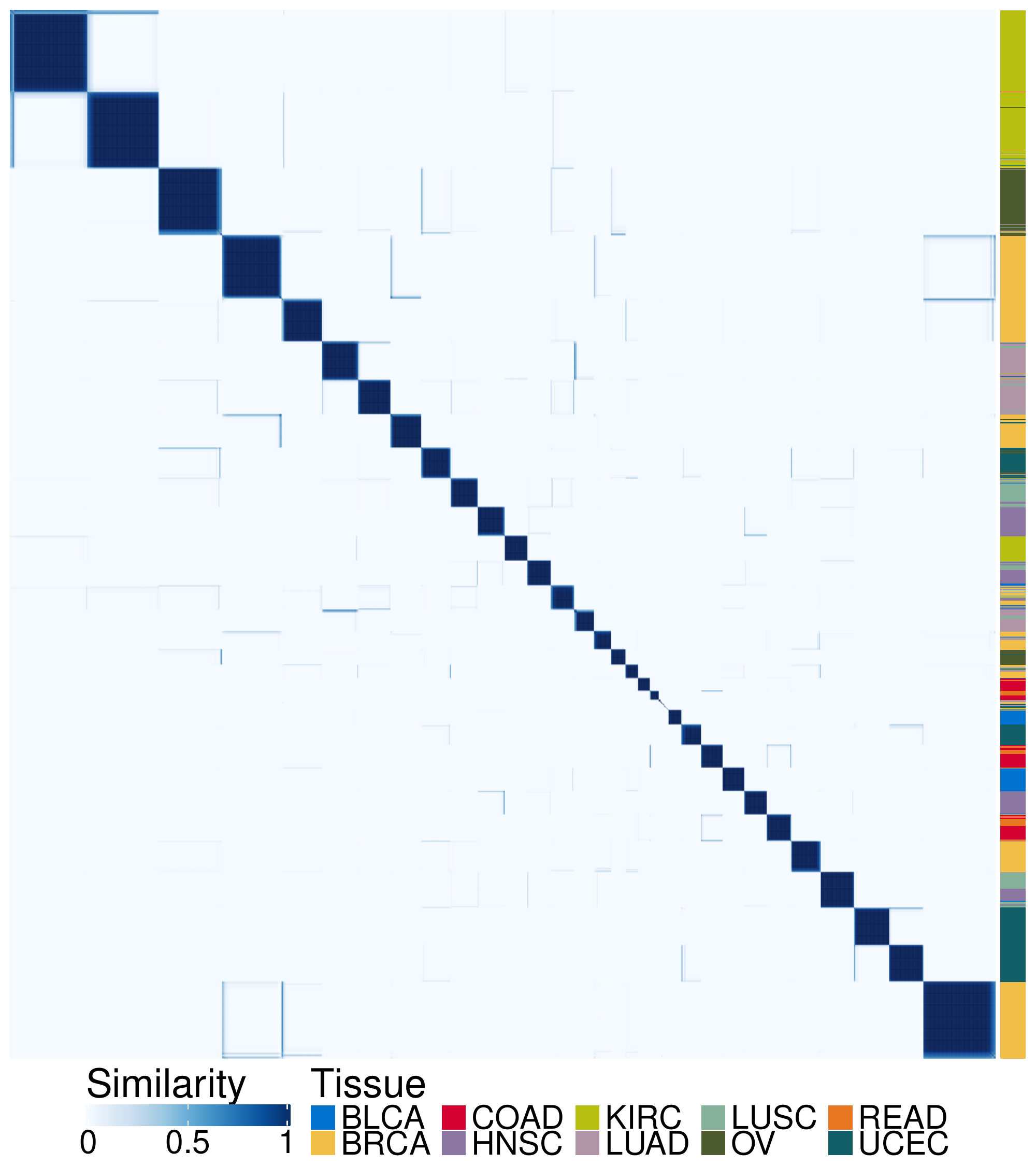}
	\includegraphics[width=.36\linewidth]{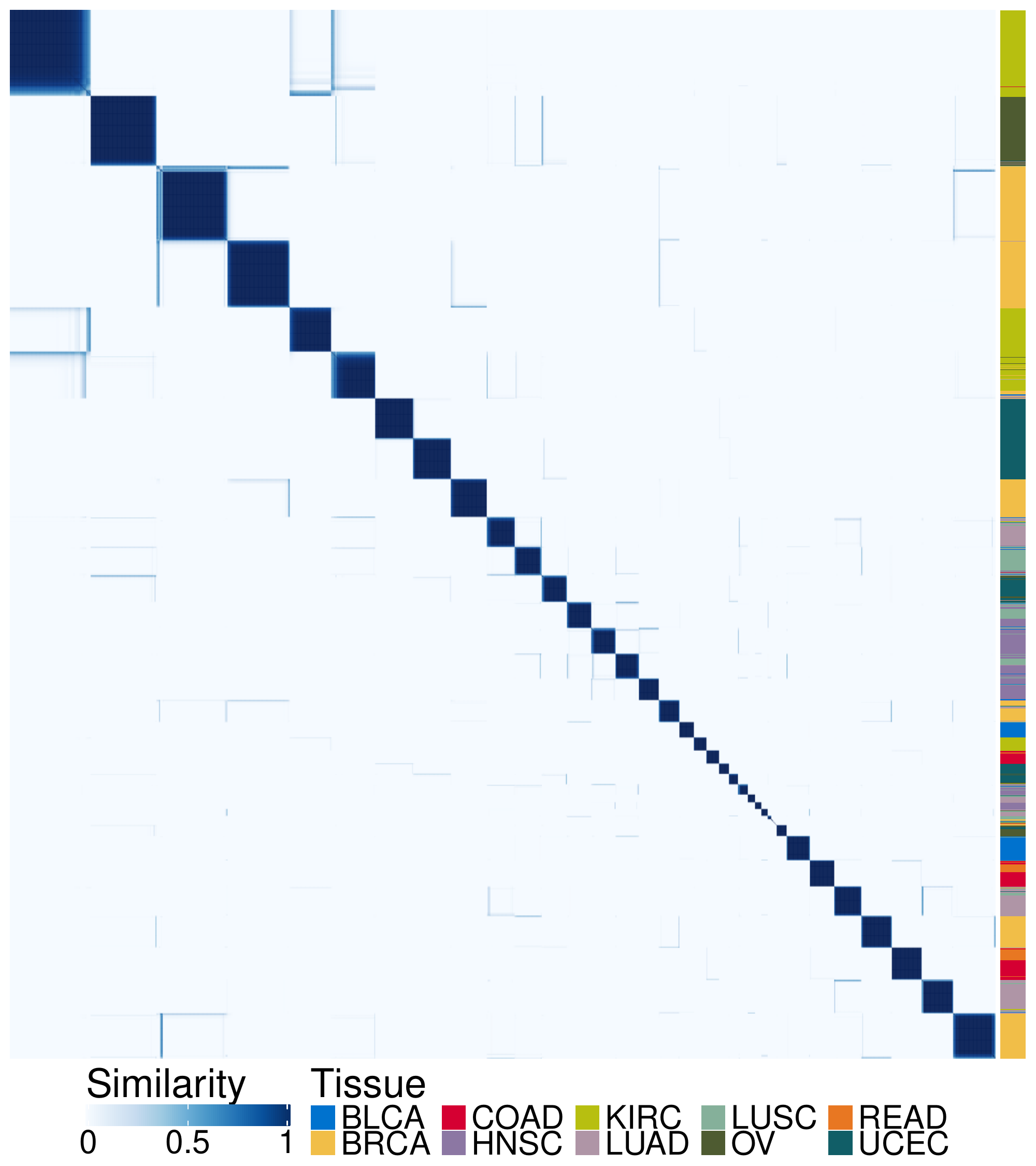}
	\includegraphics[width=.36\linewidth]{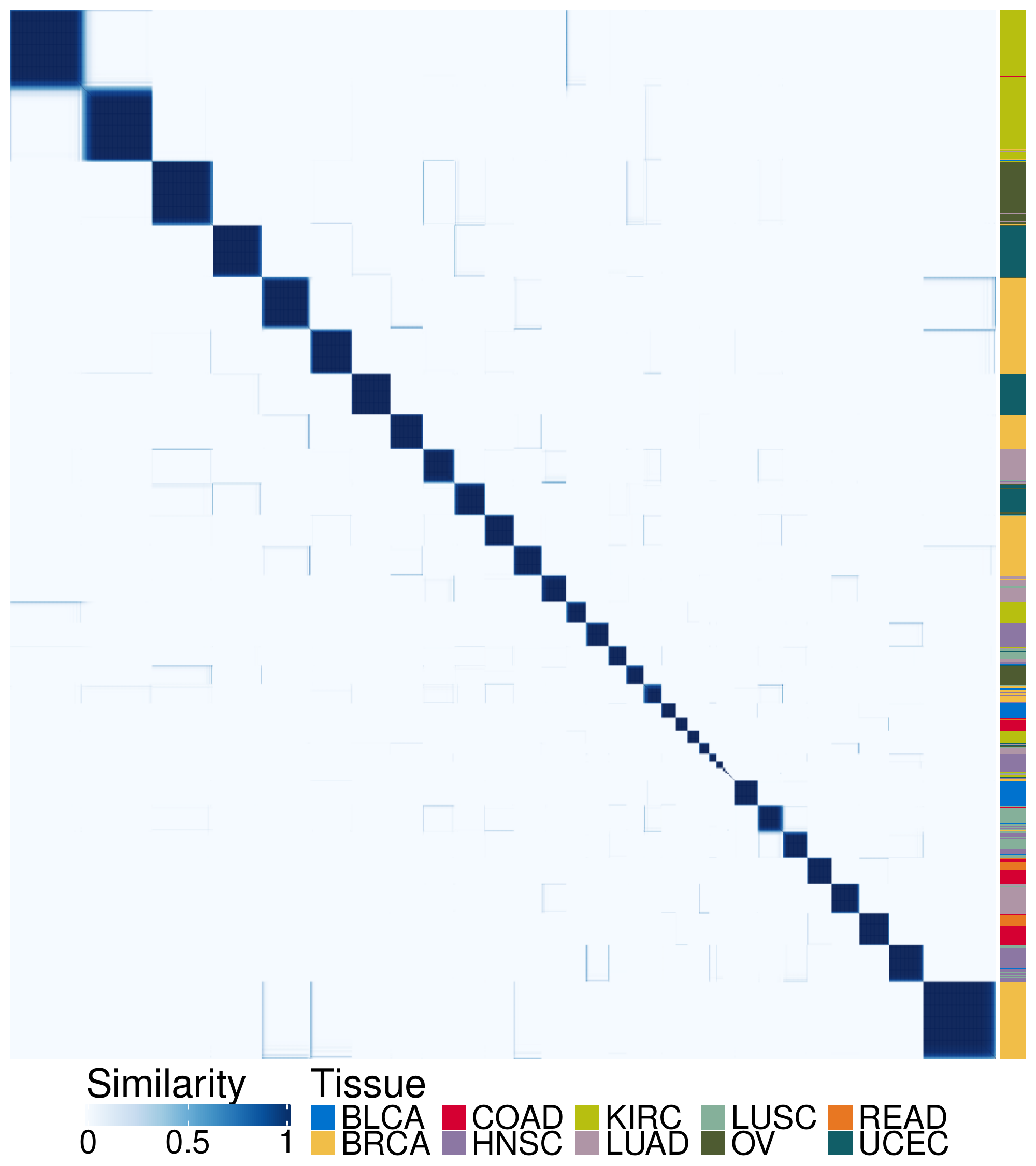}
	\includegraphics[width=.36\linewidth]{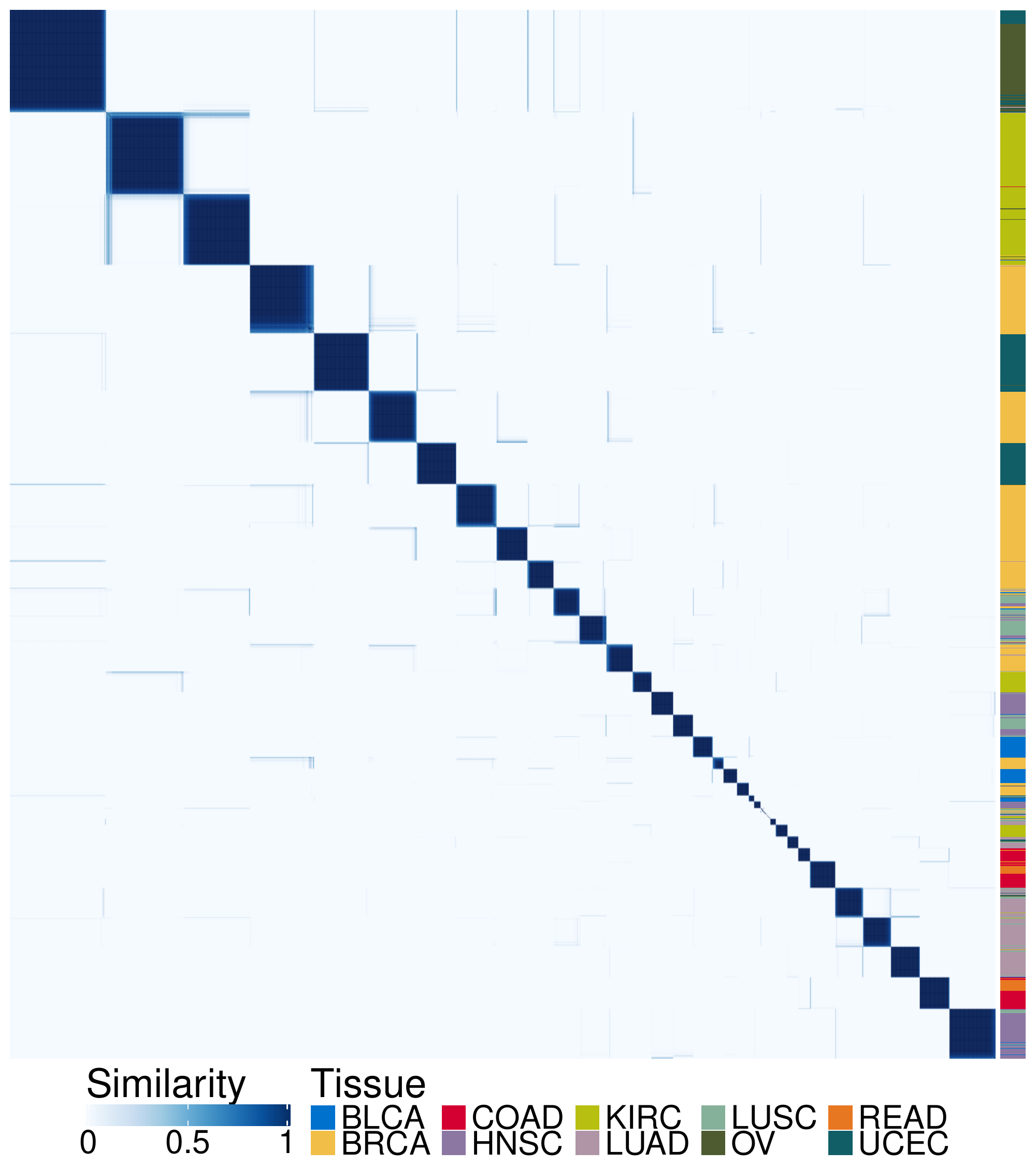}
	\includegraphics[width=.36\linewidth]{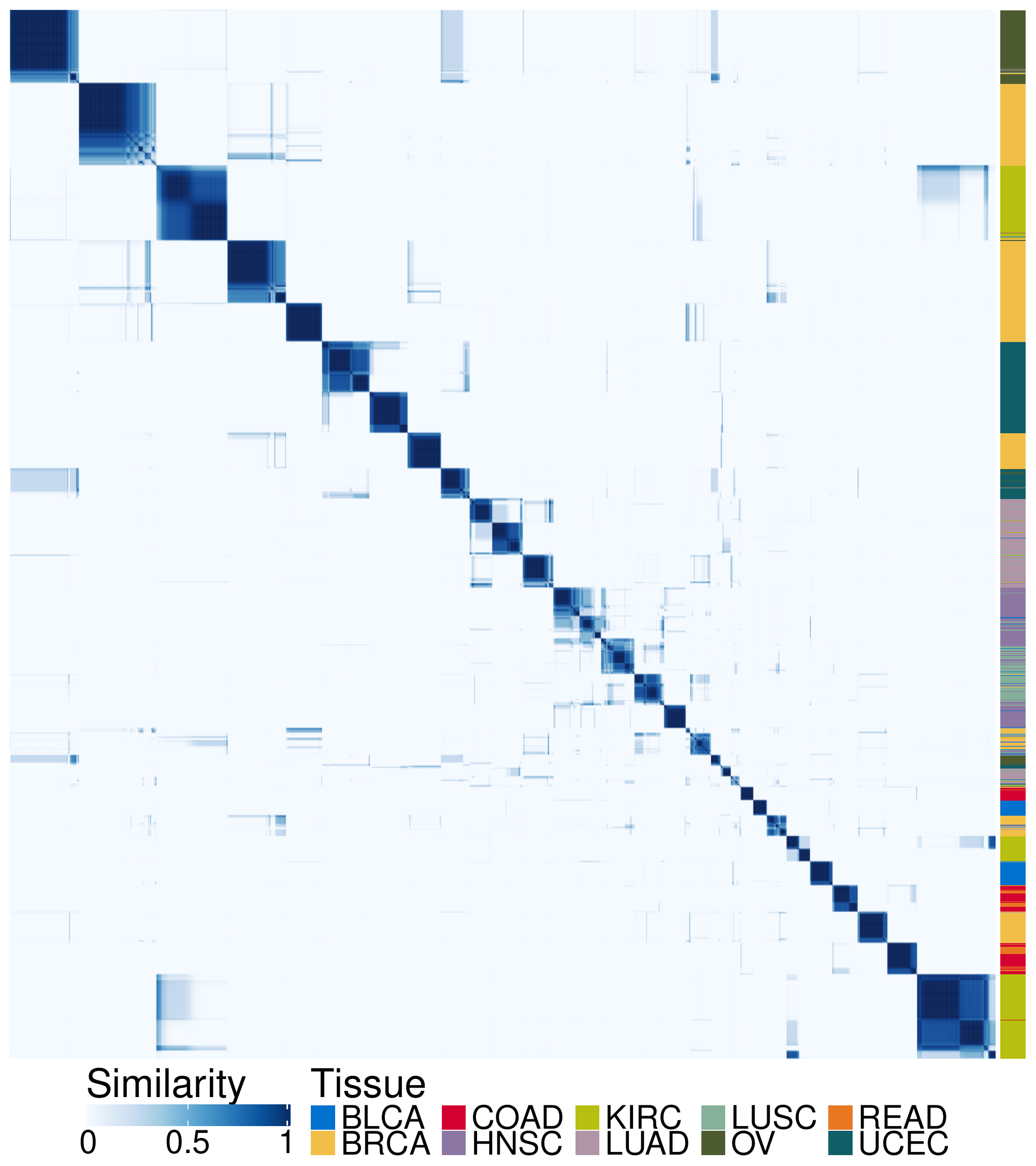}
	\caption{PSMs of the methylation data. $\alpha=1$.}
\end{figure}

\begin{table}[H]
\centering
\begin{tabular}{l c c c c}
& \textbf{Chain 2} & \textbf{Chain 3} & \textbf{Chain 4} & \textbf{Chain 5} \\
\hline
\textbf{Chain 1} & 0.81 & 0.70 & 0.83 & 0.78 \\
\textbf{Chain 2} &1 & 0.61 & 0.76 & 0.67 \\
\textbf{Chain 3} && 1 &  0.67 & 0.60 \\
\textbf{Chain 4} && & 1 & 0.76 \\
\hline\\
\end{tabular}
\caption{ARI between the clusterings found on the PSMs of different chains with the number of clusters that maximises the silhouette.}
\end{table} 

\begin{figure}[H]
\centering
\includegraphics[width=.45\linewidth]{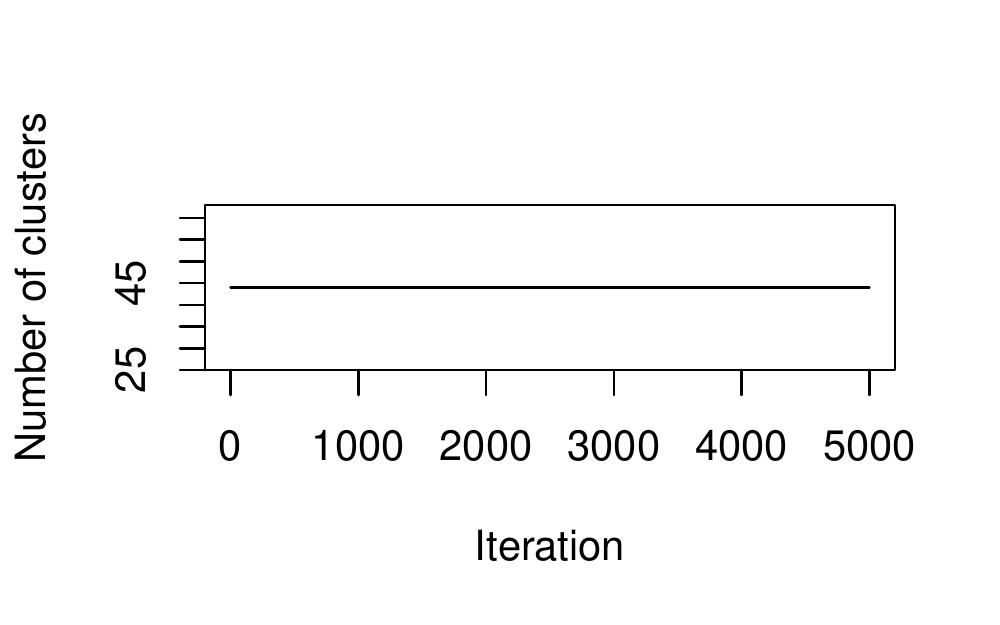}
\includegraphics[width=.45\linewidth]{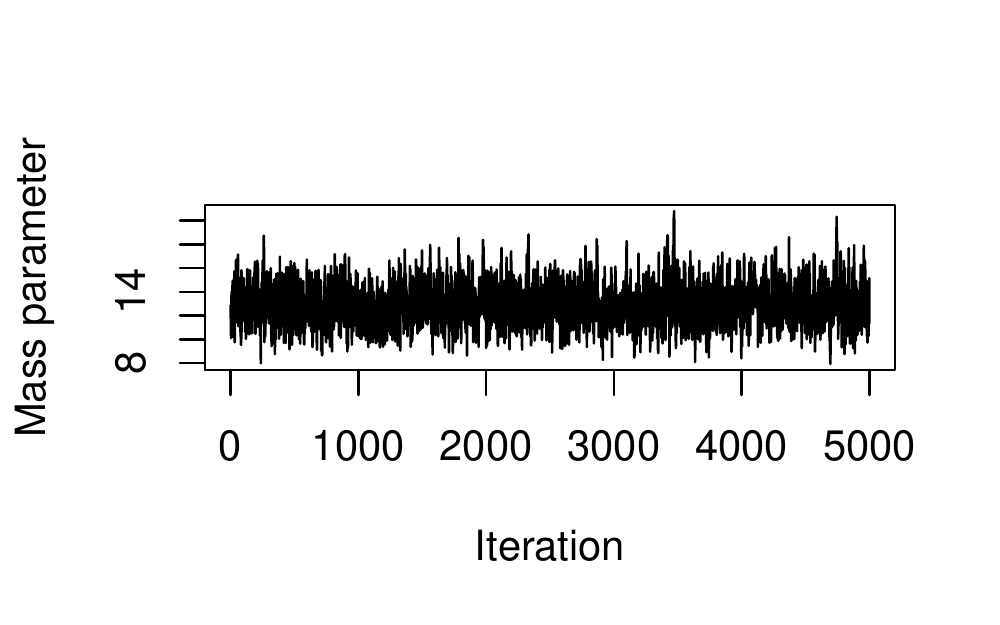}
\vspace{-1.6cm}

\includegraphics[width=.45\linewidth]{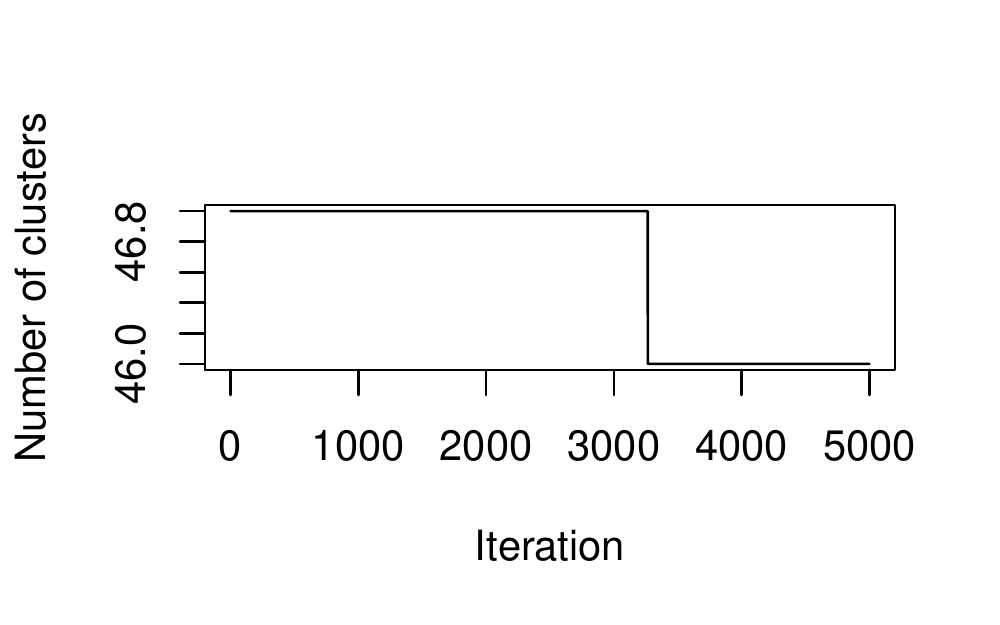}
\includegraphics[width=.45\linewidth]{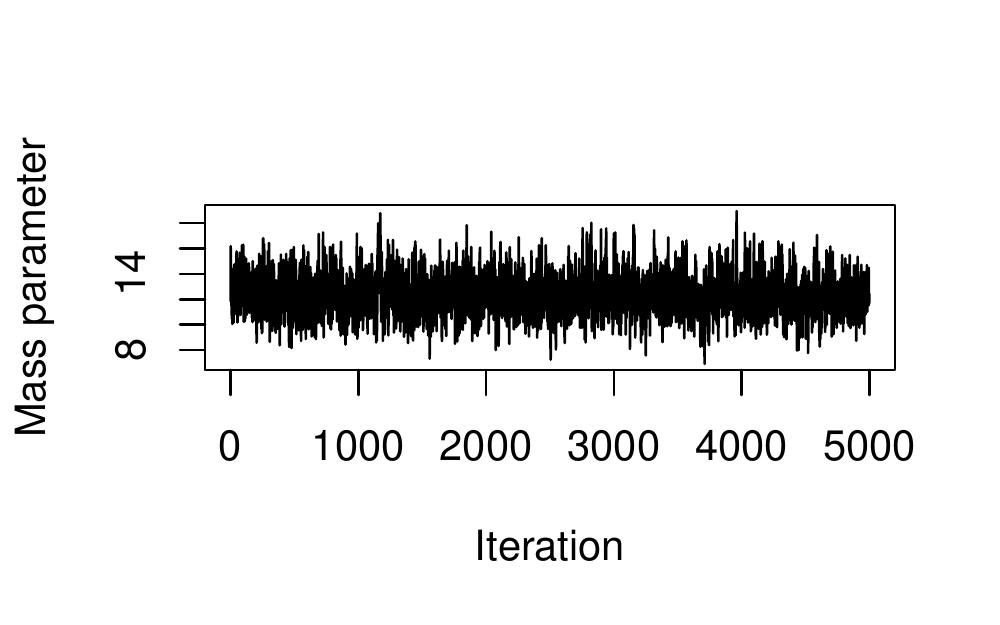}
\vspace{-1.6cm}

\includegraphics[width=.45\linewidth]{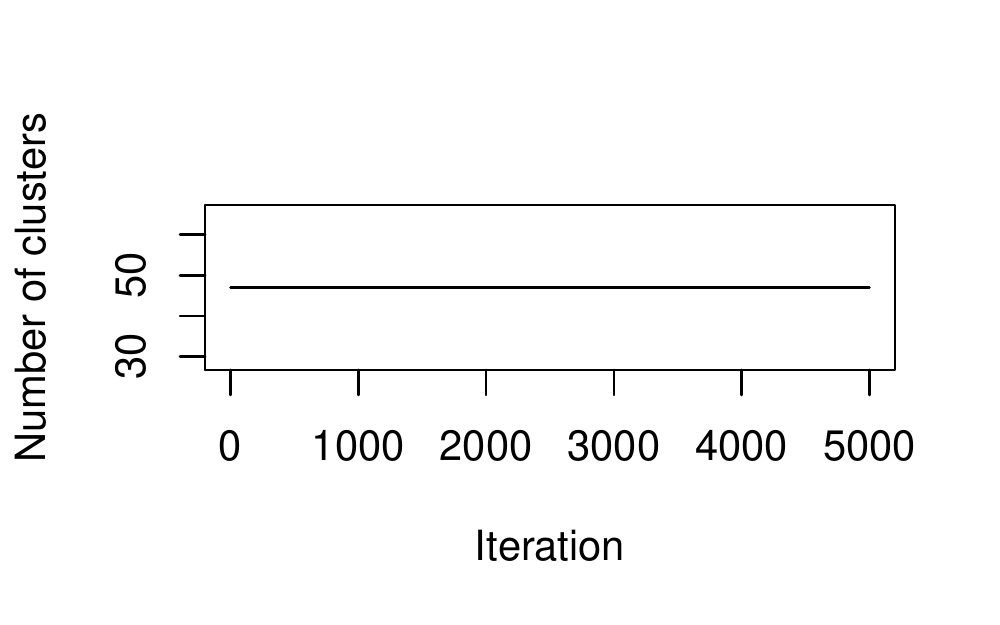}
\includegraphics[width=.45\linewidth]{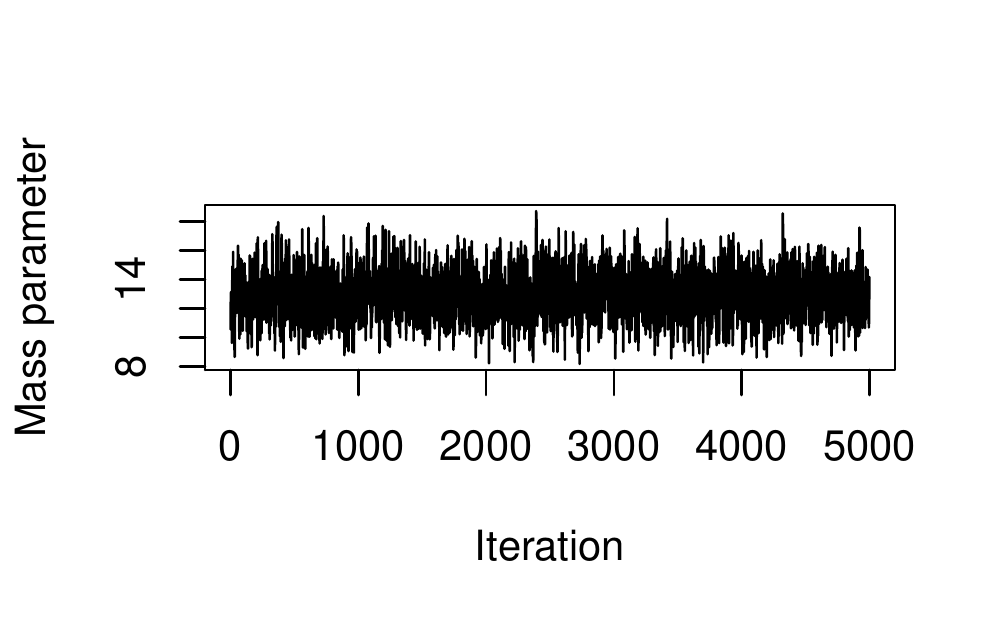}
\vspace{-1.6cm}

\includegraphics[width=.45\linewidth]{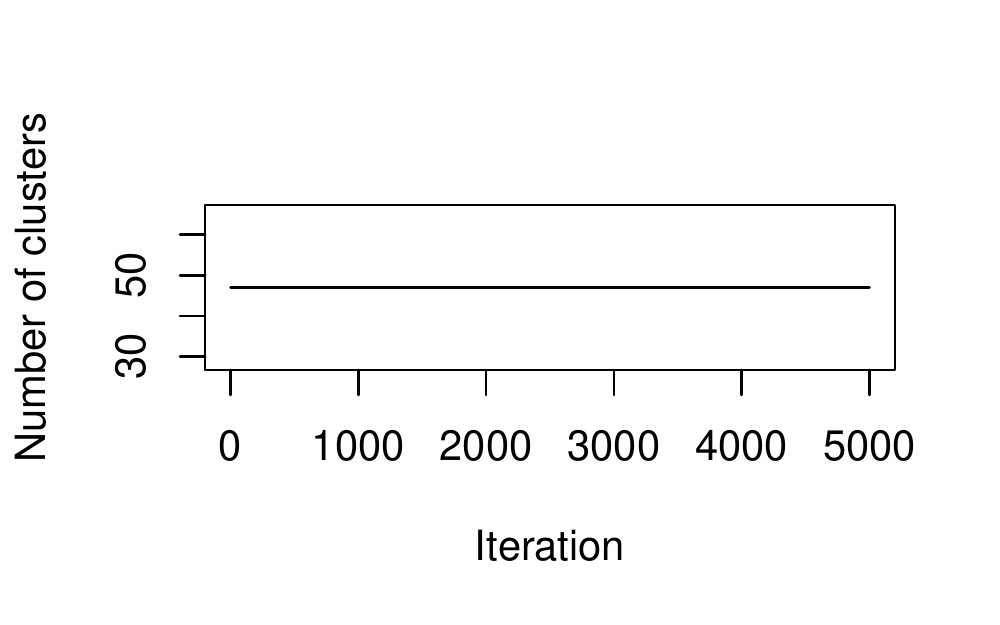}
\includegraphics[width=.45\linewidth]{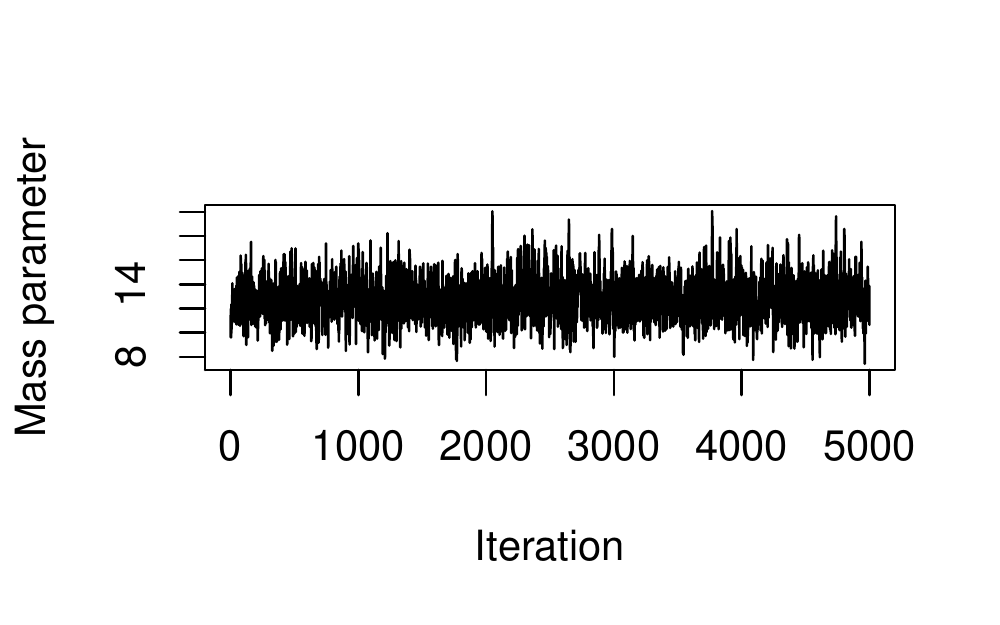}
\vspace{-1.6cm}

\includegraphics[width=.45\linewidth]{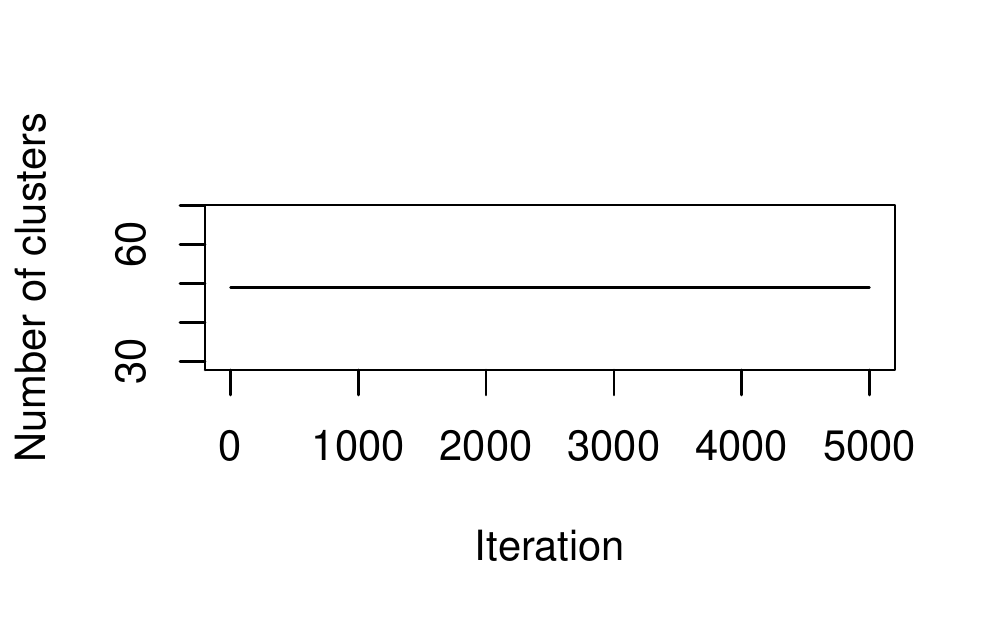}
\includegraphics[width=.45\linewidth]{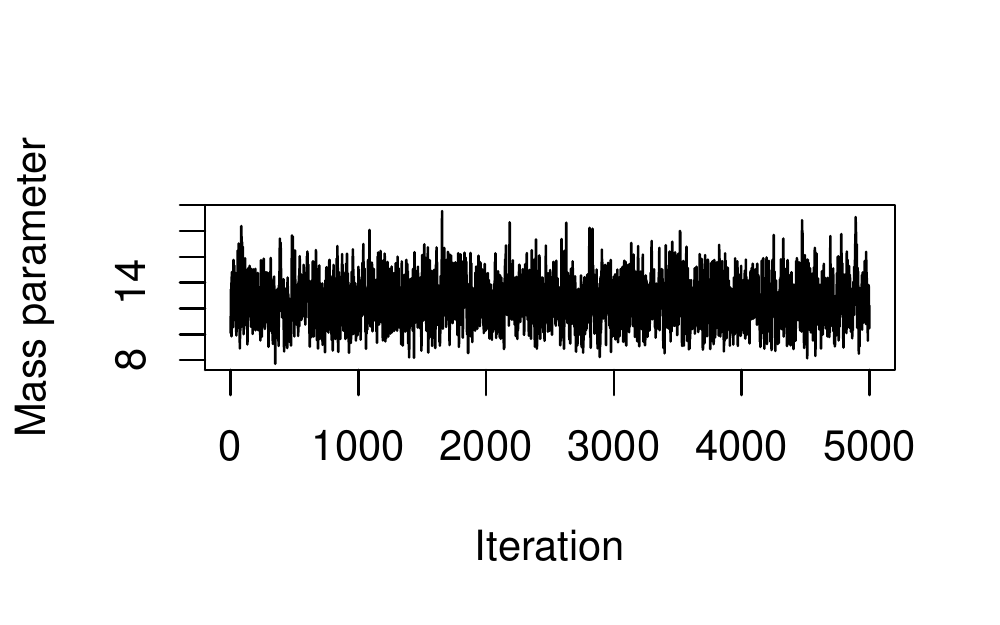}
\caption{MCMC convergence assessment, methylation data.}
\end{figure}

\clearpage

\begin{figure}[H]
	\centering
	\includegraphics[width=.36\linewidth]{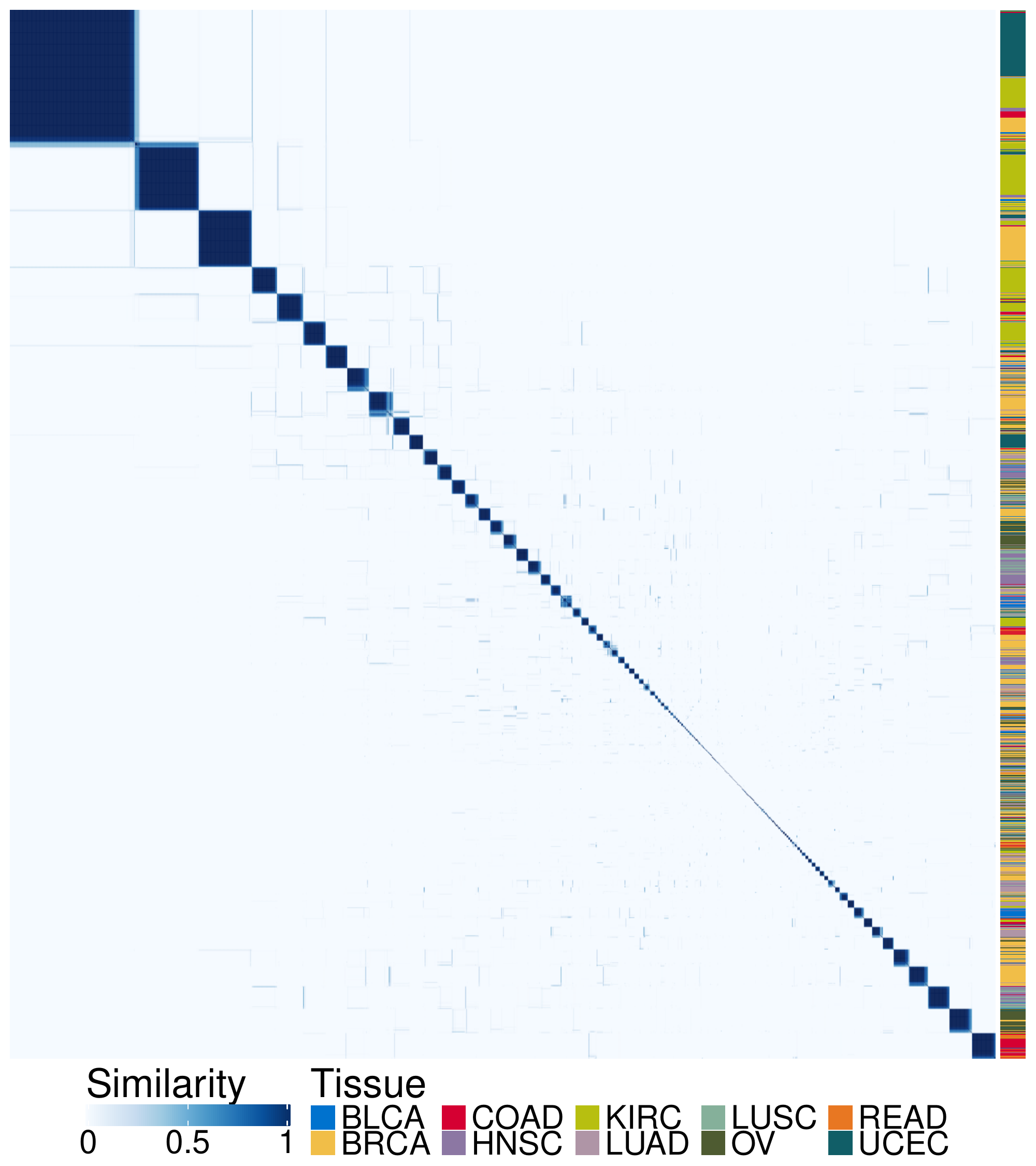}
	\includegraphics[width=.36\linewidth]{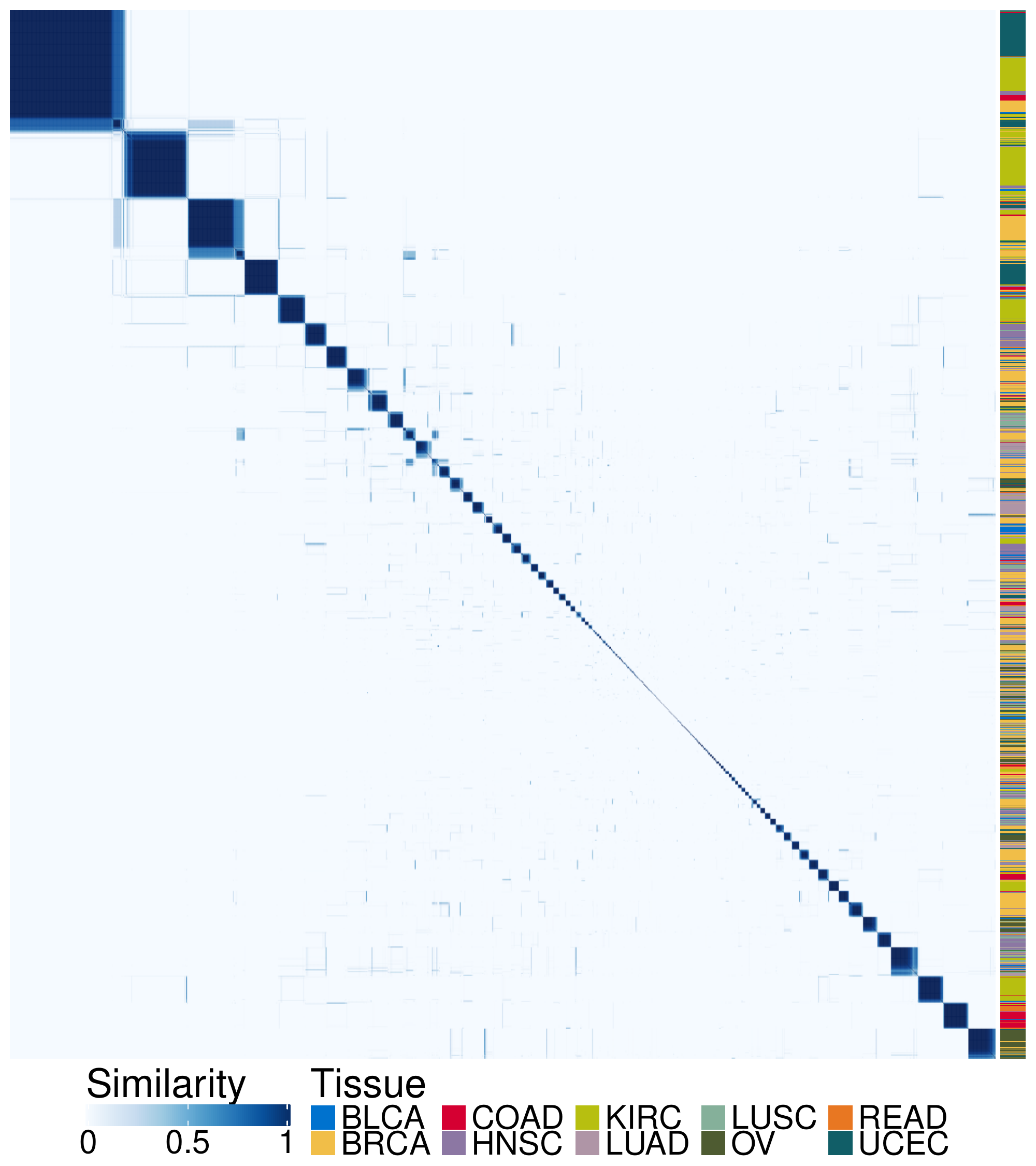}
	\includegraphics[width=.36\linewidth]{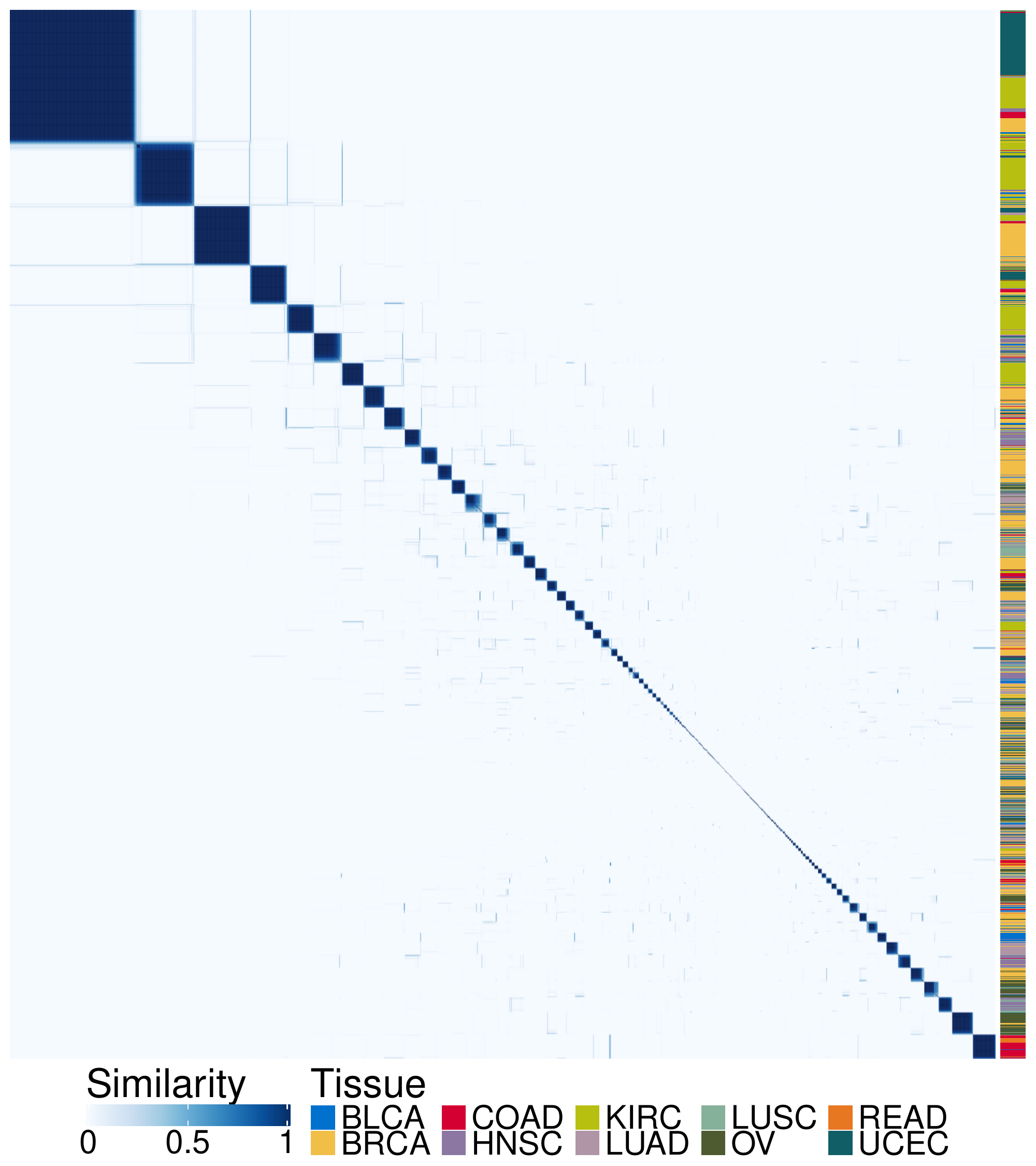}
	\includegraphics[width=.36\linewidth]{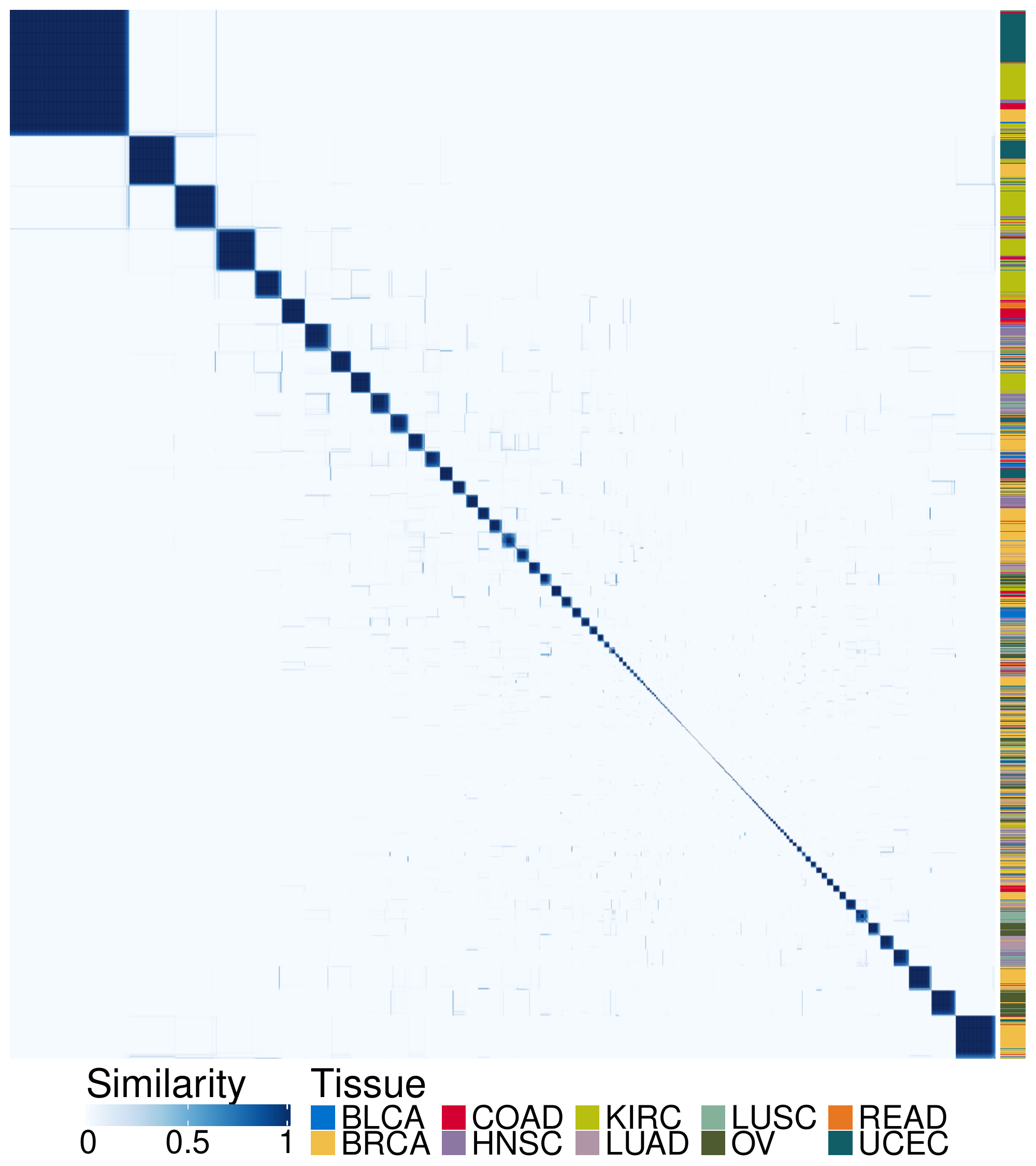}
	\includegraphics[width=.36\linewidth]{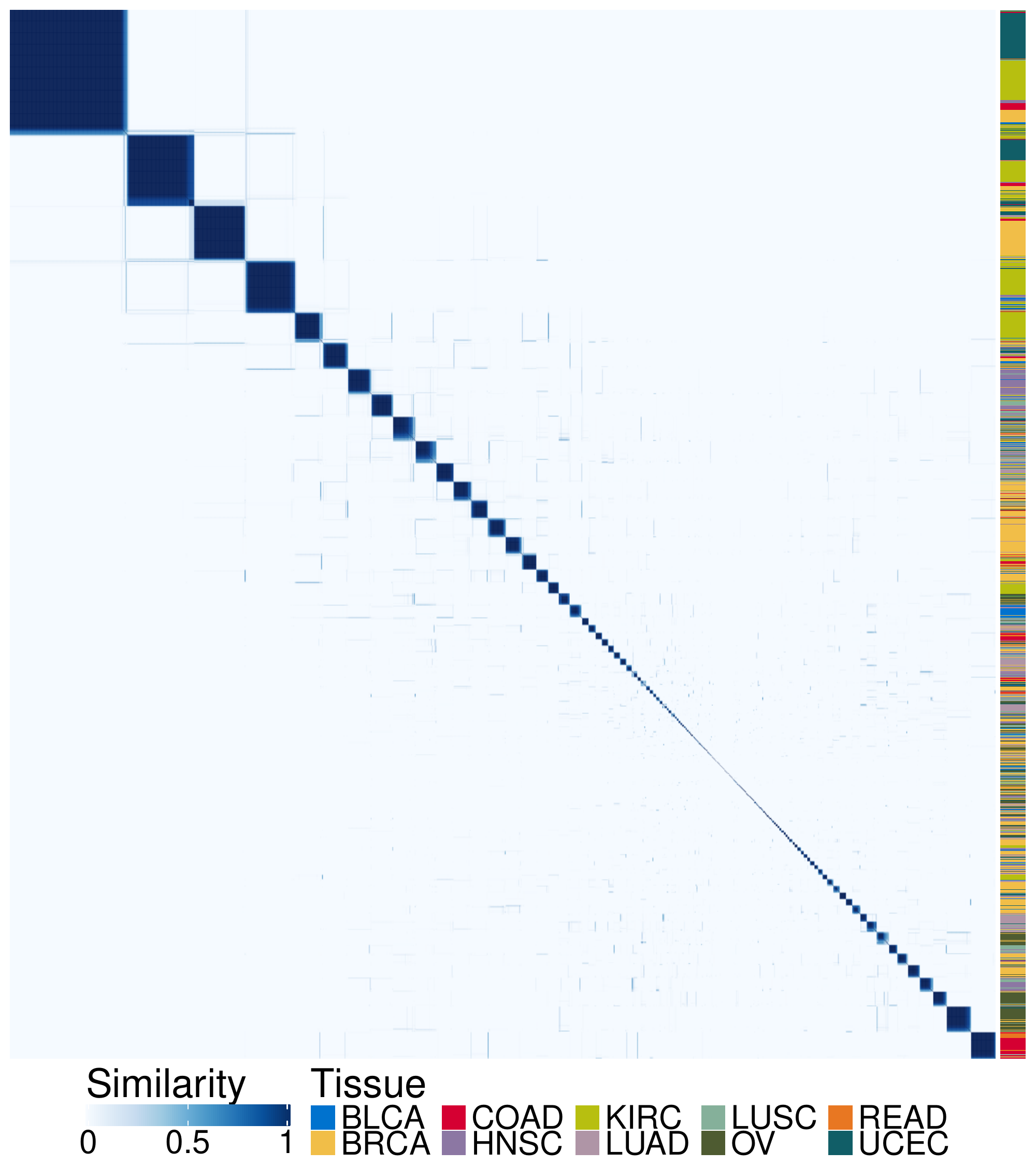}
	\includegraphics[width=.36\linewidth]{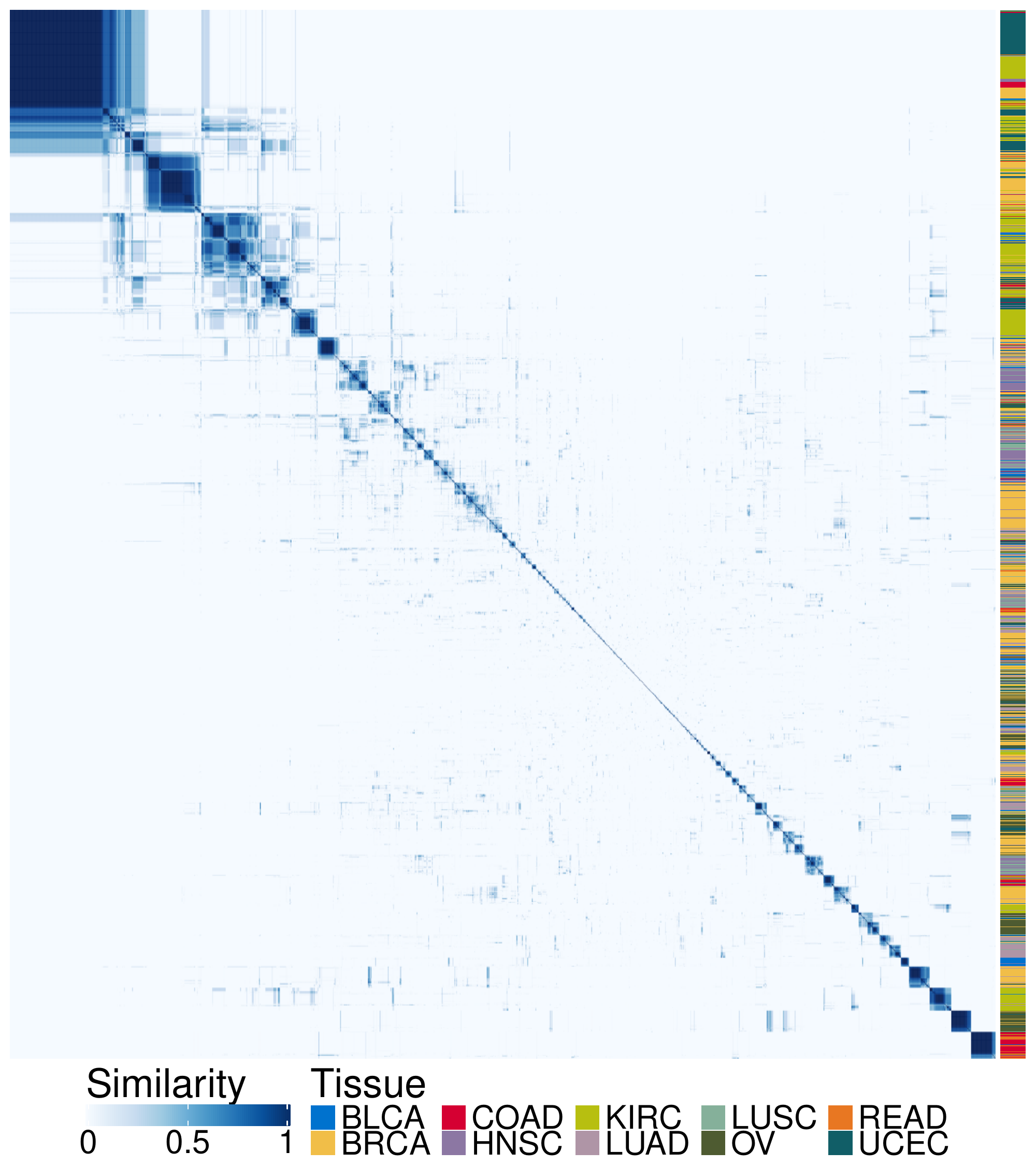}
	\caption{PSMs of the DNA copy number data.}
\end{figure}

\begin{table}[H]
\centering
\begin{tabular}{l c c c c}
& \textbf{Chain 2} & \textbf{Chain 3} & \textbf{Chain 4} & \textbf{Chain 5} \\
\hline
\textbf{Chain 1} & 0.39 & 0.45 & 0.35 & 0.35 \\
\textbf{Chain 2} &1 & 0.58 & 0.60 & 0.62 \\
\textbf{Chain 3} && 1 &  0.59 & 0.57 \\
\textbf{Chain 4} && & 1 & 0.54 \\
\hline\\
\end{tabular}
\caption{ARI between the clusterings found on the PSMs of different chains with the number of clusters that maximises the silhouette.}
\end{table} 

\begin{figure}[H]
\centering
\includegraphics[width=.45\linewidth]{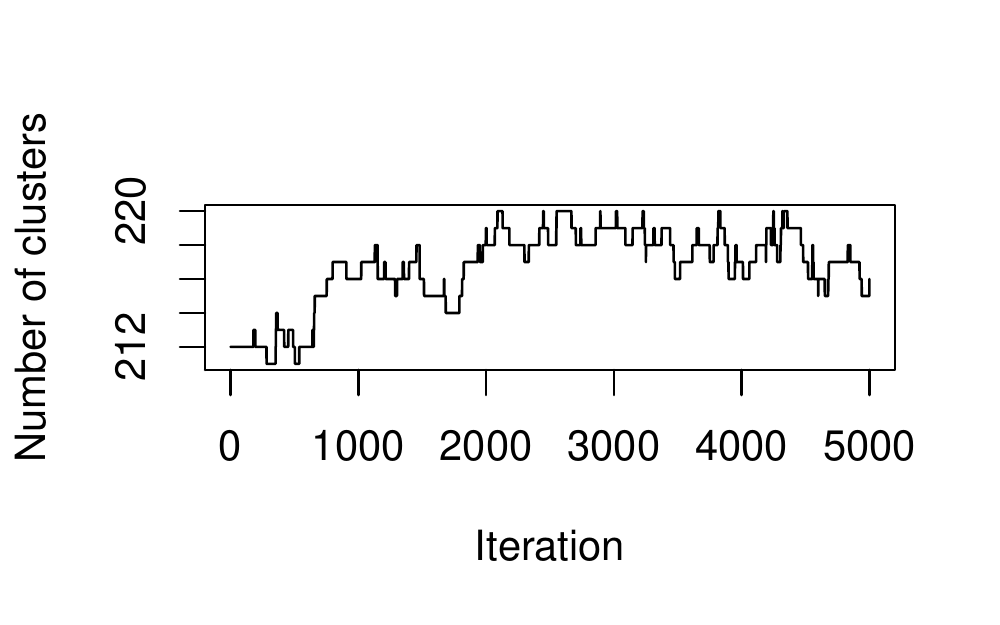}
\includegraphics[width=.45\linewidth]{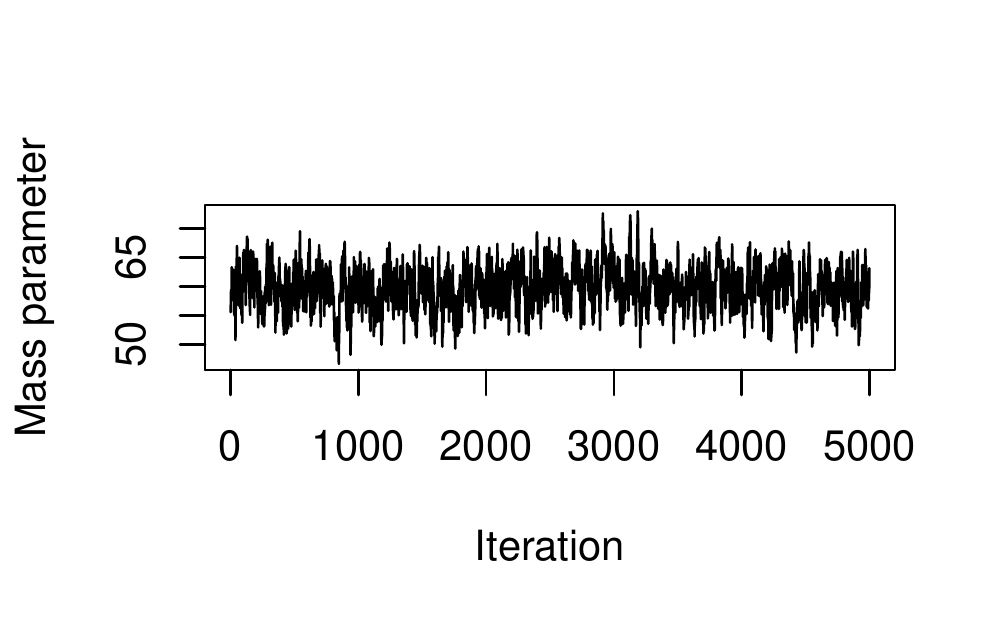}
\vspace{-1.6cm}

\includegraphics[width=.45\linewidth]{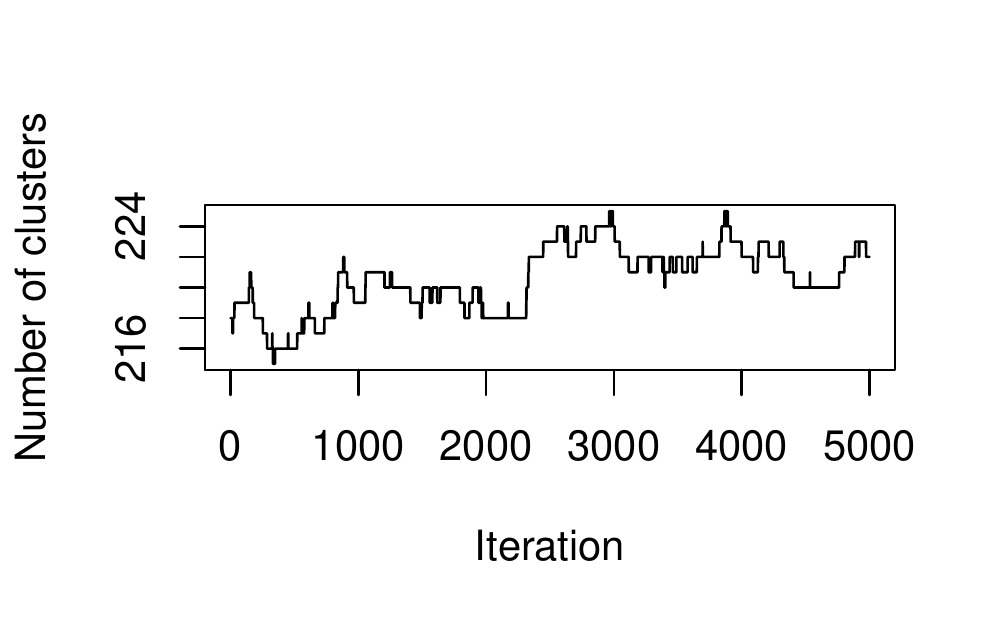}
\includegraphics[width=.45\linewidth]{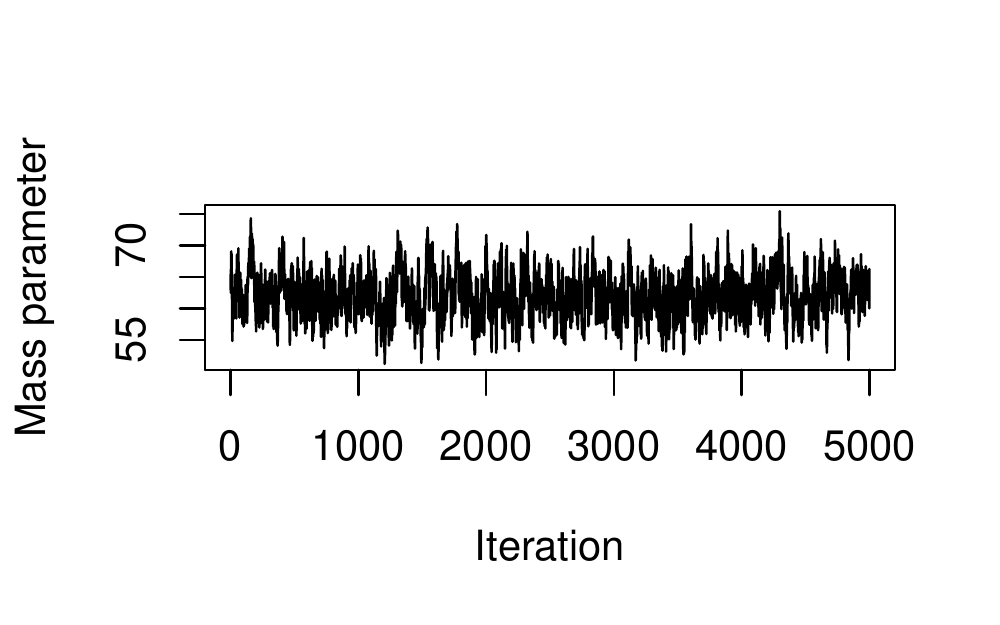}
\vspace{-1.6cm}

\includegraphics[width=.45\linewidth]{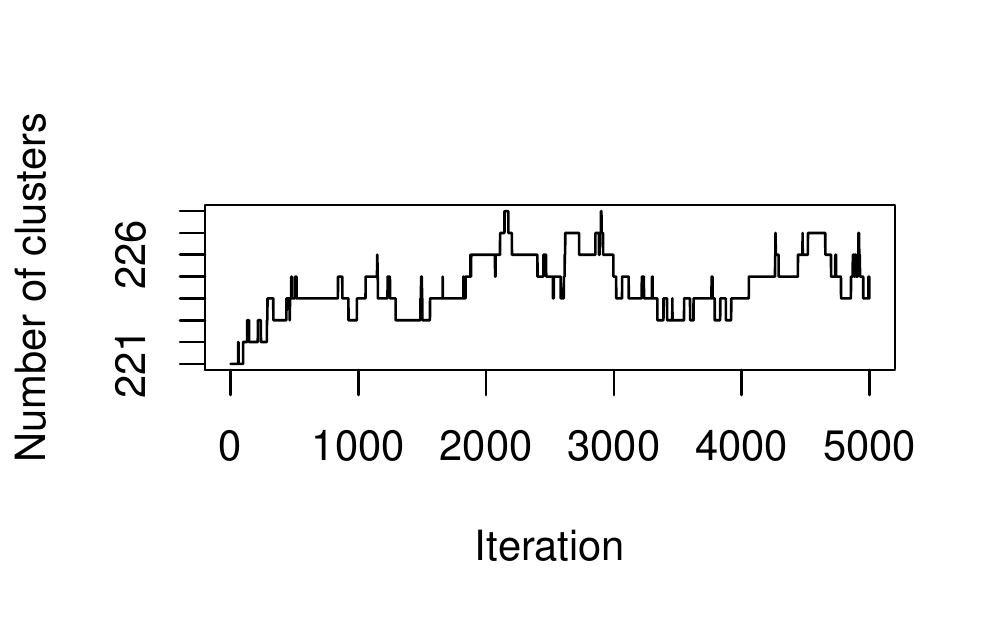}
\includegraphics[width=.45\linewidth]{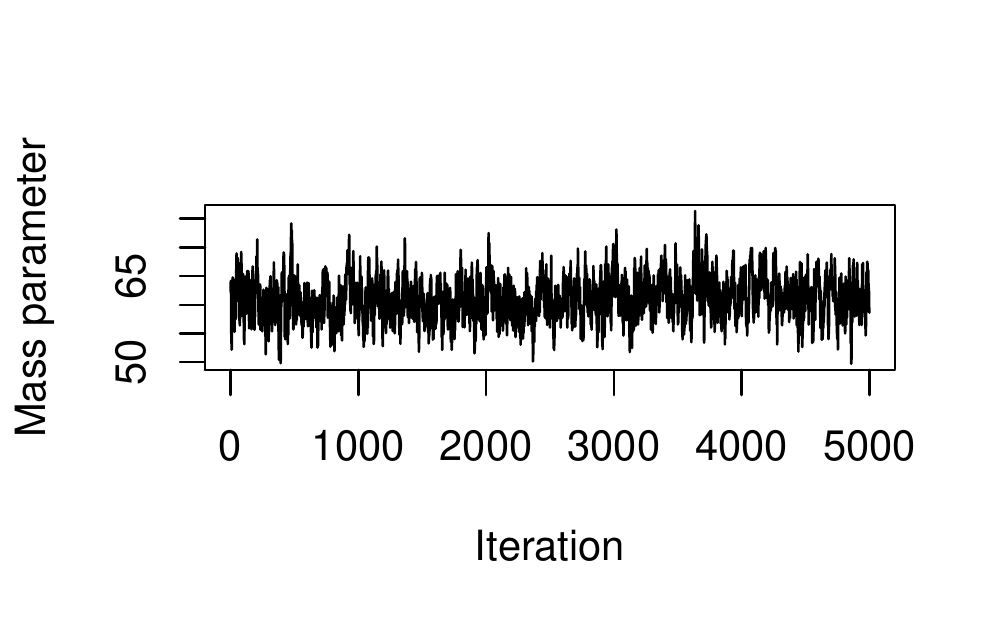}
\vspace{-1.6cm}

\includegraphics[width=.45\linewidth]{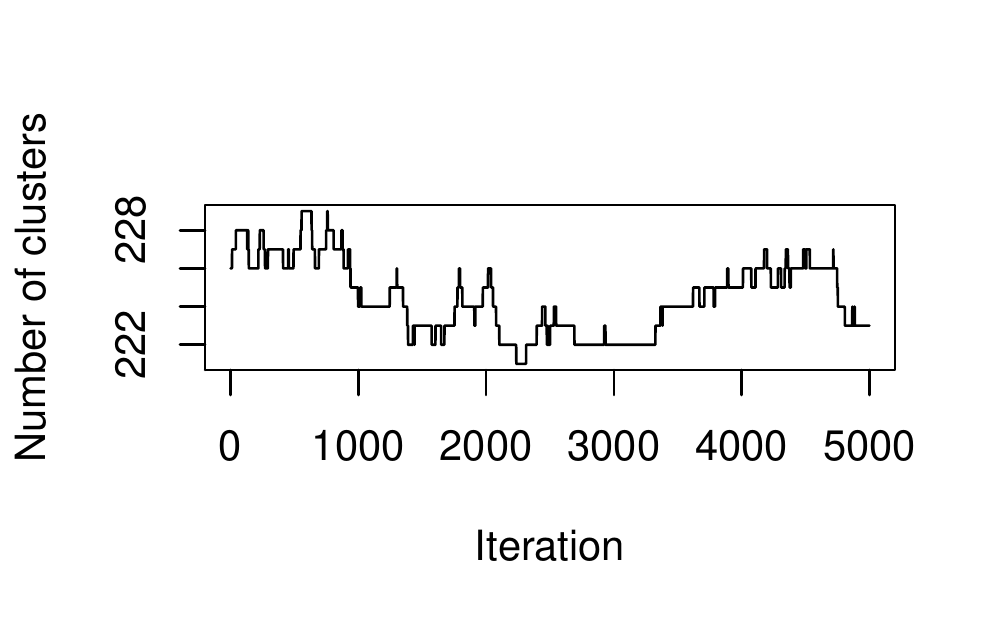}
\includegraphics[width=.45\linewidth]{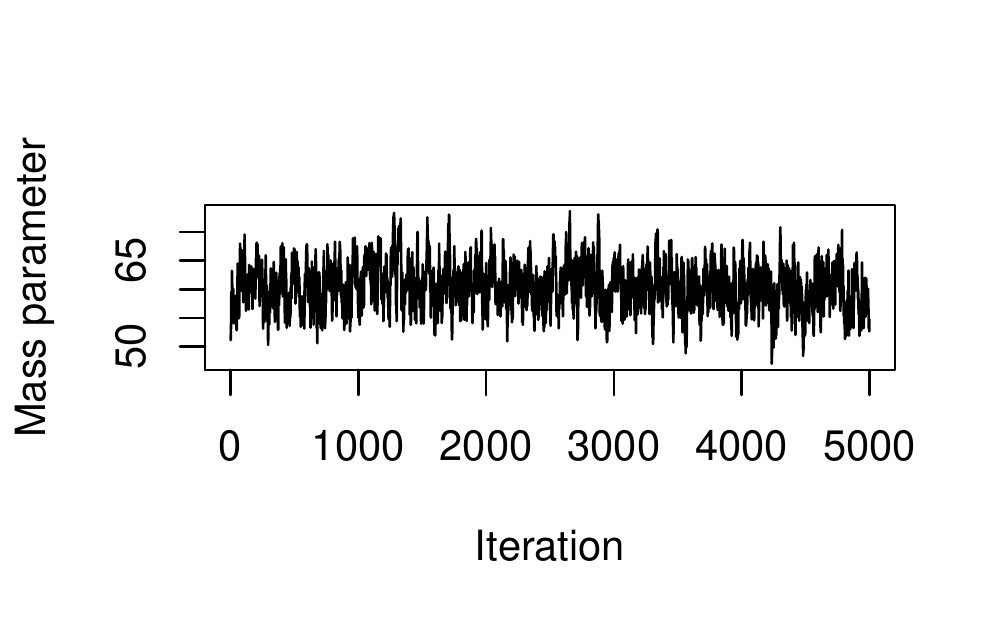}
\vspace{-1.6cm}

\includegraphics[width=.45\linewidth]{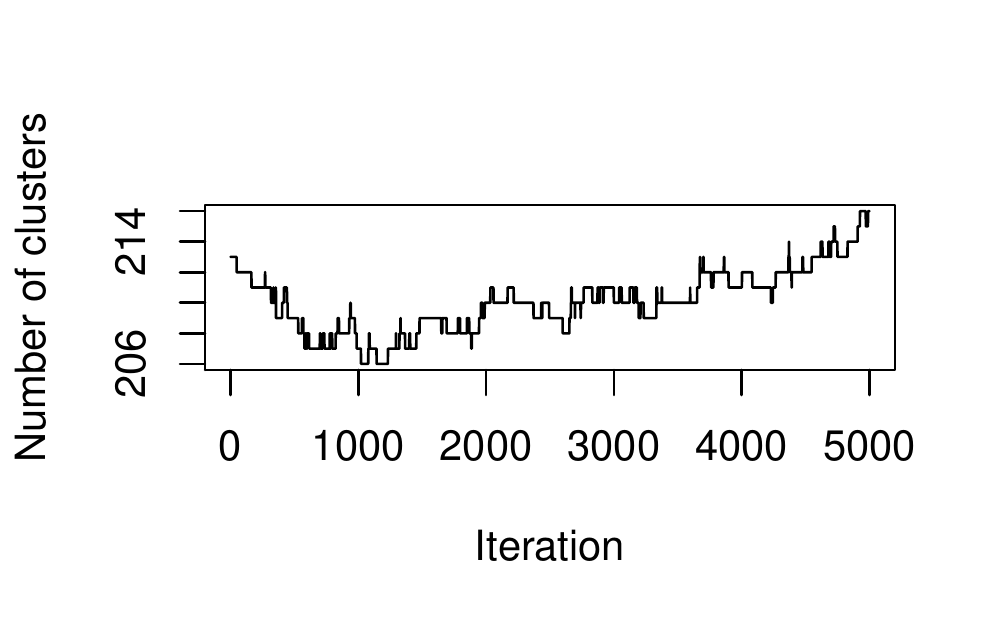}
\includegraphics[width=.45\linewidth]{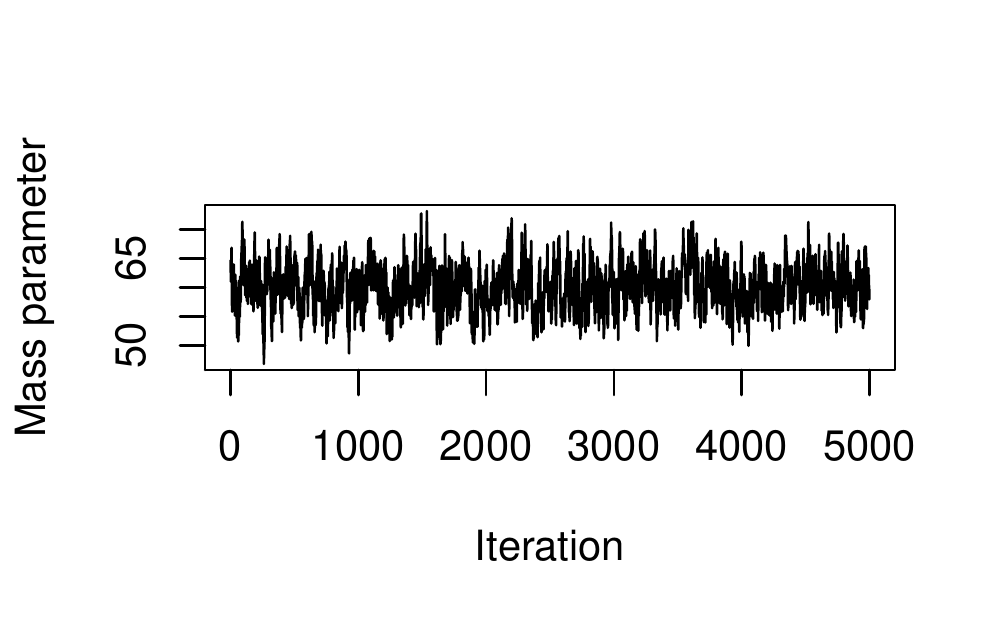}
\caption{MCMC convergence assessment, DNA copy number data.}
\end{figure}

\clearpage

\begin{figure}[H]
	\centering
	\includegraphics[width=.36\linewidth]{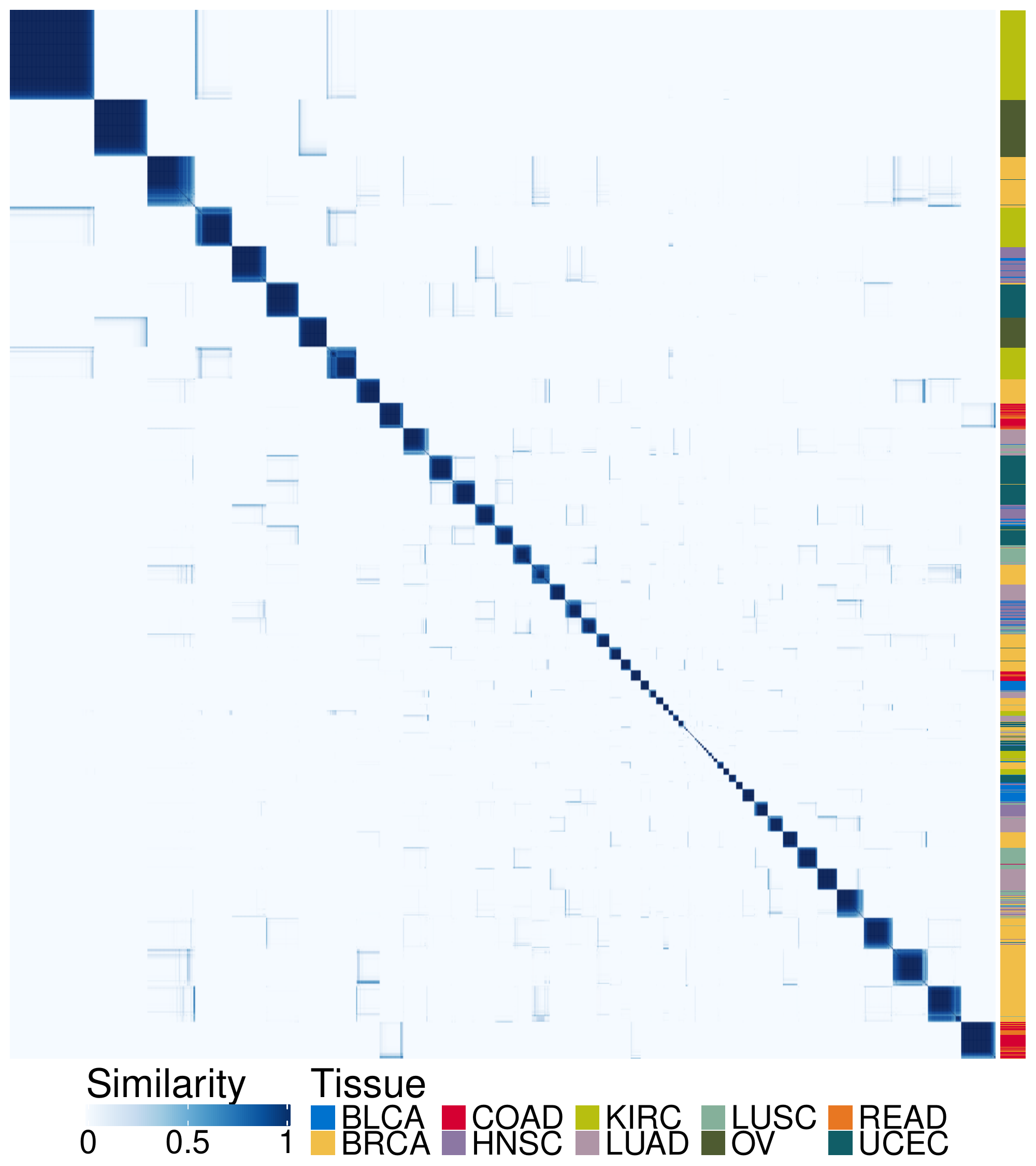}
	\includegraphics[width=.36\linewidth]{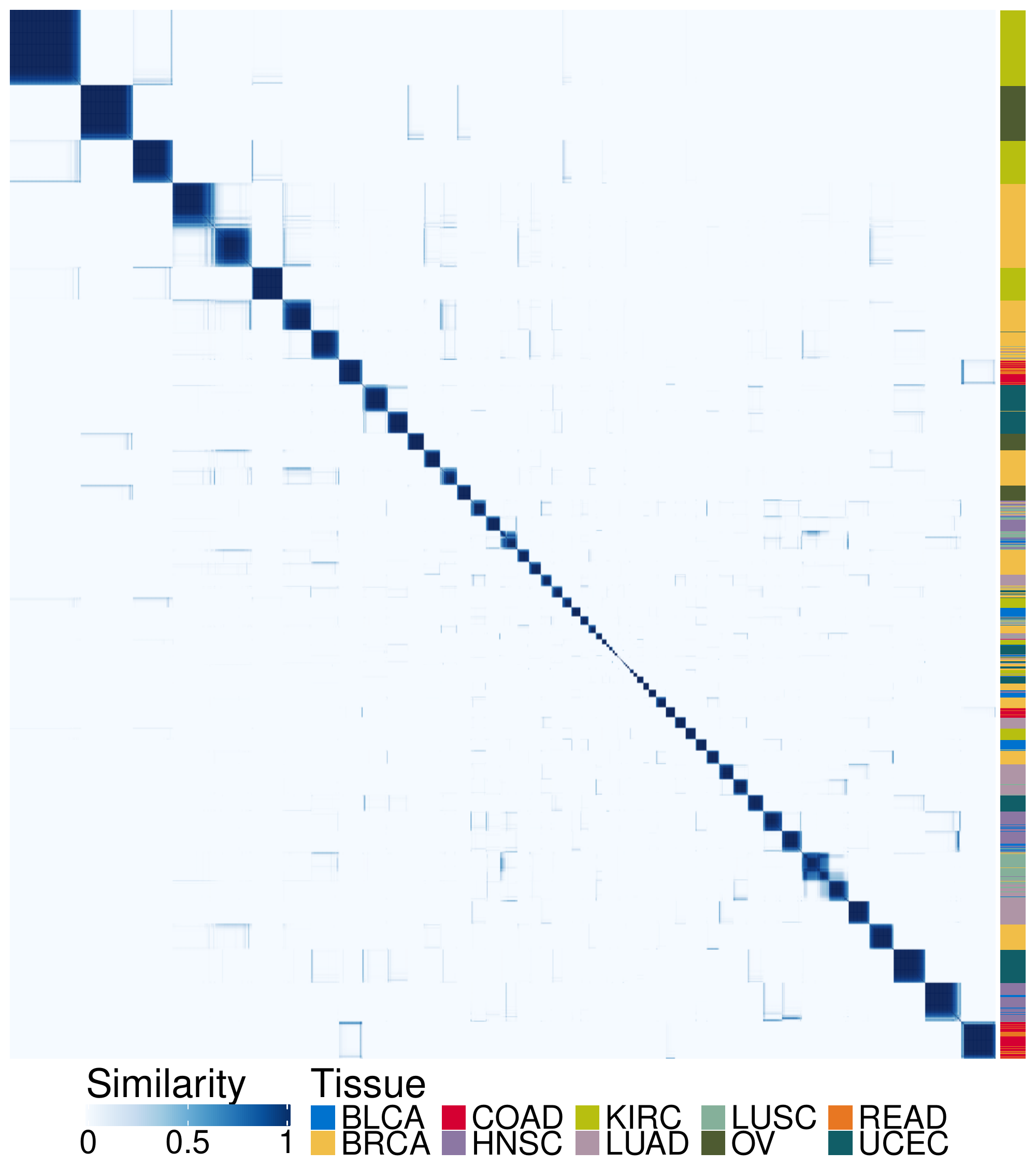}
	\includegraphics[width=.36\linewidth]{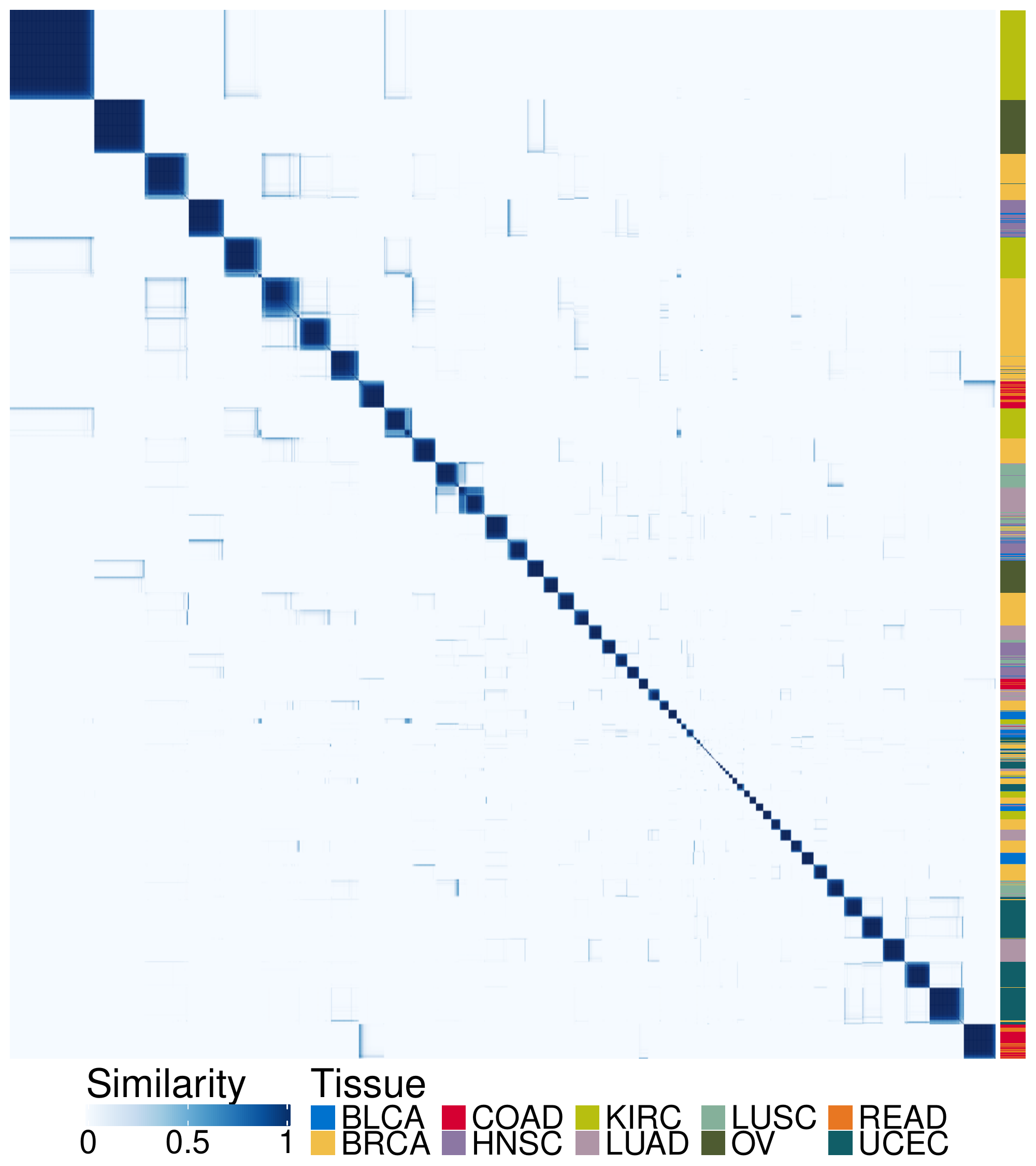}
	\includegraphics[width=.36\linewidth]{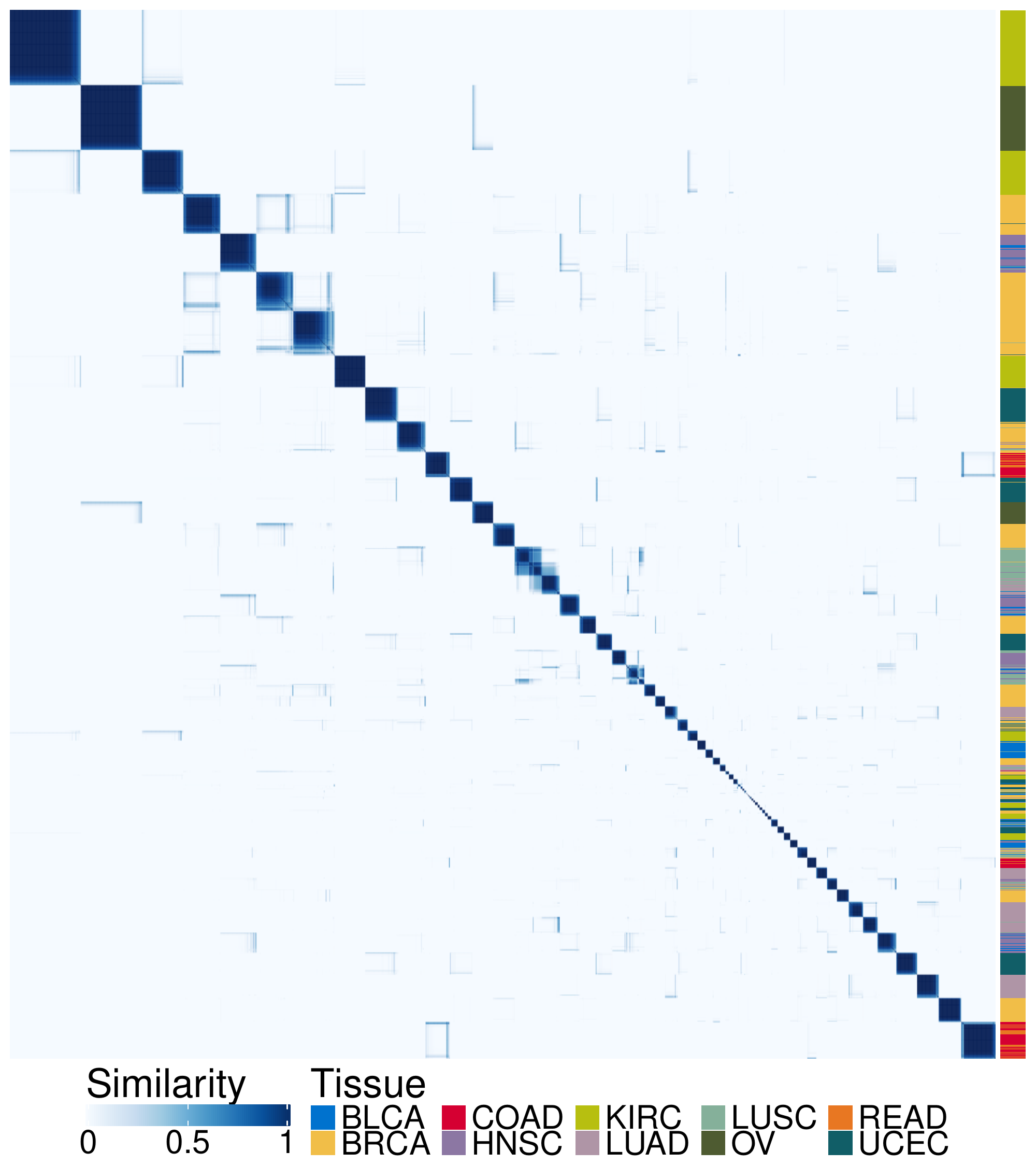}
	\includegraphics[width=.36\linewidth]{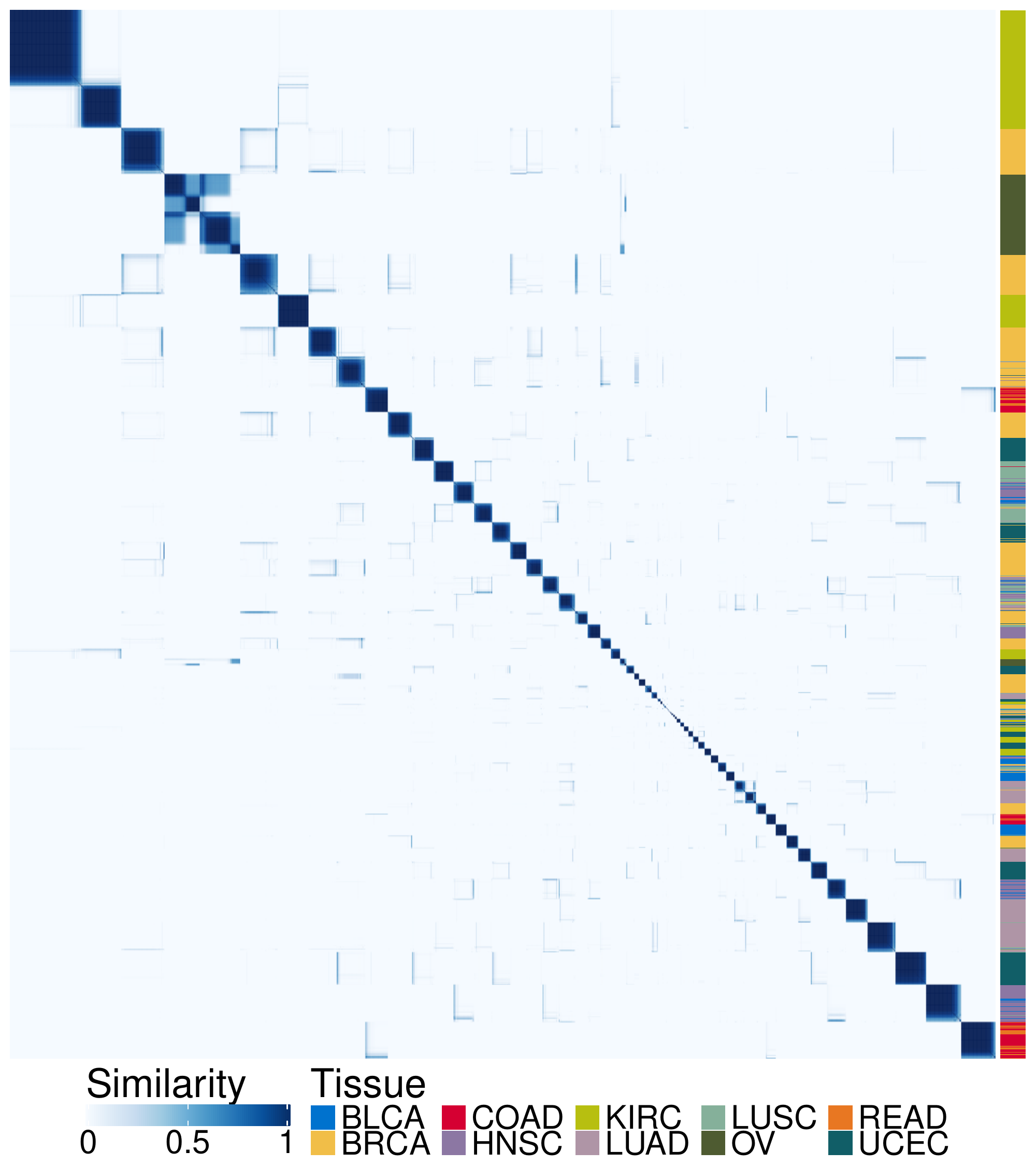}
	\includegraphics[width=.36\linewidth]{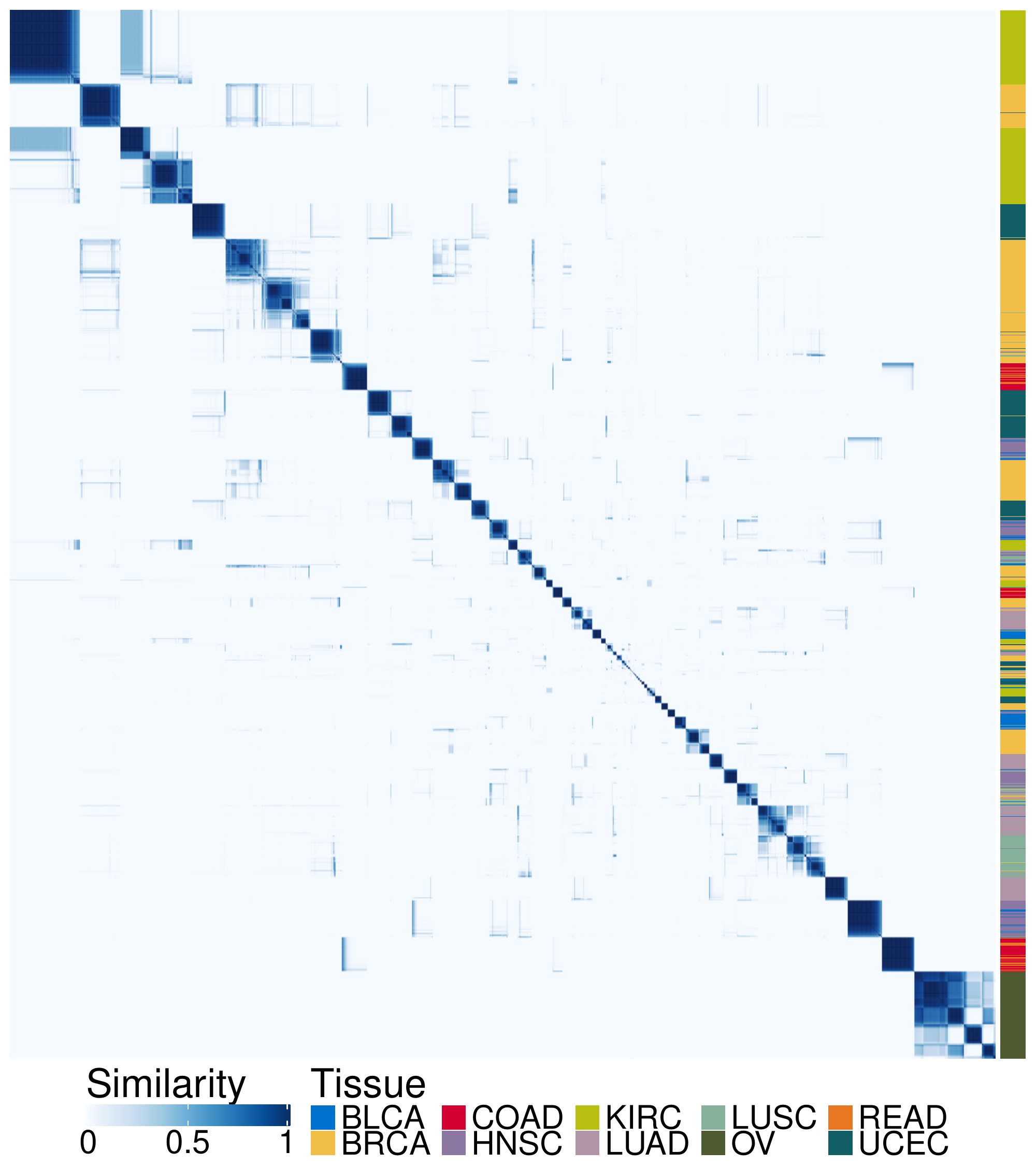}
	\caption{PSMs of the microRNA data. $\lambda=0$, $\alpha=0.1, 0.5$.}
\end{figure}

\begin{table}[H]
\centering
\begin{tabular}{l c c c c}
& \textbf{Chain 2} & \textbf{Chain 3} & \textbf{Chain 4} & \textbf{Chain 5} \\
\hline
\textbf{Chain 1} & 0.63 & 0.75 & 0.59 & 0.64 \\
\textbf{Chain 2} &1 & 0.74 & 0.82 & 0.82 \\
\textbf{Chain 3} && 1 &  0.68 & 0.69 \\
\textbf{Chain 4} && & 1 & 0.75 \\
\hline\\
\end{tabular}
\caption{ARI between the clusterings found on the PSMs of different chains with the number of clusters that maximises the silhouette.}
\end{table} 

\begin{figure}[H]
\centering
\includegraphics[width=.45\linewidth]{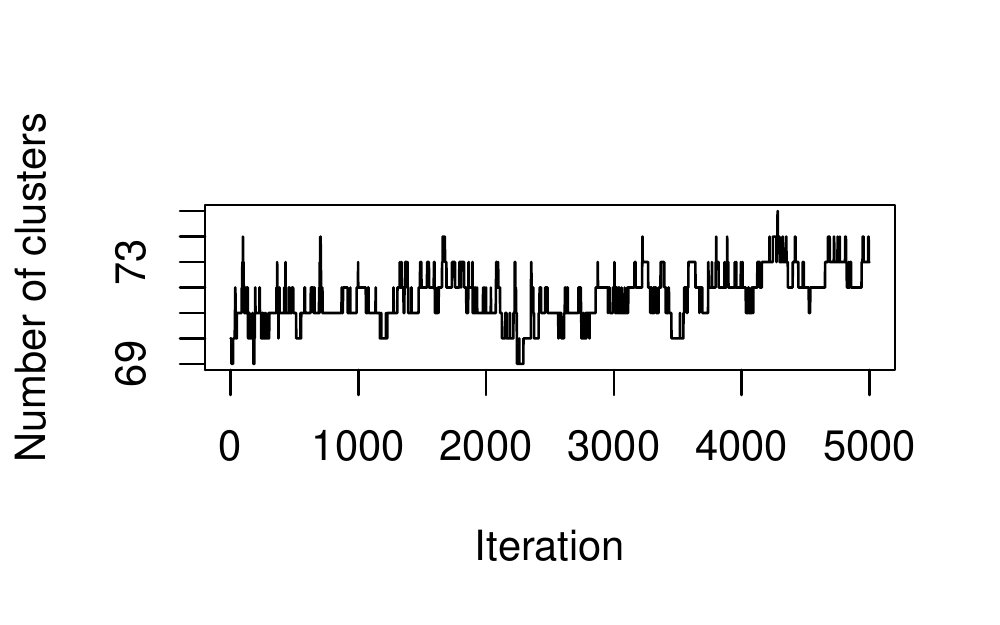}
\includegraphics[width=.45\linewidth]{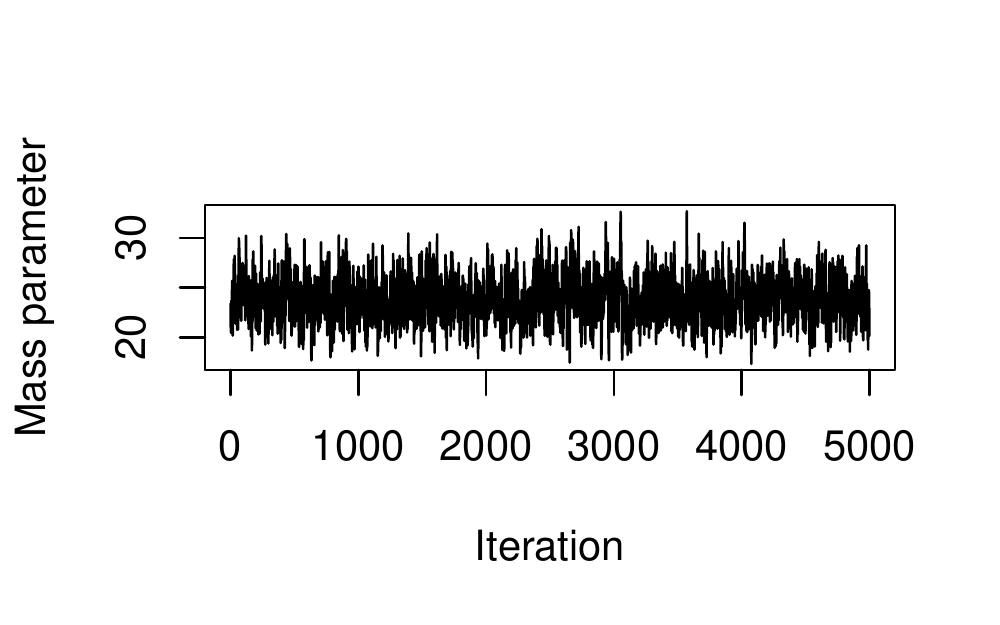}
\vspace{-1.6cm}

\includegraphics[width=.45\linewidth]{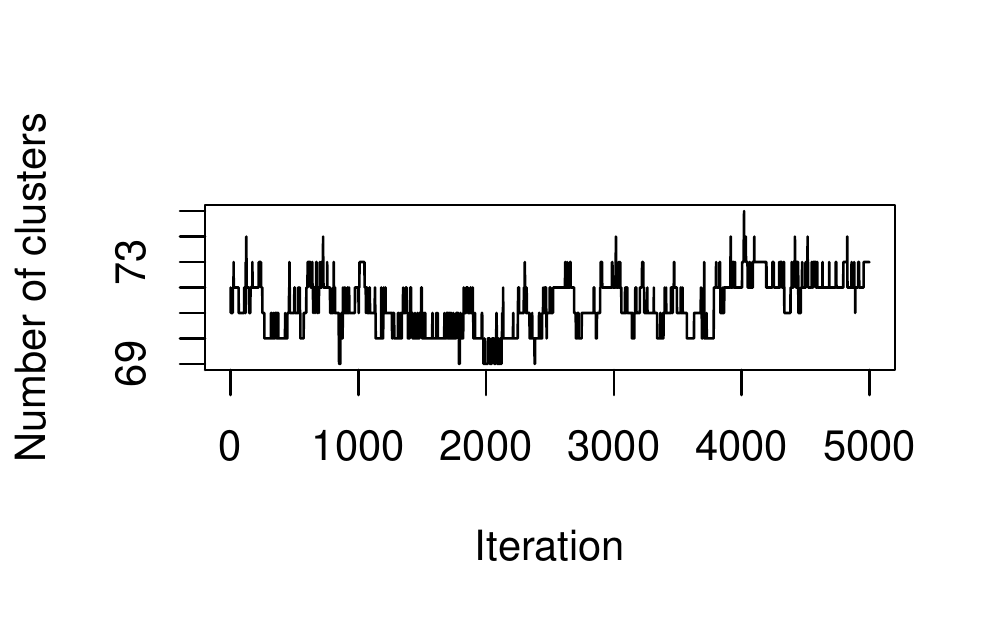}
\includegraphics[width=.45\linewidth]{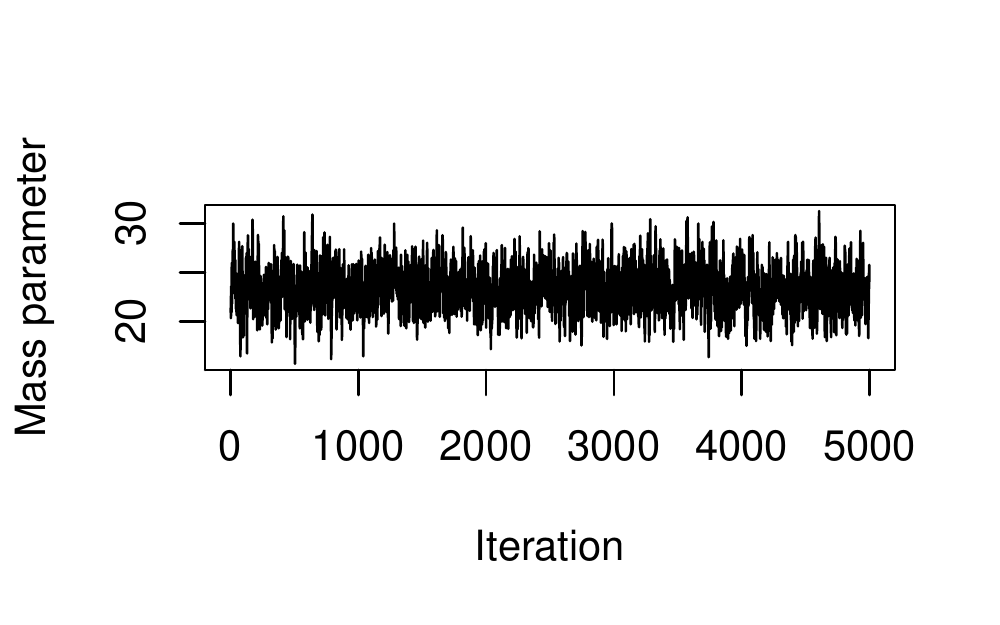}
\vspace{-1.6cm}

\includegraphics[width=.45\linewidth]{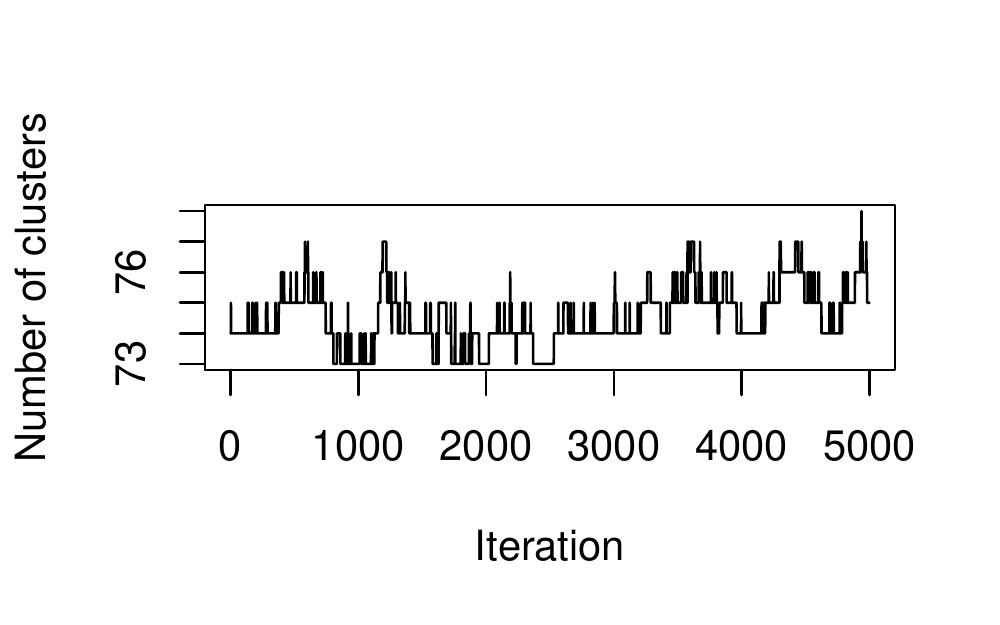}
\includegraphics[width=.45\linewidth]{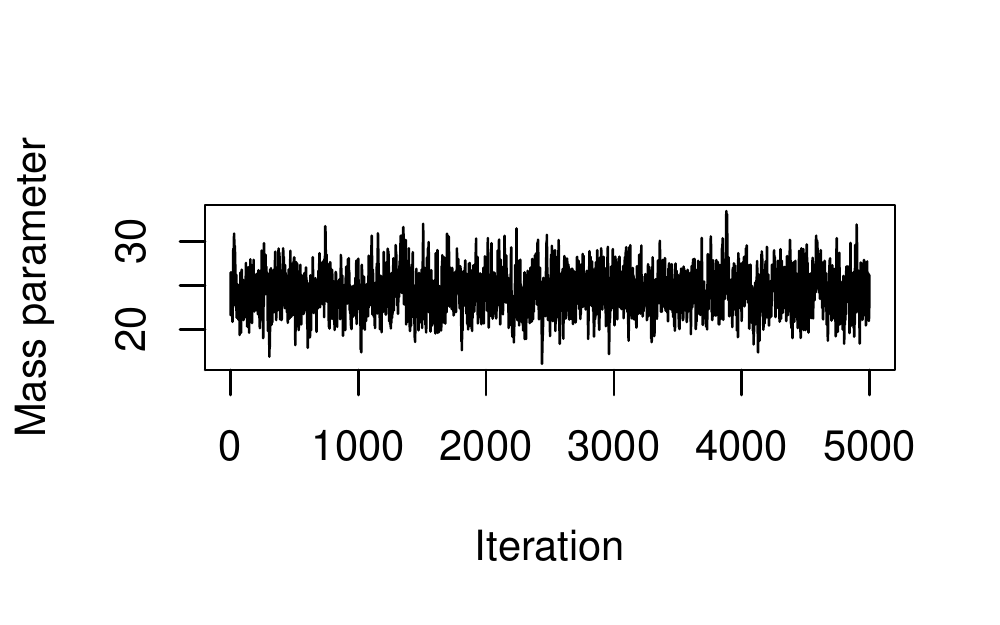}
\vspace{-1.6cm}

\includegraphics[width=.45\linewidth]{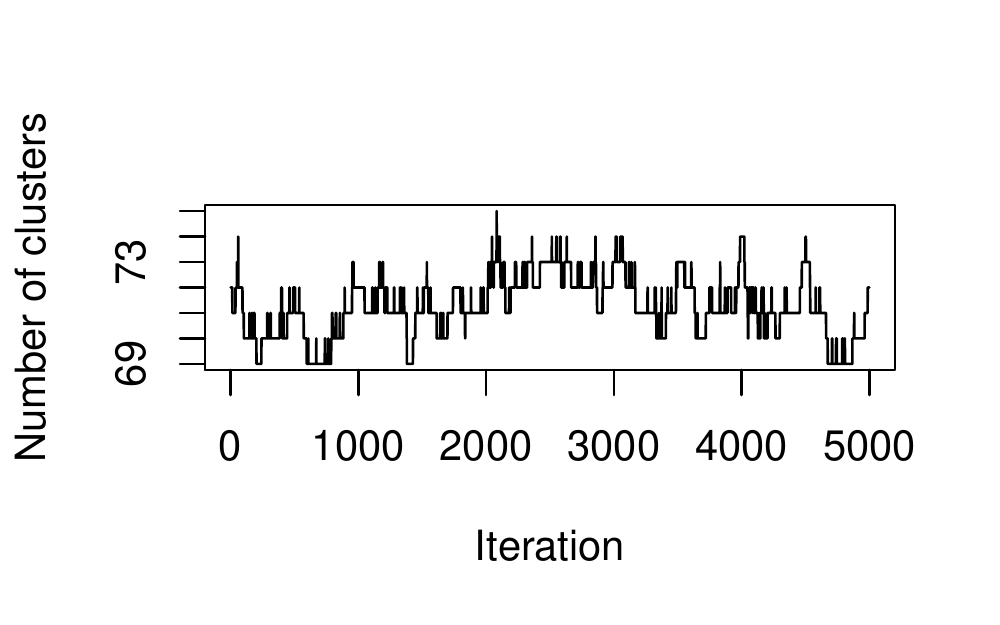}
\includegraphics[width=.45\linewidth]{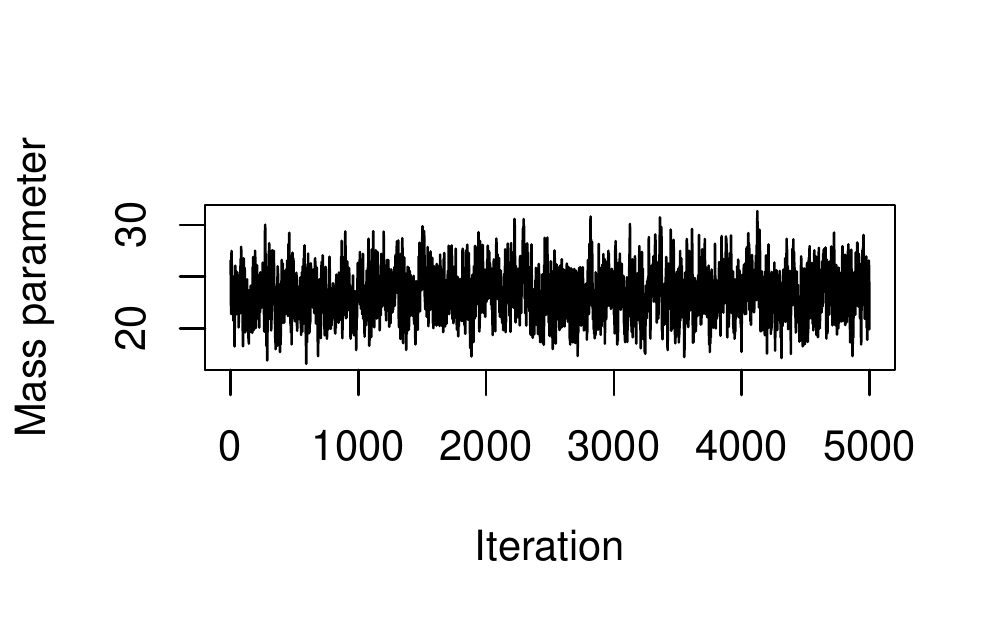}
\vspace{-1.6cm}

\includegraphics[width=.45\linewidth]{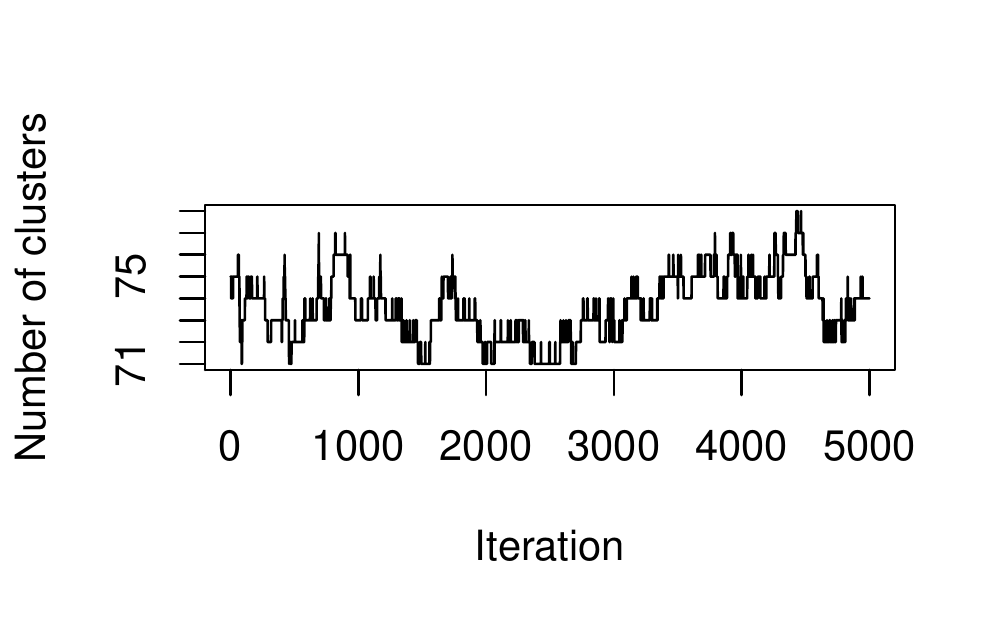}
\includegraphics[width=.45\linewidth]{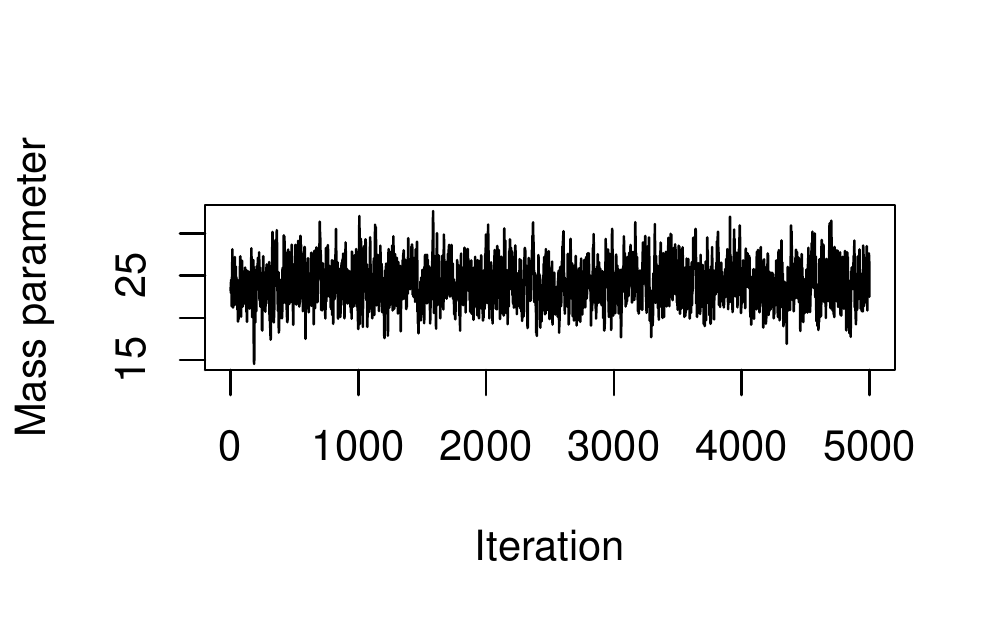}
\caption{MCMC convergence assessment, microRNA data. $\lambda=0$, $\alpha=0.1, 0.5$.}
\end{figure}

\clearpage

\begin{figure}[H]
	\centering
	\includegraphics[width=.36\linewidth]{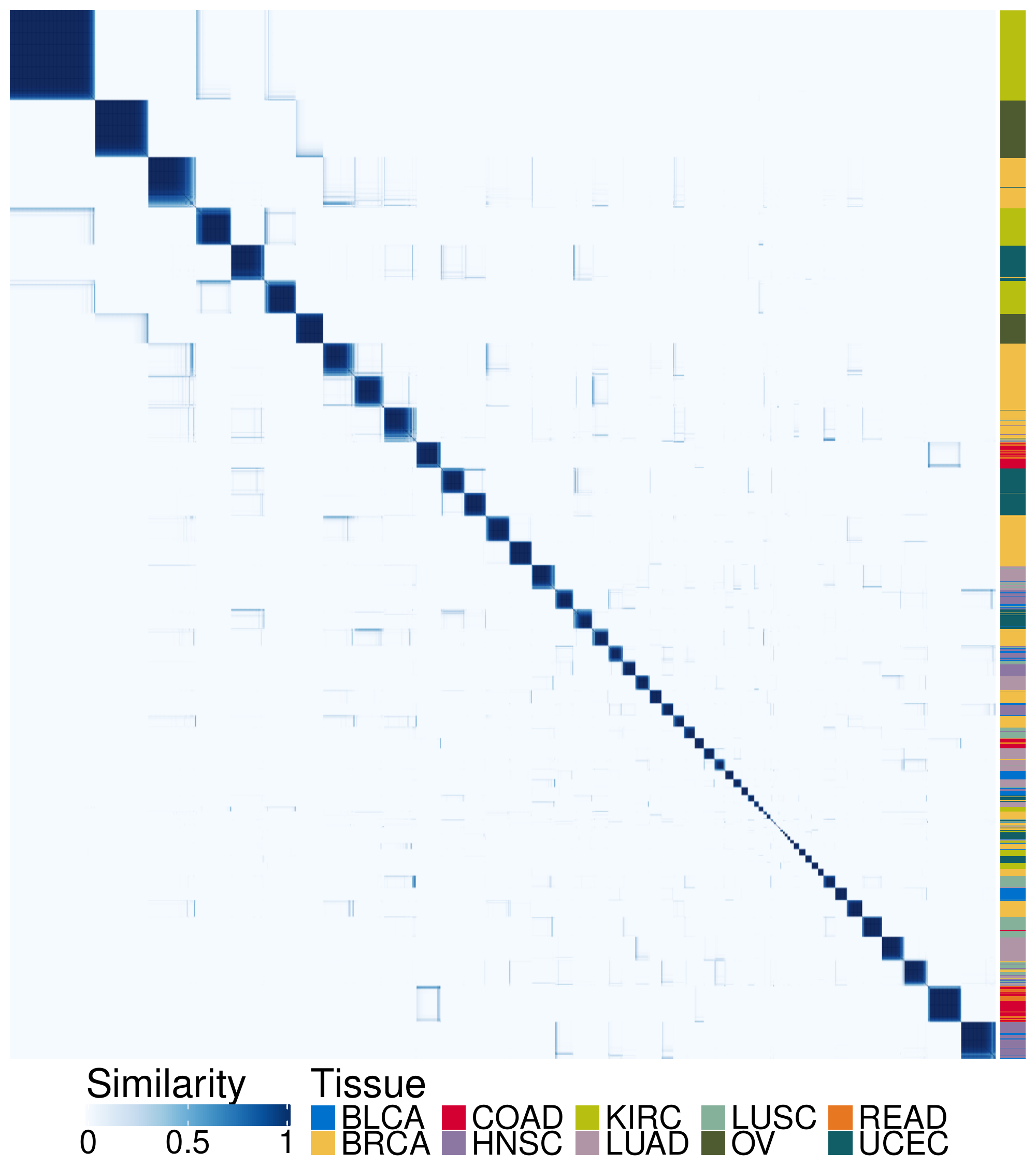}
	\includegraphics[width=.36\linewidth]{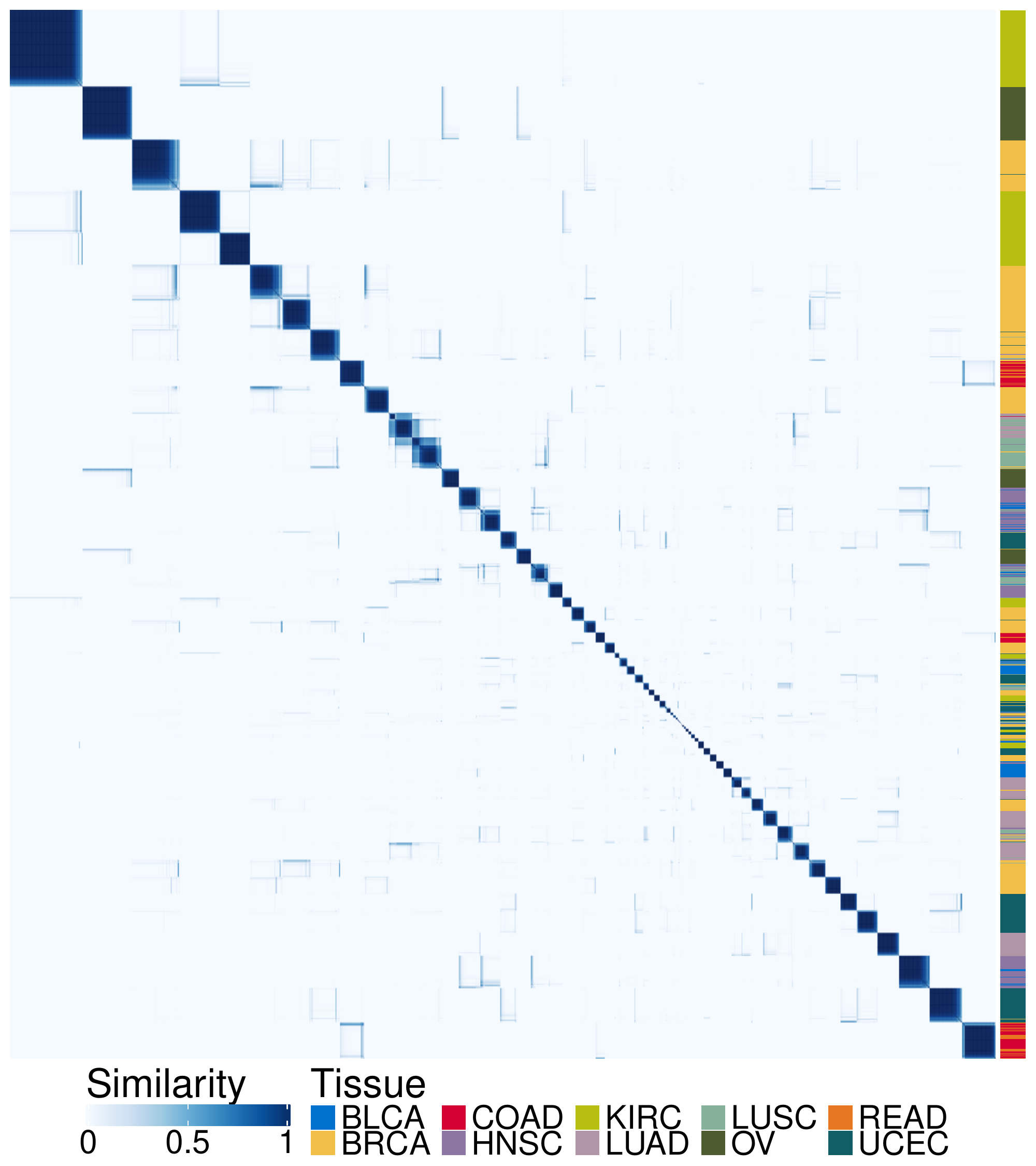}
	\includegraphics[width=.36\linewidth]{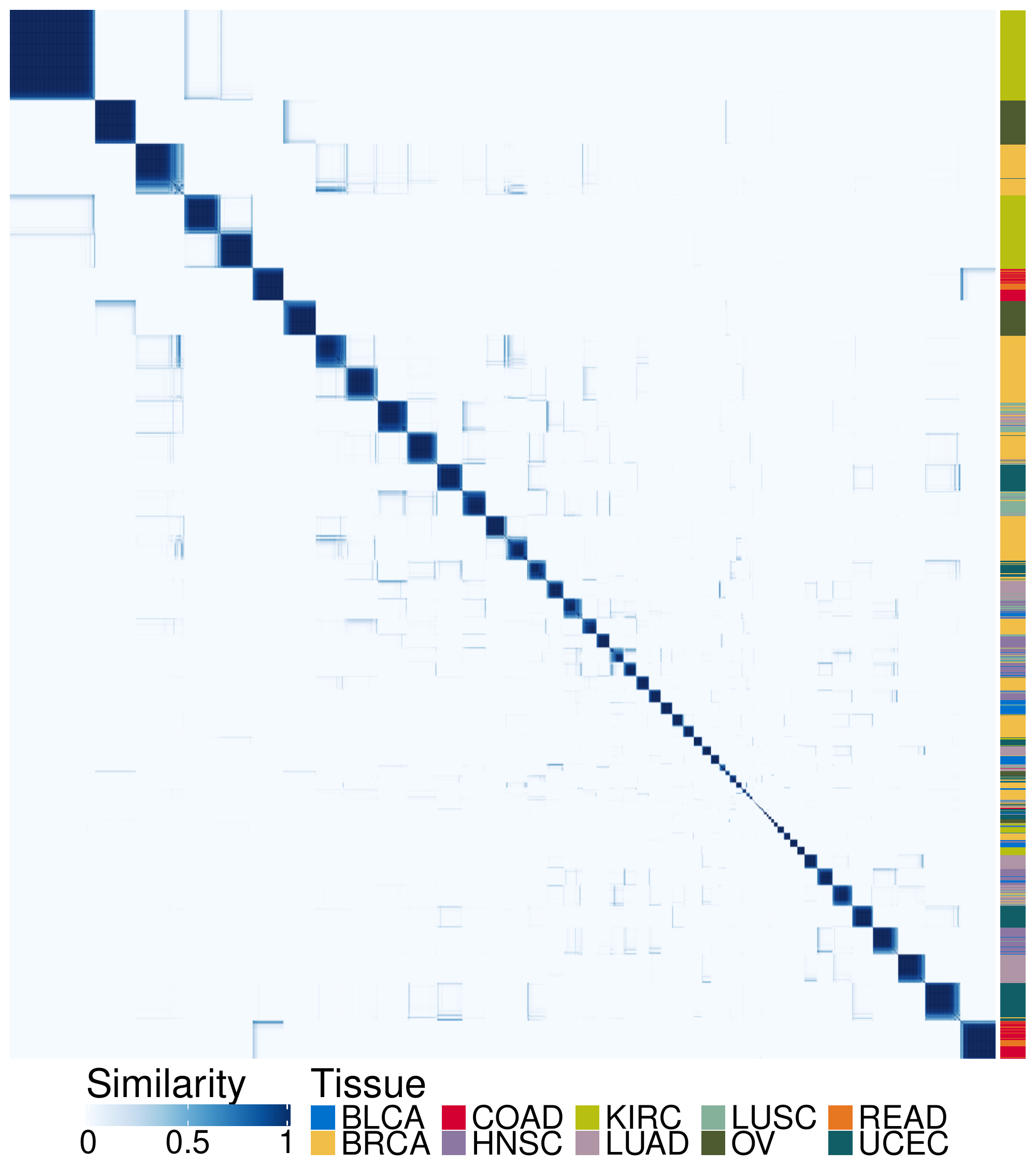}
	\includegraphics[width=.36\linewidth]{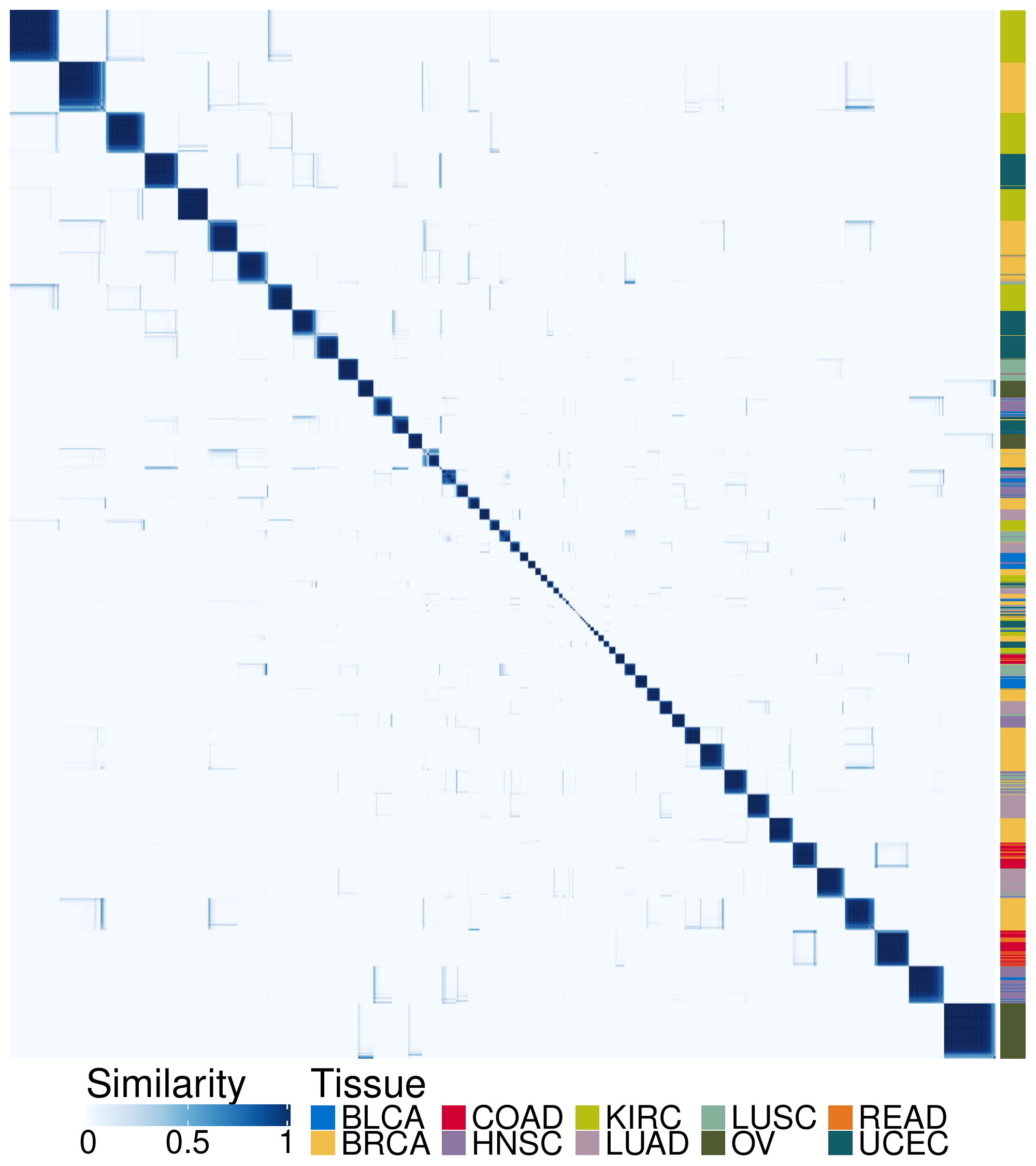}
	\includegraphics[width=.36\linewidth]{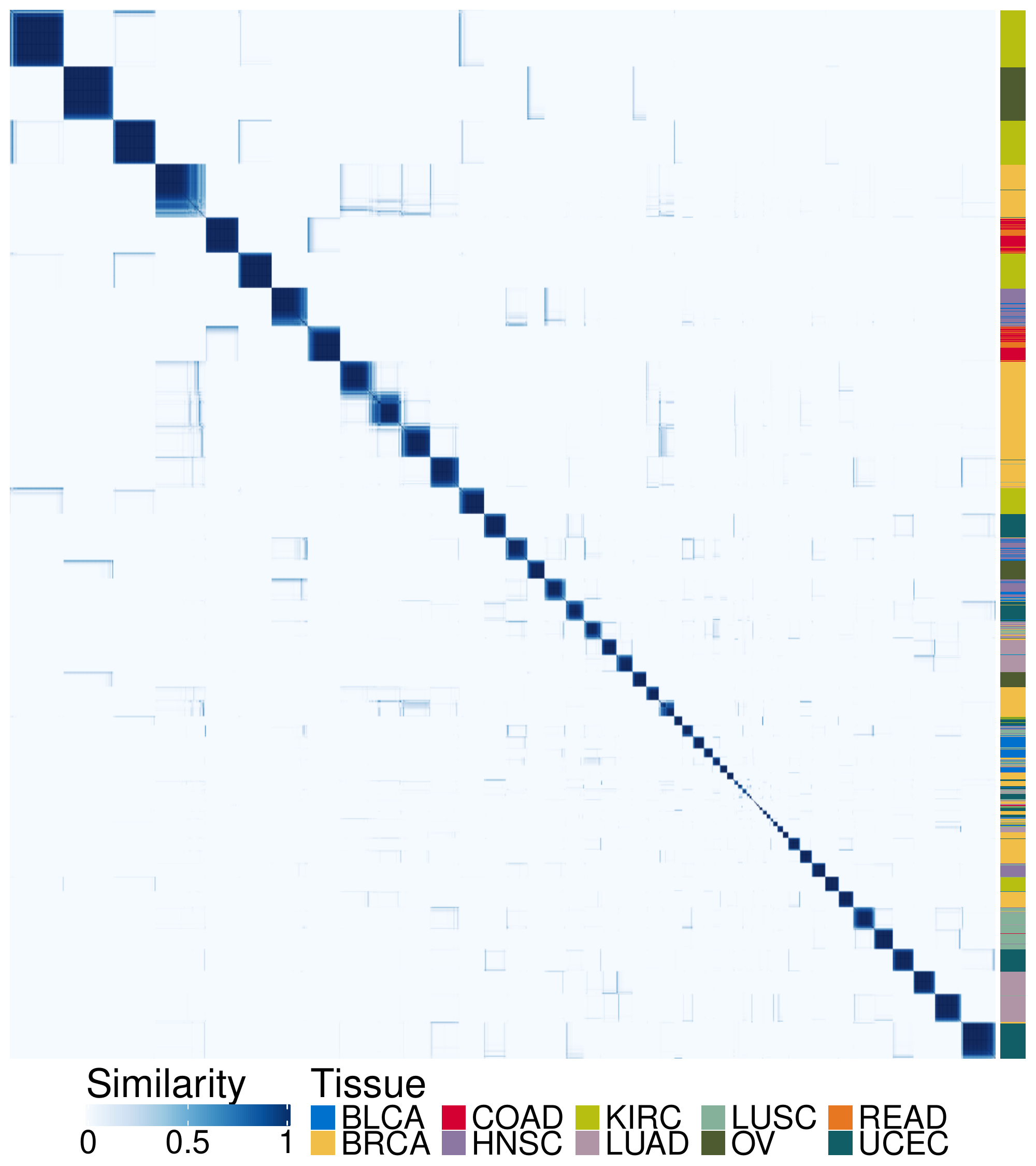}
	\includegraphics[width=.36\linewidth]{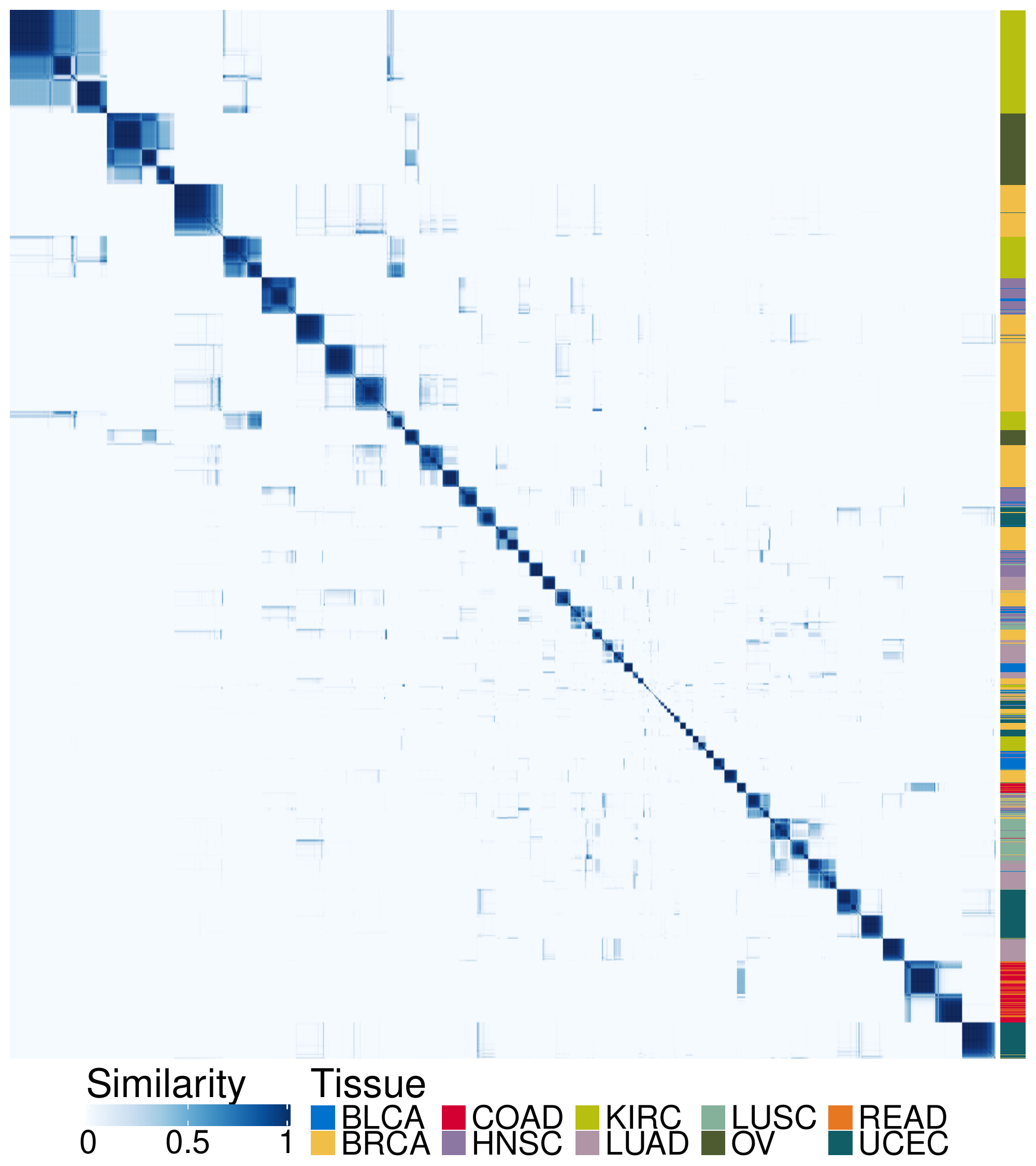}
	\caption{PSMs of the microRNA data. $\alpha=1$.}
\end{figure}

\begin{table}[H]
\centering
\begin{tabular}{l c c c c}
& \textbf{Chain 2} & \textbf{Chain 3} & \textbf{Chain 4} & \textbf{Chain 5} \\
\hline
\textbf{Chain 1} & 0.66 & 0.72 & 0.70 & 0.57 \\
\textbf{Chain 2} & 1 & 0.59 & 0.76 & 0.69 \\
\textbf{Chain 3} && 1 &  0.55 & 0.56 \\
\textbf{Chain 4} && & 1 & 0.72 \\
\hline\\
\end{tabular}
\caption{ARI between the clusterings found on the PSMs of different chains with the number of clusters that maximises the silhouette.}
\end{table} 

\begin{figure}[H]
\centering
\includegraphics[width=.45\linewidth]{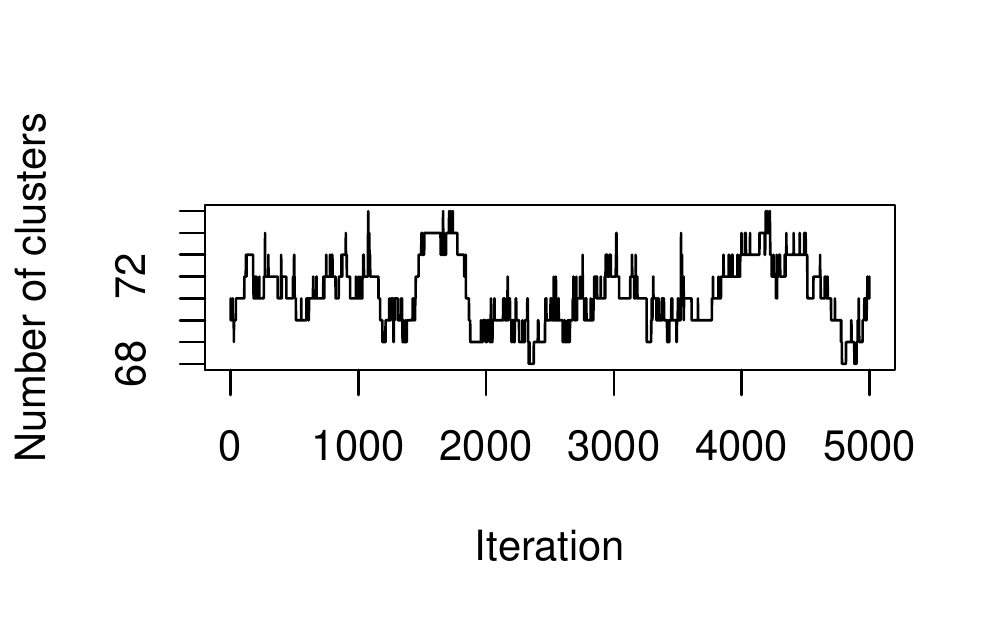}
\includegraphics[width=.45\linewidth]{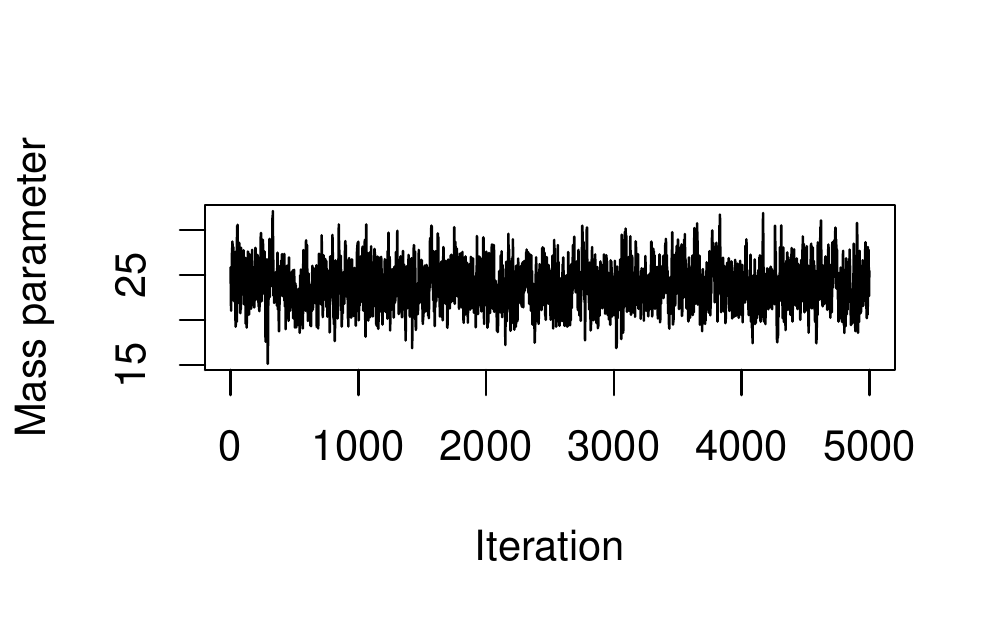}
\vspace{-1.6cm}

\includegraphics[width=.45\linewidth]{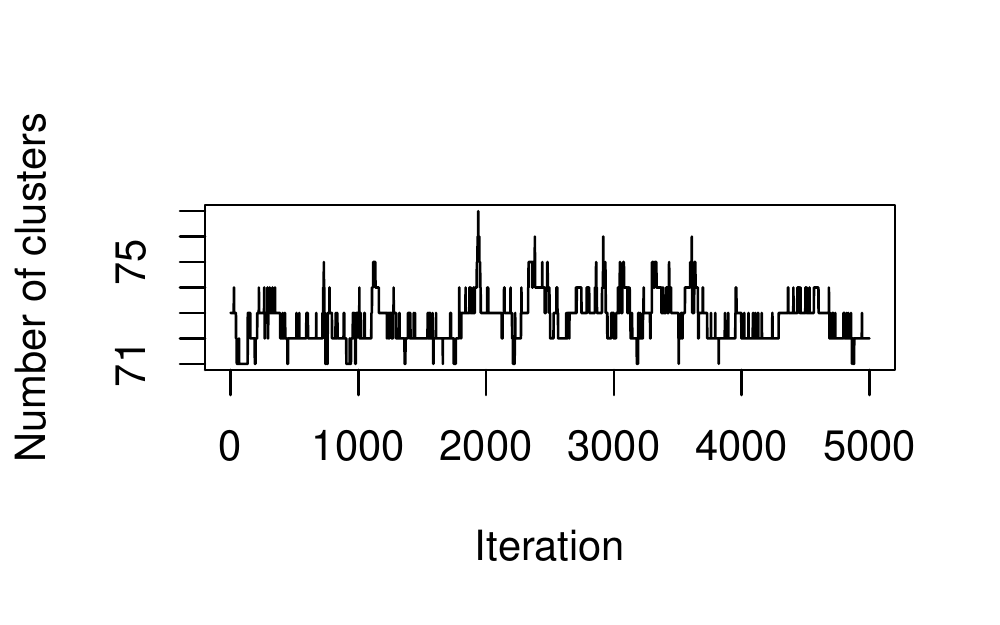}
\includegraphics[width=.45\linewidth]{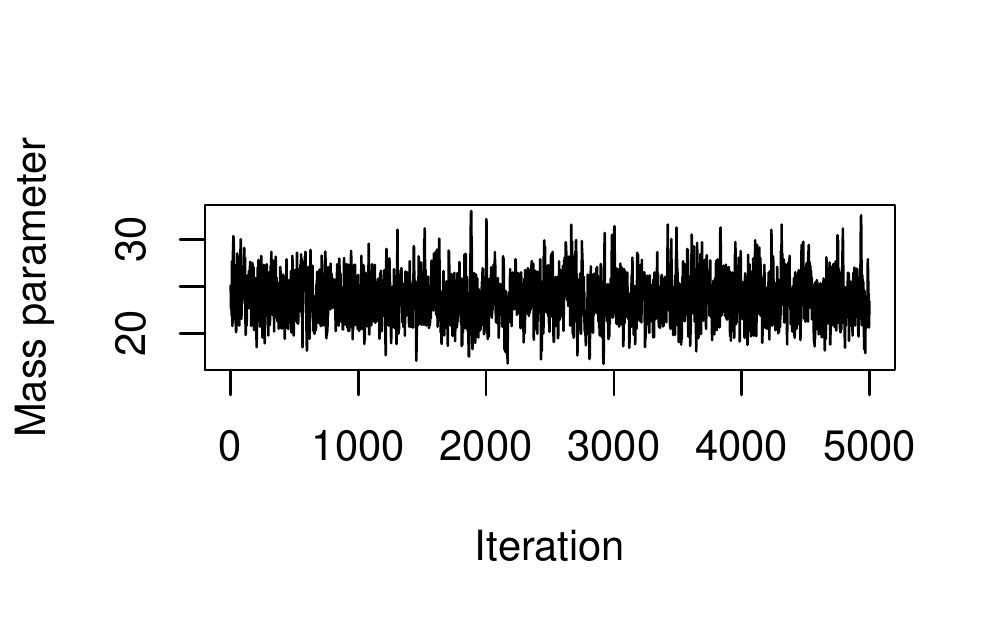}
\vspace{-1.6cm}

\includegraphics[width=.45\linewidth]{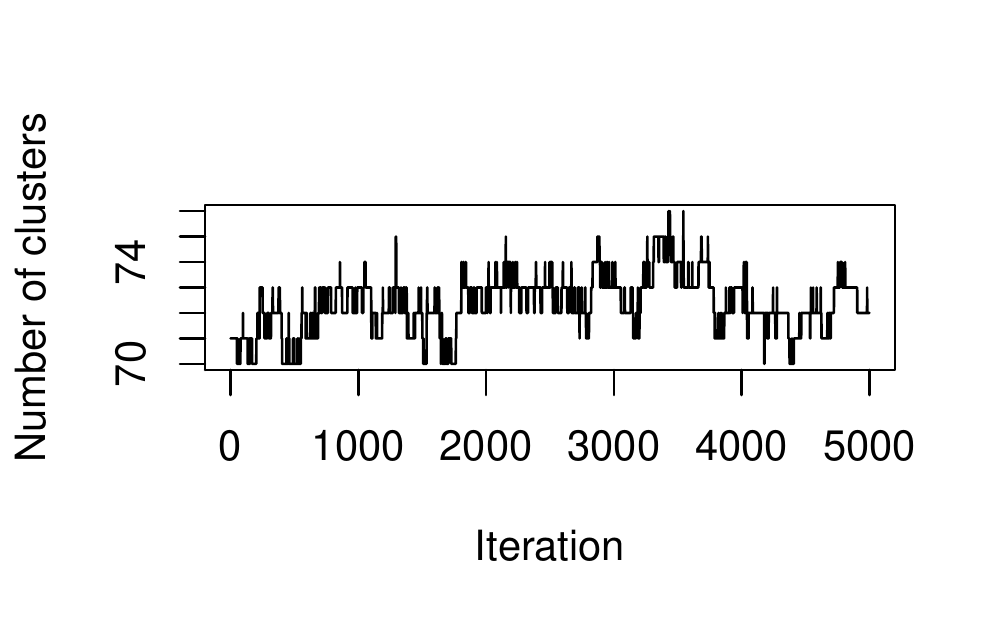}
\includegraphics[width=.45\linewidth]{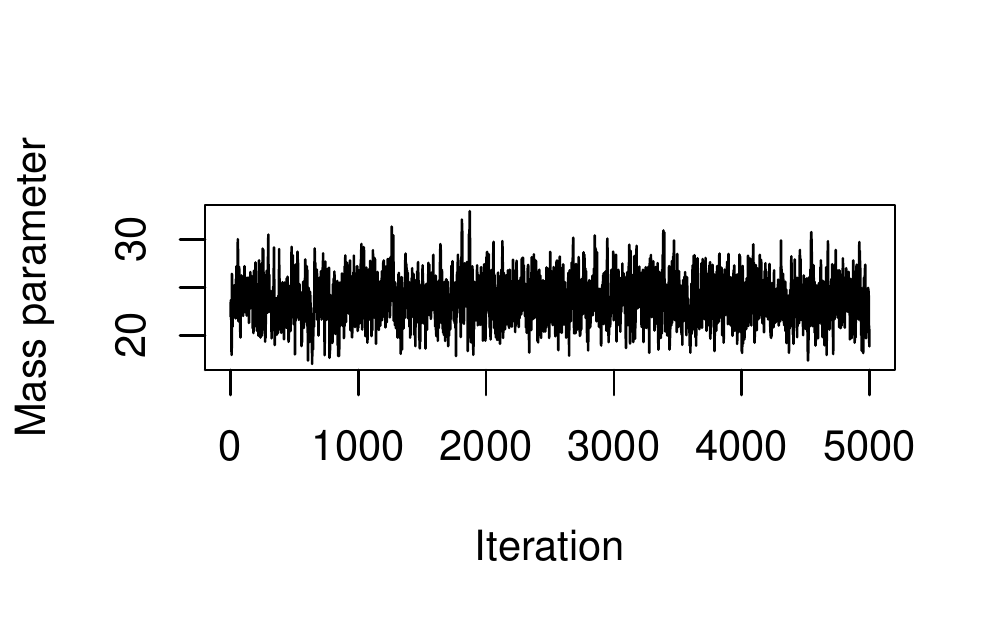}
\vspace{-1.6cm}

\includegraphics[width=.45\linewidth]{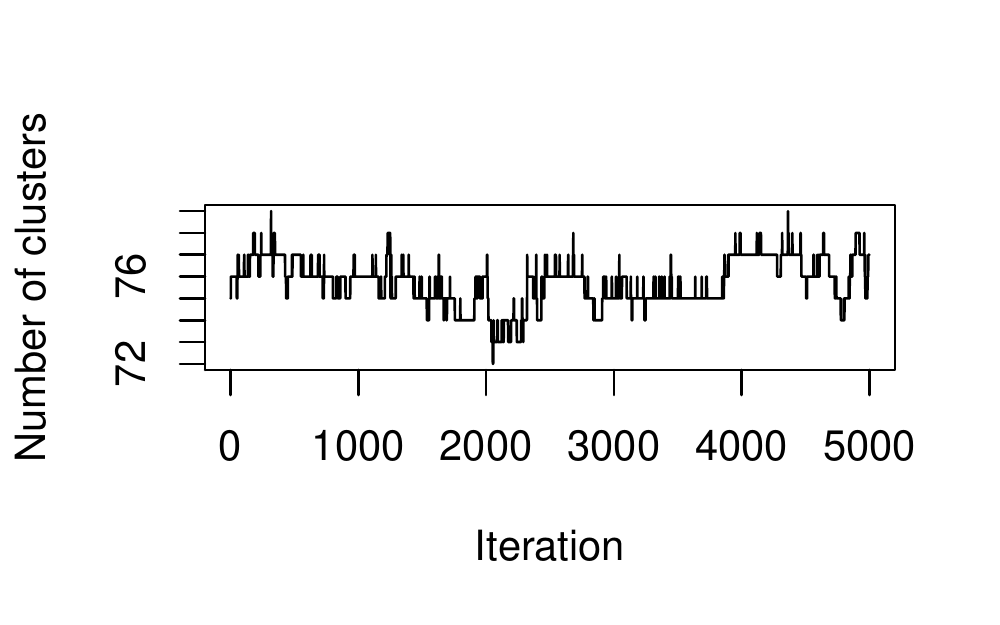}
\includegraphics[width=.45\linewidth]{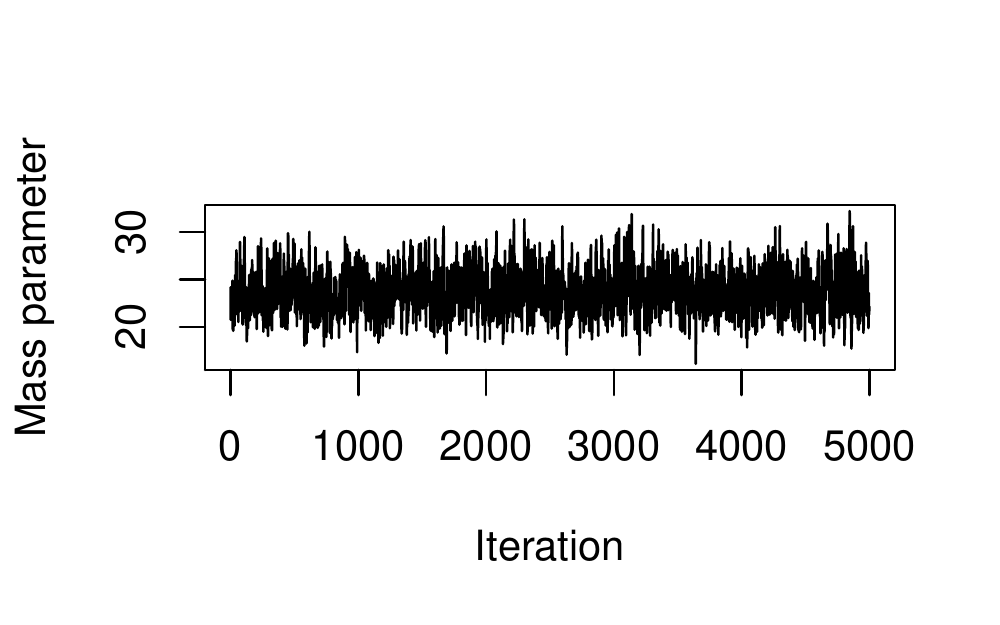}
\vspace{-1.6cm}

\includegraphics[width=.45\linewidth]{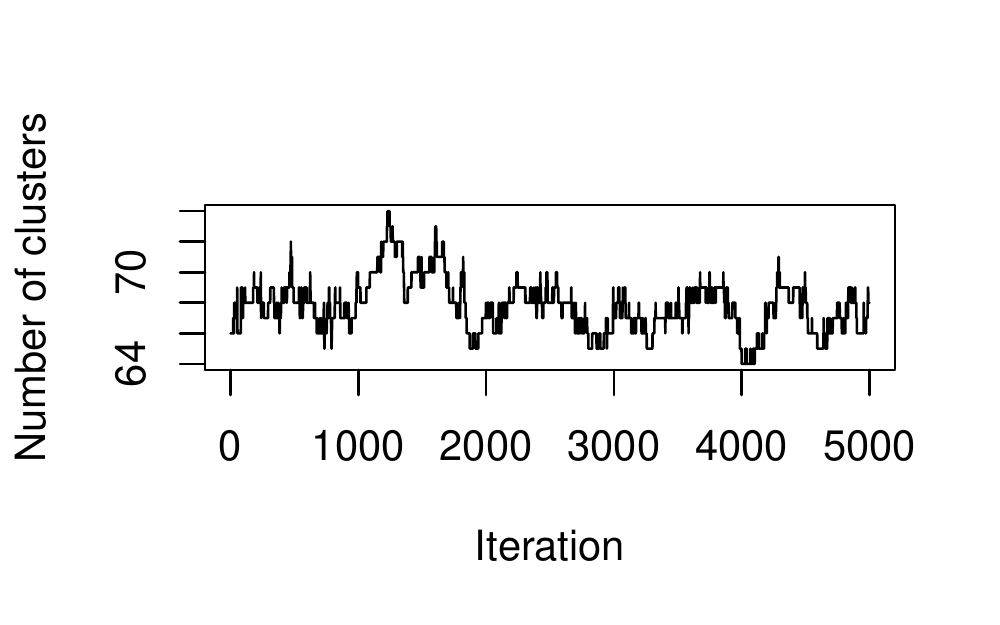}
\includegraphics[width=.45\linewidth]{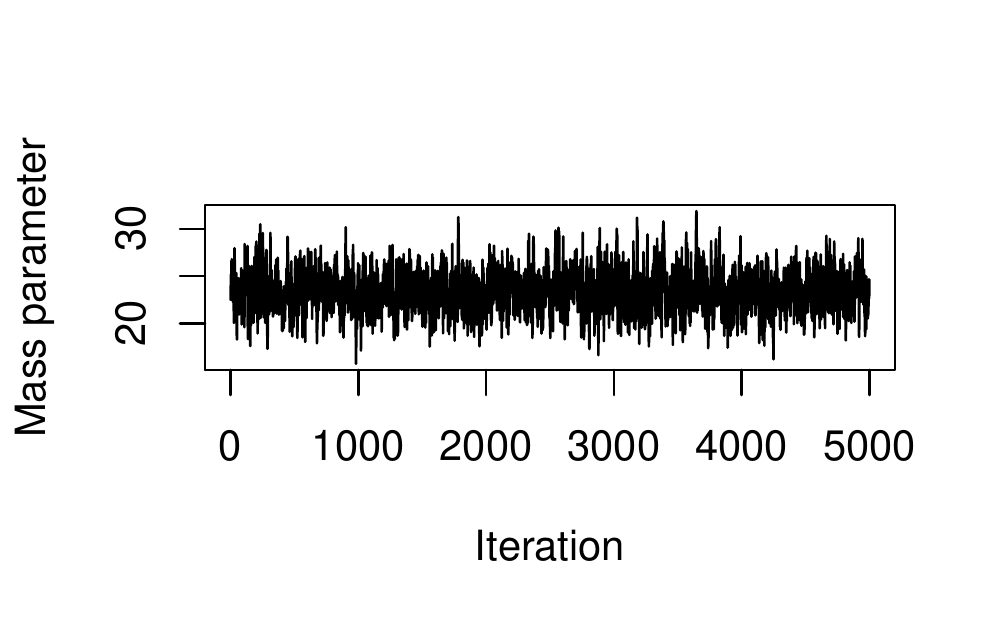}
\caption{MCMC convergence assessment, microRNA data. $\alpha=1$.}
\end{figure}

\clearpage

\subsubsection{Unsupervised integration: additional figures}

\begin{figure}[H]
	\centering
	\includegraphics[width=.675\textwidth]{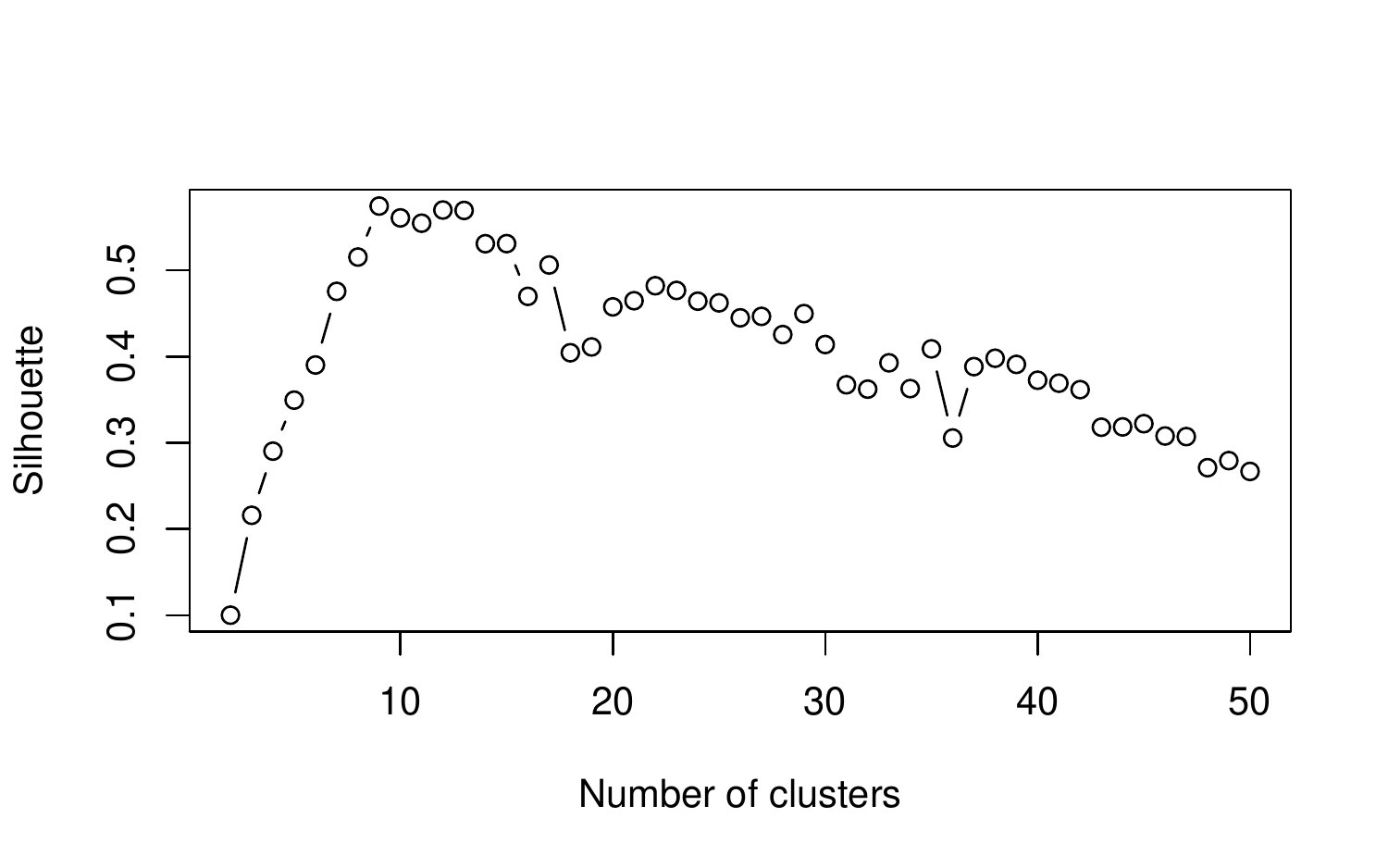}
	\caption{Average silhouette.}
	\label{fig:pancan-10-silhouette-unsupervised}
\end{figure}

\paragraph{Choice of the number of clusters} In Figure \ref{fig:pancan-10-silhouette-unsupervised} are reported the average values of the silhouette when the number of clusters goes from 2 to 50. The maximum is at $K=15$.

\begin{figure}[H]
	\centering
	\includegraphics[width=.6\textwidth]{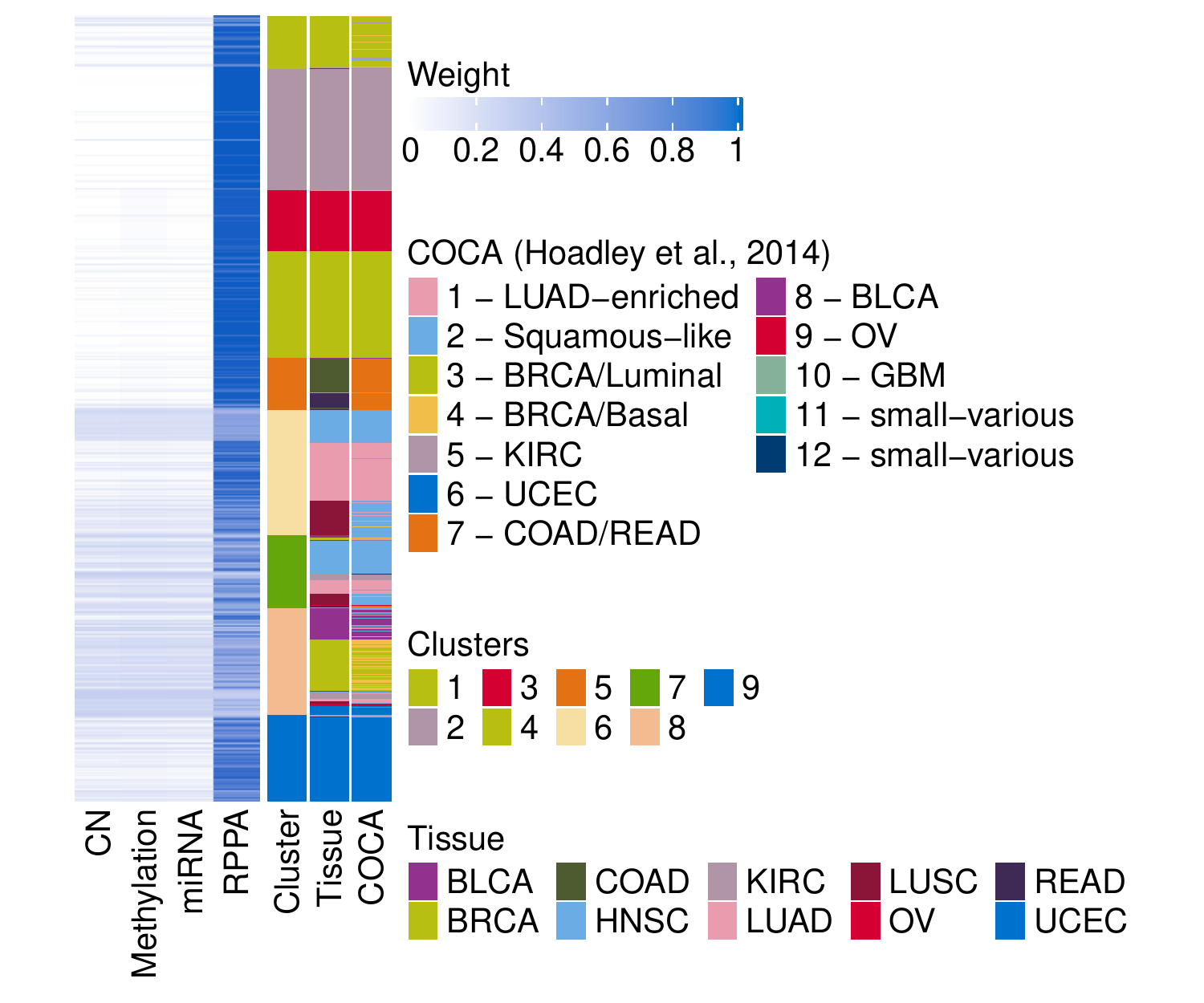}
	\caption{Weights assigned by the multiple kernel $k$-means algorithm to each observations in each layer, where ``CN'' stands for copy number and ``RPPA'' for reverse phase protein array.}
	\label{fig:pancan-10-weights}
\end{figure}

\paragraph{Kernel weights} Figure \ref{fig:pancan-10-weights} shows the weights assigned to each PSM by the multiple kernel $k$-means algorithm. 

\clearpage

\begin{figure}[H]
	\centering
	\includegraphics[width=.48\textwidth]{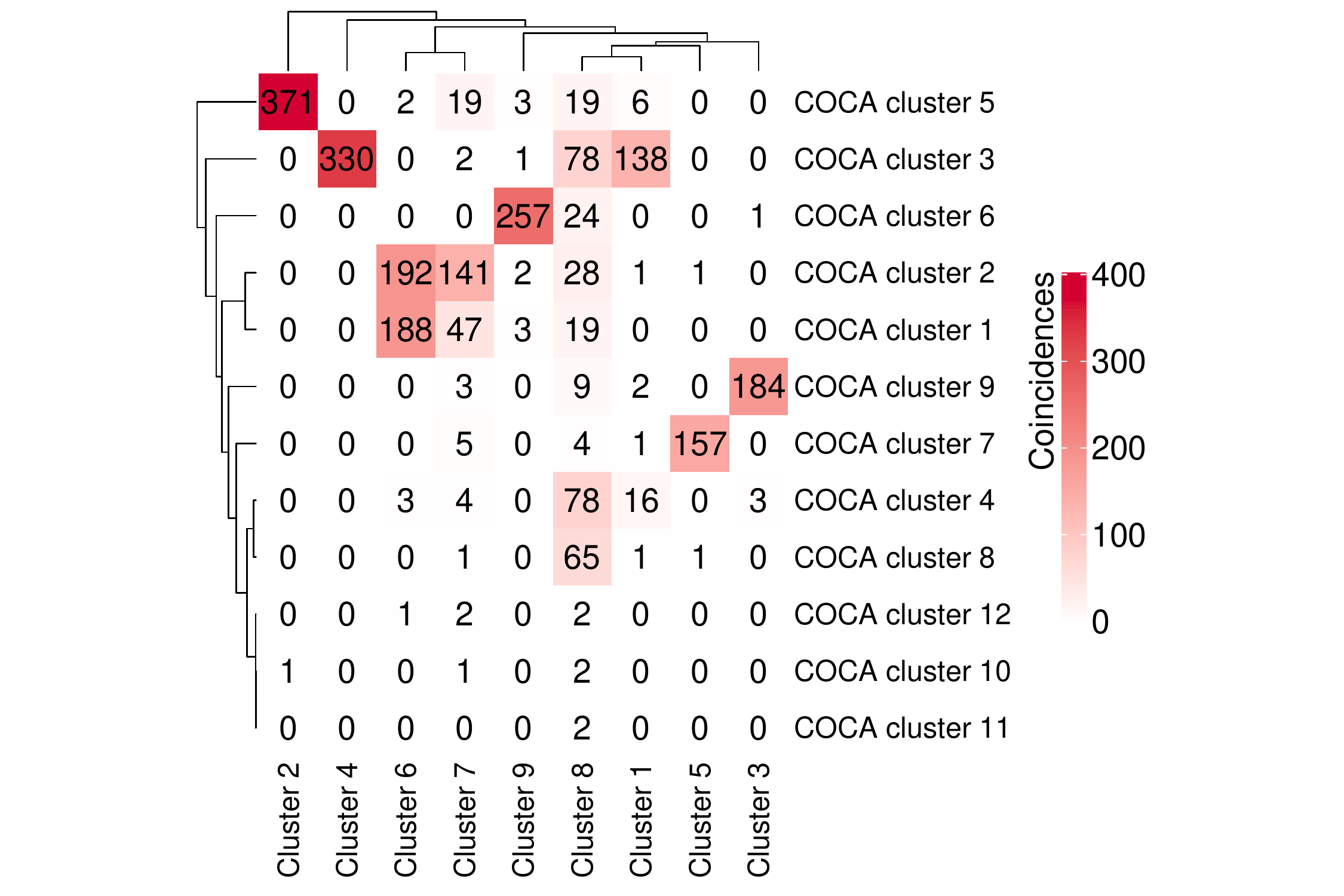}
	\caption{Comparison between the clusters found combining the PSMs of each layer using multiple kernel learning and those identified by \citet{hoadley2014multiplatform} using COCA.}
	\label{fig:pancan-10-comparison-hoadley-et-al-unsupervised}
\end{figure}

\paragraph{Comparison with the clusters identified by \citet{hoadley2014multiplatform}}
In Figure \ref{fig:pancan-10-comparison-hoadley-et-al-unsupervised} are shown the correspondences between the clusters found in the main paper and the clusters identified by \citet{hoadley2014multiplatform} using Cluster-Of-Clusters Analysis (COCA).

\begin{figure}[H]
	\centering
	\includegraphics[width=.8\textwidth]{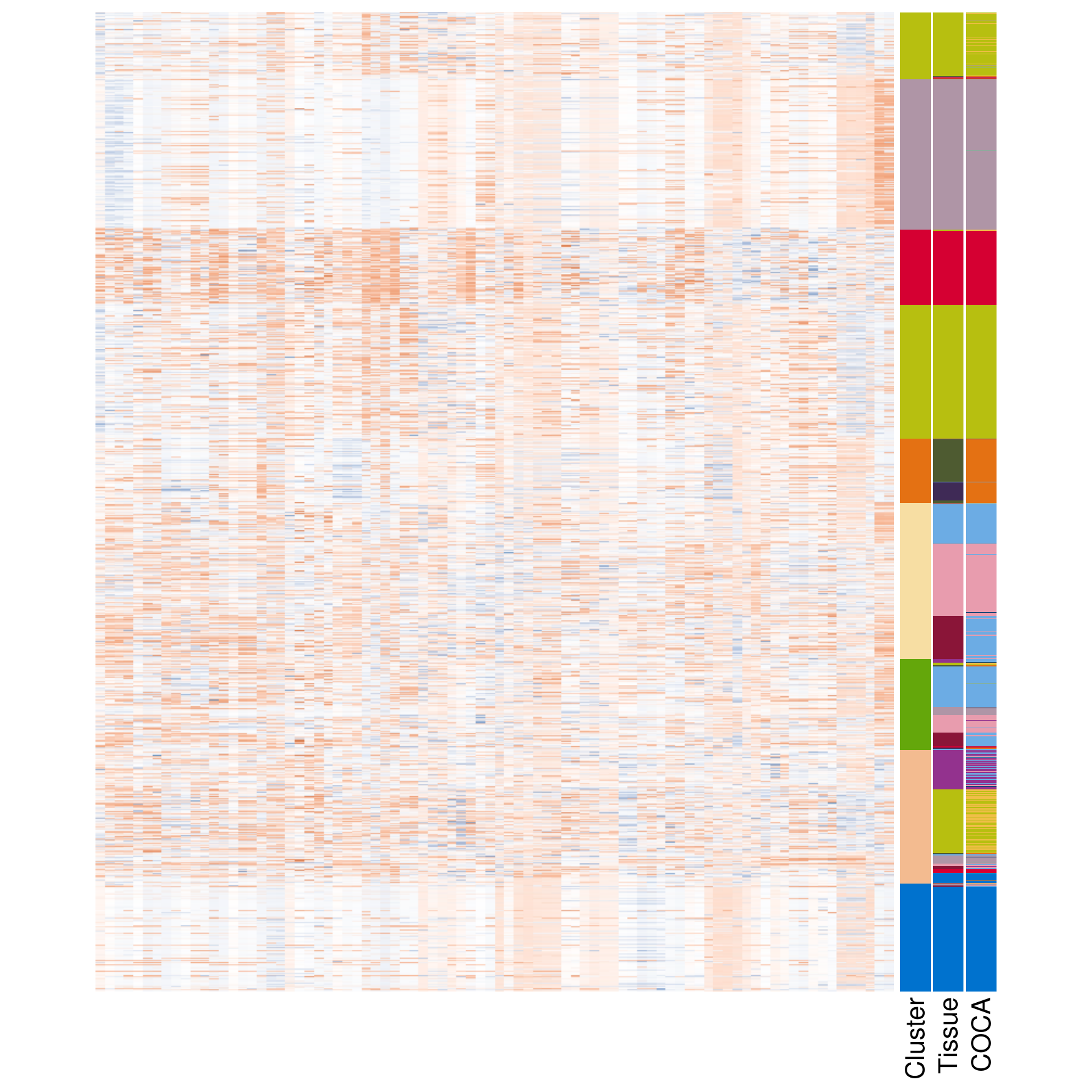}
	\caption{Copy number data and final clusters.}
	\label{fig:pancan-cn-unsupervised-finalclusters}
\end{figure}

\begin{figure}[H]
	\centering
	\includegraphics[width=.8\textwidth]{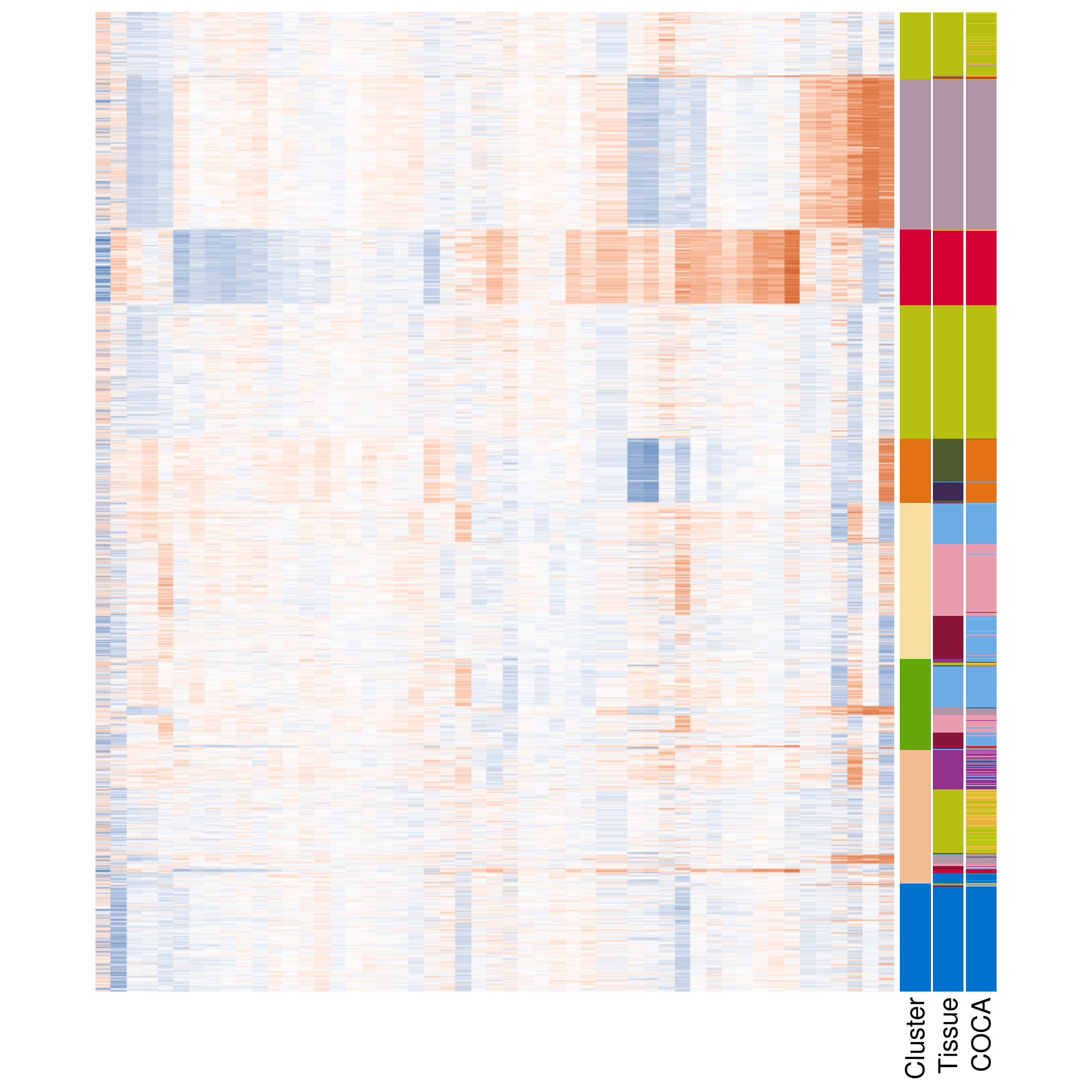}
	\caption{microRNA expression data and final clusters.}
	\label{fig:pancan-mirna-unsupervised-finalclusters}
\end{figure}

\begin{figure}[H]
	\centering
	\includegraphics[width=.8\textwidth]{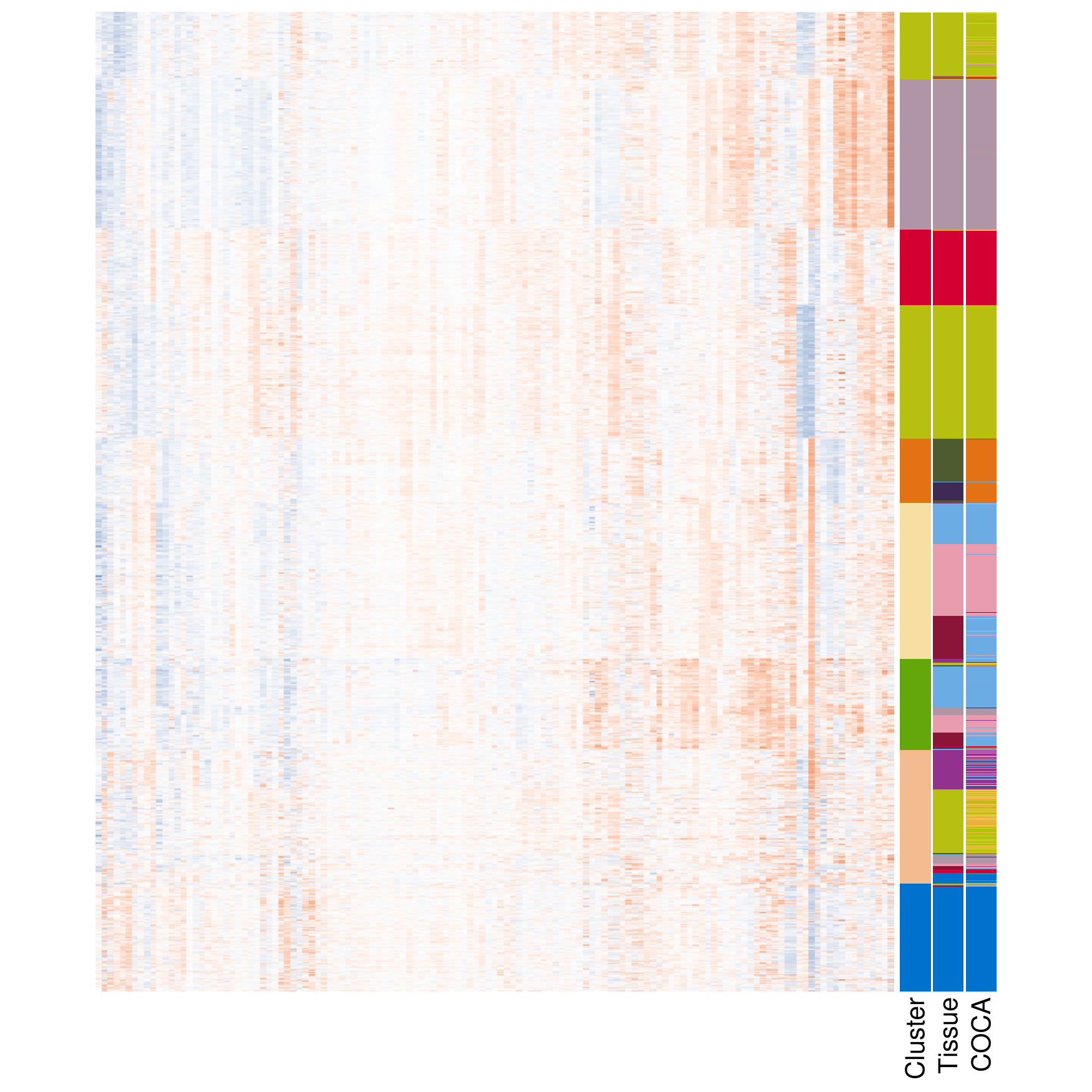}
	\caption{Protein expression data and final clusters.}
	\label{fig:pancan-protein-unsupervised-finalclusters}
\end{figure}

\clearpage
\paragraph{Clustering structure in the data}
Figures \ref{fig:pancan-cn-unsupervised-finalclusters}, \ref{fig:pancan-mirna-unsupervised-finalclusters}, \ref{fig:pancan-mirna-unsupervised-finalclusters}, and \ref{fig:pancan-protein-unsupervised-finalclusters} show the four data layers where the rows have been sorted by final cluster.

\begin{figure}[H]
	\centering
	\includegraphics[width=.8\textwidth]{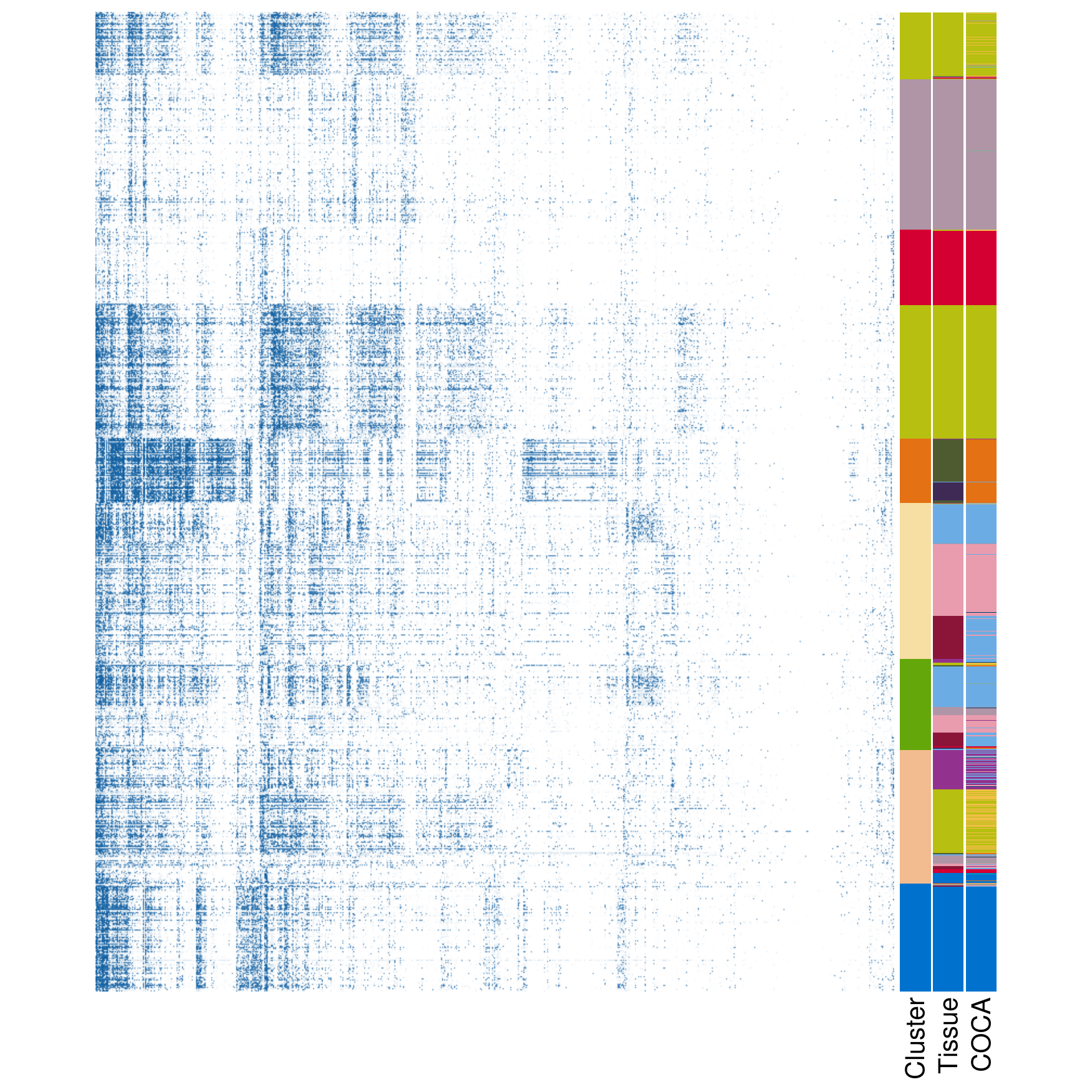}
	\caption{Methylation data and final clusters.}
	\label{fig:pancan-methylation-unsupervised-finalclusters}
\end{figure}

\subsubsection{Unsupervised integration after variable selection, $\alpha=0.1$}

\begin{figure}[H]
	\centering
	\begin{subfigure}[t]{\textwidth}
		\includegraphics[width=\textwidth]{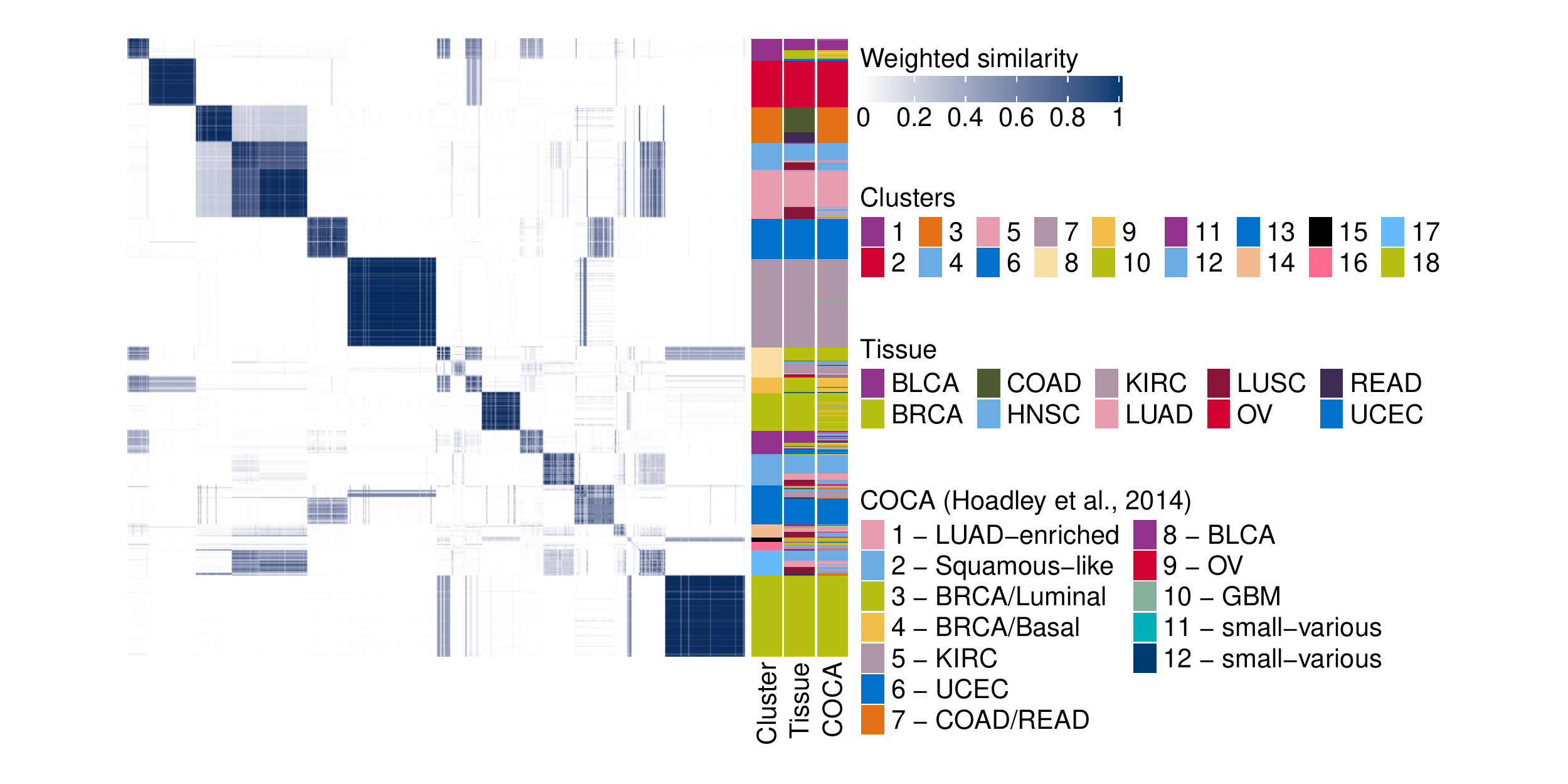}
		\caption{Weighted kernel and clusters.}
	\end{subfigure}
	\caption{Unsupervised multiplatform analysis of ten cancer types. Weighted kernel, final clusters, tissues of origin, and COCA clusters.}
\end{figure}
\begin{figure}
	\begin{subfigure}[t]{\textwidth}
		\includegraphics[width=\textwidth]{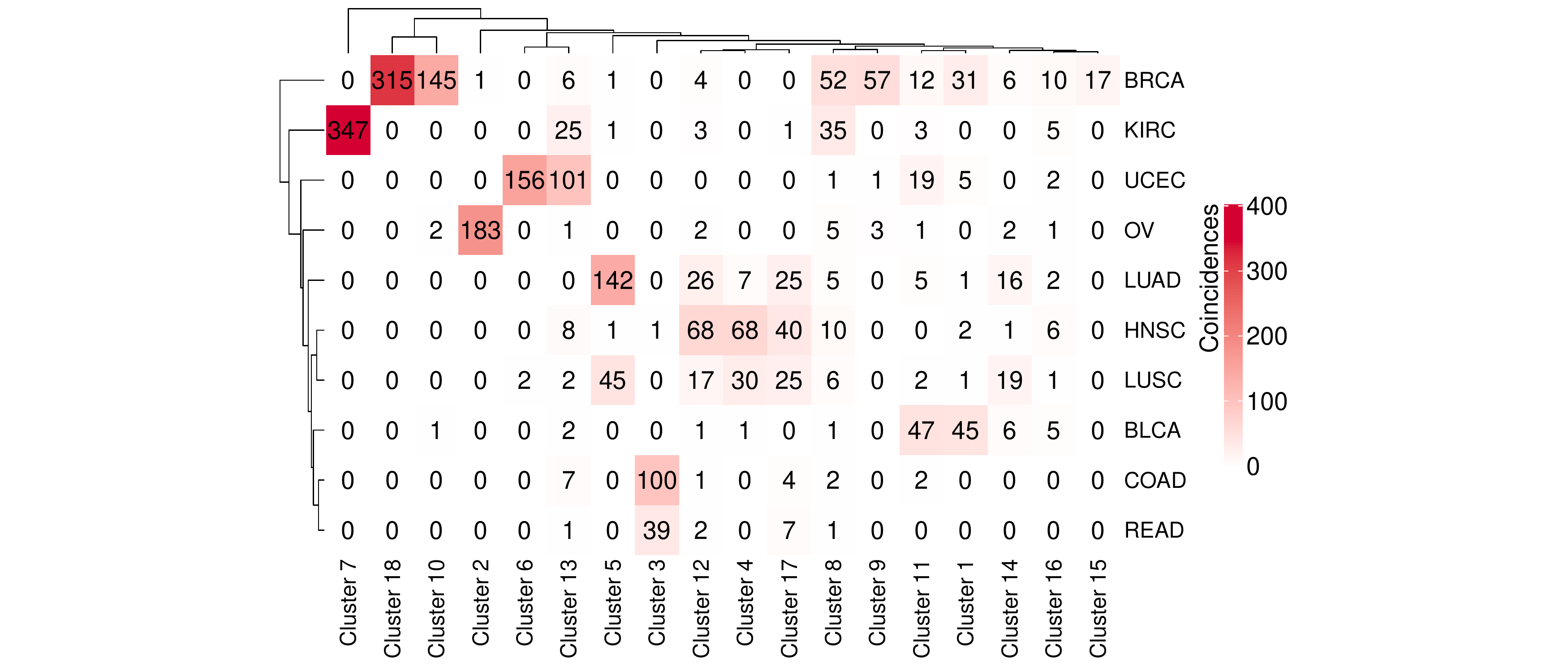}
		\caption{Comparison to tissue of origin.}
	\end{subfigure}
	\vspace{1cm}
	
	\begin{subfigure}[t]{\textwidth}
		\includegraphics[width=\textwidth]{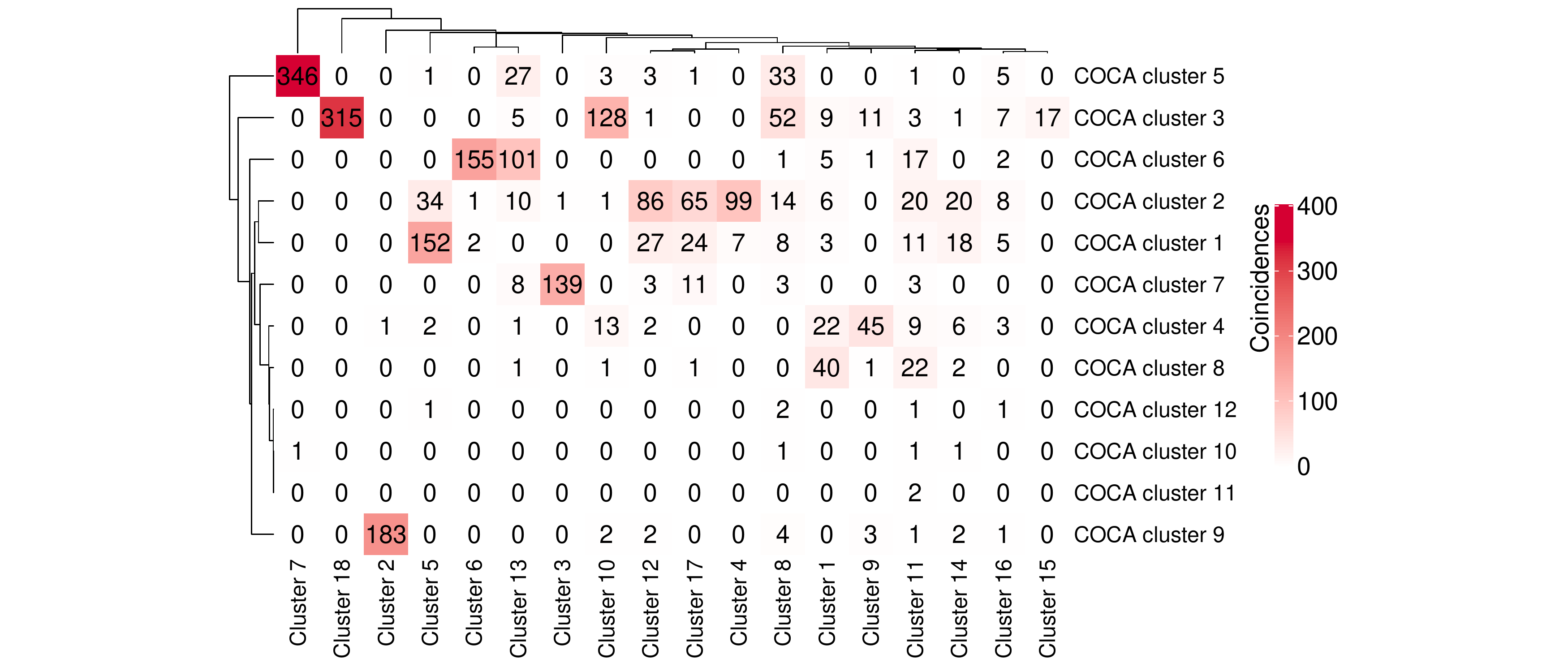}
		\caption{Comparison to COCA clusters.}
	\end{subfigure}
	\caption{Unsupervised multiplatform analysis of ten cancer types. \textbf{(a)} Coincidence matrix comparing the tissue of origin of the tumour samples to the new clusters. \textbf{(b)} Coincidence matrix comparing the COCA clusters of Hoadley \emph{et al.} to the new clusters.}
	\label{fig:unsupervised-pancan10-alpha01}
\end{figure}

\begin{figure}[H]
	\centering
	\includegraphics[width=.675\textwidth]{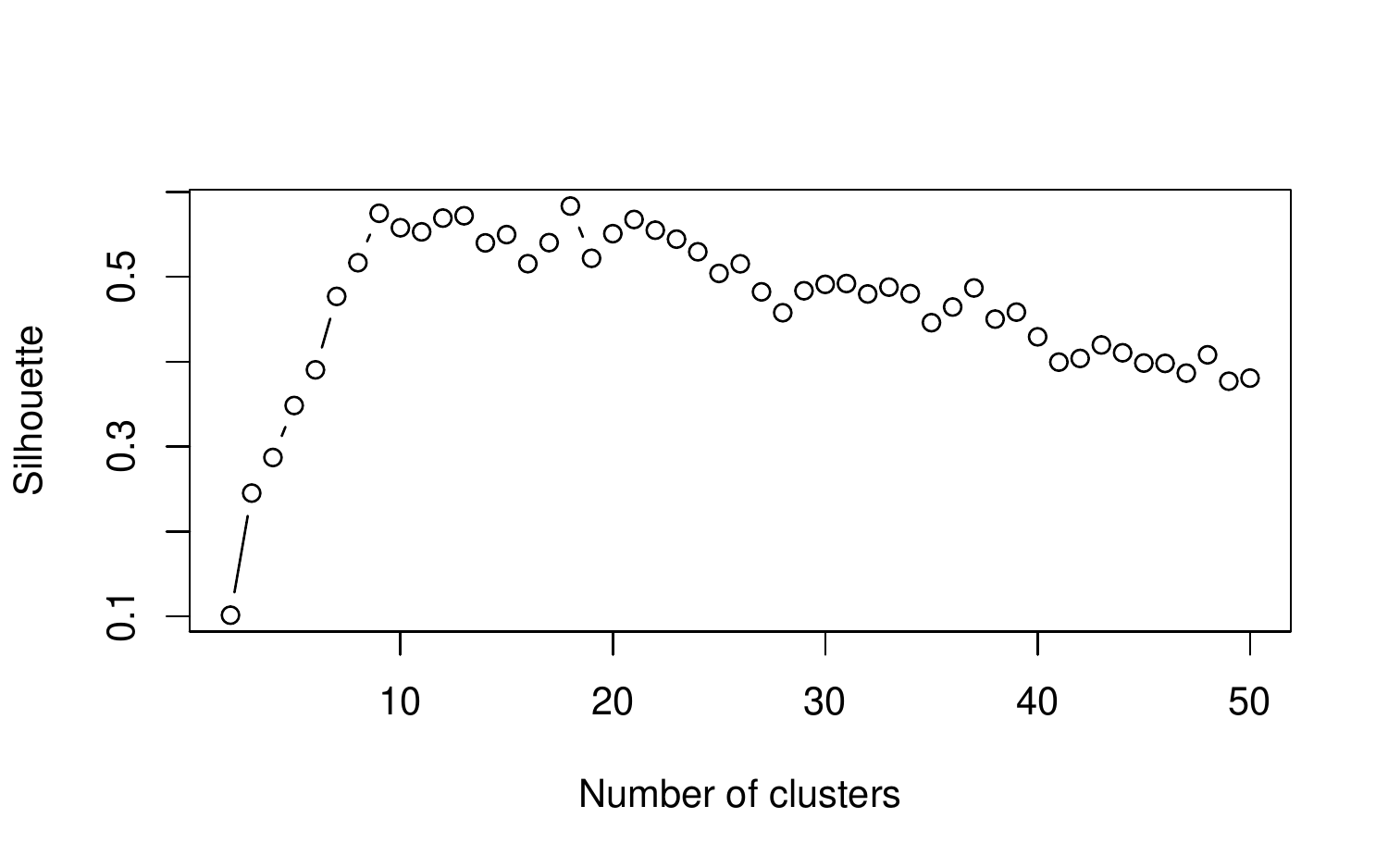}
	\caption{Average silhouette.}
	\label{fig:pancan-10-silhouette-unsupervised-alpha01}
\end{figure}

\begin{figure}[H]
	\centering
	\includegraphics[width=.6\textwidth]{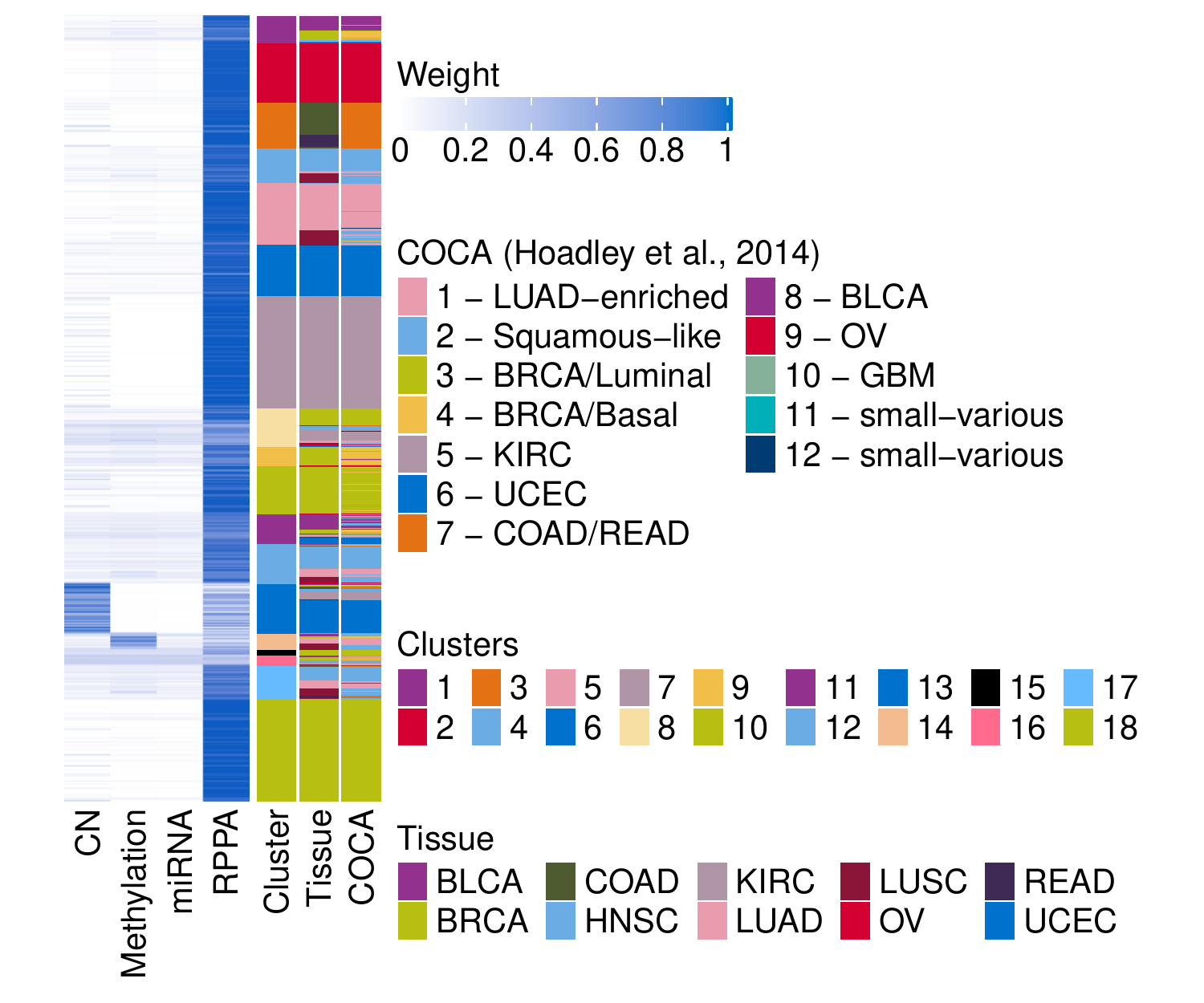}
	\caption{Weights assigned by the multiple kernel $k$-means algorithm to each observations in each layer, where ``CN'' stands for copy number and ``RPPA'' for reverse phase protein array. The weights assigned on average to the tumour samples in each layer are: copy number 7.9\%, methylation 4.6\%,  miRNA 3.2\%,  protein 84.3\%.}
	\label{fig:pancan-10-weights-alpha01}
\end{figure}

\clearpage

\subsubsection{Unsupervised integration after variable selection, $\alpha=0.5$}

\begin{figure}[H]
	\centering
	\begin{subfigure}[t]{\textwidth}
		\includegraphics[width=\textwidth]{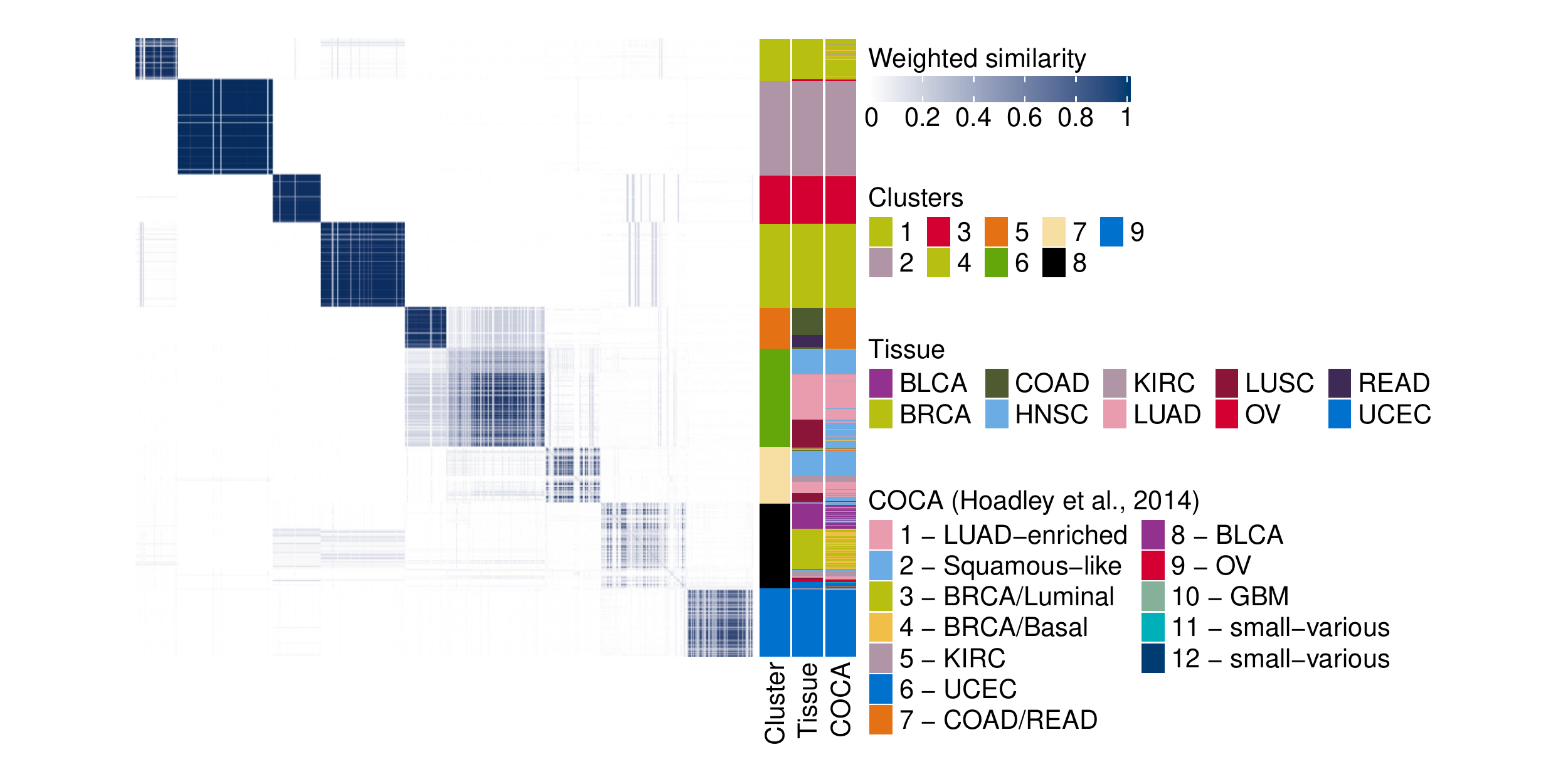}
		\caption{Weighted kernel and clusters.}
	\end{subfigure}
	\begin{subfigure}[t]{.46\textwidth}
		\includegraphics[width=\textwidth]{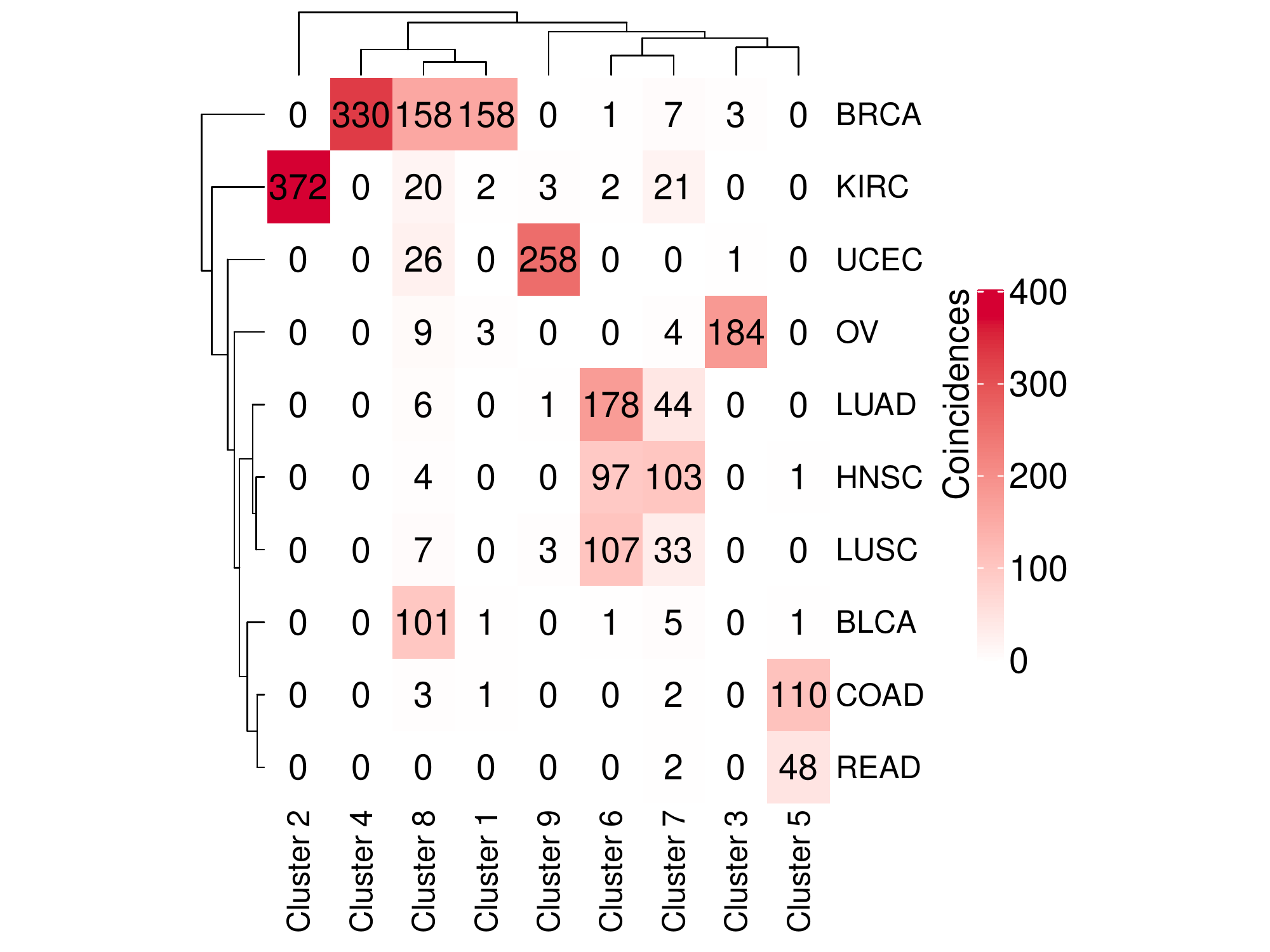}
		\caption{Comparison to tissue of origin.}
	\end{subfigure}
	\begin{subfigure}[t]{.51\textwidth}
		\includegraphics[width=\textwidth]{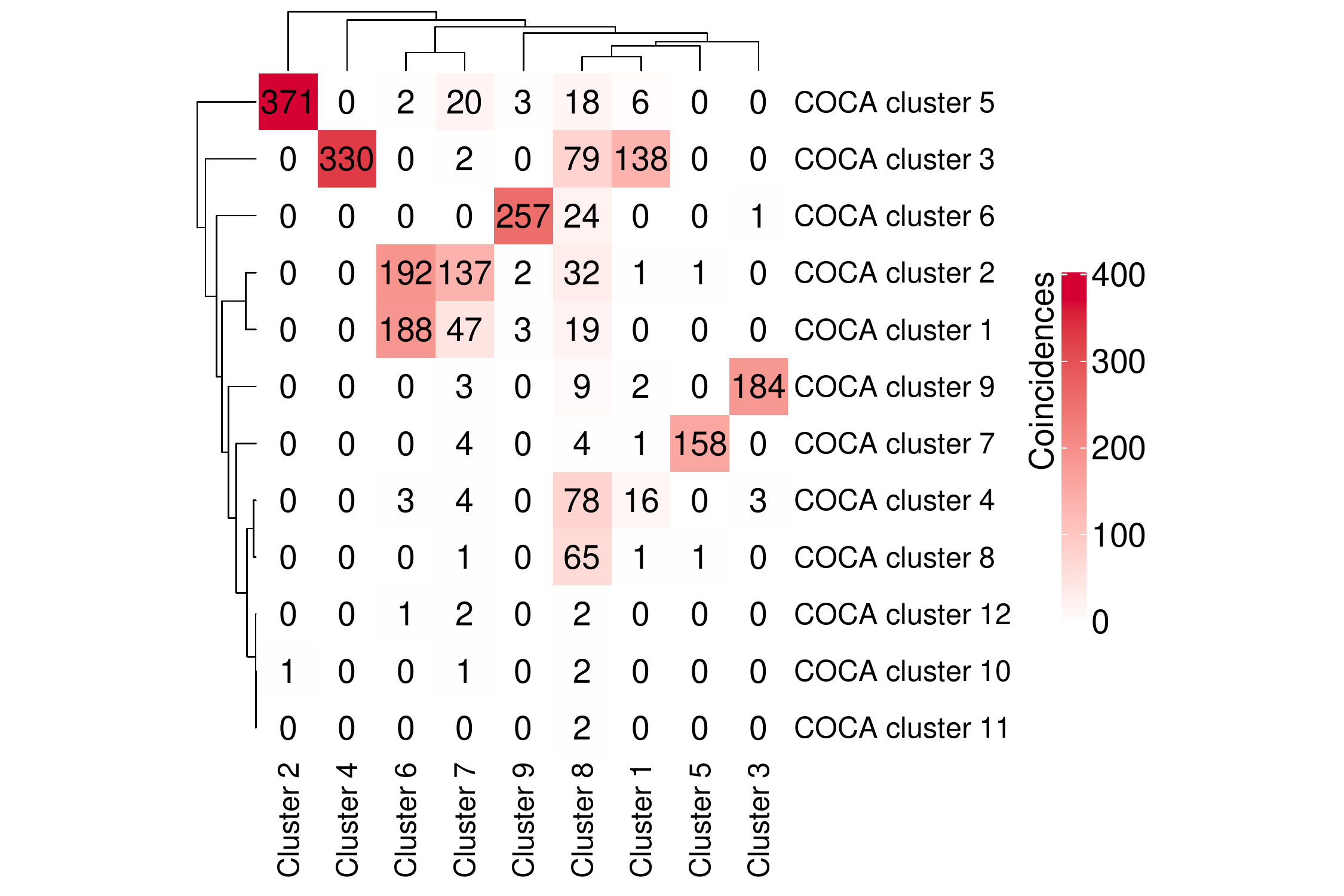}
		\caption{Comparison to COCA clusters.}
	\end{subfigure}
	\caption{Unsupervised multiplatform analysis of ten cancer types. \textbf{(a)} Weighted kernel, final clusters, tissues of origin, and COCA clusters. \textbf{(b)} Coincidence matrix comparing the tissue of origin of the tumour samples to the new clusters. \textbf{(c)} Coincidence matrix comparing the COCA clusters of Hoadley \emph{et al.} to the new clusters.}
	\label{fig:unsupervised-pancan10-alpha05}
\end{figure}

\begin{figure}[H]
	\centering
	\includegraphics[width=.675\textwidth]{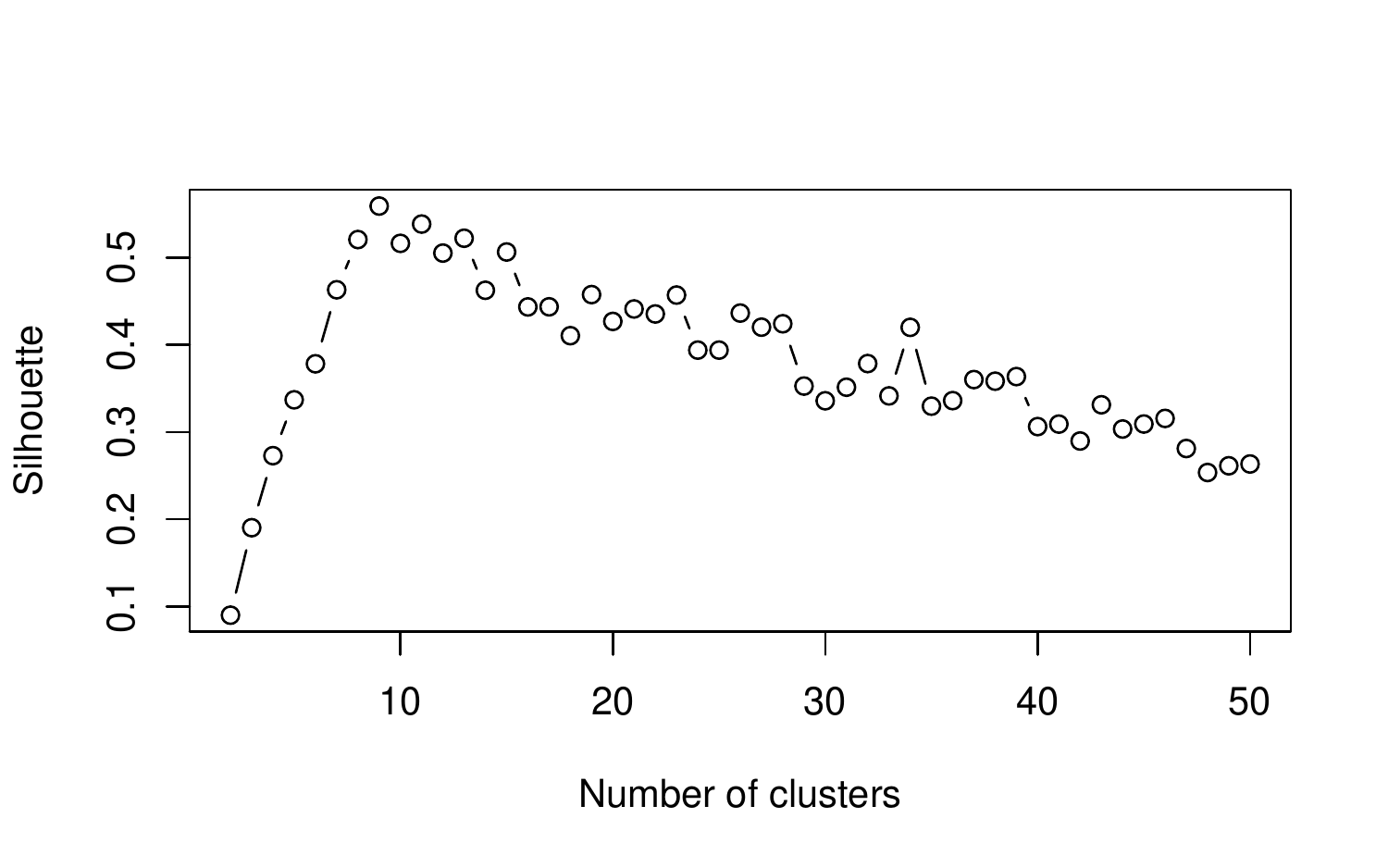}
	\caption{Average silhouette.}
	\label{fig:pancan-10-silhouette-unsupervised-alpha05}
\end{figure}

\begin{figure}[H]
	\centering
	\includegraphics[width=.6\textwidth]{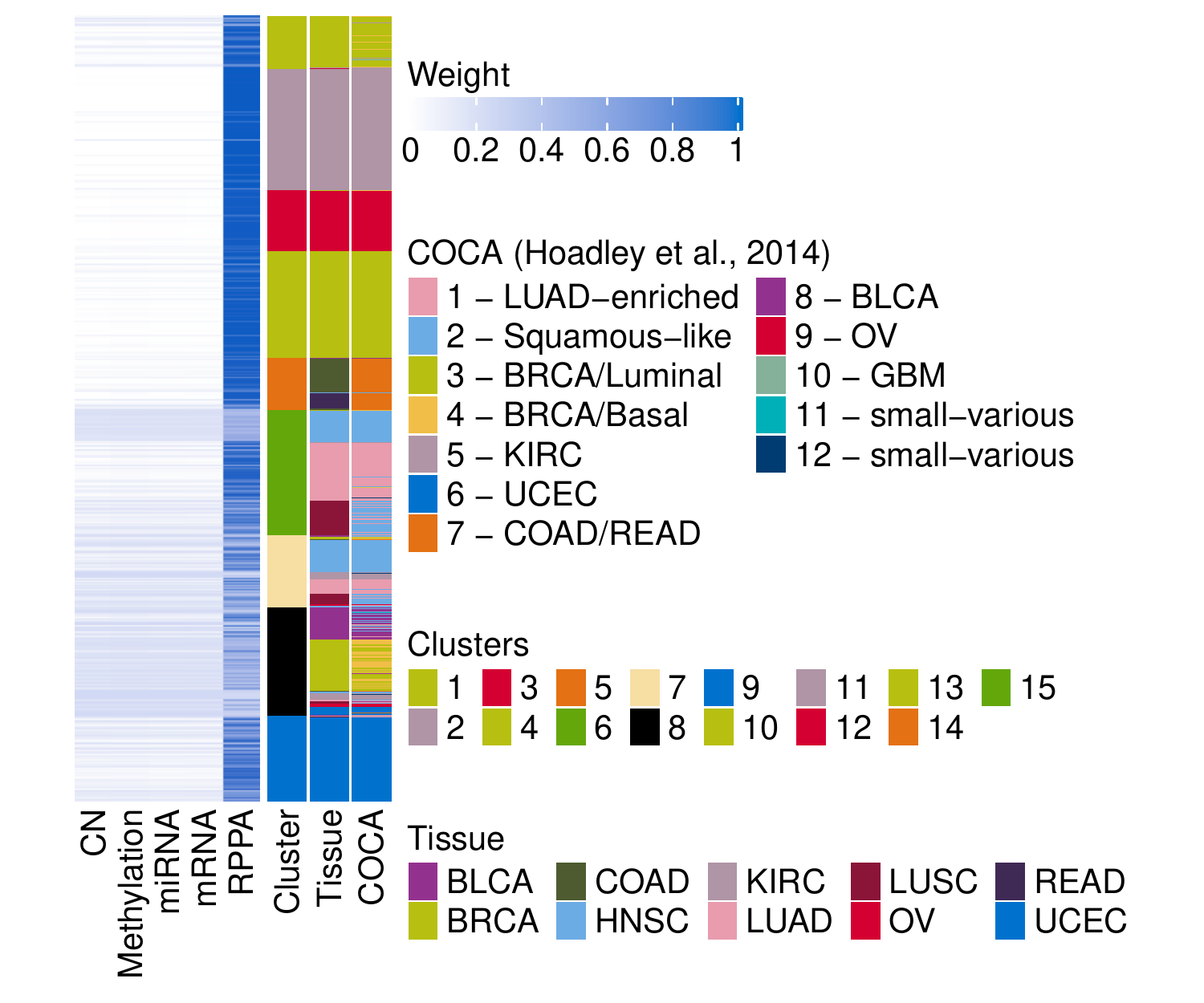}
	\caption{Weights assigned by the multiple kernel $k$-means algorithm to each observations in each layer, where ``CN'' stands for copy number and ``RPPA'' for reverse phase protein array.  The weights assigned on average to the tumour samples in each layer are: copy number 5.8\%, methylation 6\%, microRNA 5.9\%, mRNA 5.9\%, protein 76.4\%.}
	\label{fig:pancan-10-weights-alpha05}
\end{figure}

\clearpage

\subsubsection{Unsupervised integration after variable selection, $\alpha=1$}

\begin{figure}[H]
	\centering
	\begin{subfigure}[t]{\textwidth}
		\includegraphics[width=\textwidth]{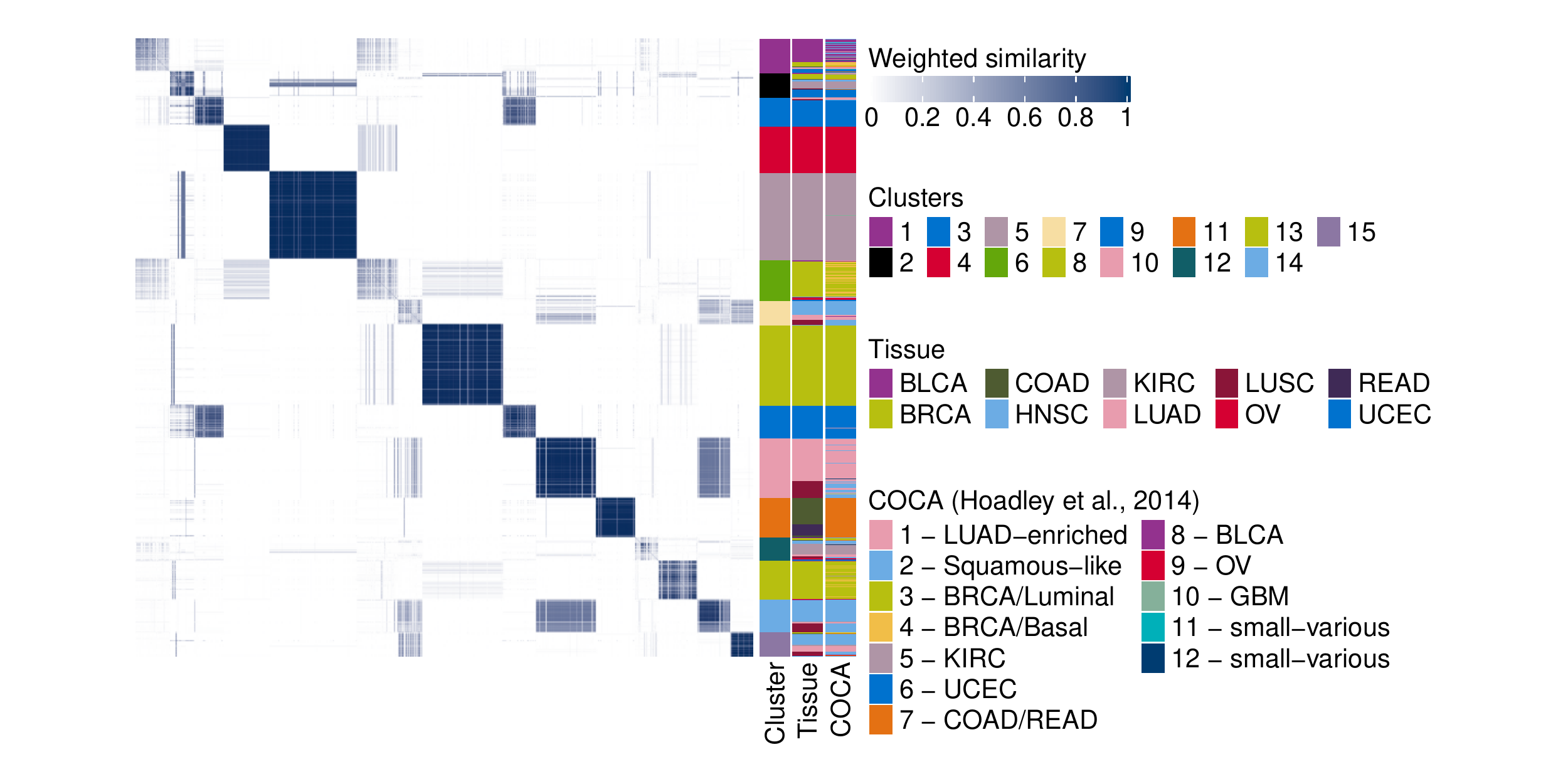}
		\caption{Weighted kernel and clusters.}
	\end{subfigure}
	\begin{subfigure}[t]{.46\textwidth}
		\includegraphics[width=\textwidth]{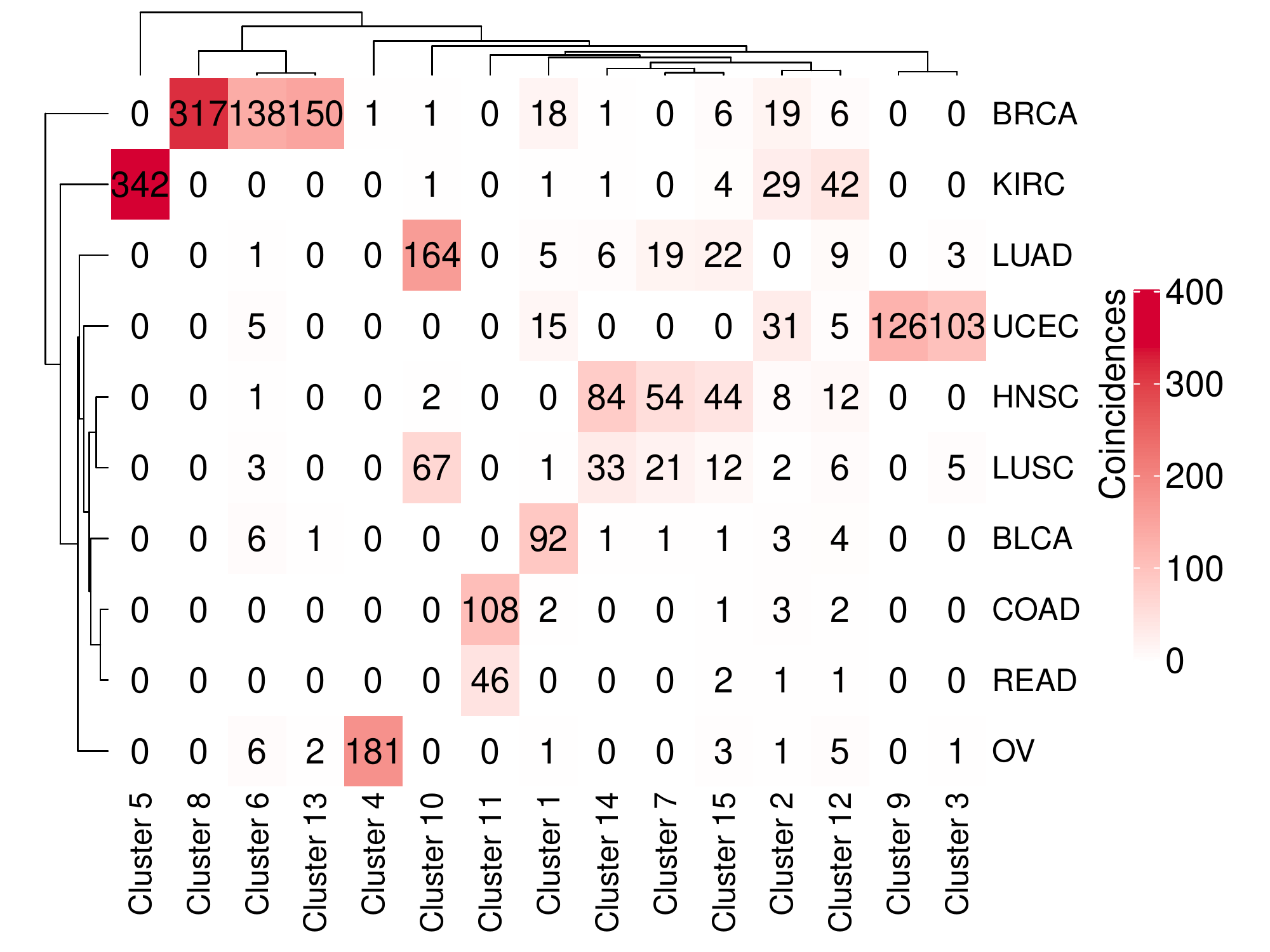}
		\caption{Comparison to tissue of origin.}
	\end{subfigure}
	\begin{subfigure}[t]{.51\textwidth}
		\includegraphics[width=\textwidth]{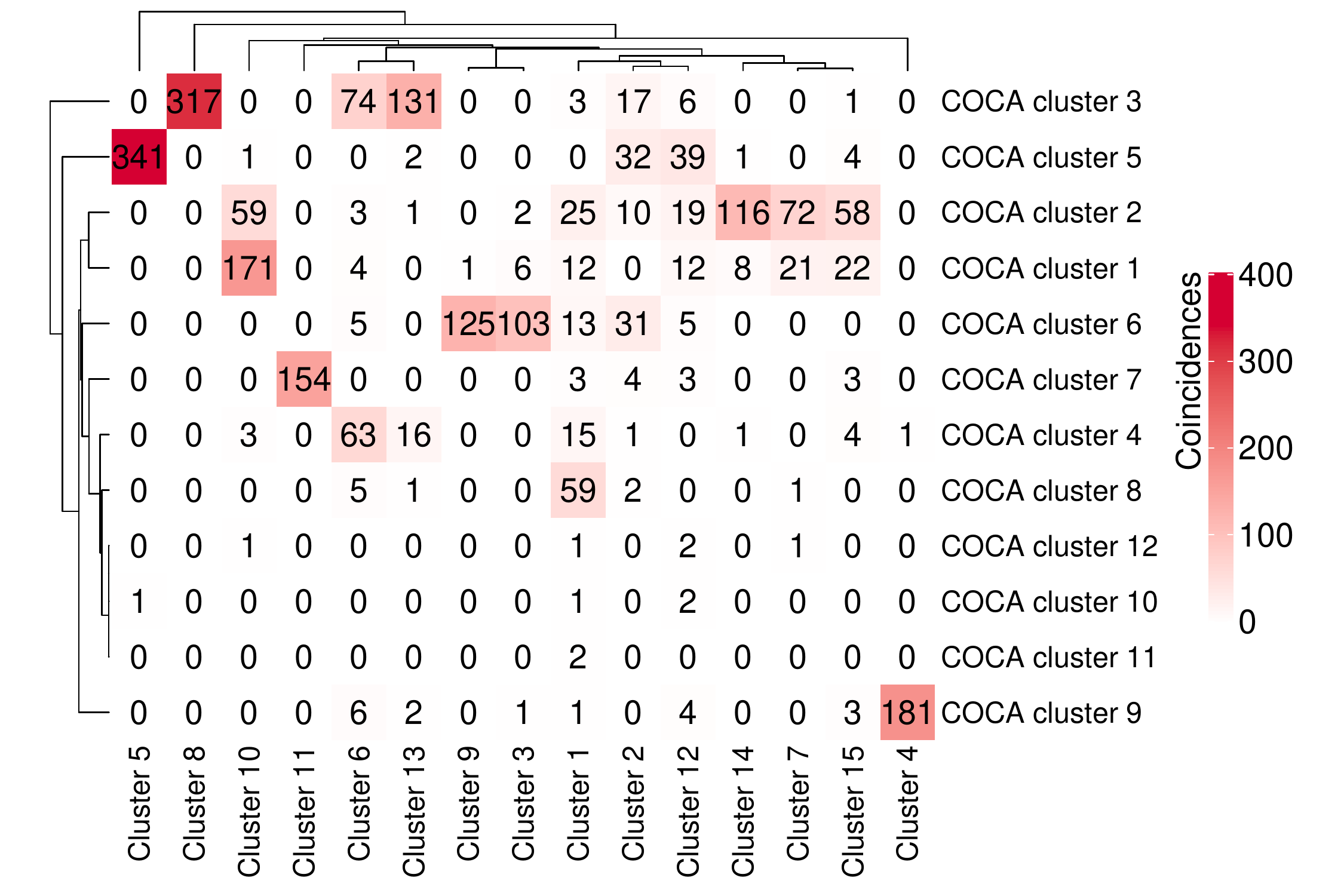}
		\caption{Comparison to COCA clusters.}
	\end{subfigure}
	\caption{Unsupervised multiplatform analysis of ten cancer types. \textbf{(a)} Weighted kernel, final clusters, tissues of origin, and COCA clusters. \textbf{(b)} Coincidence matrix comparing the tissue of origin of the tumour samples to the new clusters. \textbf{(c)} Coincidence matrix comparing the COCA clusters of Hoadley \emph{et al.} to the new clusters.}
	\label{fig:unsupervised-pancan10-alpha1}
\end{figure}

\begin{figure}[H]
	\centering
	\includegraphics[width=.675\textwidth]{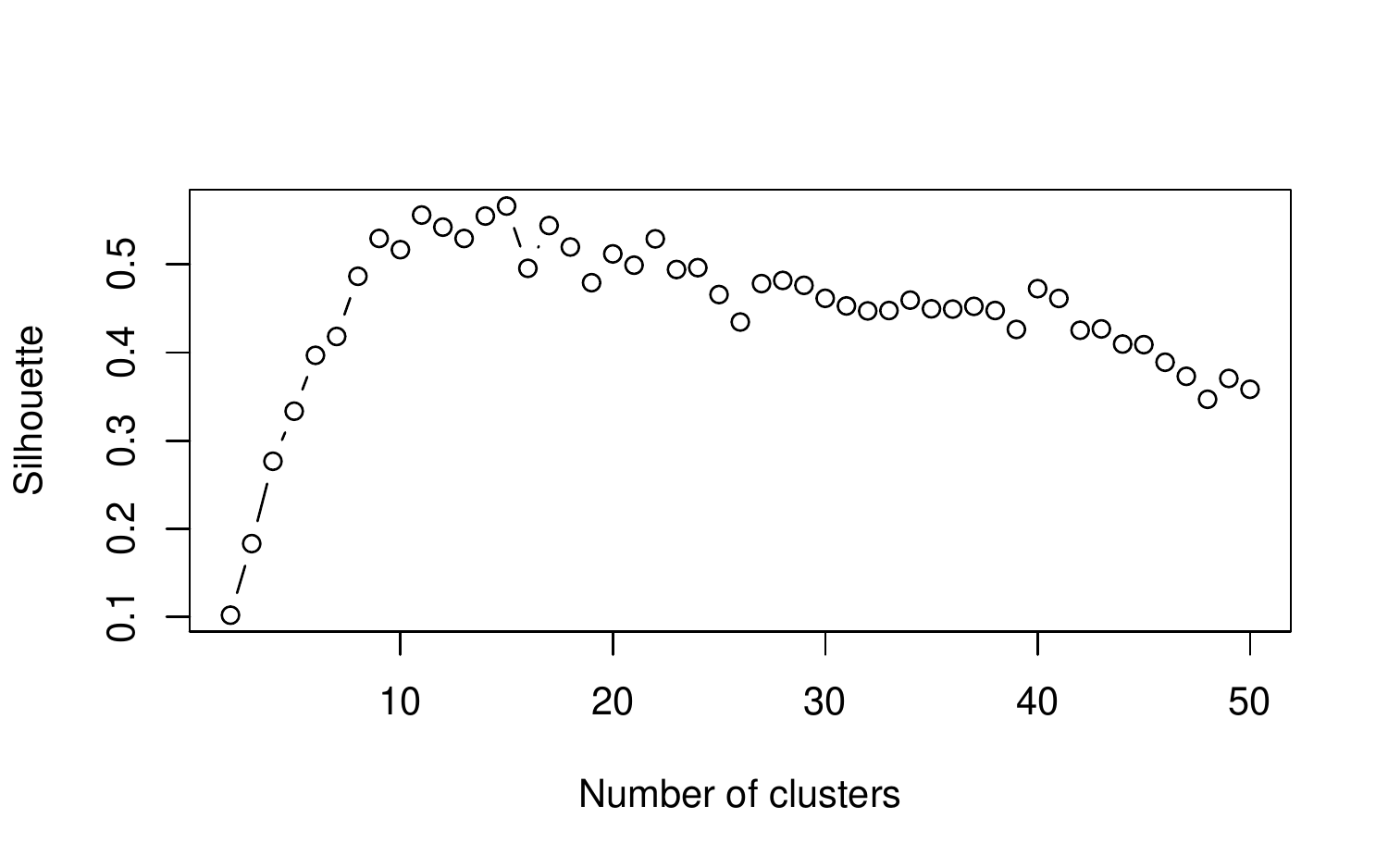}
	\caption{Average silhouette.}
	\label{fig:pancan-10-silhouette-unsupervised-alpha1}
\end{figure}

\begin{figure}[H]
	\centering
	\includegraphics[width=.6\textwidth]{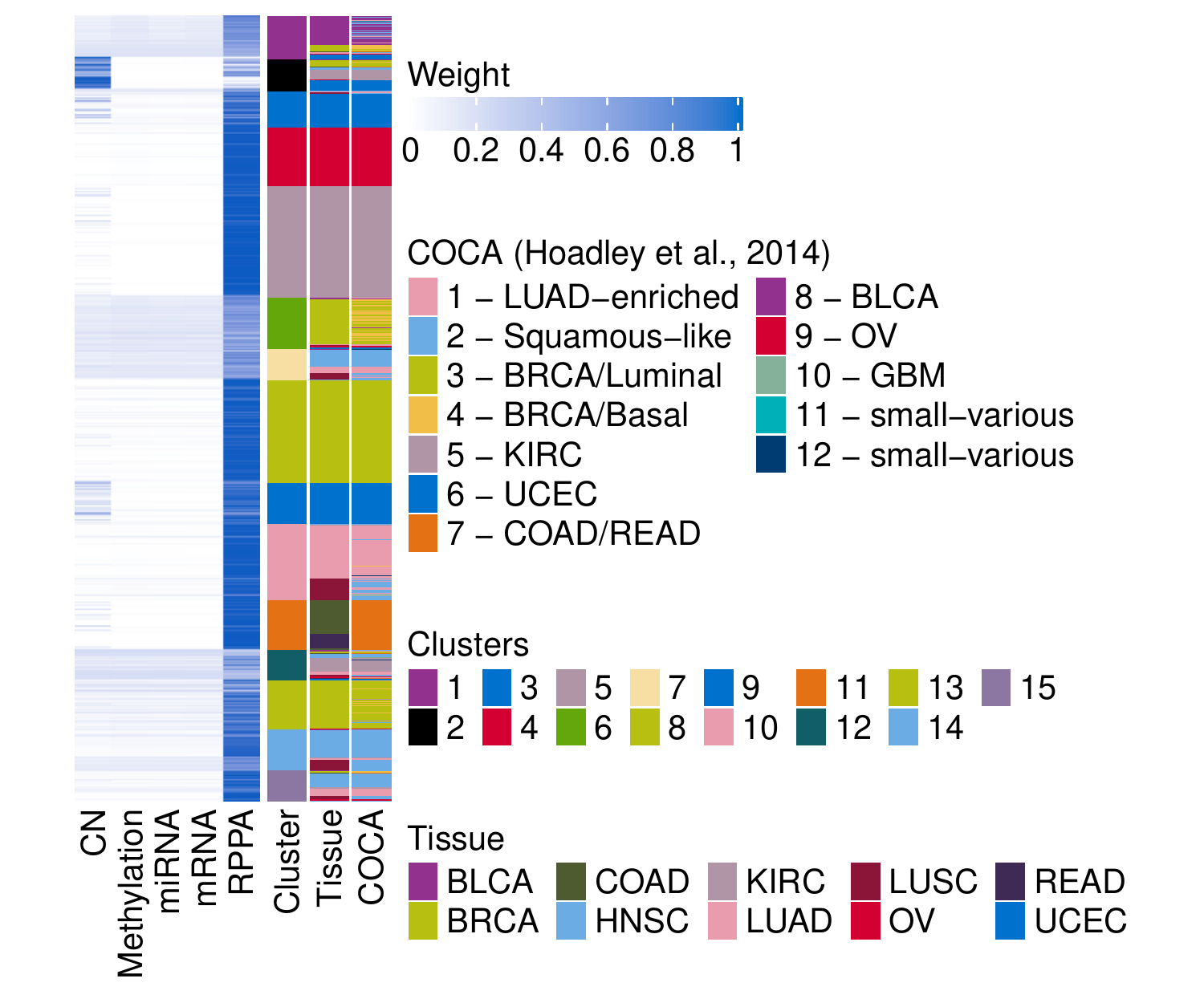}
	\caption{Weights assigned by the multiple kernel $k$-means algorithm to each observations in each layer, where ``CN'' stands for copy number and ``RPPA'' for reverse phase protein array. The weights assigned on average to the tumour samples in each layer are: copy number 8.1\%, methylation 3.8\%, microRNA 3.6\%, mRNA 3.7\%, protein 80.8\%.}
	\label{fig:pancan-10-weights-alpha1}
\end{figure}

\clearpage

\subsubsection{Outcome-guided integration: additional figures}

\begin{figure}[H]
	\centering
	\includegraphics[width=.675\textwidth]{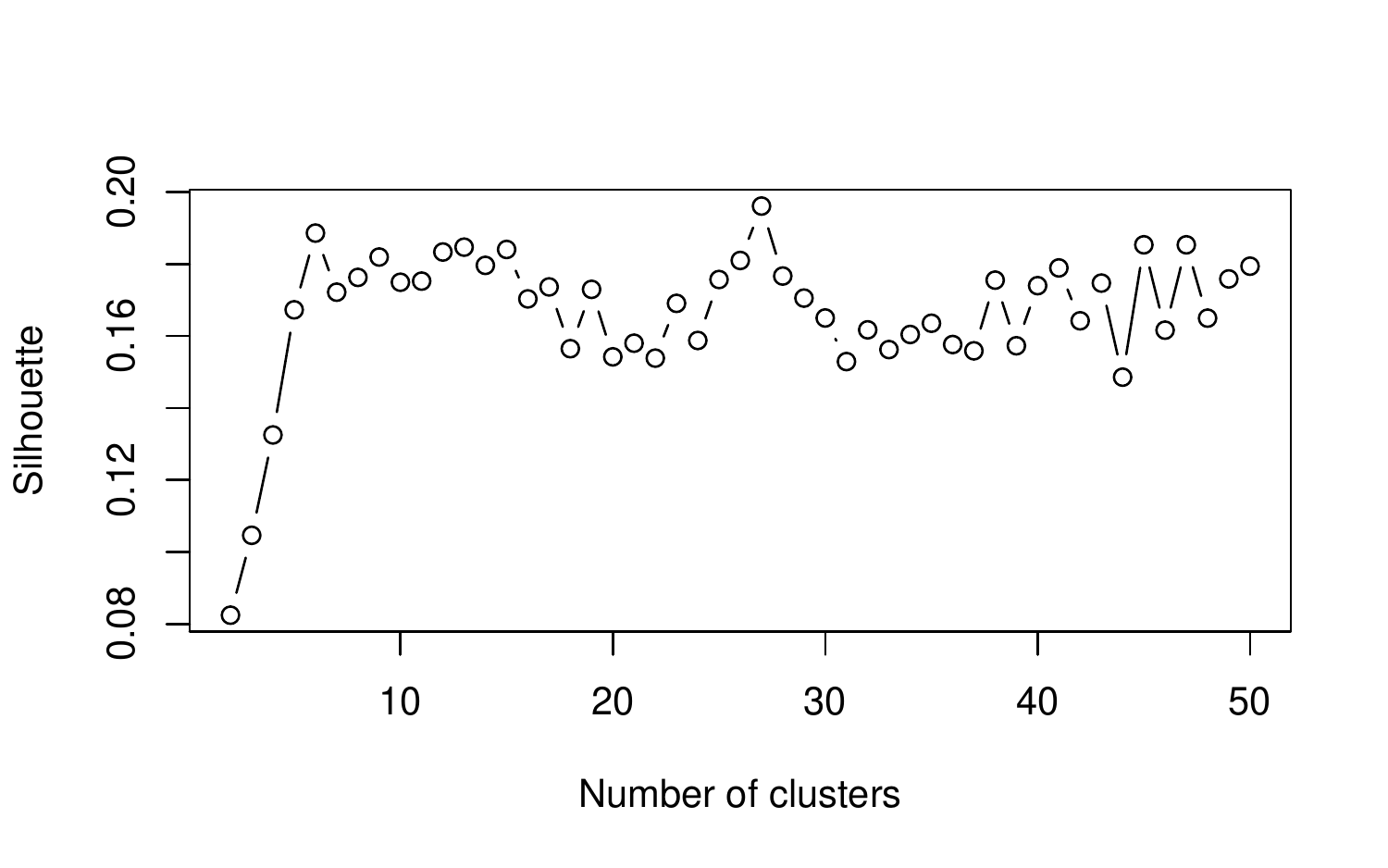}
	\caption{Average silhouette.}
	\label{fig:pancan-10-silhouette-outcome-guided}
\end{figure}

\paragraph{Choice of the number of clusters} In Figure \ref{fig:pancan-10-silhouette-outcome-guided} are reported the average values of the silhouette when the number of clusters goes from 2 to 50. The maximum is at $K=27$.

\begin{figure}[H]
	\centering
	\includegraphics[width=\textwidth]{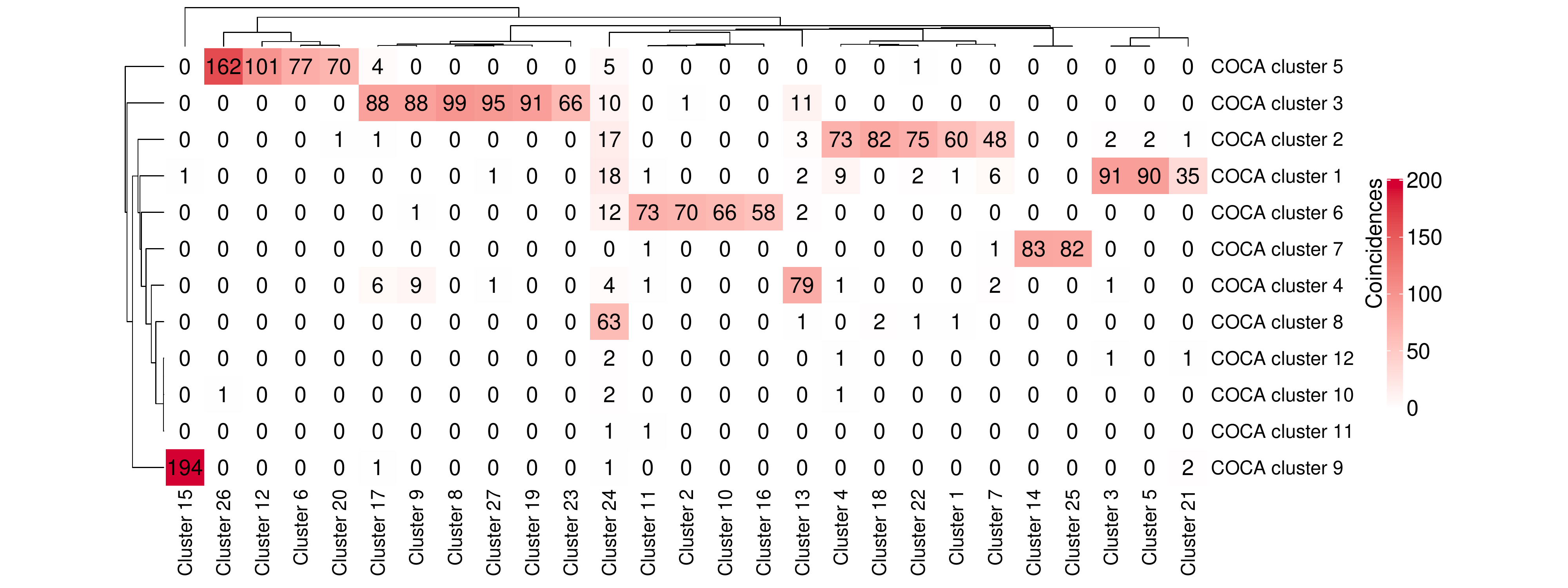}
	\caption{Comparison between the clusters found combining the PSMs of each layer using the outcome-guided approach and those identified by \citet{hoadley2014multiplatform} using COCA.}
	\label{fig:pancan-10-comparison-hoadley-et-al-outcome-guided}
\end{figure}

\paragraph{Comparison with the clusters identified by \citet{hoadley2014multiplatform}}
In Figure \ref{fig:pancan-10-comparison-hoadley-et-al-outcome-guided} are shown the correspondences between the clusters found in the main paper in the outcome-guided case and the clusters identified by \citet{hoadley2014multiplatform} using Cluster-Of-Clusters Analysis (COCA).

\begin{figure}[H]
	\centering
	\includegraphics[width=.8\textwidth]{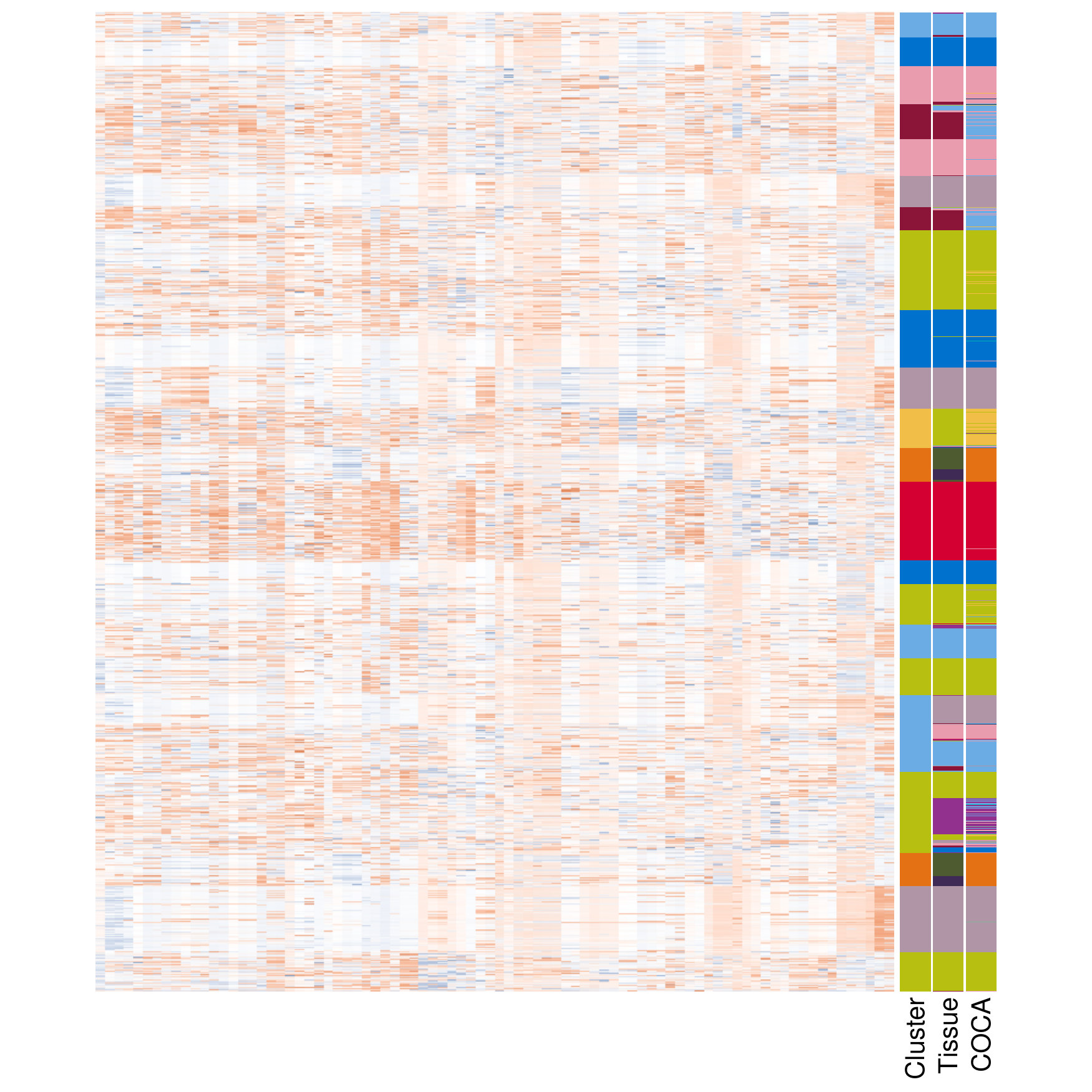}
	\caption{Copy number data and final clusters.}
	\label{fig:pancan-cn-outcomeguided-finalclusters}
\end{figure}

\begin{figure}[H]
	\centering
	\includegraphics[width=.8\textwidth]{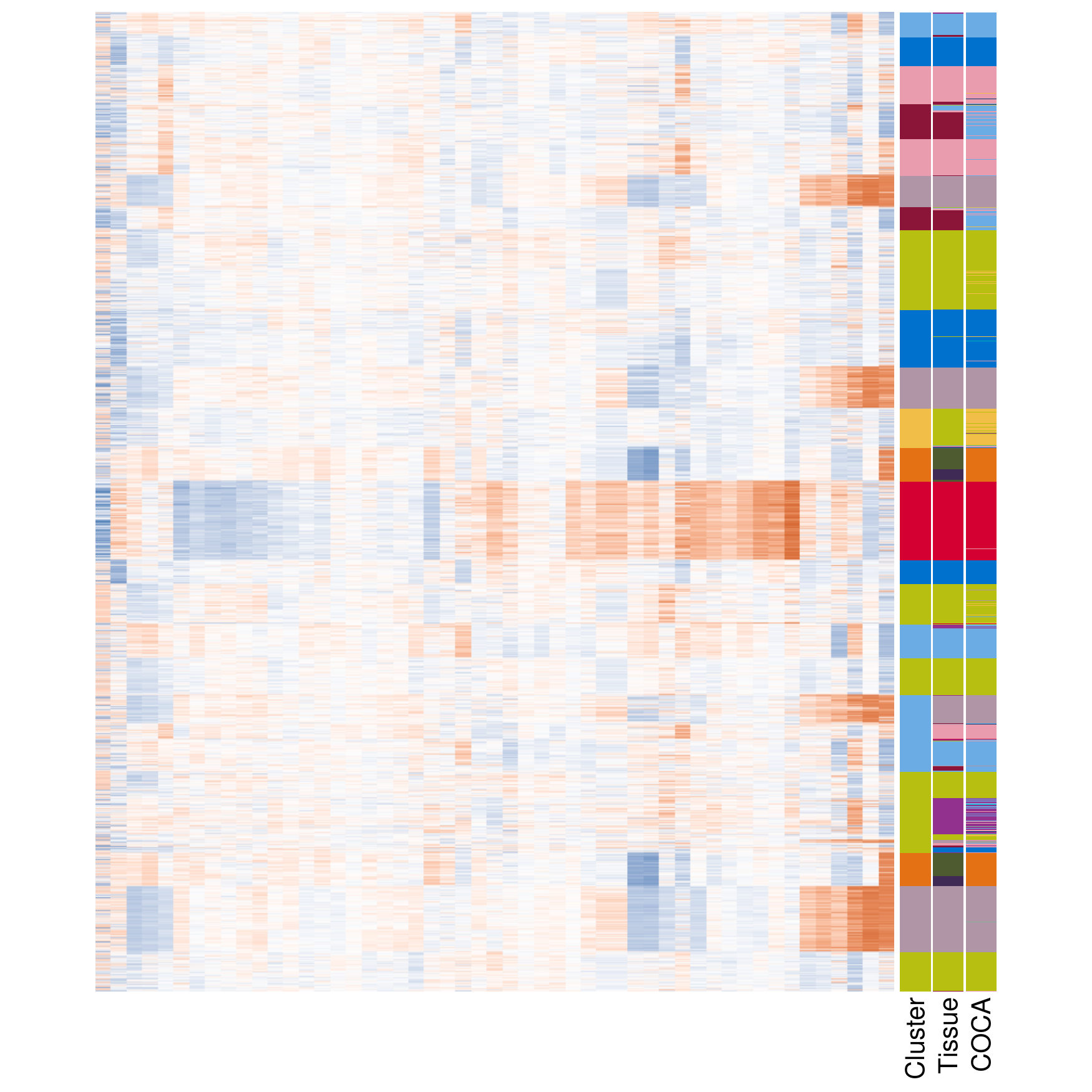}
	\caption{microRNA expression data and final clusters.}
	\label{fig:pancan-mirna-outcomeguided-finalclusters}
\end{figure}

\begin{figure}[H]
	\centering
	\includegraphics[width=.8\textwidth]{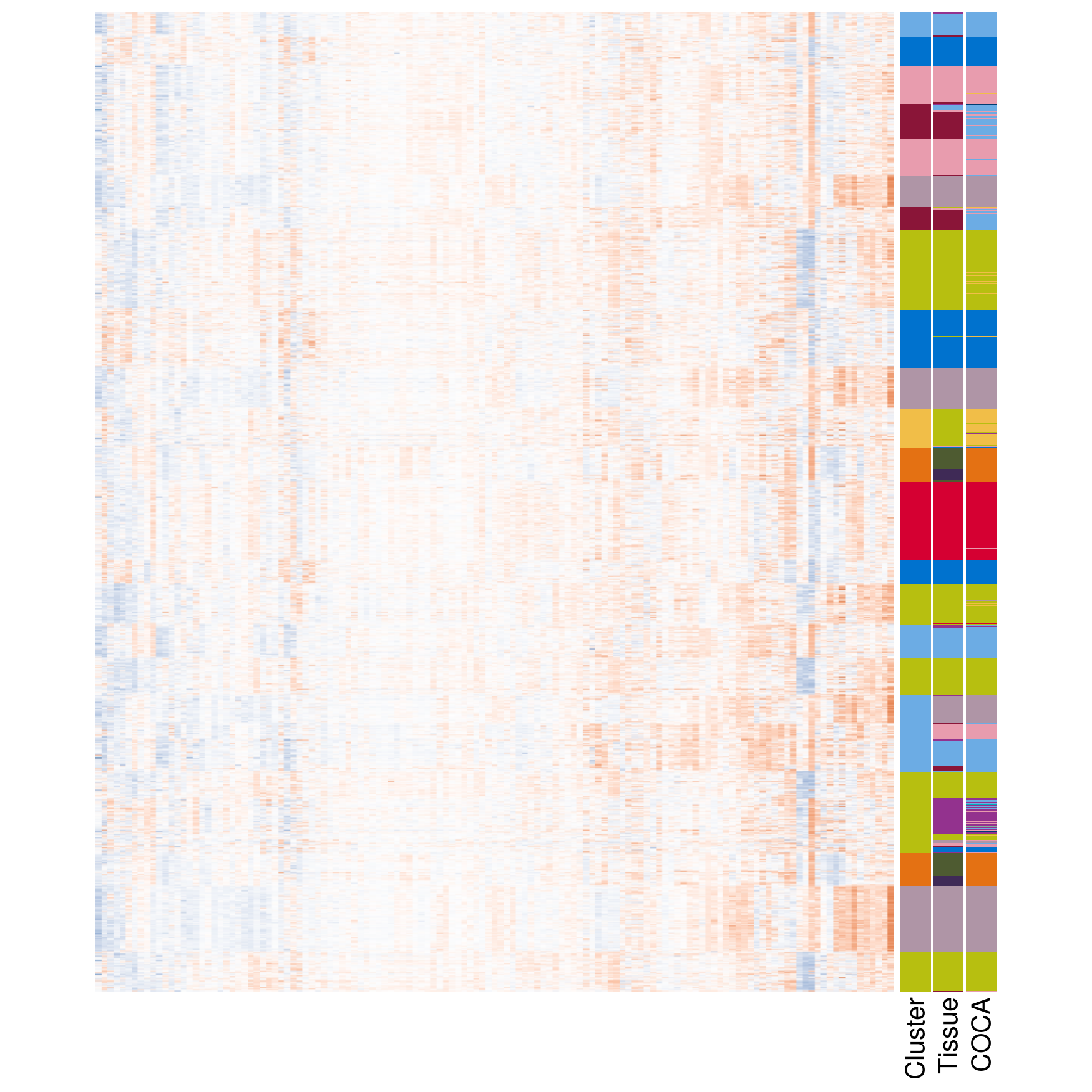}
	\caption{Protein expression data and final clusters.}
	\label{fig:pancan-protein-outcomeguided-finalclusters}
\end{figure}

\clearpage
\paragraph{Clustering structure in the data}
Figures \ref{fig:pancan-cn-outcomeguided-finalclusters}, \ref{fig:pancan-mirna-outcomeguided-finalclusters}, \ref{fig:pancan-mirna-outcomeguided-finalclusters}, and \ref{fig:pancan-protein-outcomeguided-finalclusters} show the four data layers where the rows have been sorted by final cluster.

\begin{figure}[H]
	\centering
	\includegraphics[width=.8\textwidth]{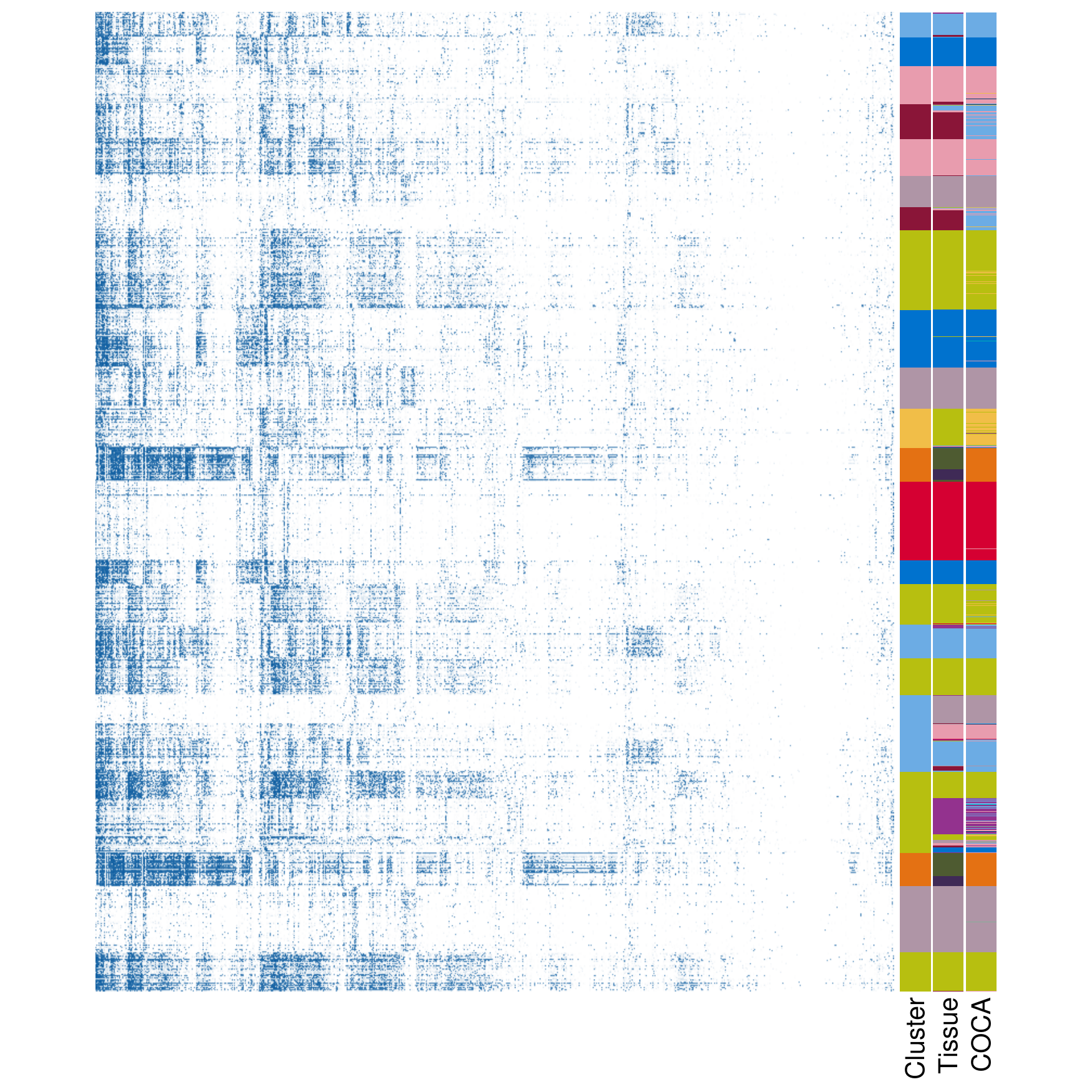}
	\caption{Methylation data and final clusters.}
	\label{fig:pancan-methylation-outcomeguided-finalclusters}
\end{figure}

\clearpage

\subsubsection{Outcome-guided integration after variable selection, $\alpha=0.1$}

The weights assigned on average to the tumour samples in each layer are: copy number 33.1\%,  methylation  16.3\%,    microRNA 34.1\%, and protein 16.5\%.

\begin{figure}[H]
	\centering
	\begin{subfigure}[t]{\textwidth}
		\includegraphics[width=\textwidth]{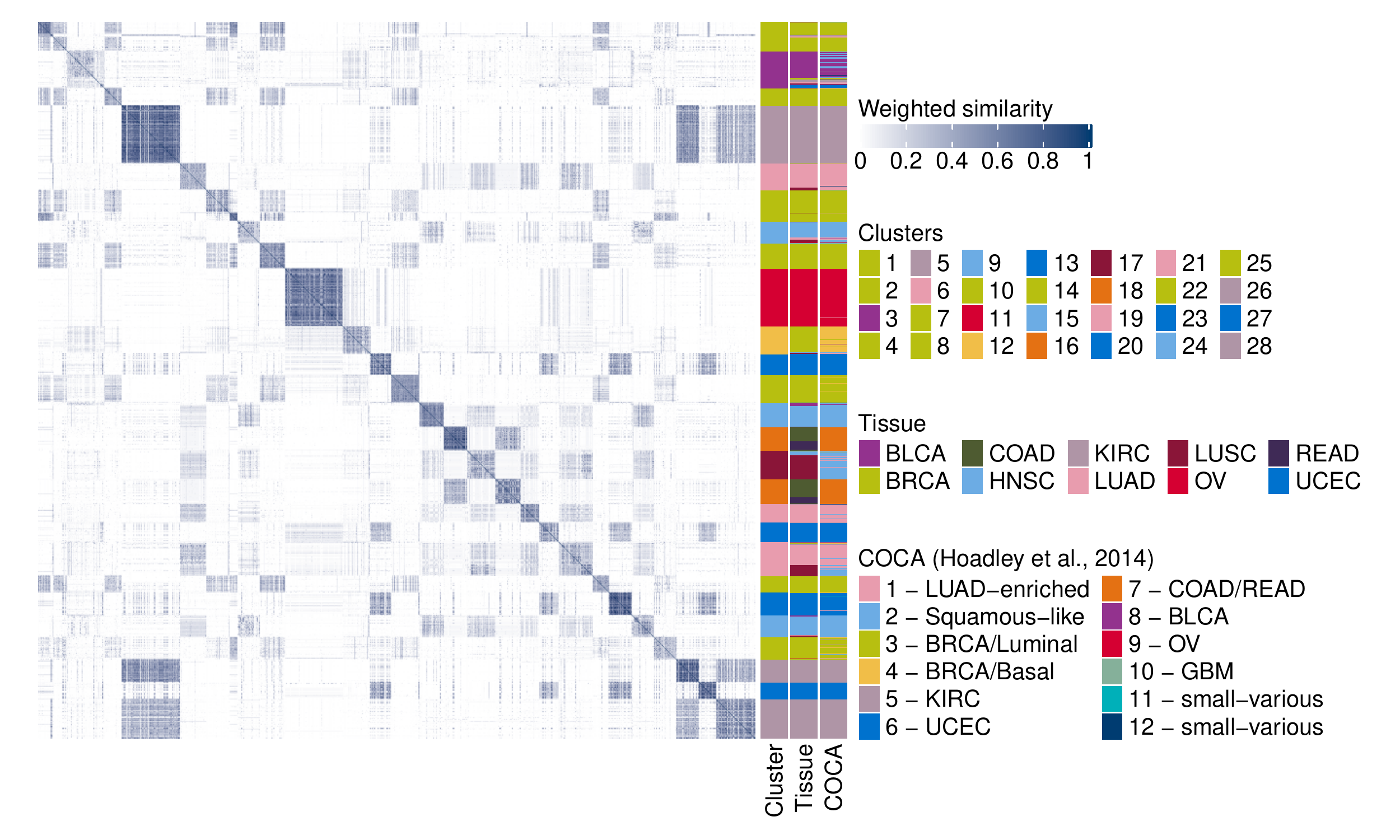}
		\caption{Clusters and weighted kernel.}
	\end{subfigure}
	\begin{subfigure}[t]{\textwidth}
		\includegraphics[width=\textwidth]{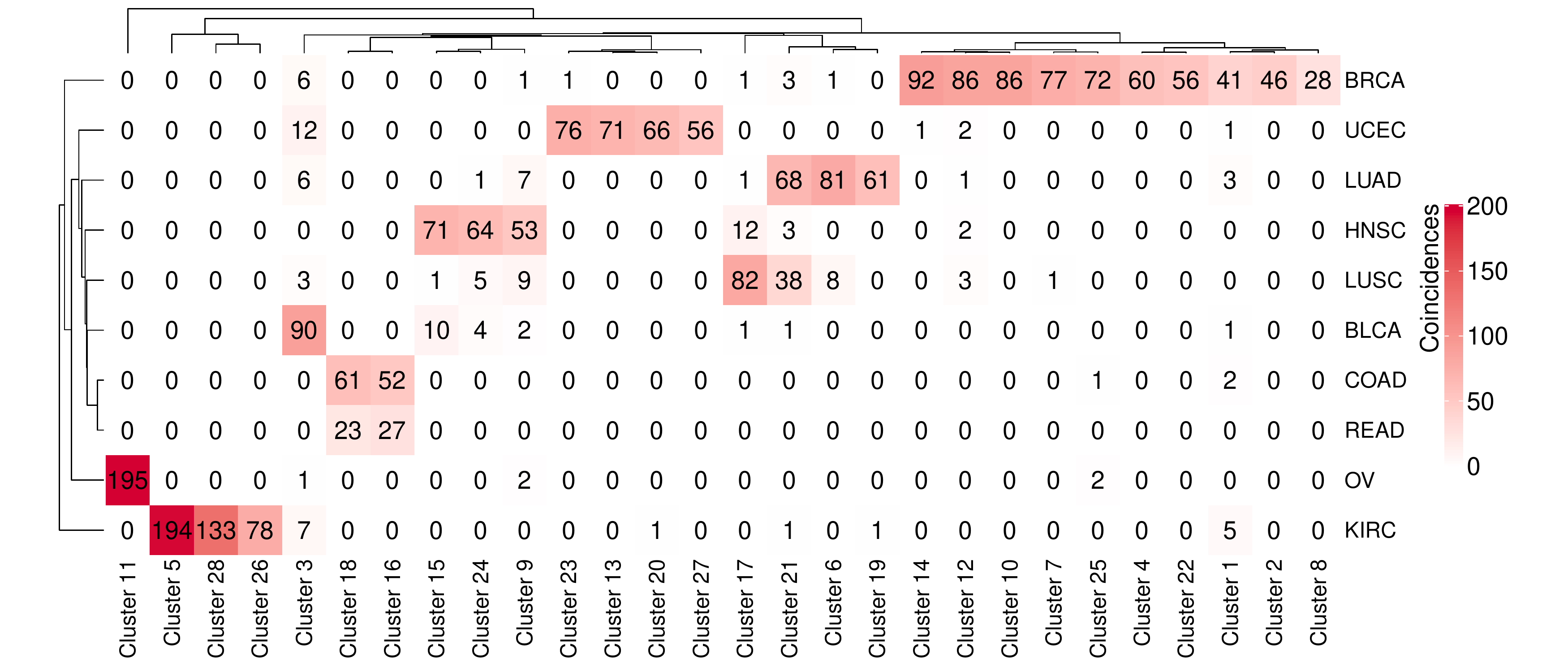}
		\caption{Coincidence matrix.}
	\end{subfigure}
	\caption{Outcome-guided multiplatform analysis of ten cancer types. \textbf{(a)} Weighted kernel, final clusters, tissues of origin, and COCA clusters. \textbf{(b)} Coincidence matrix comparing the tissue of origin of the tumour samples with the clusters.}
	\label{fig:outcome-guided-pancan10-alpha05}
\end{figure}

\begin{figure}[H]
	\centering
	\includegraphics[width=.675\textwidth]{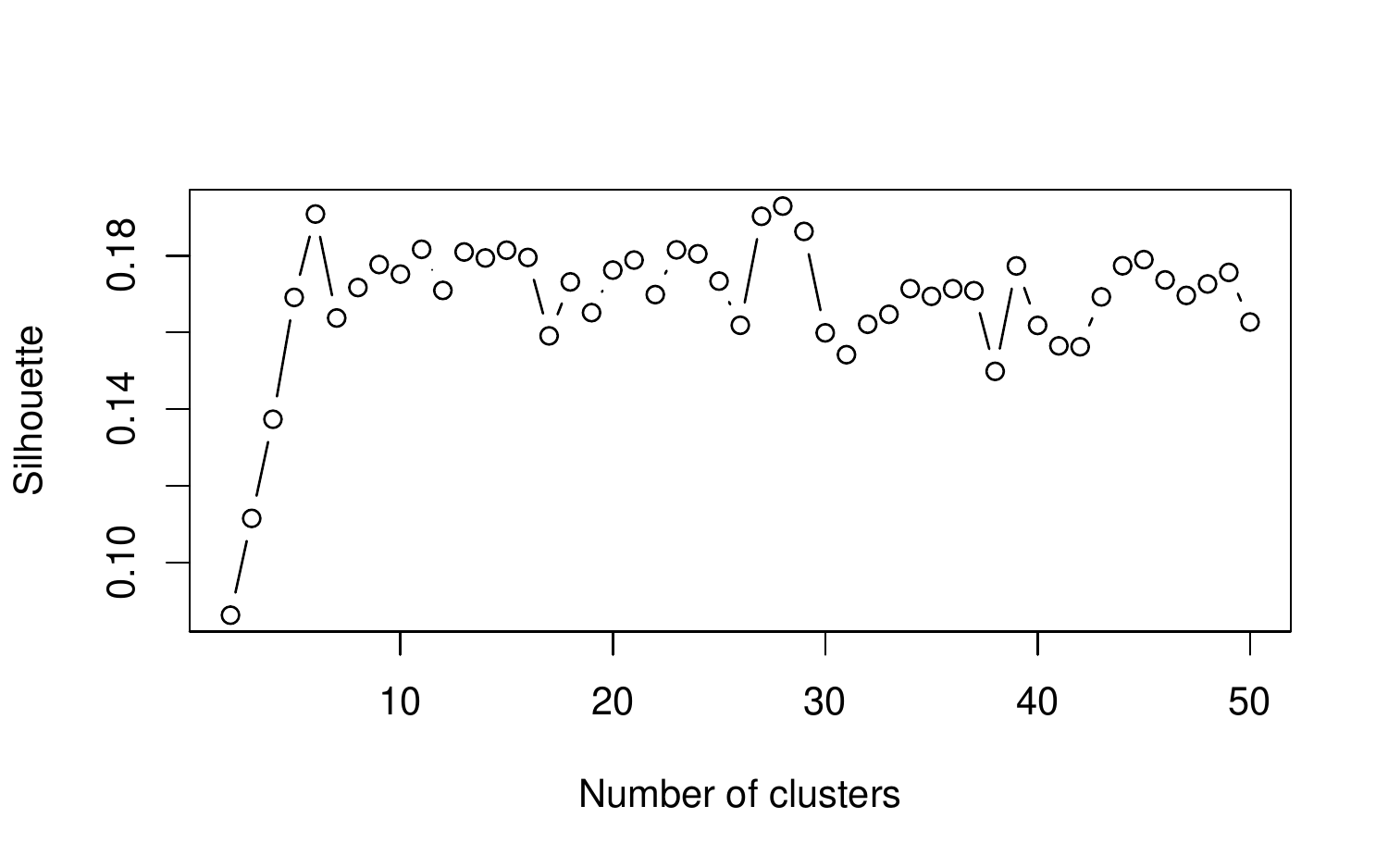}
	\caption{Average silhouette.}
	\label{fig:pancan-10-silhouette-outcome-guided-alpha05}
\end{figure}

\begin{figure}[H]
	\centering
	\includegraphics[width=\textwidth]{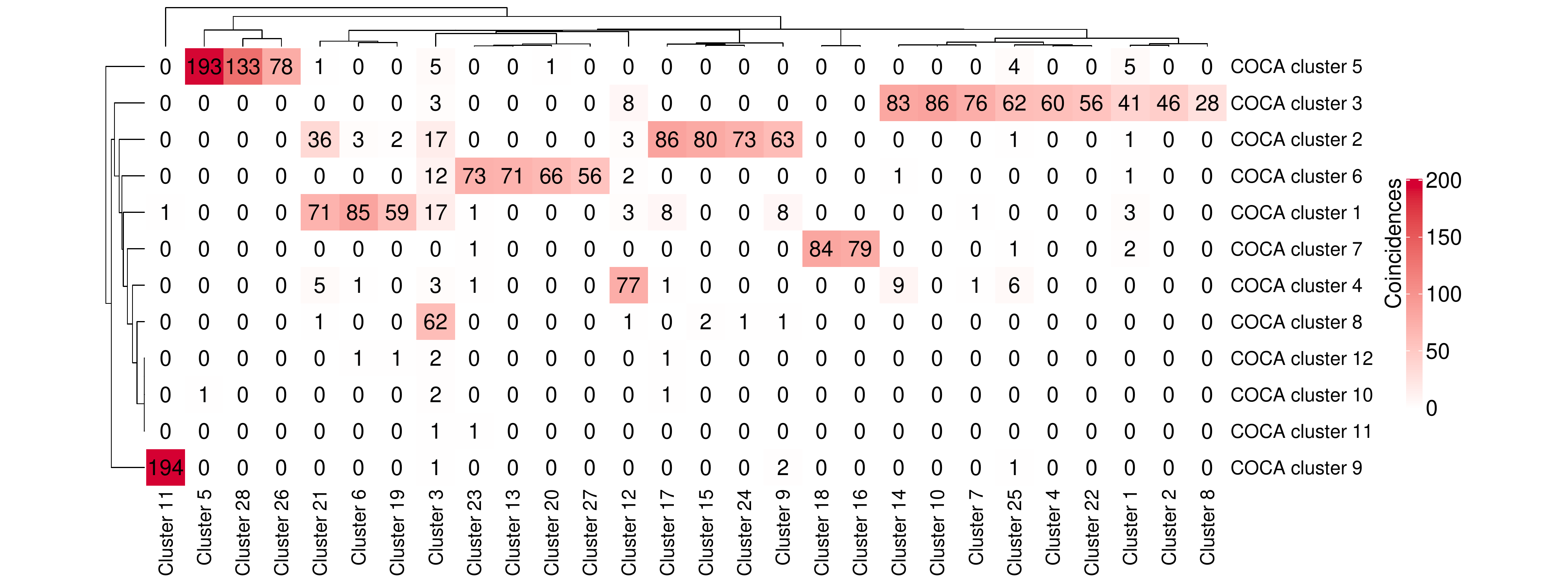}
	\caption{Comparison between the clusters found combining the PSMs of each layer using the outcome-guided approach and those identified by \citet{hoadley2014multiplatform} using COCA.}
	\label{fig:pancan-10-comparison-hoadley-et-al-outcome-guided-alpha05}
\end{figure}
\clearpage

\clearpage

\subsubsection{Outcome-guided integration after variable selection, $\alpha=0.5$}

The weights assigned on average to the tumour samples in each layer are: copy number 34.2\%, mRNA  17.7\%, methylation 7.2\%, miRNA 27.4\%, protein 13.5.

\begin{figure}[H]
	\centering
	\begin{subfigure}[t]{\textwidth}
		\includegraphics[width=\textwidth]{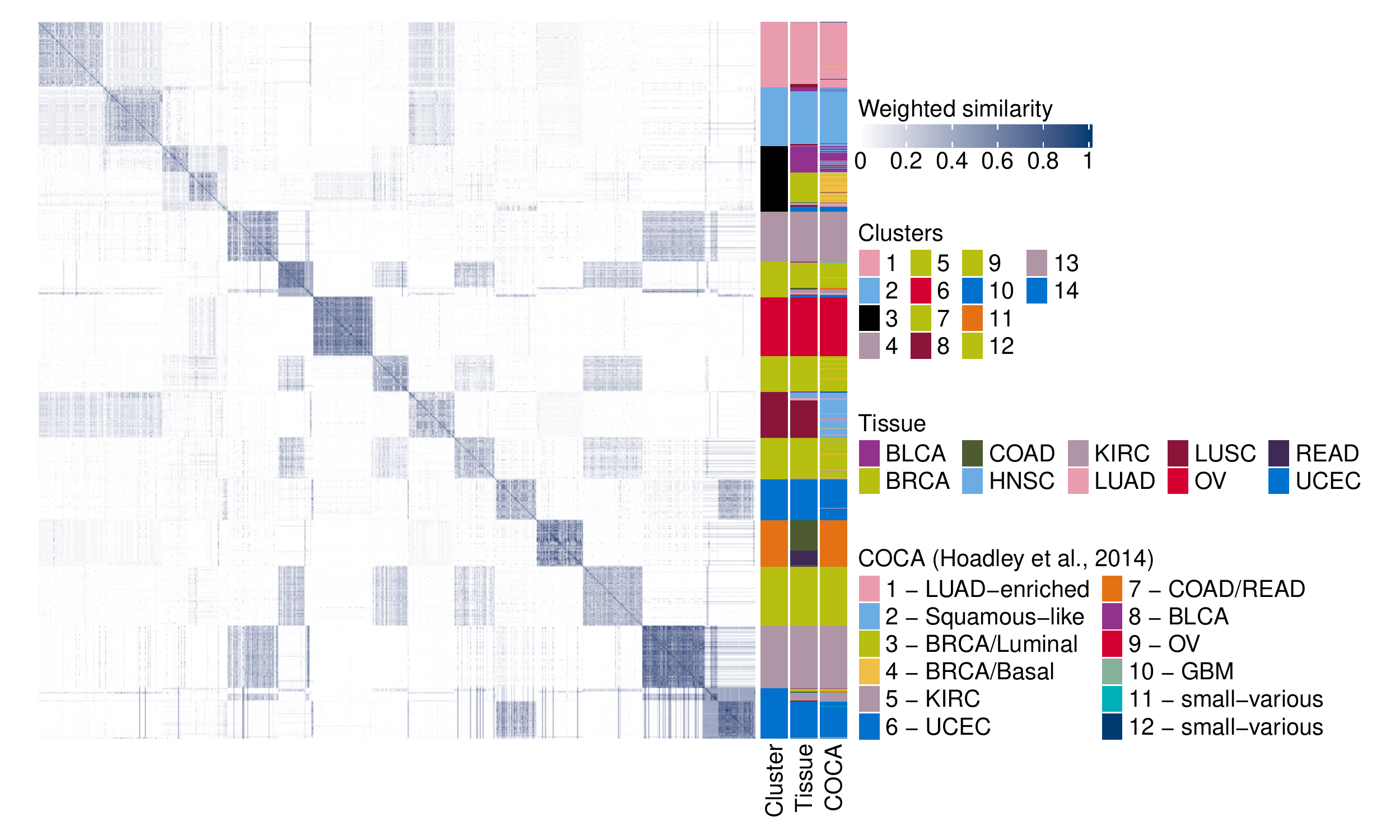}
		\caption{Clusters and weighted kernel.}
	\end{subfigure}
	\begin{subfigure}[t]{\textwidth}
		\includegraphics[width=\textwidth]{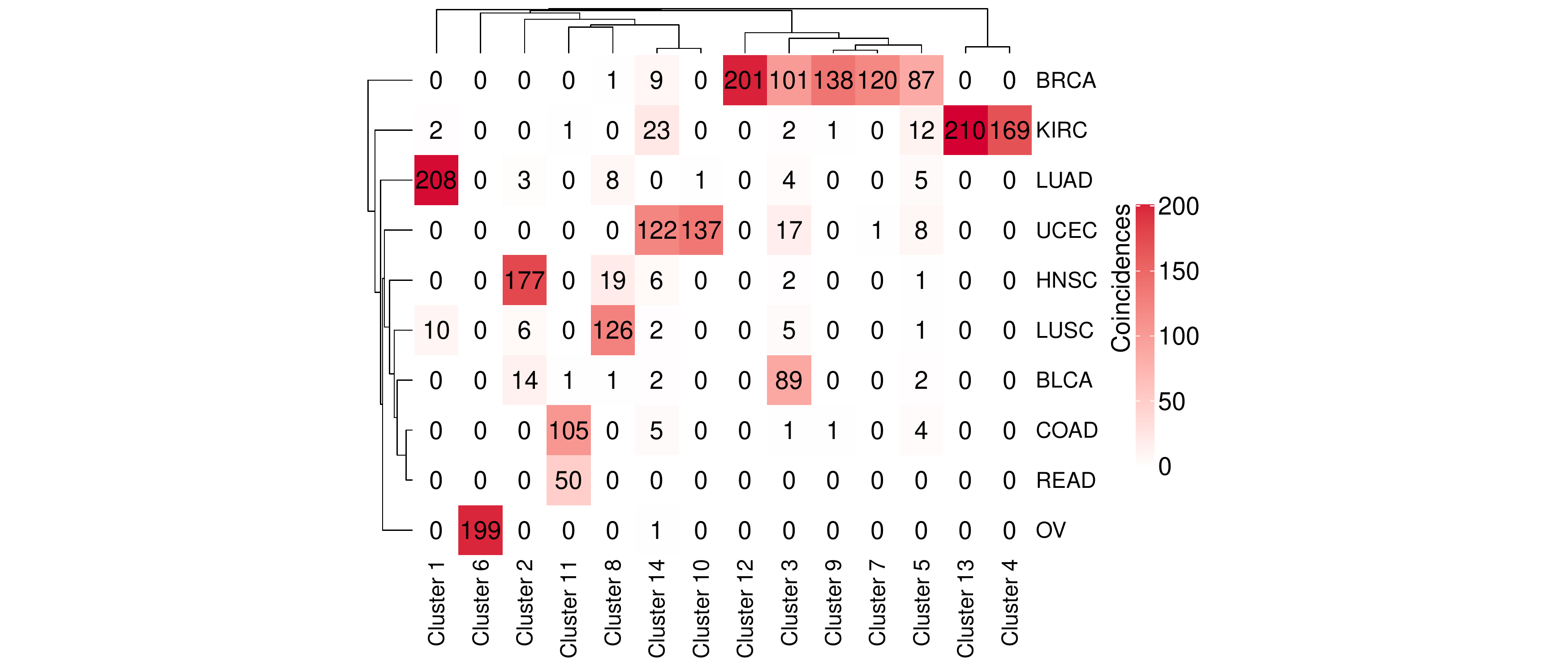}
		\caption{Coincidence matrix.}
	\end{subfigure}
	\caption{Outcome-guided multiplatform analysis of ten cancer types. \textbf{(a)} Weighted kernel, final clusters, tissues of origin, and COCA clusters. \textbf{(b)} Coincidence matrix comparing the tissue of origin of the tumour samples with the clusters.}
	\label{fig:outcome-guided-pancan10-alpha05}
\end{figure}

\begin{figure}[H]
	\centering
	\includegraphics[width=.675\textwidth]{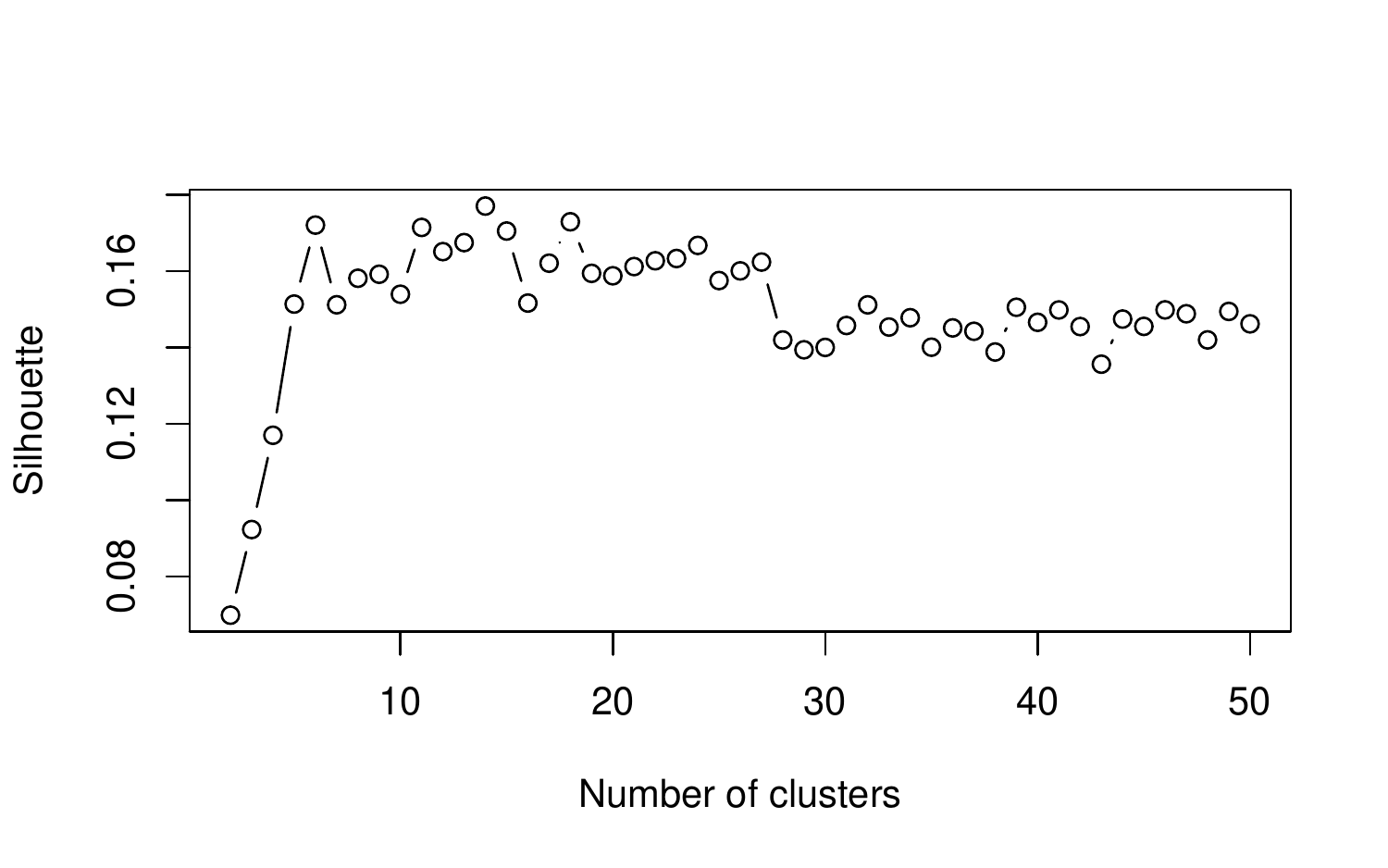}
	\caption{Average silhouette.}
	\label{fig:pancan-10-silhouette-outcome-guided-alpha05}
\end{figure}

\begin{figure}[H]
	\centering
	\includegraphics[width=\textwidth]{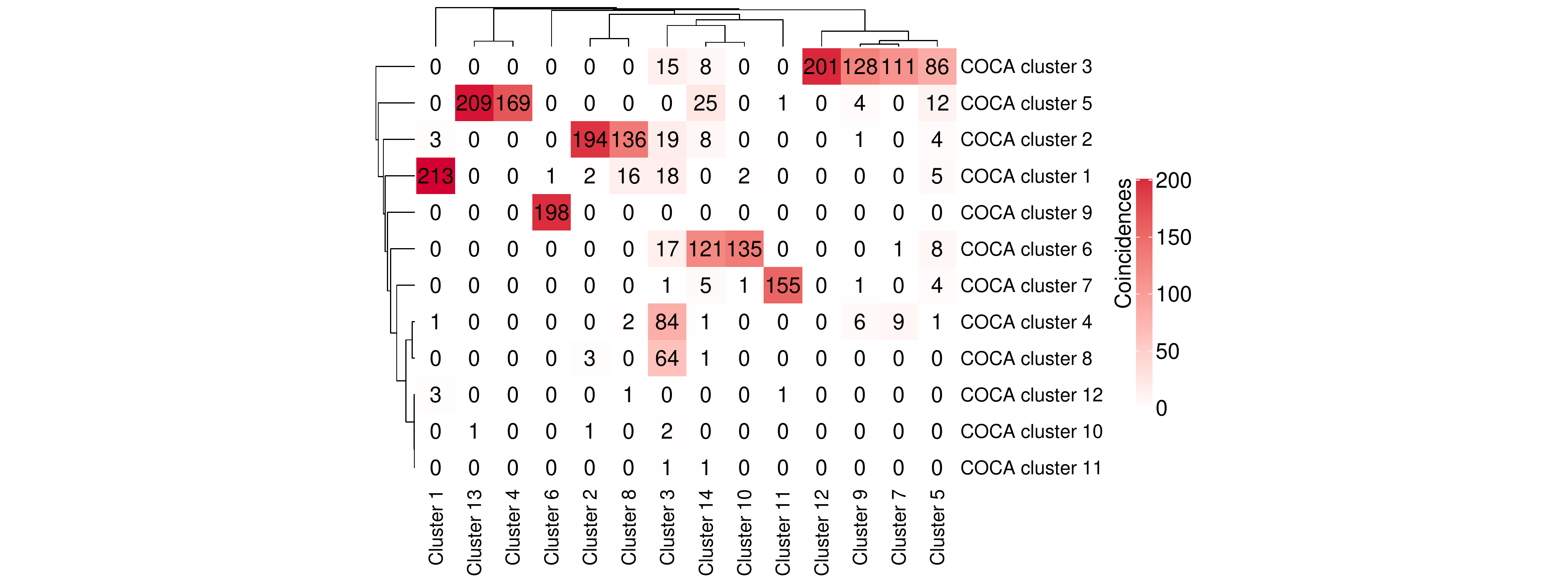}
	\caption{Comparison between the clusters found combining the PSMs of each layer using the outcome-guided approach and those identified by \citet{hoadley2014multiplatform} using COCA.}
	\label{fig:pancan-10-comparison-hoadley-et-al-outcome-guided-alpha05}
\end{figure}
\clearpage

\subsubsection{Outcome-guided integration after variable selection, $\alpha=1$}

The weights assigned on average to the tumour samples in each layer are: copy number 0\%, mRNA 37.6\%, methylation 25.2\%, miRNA 24.5\%, protein 12.7\%.

\begin{figure}[H]
	\centering
	\begin{subfigure}[t]{\textwidth}
		\includegraphics[width=\textwidth]{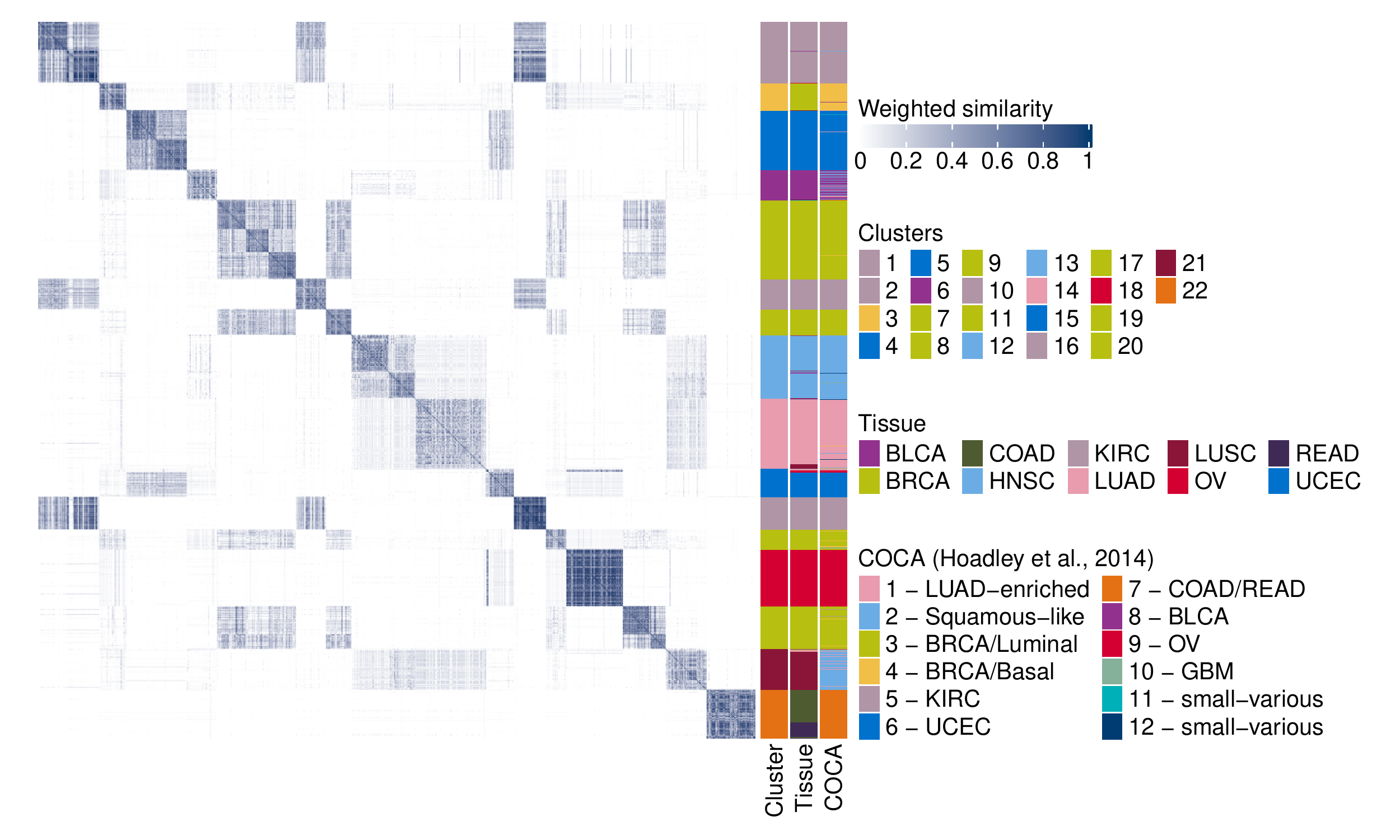}
		\caption{Clusters and weighted kernel.}
	\end{subfigure}
	\begin{subfigure}[t]{\textwidth}
		\includegraphics[width=\textwidth]{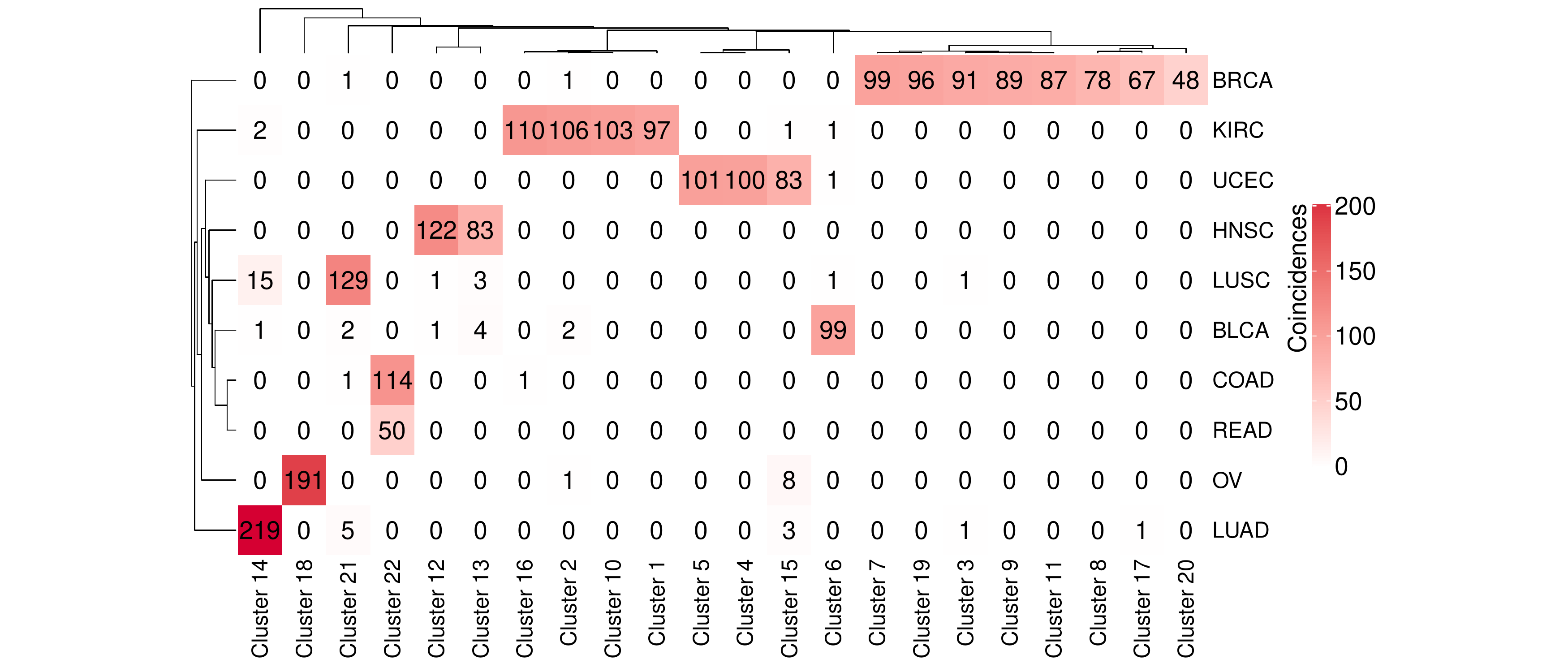}
		\caption{Coincidence matrix.}
	\end{subfigure}
	\caption{Outcome-guided multiplatform analysis of ten cancer types. \textbf{(a)} Weighted kernel, final clusters, tissues of origin, and COCA clusters. \textbf{(b)} Coincidence matrix comparing the tissue of origin of the tumour samples with the clusters.}
	\label{fig:outcome-guided-pancan10-alpha1}
\end{figure}

\begin{figure}[H]
	\centering
	\includegraphics[width=.675\textwidth]{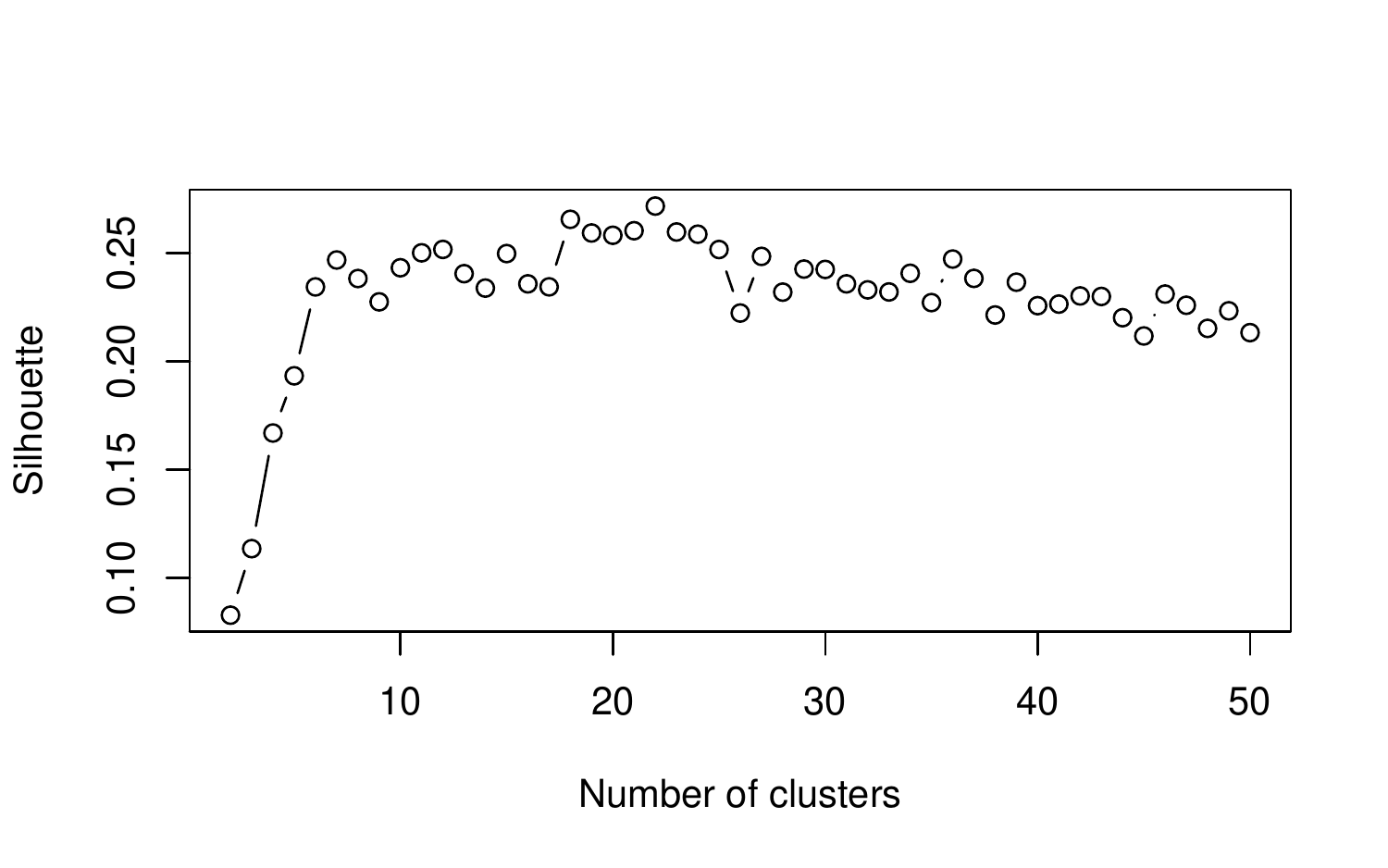}
	\caption{Average silhouette.}
	\label{fig:pancan-10-silhouette-outcome-guided-alpha1}
\end{figure}

\begin{figure}[H]
	\centering
	\includegraphics[width=\textwidth]{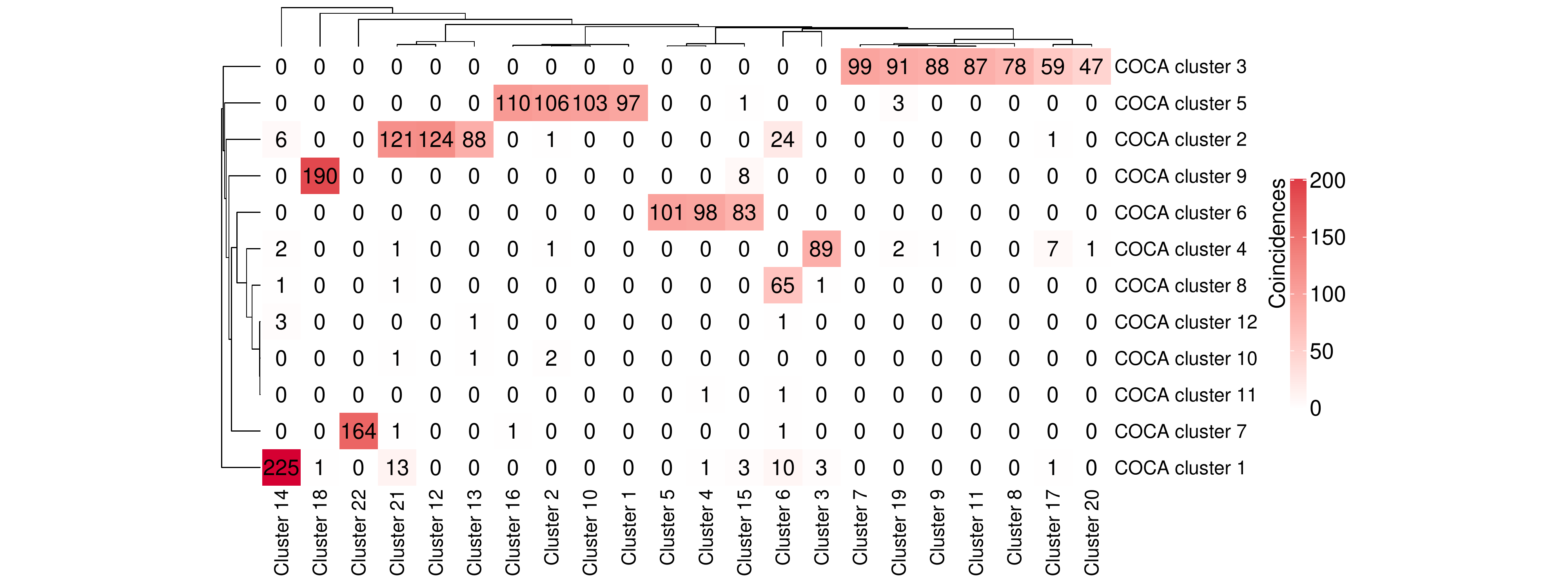}
	\caption{Comparison between the clusters found combining the PSMs of each layer using the outcome-guided approach and those identified by \citet{hoadley2014multiplatform} using COCA.}
	\label{fig:pancan-10-comparison-hoadley-et-al-outcome-guided-alpha1}
\end{figure}

\clearpage

\subsection{Transcriptional module discovery}

\subsubsection{Additional figures}

\begin{figure}[h]
	\centering
	\begin{subfigure}[b]{.49\textwidth}
		\centering
		\includegraphics[width =\textwidth]{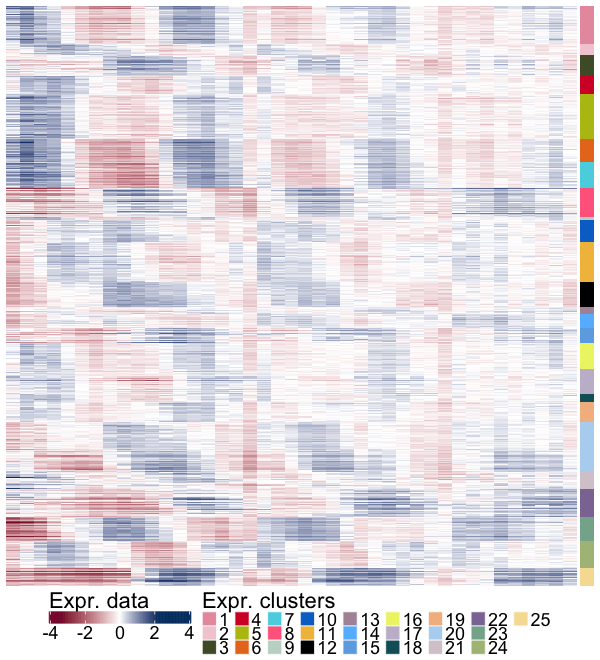}
		\caption{Expression data.}
	\end{subfigure}
	\begin{subfigure}[b]{.49\textwidth}
		\centering
		\includegraphics[width = \textwidth]{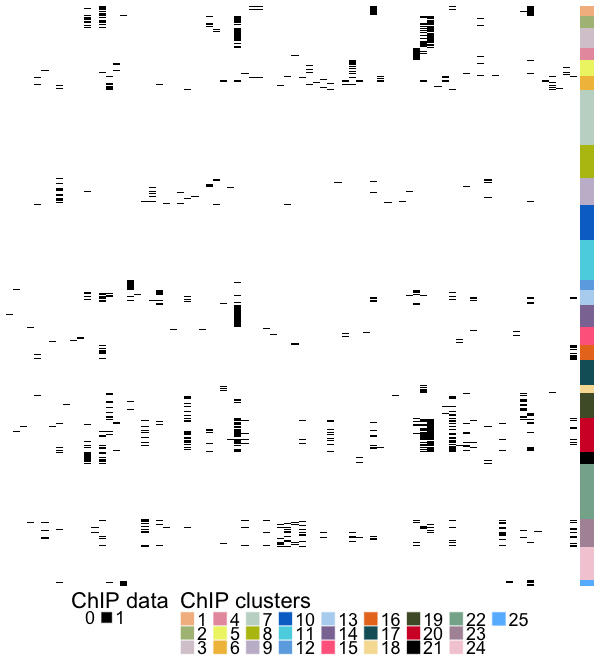}
		\caption{ChIP data.}
	\end{subfigure}
	\caption{Clusters obtained on each dataset separately. The ordering of the rows is different in the two  figures.}
	\label{fig:initial-clusters-second-set-of-data}
\end{figure}

\begin{figure}[h]
	\centering
	\begin{subfigure}{.35\textwidth}
		\includegraphics[width=\textwidth]{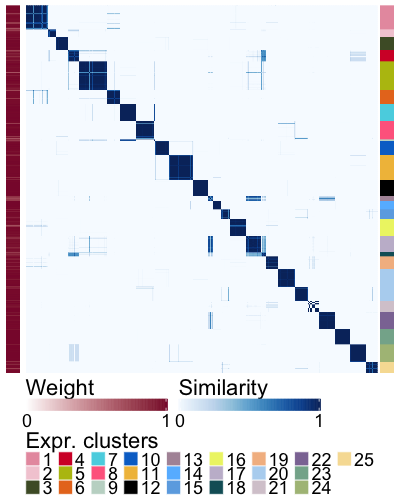}
		\caption{Expression data.}
	\end{subfigure}
	\hspace{1cm}
	\begin{subfigure}{.35\textwidth}
		\includegraphics[width=\textwidth]{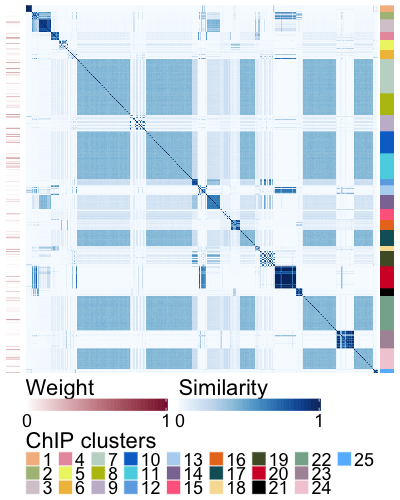}
		\caption{ChIP data.}
	\end{subfigure}
	\caption{Posterior similarity matrices and clusterings obtained via kernel $k$-means on each dataset separately.  The ordering of the observations is different in the two  figures.}
	\label{fig:psms-second-set-of-data}
\end{figure}

Figures \ref{fig:initial-clusters-second-set-of-data} and \ref{fig:psms-second-set-of-data} show the initial data, the clusterings obtained on each dataset individually and the PSMs.
The cophenetic correlation coefficients are 0.953685 for the expression data and 0.9841434 for the ChIP data.
3.5\% of the weight is assigned on average to the ChIP data, abd the remaining 96.5\% to the expression data.
 
\begin{figure}[h]
	\centering
	\includegraphics[width=.49\linewidth]{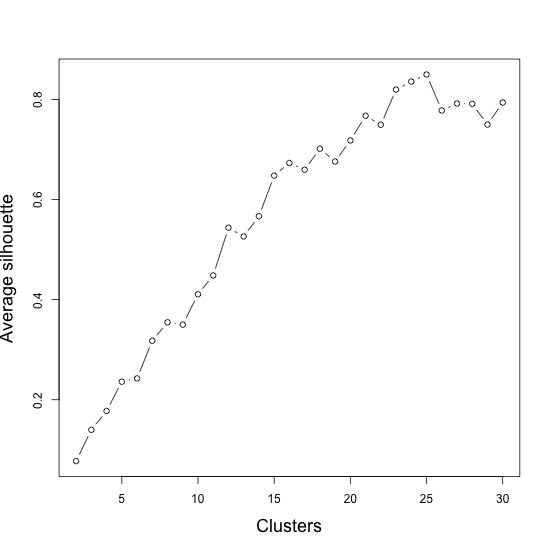}
	\caption{Plot of silhouette for different numbers of clusters for the integration of the datasets of Harbison {\em et al.} and Granovskaia {\em et al.}. The maximum value is attained at $K = 25$.}
	\label{fig:silhouette-yeast-second-set-of-data}
\end{figure}

In Figure \ref{fig:silhouette-yeast-second-set-of-data}  are reported the values of the average silhouette for different values of the number of clusters $K$. We choose $K=25$, which gives the highest value of the silhouette.

%\begin{figure}
%	\centering
%%	\includegraphics[width=.4\linewidth]{onlyPSMs_weights.png}
%	\caption{Matrix of the weights $\Theta$ assigned by the localised kernel $k$-means algorithm to each data point. Each row corresponds to a patient and each column to a data set: gene expression data, ChIP-chip data.}
%	\label{fig:weight-matrix-onlyPSMs}
%\end{figure}
%
%\clearpage

%\subsubsection{Posterior similarity matrices}

%\subsubsection{MCMC convergence assessment}

\clearpage
\subsubsection{A different set of data}

Here we combine the expression dataset of \cite{ideker2001integrated} with the ChIP-chip dataset of \citet{harbison2004transcriptional}, which provides binding information for 117 transcriptional regulators. Both datasets are discretised as in \citet{savage2010discovering} and \citet{kirk2012bayesian}. The dataset of \citet{ideker2001integrated} contains measurements related to 205 genes whose expression patterns reflect four functional categories based on gene ontology annotations.

\begin{figure}[h]
	\centering
	\begin{subfigure}{.35\textwidth}
		\includegraphics[width=\textwidth]{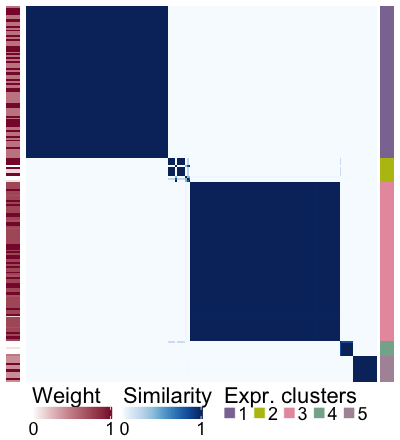}
		\caption{Expression data.}
	\end{subfigure}
	\hspace{1cm}
	\begin{subfigure}{.35\textwidth}
		\includegraphics[width=\textwidth]{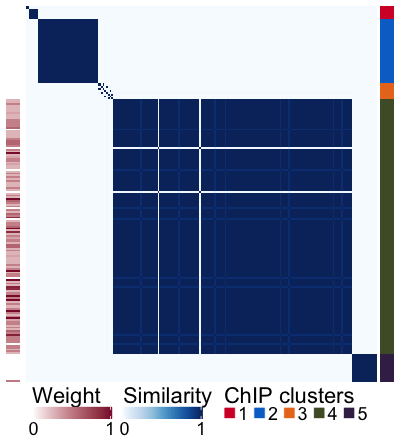}
		\caption{ChIP data.}
	\end{subfigure}
	\caption{Posterior similarity matrices and clusterings obtained via kernel $k$-means on each dataset separately.  The ordering of the observations (i.e. rows) is different in the two  figures.}
	\label{fig:psms-first-set-of-data}
\end{figure}

Figure \ref{fig:psms-first-set-of-data} shows the PSMs obtained for each dataset. In Figure \ref{fig:psms-first-set-of-data}, the cophenetic correlation coefficient is 0.9999932 for the ChIP data and 0.9999184 for the PSM of the expression data.

\begin{figure}[h]
	\centering
	\begin{subfigure}[b]{.35\textwidth}
		\centering
		\includegraphics[width =\textwidth]{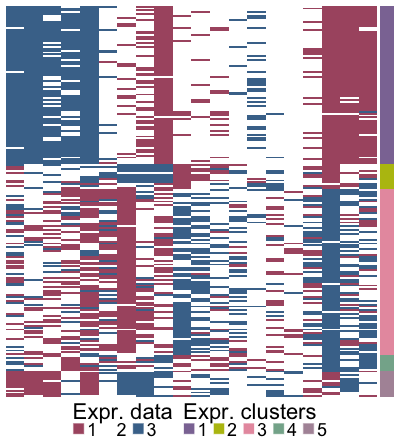}
		\caption{Expression data.}
	\end{subfigure}
	\hspace{1cm}
	\begin{subfigure}[b]{.35\textwidth}
		\centering
		\includegraphics[width = \textwidth]{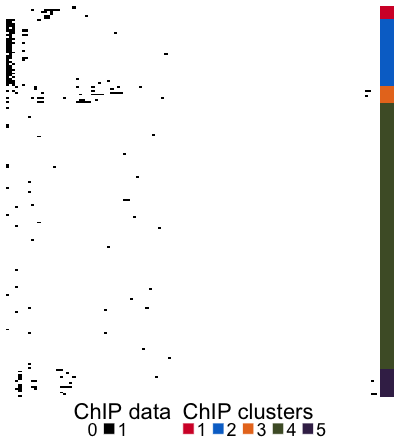}
		\caption{ChIP data.}
	\end{subfigure}
	\caption{Clusters obtained on each dataset separately. The ordering of the observations (i.e. rows) is different in the two  figures.}
	\label{fig:transcriptional-psms-first-set-of-data}
\end{figure} 

\clearpage

\begin{figure}[h]
	\centering
	\includegraphics[width=.49\linewidth]{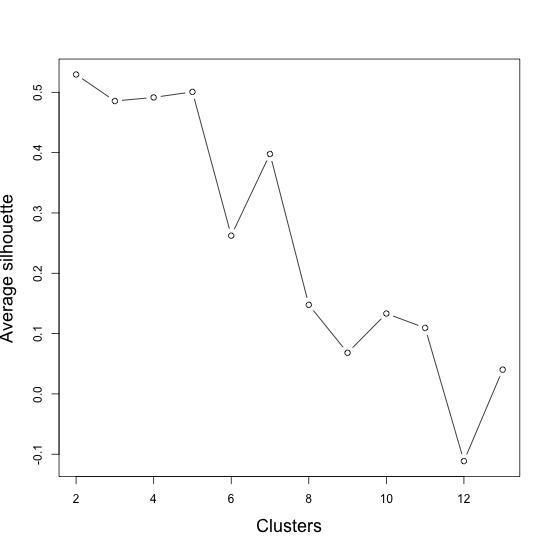}
	\caption{Plot of silhouette for different numbers of clusters for the integration of the datasets of Harbison {\em et al.} and Ideker {\em et al.}. The maximum value is attained for $K=2$.}
	\label{fig:silhouette-yeast-first-set-of-data}
\end{figure}

In Figure \ref{fig:silhouette-yeast-first-set-of-data} are reported the values of the average silhouette for different values of the number of clusters $K$. 
The number of clusters chosen for the data analysis is 5, since it has a similar value of the average silhouette to 2.

The final clusters are shown in Figure \ref{fig:transcriptional-psms-first-set-of-data} next to the initial datasets and combined PSM. 28.7\% of the weight is assigned on average to the ChIP data, the remaining 71.3\% to the expression data.

\begin{figure}[h]
	\centering
	\begin{subfigure}[b]{.32\textwidth}
		\centering
		\includegraphics[width =\textwidth]{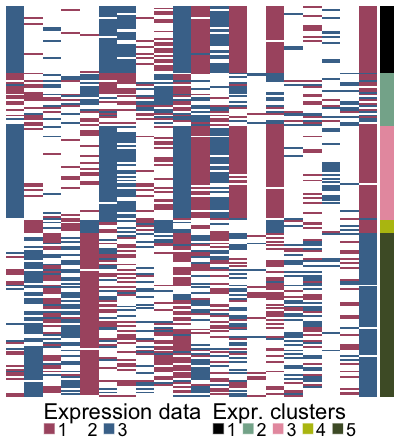}
		\caption{Expression data.}
	\end{subfigure}
	\hspace{1cm}
	\begin{subfigure}[b]{.32\textwidth}
		\centering
		\includegraphics[width = \textwidth]{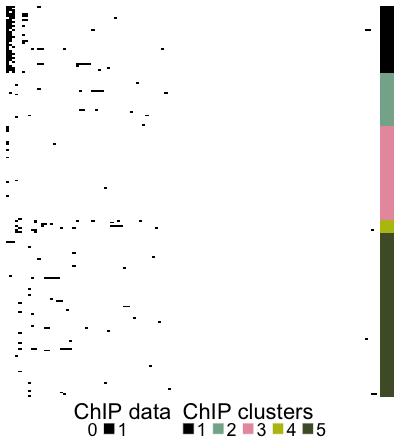}
		\caption{ChIP data.}
	\end{subfigure}
	\vspace{1em}
	\begin{subfigure}[b]{.49\textwidth}
		\centering
				\includegraphics[width = \textwidth]{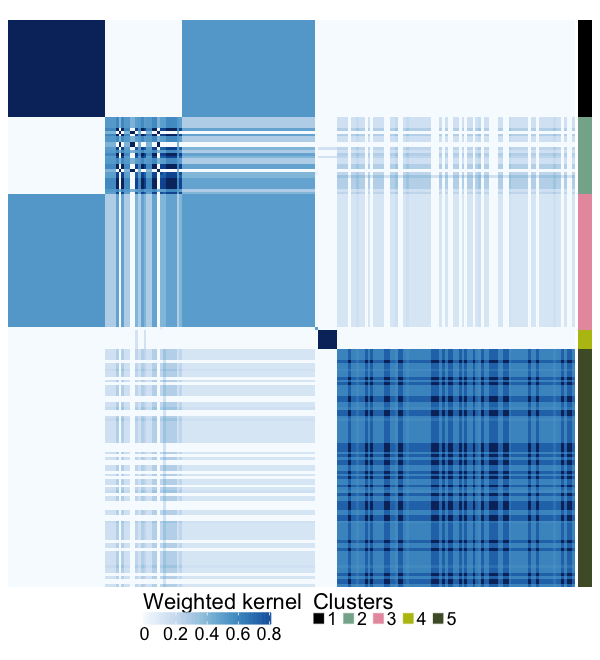}
		\caption{Weighted kernel.}
	\end{subfigure}
	\caption{Transcriptional module discovery, integration of the \citet{harbison2004transcriptional} and \citet{ideker2001integrated} datasets.}
	\label{fig:transcriptional-psms-first-set-of-data}
\end{figure} 

\begin{table}[h]
\centering
\begin{tabular}{l c c c }
Dataset(s) &GOTO BP & GOTO MF & GOTO CC \\
\hline
ChIP (Harbison {\em et al.}) &  13.36 & 1.37  & 12.26  \\
Expr. (Ideker {\em et al.}) & \textbf{17.34} &  \textbf{2.52} & \textbf{15.38} \\
Both & 16.51 & 2.15 & 14.84\\
\hline\\
\end{tabular}
\caption{Gene Ontology Term Overlap scores. ``BP'' stands for Biological Process ontology, ``MF'' for Molecular Function, and ``CC'' for Cellular Component.
The number of clusters used to combine the datasets of Harbison {\em et al.} and Ideker {\em et al.} is 5.}
\label{table:goto-scores-first-set-of-data}
\end{table}

\clearpage
\bibliographystyle{apalike}
\addcontentsline{toc}{section}{Bibliography}
\bibliography{supplement}